# Gravitational Lensing of the supernovae from the Supernova Legacy Survey (SNLS)

Ayan Mitra

**THÈSE DE DOCTORAT DE L'UNIVERSITÉ PIERRE ET MARIE CURIE**

Spécialité Astrophysique et Cosmologie

Particules, Noyaux et Cosmos (ED 517)

Présentée par

M. Ayan MITRA

En vue de l'obtention du grade de

**DOCTEUR de l'UNIVERSITÉ PIERRE et MARIE CURIE**

# Gravitational lensing of the supernovae from the

# SuperNova Legacy Survey

Date de la soutenance : le 30 septembre 2016

devant le jury composé de :

| | |
|---|---|
| M. Dominique FOUCHEZ | *Examinateur* |
| Mme Delphine HARDIN | *Directrice de thèse* |
| M. Michael JOYCE | *Examinateur* |
| Mme Simona MEI | *Rapporteur* |
| Mme Cécile RENAULT | *Rapporteur* |
| Mme Vanina RUHLMANN-KLEIDER | *Examinateur* |

# Contents

# List of Figures

# List of Tables

# Introduction

Towards the end of the last century the discovery of the present time acceleration of the expansion of the Universe [Per97], [Rie98],[Sch98] changed our perspective of modern cosmology. This discovery was achieved by using distant type Ia supernovae (SNe Ia) explosions as standard candles to estimate their distances and then build a Hubble diagram : the luminosity distance, as a function of the redshift, which is related to the Universe's expansion scale factor at the time of explosion. The observed departure from the Hubble law (valid at small distances) was not compatible with a matter filled Universe, and required an additional constituent of the Universe : the dark energy.

Dark energy can be modeled as a perfect fluid, with an equation of state parameter $w = p/\rho$, where $\rho$ is the density and p the fluid pressure. It can be equivalent to a cosmological constant $\Lambda$ – then $w = 1$ – and is believed to be the responsible factor for the present accelerating expansion. Current observational constraints are in favor of a flat Universe with a cosmological constant $\Lambda$, filling up roughly 70% of the Universe.

SNe Ia exhibit homogenous brightness properties. When the intrinsic sources of variations for the peak luminosity of SN Ia are corrected, their corresponding brightness distribution are almost perfectly standard with only 15% peak luminosity variations.

The Supernovae Legacy Survey (SNLS) is a project which main objective was to pursue the nature of dark energy and put a better constraint on the corresponding equation of state parameter $w$, using high redshift (z) supernova type Ia as a probe. It drew it's data between $2003 - 2008$ from the CFHTLS [1]. Also a joint analysis was carried out using the SNLS 3 years sample (high-z samples) and SDSS II sample (z<0.4 samples) SNe Ia, called the Joint Light Curve Analysis (JLA). With 740 spectroscopically confirmed SN Ia, the JLA result was consistent with other probe's ( e.g. CMB observations by the Planck (and WMAP), the BAO) results of a constant dark energy equation of state parameter in a flat Universe. The best fit value for the equation of state parameter obtained was $w = -1.018 \pm 0.057$ for a flat $w$-CDM cosmological model [Bet14]. The full SNLS 5 year sample analysis, which is underway, will provide $\sim 150$ additional spectroscopic SNe Ia.

In this thesis we carried out weak gravitational lensing analysis using SN Ia of the full 5 years data sample of the SNLS. Gravitational lensing is the bending of light rays in the presence of massive gravitational field. It leads to an isotropic alteration in the measured brightness of the source. As a consequence, the supernova light can be either magnified or de-magnified by the presence of dark matter galaxy haloes on their line-of-sight. In relation to the Hubble diagram, this effect will increase it's dispersion.

Our lensing magnification signal detection algorithm was thus based on the computation of the correlation of SN luminosity which is calculated using two separate (and independent) methods i.e. the lensing magnification and the Hubble residual.

The lensing magnification is computed by modelling the dark matter halo (with a prior on the mass-luminosity relation) embedded on foreground galaxies along the line-of-sight. Due to the lensing effect, the SN flux will be slightly increased or decreased with respect to its expected value, leading to a residual in the Hubble diagram, between the distance modulus (or SN flux) obtained from the SN magnitudes and its value predicted from the cosmological model using the SN redshift.

[1] http://www.cfht.hawaii.edu/Science/CFHTLS/

Thus, the detection of the lensing signal relies on the measurement of a positive correlation between the Hubble residual and the lensing magnification. This method was first developed and carried out on SNLS3 data by [Kro10] and in this thesis we present an updated analysis performed with the full spectroscopic sample of the SNLS 5 years data.

In the near future upcoming large scale high-z (deep field) surveys like the LSST, are expected to detect thousands of SNe. With such a statistics at high redshift, SNe gravitational lensing will help probe the dark matter distribution at small scale in a complementary way to the main cosmological probes, such as galaxy-galaxy lensing, carried out by these projects.

The chronology of this monograph is broadly divided into two sections. The first part (chapters 1-3) presents the theoretical framework and the observational background of the subject. In chapter 1, we recap the cosmological background and the recent results relevant to our analysis. The following chapter 2 discusses the supernovae categories and the advantages of the type Ia as a standard candle, along with the description of the SNLS project. Chapter 3 presents the mathematical frameworks of gravitational lensing and the modeling of the galaxy dark matter haloes, along with the previous results obtained in detecting lensing on SNe Ia.

The remaining last two chapters of the thesis are directed towards the analysis and results obtained from the SNLS5 data. Chapter 4 explains the SNe line of sight modeling and selection. We present the galaxy catalogue used to select the line of sight lenses and present the computation process of their photometric redshifts. The identification of the contaminated regions (luminous pollution) by bright stars are described, which determines the selection of the line-of-sight suitable for the analysis. An important step for the SNe selection is the identification of the host galaxy.

In chapter 5 we present the magnification analysis for the SNLS5 sample and the sensitivity to two different halo models (SIS and NFW) and SNe selection. We discuss on the prospects of lensing signal expected with the addition of a complementary $\sim 300$ photometric SNe in future. The appendix gives the list of all supernovae selected for the lensing signal detection process and their lensing magnification estimated with two different halo models.

# Chapter 1

# Cosmology

The twentieth century saw the building of cosmology theoretical frame : Einstein general relativity [Ein52], its solution the Friedman's equation in 1922 [Fri99], the observational discovery of redshift by V. Slipher[Sli13], the proposal of the Big Bang model by Lemaître [Lem33] in 1933 and the establishing by Edwin Hubble of the Hubble's law ([Hub26], [Hub36]). Subsequently is modern cosmology based on the current standard Hot Big Bang Model which states that the universe expanded from an initially hot and dense state, and that it is ever-expanding today.

Most observational evidence agree with the concordance ΛCDM model : a flat universe, where dark energy makes up approximately 70% of the energy density, dark matter about 25%, which leaves 5% for atoms, outnumbered, in 1:6 $10^{10}$ by the cold relic black-body photons of the Cosmic Microwave Background radiation at $T \simeq 2.7255$ K[1].

We present here the cosmological standard model frame, and some of the observational evidences that sustain it. This chapter is based in particular on [Ryd03], [Ric09], and [Har16a] notes.

## 1.1 A homogeneous expanding universe

The Cosmological Principle states that "the universe looks the same whoever and wherever you are" ([Lid03]) or that "there is nothing special about our location in the universe" ([Ryd03]): this implies that is is homogeneous and isotropic. Isotropy refers to the fact that there are no preferred directions, which is sustained by the exceptional isotropy observed by the COBE mission of the cosmic microwave background (CMB) temperature $T_0 = 2.7255 \pm 0.0006$ K ([Fix96], [Fix09]), to roughly one part in 100 000, once the dipole due to the earth motion is subtracted ([Smo92]). Homogeneity means that there are no preferred location. This holds at large scales[2], of about 100Mpc ([Hog05]), corresponding to a scale larger than super-clusters and voids (see figure 1.1).

The universe geometry is described through its metric, and the homogeneity and isotropy hypothesis lead to the Robertson-Walker (RW) metric :

$$ds^2 = dt^2 - R^2(t) \left( \frac{dr^2}{1 - kr^2} + r^2(d\theta^2 + \sin^2\theta d\phi^2) \right) \tag{1.1}$$

where $t$ is the cosmic time, $R(t)$ is the cosmic scale factor, the coordinate system for a spatial location is given by the spherical coordinates $(r, \theta, \phi)$, while $k = -1, 0, +1$ is the sign of the

[1] ... and relic neutrinos

[2] And obviously not at scales as small as our Galaxy or our Local Group.

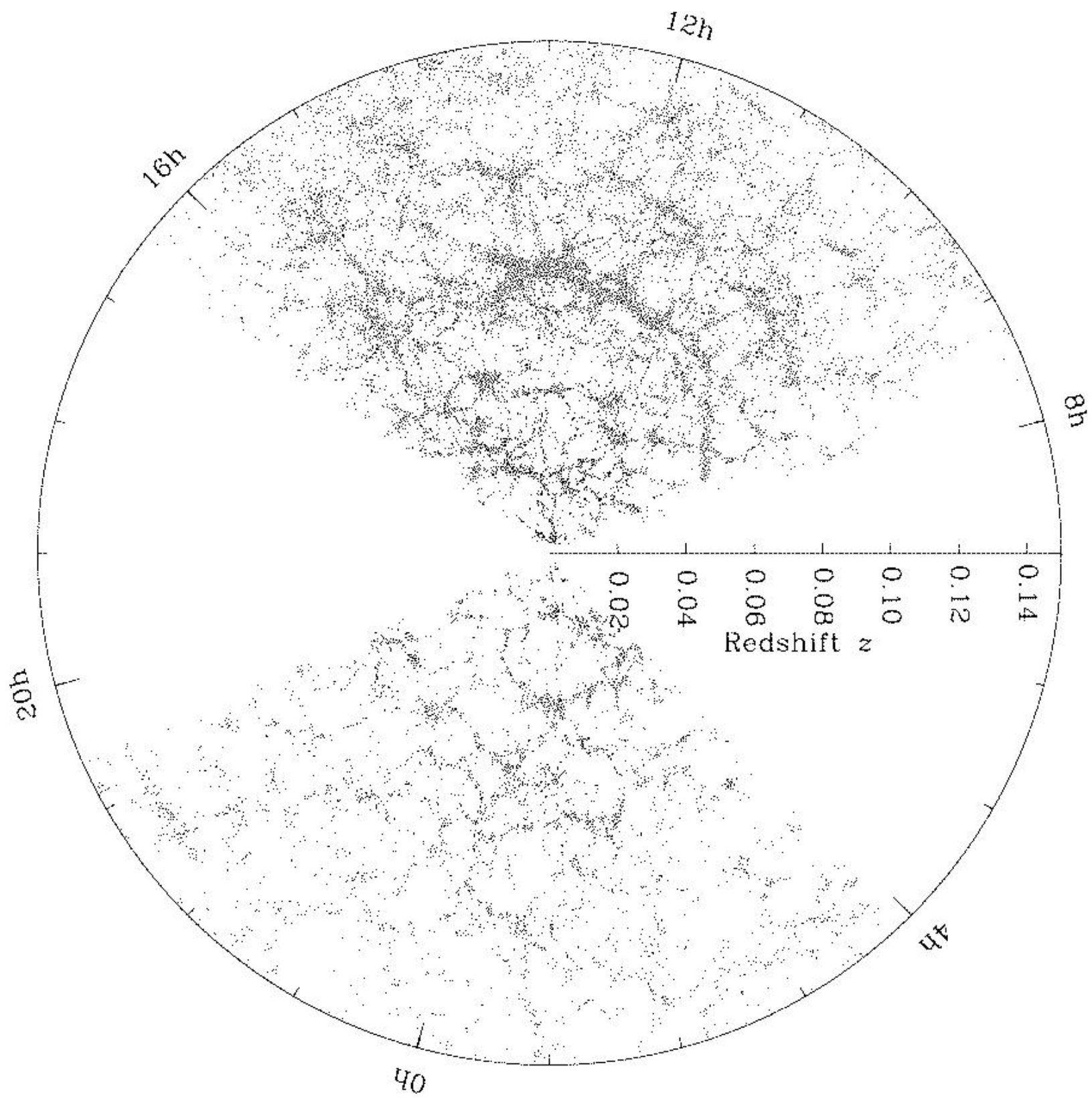

**Figure 1.1**: Slice of the 3D map of the distribution of the galaxies in the universe from the Sloan Digital Sky Survey III. The earth is situated at the center and each point represents a galaxy. The radius of the circle is roughly 0.5 Gpc. Super-cluster and voids occur up to a few hundred Mpc, which is the scale on which the cosmological principle can be applied in the universe. The thinning out of the galaxies near the edges which are farthest from us represents the sparser sampling of these faint objects in the survey [Kee07]. More extensive mapping work will be done in future, for example by the Square Kilometre Array (SKA) [Sch04], [San15][Maa15].

uniform curvature radius of spatial sections and sets the universe geometry :

$$k = \begin{cases} -1 & \text{closed} \\ 0 & \text{flat} \\ +1 & \text{open} \end{cases}$$

There are compelling observational evidences that tell us that the universe at present is flat with $k = 0$ ([Hin13], [Ade15]). The scale factor $R(t)$, or equivalently its normalized expression to present day $t_0$ value : $a(t) \equiv R(t)/R(t_0)$, describes how the distances expand (or contract) with time. The spatial coordinates $(r, \theta, \phi)$ are said to be comoving as they do not change with the expansion of the universe. The RW metric can also be written using the comoving spatial coordinates $(\chi, \theta, \phi)$, with $d\chi = dr/\sqrt{1 - kr^2}$ so that the relation between the coordinates $r$ and $\chi$ depends on $k$ sign : $r = S_k(\chi)$, with :

$$S_k = \begin{cases} \sin(\chi) & k = +1 \\ \chi & k = 0 \\ \sinh(\chi) & k = -1 \end{cases}$$

**Hubble Law**

The proper distance $d\ell$ traveled by a photon between two given points at comoving coordinate $\chi$ and $\chi + d\chi$ is obtained using the RW metric and setting $ds^2 = 0$ :

$$d\ell = cdt = R(t)d\chi \tag{1.2}$$

Their distance evolves trough time, due to the expansion. Their relative velocity $dv = \dot{R}(t)d\chi$ is thus proportional to their distance :

$$dv = \frac{\dot{R}(t)}{R(t)} d\ell = H(t) d\ell \tag{1.3}$$

with $H(t) \equiv \dot{R}(t)/R(t)$ being the Hubble parameter, the expansion rate of the universe at time $t$, and $H_0 = H(t_0)$ its present day value.

The Hubble law (1.3) was discovered by Edwin Hubble in 1929 ([Hub29]), who interpreted the observed redshift (see below) of nearby galaxies as a (non-relativistic) Doppler shift due to their radial velocity $v$ away from us, and established that it was proportional to their distance $d$ through the linear law : $v = H_0 \times d$. Due to the underestimation of the galaxies distances he was using, the value of $H_0 \simeq 500\text{km s}^{-1}\text{ Mpc}^{-1}$ he obtained was largely overestimated. A recent estimation is ([Ade15], see section 1.6.1):

$$H_0 = 67.51 \pm 0.64\text{km.s}^{-1}.\text{Mpc}^{-1}$$

$H_0$ value can also be expressed using the dimensionless number $h$ : $H_0 = 100h$ km s$^{-1}$ Mpc$^{-1}$. The Hubble time $H_0^{-1} \sim 14$ Gyr is the time scale of the expansion, related to the time since the primordial highly dense state of the universe. The Hubble distance $c/H_0 \sim$ provides a natural distance scale. The Hubble parameter is a function of time, and the next section will establish its dependency, determined by the composition of the universe.

**Cosmological redshift**

In such a universe, the wavelength $\lambda_{\text{emt}}$ of the light emitted at time $t$ by a given source (located at a radial position $r_E$), and received at $t_0$ by an observer (located at $r = 0$) is redshifted at a wavelength $\lambda_{\text{obs}}$, through the relation :

$$\frac{R(t_0)}{R(t)} = \frac{\lambda_{\text{obs}}}{\lambda_{\text{emt}}}$$

The redshift, defined as :

$$z \equiv \frac{\lambda_{obs} - \lambda_{emt}}{\lambda_{emt}}$$

is thus related to the change of the expansion factor between emission and observation of the light :

$$1 + z = \frac{R(t_0)}{R(t)}$$

Vesto Slipher in 1912 was the first to observe this effect on the light emitted by spiral nebulae [Sli15]. It is however not identical to Doppler shift[3].

## 1.2 The Friedman universe

Einstein's equation relates the geometry and the matter and energy content of the universe:

$$G_{\mu\nu} = \mathcal{R}_{\mu\nu} - \frac{1}{2}\mathcal{R}g_{\mu\nu} - \Lambda g_{\mu\nu} = \frac{8\pi G}{c^4}T_{\mu\nu} \tag{1.4}$$

In this famous equation, $G_{\mu\nu}$ is the Einstein tensor and $g_{\mu\nu}$ , $\mathcal{R}_{\mu\nu}$ , $\mathcal{R}$, $G$, $\Lambda$, $T_{\mu\nu}$ are respectively the metric tensor, the Ricci tensor, the Ricci scalar, the gravitational constant, the cosmological constant, and the energy-momentum tensor, describing the matter-energy contents of the universe.

For a perfect fluid the energy-momentum tensor is given by:

$$T^{\mu}_{\nu} = pg^{\mu}_{\nu} + (p + \rho)\mathcal{U}^{\mu}\mathcal{U}_{\nu} \tag{1.5}$$

where $\mathcal{U}^{\nu}$ is the fluid four velocity, $p$ the pressure and $\rho$ the energy density.

The cosmological constant $\Lambda$ was first introduced in 1915 by Einstein to force the existence of solutions describing a static universe. As it came to knowledge that the universe was actually expanding, Einstein would famously call it his 'greatest blunder'. In the late 90's, the cosmological constant came in favor again, as its presence in equation 1.4 could account for the acceleration of the cosmic expansion observed by supernovae cosmology surveys ([Per99], [Sch98]). Equivalently, the cosmological constant can be taken into account by adding a supplementary component entering the energy-momentum tensor, with a constant energy density $\rho_{\Lambda} \equiv c^2\Lambda/(8\pi G)$ and a negative pressure $p_{\Lambda} = -\rho_{\Lambda}$. Physically it can be interpreted as a "zero point energy" equivalent to the quantum vacuum energy. There is a big mismatch though between its estimated value from the observations of the universe expansion at large scales, and its natural value as suggested by particles physics ([Rug00], [Car92], [Pad03]), with a difference in order of $10^{120}$.

Given the FRW metric (1.4), the $0 - 0$ component yields the Friedmann equation 1.6a (see e.g. [Ric09]). The $i - i$ component of the Einstein equation together with eq. 1.6a leads to the acceleration equation 1.6b :

$$\left(\frac{\dot{a}}{a}\right)^2 = H^2 = \frac{8\pi G}{3}\rho - \frac{k}{R^2} \tag{1.6a}$$

$$\frac{\ddot{a}}{a} = -\frac{4\pi G}{3}(\rho + 3p) + \frac{\Lambda}{3} \tag{1.6b}$$

The Friedmann's equation 1.6a relates the energy content of the universe to it's geometry. The acceleration equation 1.6b shows that gravitational forces slow down the expansion while a positive cosmological constant can lead to an accelerating expansion, as observed today. Setting

[3] Doppler shift depends on the speed of the emitter while in cosmological redshift the stretching of wavelength depends on the expansion of the intermediate space. In a locally flat space, at a small distance ($z \ll 1$) it will be a consequence of the Doppler shift and $z = \frac{v}{c}$. With space time curvature, gravitational shift comes into play and the redshift caused is not due to Doppler shift [Bun09]

$k = 0$ in 1.6a defines the critical density corresponding to a flat universe :

$$\rho_{cr}(t) = \frac{3H^2}{8\pi G} \tag{1.7}$$

Numerically, the critical density value is today $\rho_{cr}(t_0) = 1.88\, h^2 \times 10^{-26}$ kg.m$^{-3}$. The $\Omega$'s parameters are defined as the ratio to the critical density of each contribution:

$$\Omega_m = \frac{\rho_m(t_0)}{\rho_{cr}(t_0)} \quad \Omega_r = \frac{\rho_r(t_0)}{\rho_{cr}(t_0)} \quad \Omega_\Lambda = \frac{\Lambda}{3H_0^2} \quad \Omega_k = \frac{-k}{R_0^2 H_0^2}$$

where the subscript $m$ stands for matter and $r$ for radiation. Although they can be defined at anytime $t$, we refer here and in the following (if not otherwise stated) to their present day value at $t = t_0$.

Energy conservation is expressed by the vanishing of the covariant divergence of the energy-momentum tensor, leading to equation 1.8 :

$$\dot{\rho} + 3\frac{\dot{a}}{a}\left(\rho + \frac{p}{c^2}\right) = 0 \tag{1.8}$$

which is equivalent to thermodynamics first law $d(\rho a^3) = -pd(a^3)$. The composition of the matter-energy content of the universe is described as an ideal fluid with its density $\rho$ and its pressure $p$ related trough its equation of state parametrised by $w$ :

$$w = \frac{p}{\rho} \tag{1.9}$$

For non-relativistic particles ($p = 0$) like dark matter, galaxies, $w = 0$, and for radiation (relativistic particles) with pressure $p = \rho/3$, $w = 1/3$. Using this relation and the fluid equation (1.8) we obtain a general solution of the fluid :

$$\rho = \rho_0 a(t)^{-3(1+w)} \tag{1.10}$$

so that $\rho_r \propto a(t)^{-4}$ for radiation and $\rho_m \propto a(t)^{-3}$ for matter. As a consequence, the universe started with an era of radiation domination, with radiation density producing almost all the gravitational force. At $a_{eq}^{-1} \equiv (1.68\Omega_m/\Omega_\gamma)^{-1} \simeq 3200$ begun the era of matter domination.

The value $w = -1$ corresponds to the cosmological constant, with a negative energy density. More generally, one can postulate the existence of a dark energy (DE) component, with no assumption on $w_{DE}$ value or even its evolution with time or redshift $w_{DE}(z)$, only needing (see equation 1.6b) that $w_{DE} < -1/3$ so that this dark energy component accelerate the expansion. A range of value of $w_{DE}$ can be obtained in the frame of canonical scalar field models also called quintessence with $-1 < w_{DE} < 1$ ([Tsu13], [Tak14]). The phantom field energy, with $w < -1$, can give rise to strange situations including a universe ending in a Big Rip [Sam04].

Integrating the Friedman equations to obtain the evolution of the expansion rate $H$ as a function of time $t$, or equivalently the redshift $z$ as $1 + z = 1/a(t)$, we obtain:

$$H(z)^2 = H_0^2\left[\Omega_r(1+z)^4 + \Omega_m(1+z)^3 + \Omega_k(1+z)^2 + \Omega_\Lambda\right] \tag{1.11}$$

leading, when taken at $t = t_0$, to the relation :

$$\Omega_\mathrm{T} = \Omega_m + \Omega_r + \Omega_\Lambda = 1 - \Omega_k \tag{1.12}$$

so that we have $\sqrt{|1 - \Omega_T|} = c/(R_0 H_0)$. Note that $\Omega_r(t) \ll \Omega_m(t)$ for $z \ll z_\mathrm{eq}$ and hence $\Omega_r(t)$ can generally be omitted at later epochs.

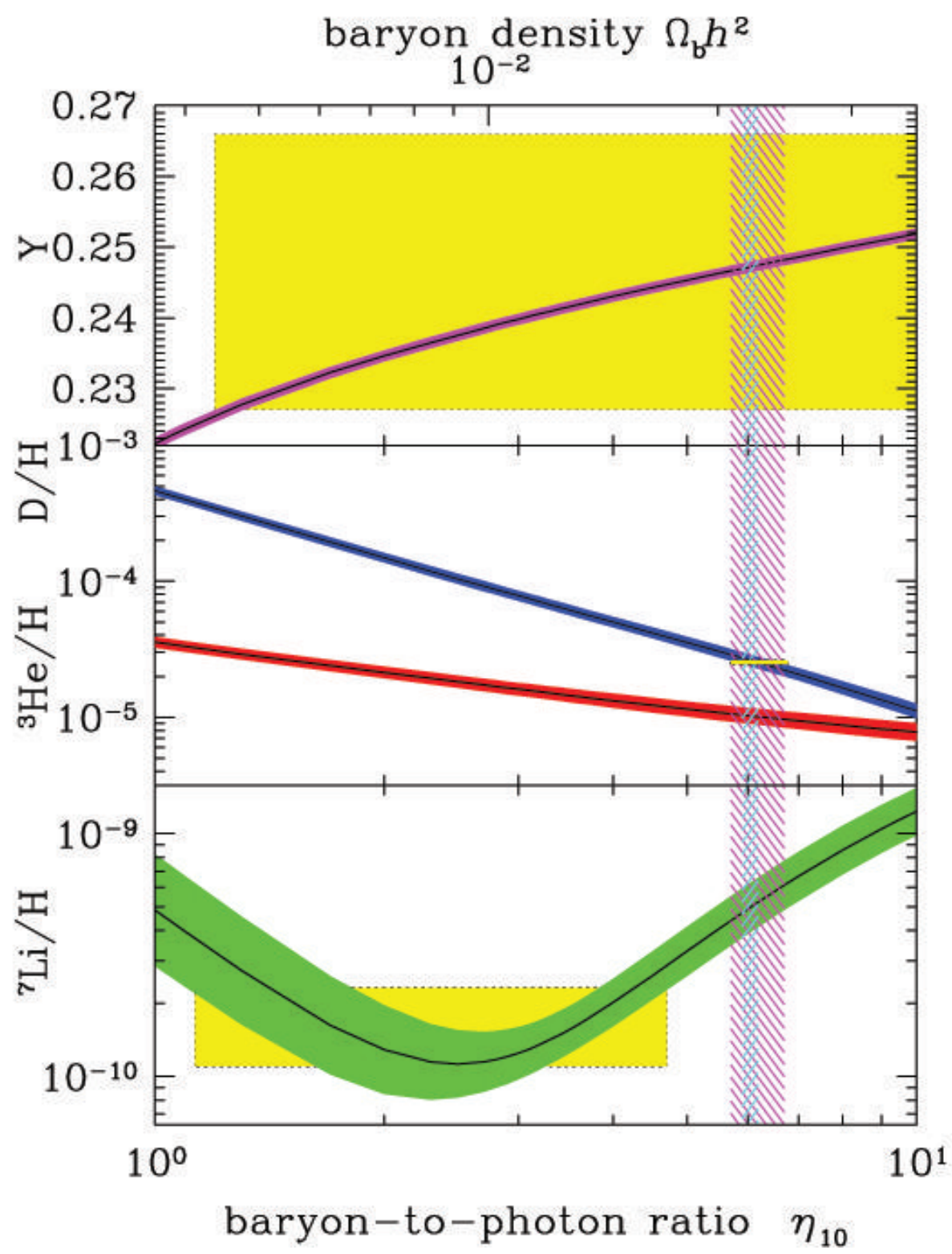

**Figure 1.2**: The abundances of light elements as predicted by the BBN as a function of the baryon-to-photon ratio $\eta \times 10^{10}$. The bands show the 95% C.L. The narrow vertical band indicates the measurement from CMB for $\Omega_B h^2 = 0.2230 \pm 0.00014$ ([Ade14]) while the wider band indicates the BBN concordance region. The 4 boxes represent Helium-4, Deuterium, Helium-3 and Lithium-7. The discrepancy between Lithium-7 and the otherwise agreeing abundances could simply reflect difficulty in determining the primordial lithium abundance. Image : [Fie14]

## 1.3 Hot Big Bang model and the thermal history of the universe

The Hot Big Bang model is based on the observation of the universe expansion, which might be traced back in time to an originating single point. In this model, the universe started off from a very hot and dense state after the initial 'Big Bang' to its current cool and tenuous state, and is still expanding today.

As the universe expanded adiabatically, its temperature $T$ cooled, scaling as $T \propto a(t)^{-1}$ : the mean energy per photons dropped from $10^{28}$ eV at the Planck time, at a time $t \sim 10^{-44}$ s before which no proven theory can correctly describes the universe, to a mean energy of $6.\,10^{-4}$ eV today. Salient phases in the universe thermal history are presented in table 1.1.

During the first few microseconds following the Planck epoch, all matter existed as a free sea of quarks and leptons. When the temperature dropped below $T \simeq 150$ MeV, the quarks bounded together into baryons, mostly protons and neutrons. The small excess of quarks over antiquarks led to a small excess of baryons over antibaryons. The radiation was in the form of extremely abundant photons, neutrinos (and anti-neutrinos). The ratio of baryons to photons numbers in the universe $\eta \equiv n_B(t)/n_\gamma(t)$ is constant as $n \propto t^{-3}$ for both species, and its value is of order $\eta \sim 6\,10^{-10}$, very much in favor of the photons. Equivalently, $\eta$ can be expressed in terms of the baryonic fraction of the critical density $\Omega_B \equiv \rho_B(t_0)/\rho_{\rm cr}(t_0)$, as $n_\gamma(t_0)$ is fixed by the present day CMB relic photons temperature $T_0 = 2.7255$ K : $\Omega_B h^2 \simeq 3.65\,10^{-3} \times (10^{10}\eta)$.

Once the temperature drops at $T \simeq 1$ MeV (at $t \sim 100$ s), deuterium nuclei could survive disruption by high-energy photons so that the fusion of neutrons and protons into nuclei heavier than hydrogen could proceed. This is the Big Bang Nucleosynthesis (BBN). The primordial nucleosynthesis stopped at $t \sim 10$ minutes. It produced essentially helium, no elements heavier than lithium, and left about 75% of hydrogen. The yields of these elements (D, $^3$He, $^4$He, and $^7$Li) depends strongly on the baryon-to-photon ration $\eta$, as the more the photons, the later the BBN can start. The measurement of the abundance of these elements today gives us a powerful tool for testing the Hot Big Bang model. Indeed (see figure 1.2), the Hot Big Bang model successfully reproduces the observed abundances of the light elements, provided $\eta$, or equivalently $\Omega_B h^2$, lies in a very narrow range.

| Temperature | Redshift | Time | Era |
|---|---|---|---|
| $\simeq 10^{19}$ GeV | | $0 - 10^{-43}s$ | Planck epoch |
| $\simeq 10^{15} - 10^{16}$ GeV | | $10^{-43} - 10^{-38}s$ | GUT scale |
| | | $10^{-34}s$ ? | Inflation? |
| $\simeq$ GeV | | $t \simeq \mu s$ | Quark-hadrons transition |
| ? | ? | ? | Dark matter freeze-out |
| $\simeq$ MeV | $z \sim 6 \times 10^9$ | 1 s | Neutrinos decouple |
| $\simeq$ 500 KeV | $z \sim 2 \times 10^9$ | 6 s | Electron-positrons annihilate |
| $\simeq$ 100 KeV | $z_{BBN} \sim 4 \times 10^8$ | $t \simeq$ 3 minutes | Big Bang Nucleosynthesis |
| $\simeq$ 0.75 eV | $z_{eq} \sim 3400$ | $t \simeq$ 55kyr | Matter radiation equality |
| $\simeq (0.25 - 0.33)$ eV | $z_{rec} \sim (1100 - 1400)$ | $t \simeq (260 - 380)$kyr | Recombination epoch |
| $\simeq (0.23 - 0.28)$ eV | $z_{rec} \sim (1000 - 1200)$ | $t \simeq$ 380kyr | Photon decoupling[4] |
| $\simeq (2.6 - 7.0)$ meV | $z \sim (11 - 30)$ | $t \simeq 100 - 400$Myr | Reionization |
| $\simeq$ 0.33 meV | $z \sim 0.4$ | $t \simeq$ 9.5 Gyr | Matter-$\Lambda$ equality |
| $\simeq$ 0.24 meV | $z = 0$ | $t \simeq$ 13.8 Gyr | Now |

**Table 1.1**: Brief history of the thermal regimes of the universe. Credits: [Bau13]

By the end of BBN, the universe is filled by a ionized plasma of nuclei (mostly protons), electrons, and photons. The high energy photons prevent the formation of neutral atoms because of photoionization, and interact primarily with the electrons through Compton scattering. So that the universe is opaque as well as ionized. At a time $t$ ~300,000 years, the temperature has cooled down to $T \simeq 3000$ K, and electrons and protons (or nuclei) can undergo radiative recombination into atoms. Since at the epoch of recombination the number of electrons drops rapidly, photons decouple from the electrons : their mean free path becomes longer than the Hubble distance $c/H$ and the universe becomes transparent. These black-body photons have been streaming (nearly) freely since the recombination-decoupling era, and are at present times observed at a wavelength redshifted into the microwave spectrum, corresponding to a temperature $T_0 \simeq 2.7255$ K (see §1.6.1). The discovery of the Cosmic Microwave Background (CMB) by Arno Penzias and Robert Wilson in 1965 ([Pen65]) established another observational pillar of the Hot Big-Bang model.

## 1.4 Distances

Locally in our Galaxy, distances can be measured using the parallax method, by taking advantage of the earth's relative motion about the Sun leading to a shift in the apparent position in the sky. Further away, in the Hubble flow, distances are related to the source redshift through the cosmological parameters. Distance measurement in the expanding universe has many categories

[4] Recombination and photon decoupling are two different events, coinciding only if recombination is instantaneous.

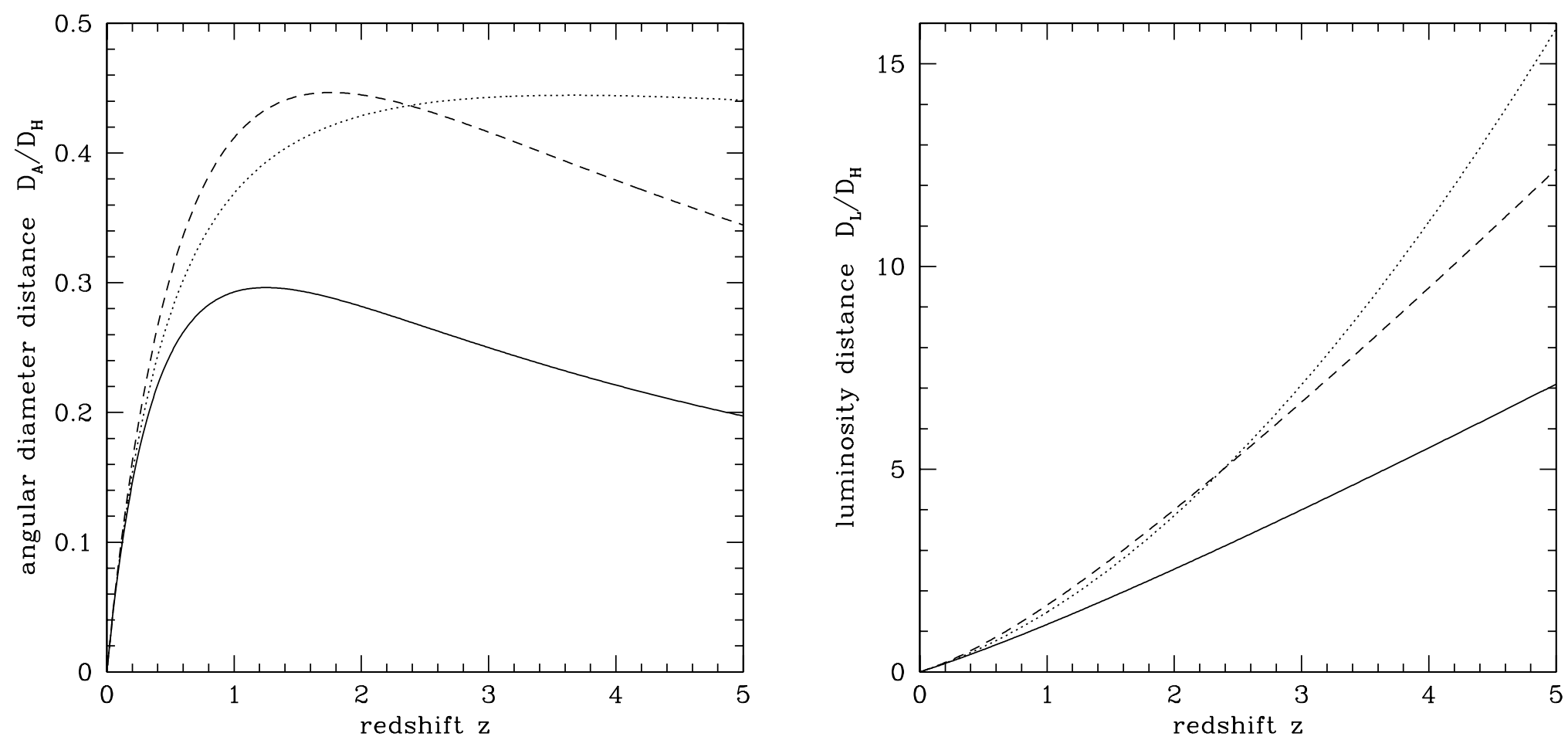

**Figure 1.3**: The angular distance $H_0 D_A/c$ and the luminosity distance $H_0 D_L/c$ as a function of redshift. The solid line shows a FCDM model ($\Omega_m = 1, \Omega_\Lambda = 0$), the dotted line a solely baryonic matter model ($\Omega_m = 0.05, \Omega_\Lambda = 0$), and the dashed line a flat $\Lambda$CDM model ($\Omega_m = 0.2, \Omega_\Lambda = 0.8$) close to the concordance ($\Omega_m = 0.3, \Omega_\Lambda = 0.7$) model. Image: [Hog99]

and we shall present here the luminosity, angular and comoving volume distances.

For most of the distances we define the common function :

$$E(z) \equiv H(z)/H_0 = \sqrt{\Omega_m(1+z)^3 + \Omega_k(1+z)^2 + \Omega_\Lambda}$$

where, as $z > z_{\rm eq}$, we have neglected $\Omega_r$. We shall first interest ourselves to the distance of a source, at a spatial position $\chi$ (or $r_E = \mathcal{S}_k(\chi)$), emitting light at $t_E$, received, redshifted by a factor $(1+z)$, at $t = t_0$ by an observer at a position $\chi = 0$.

### Comoving coordinate and proper distance

Following equation 1.2, and integrating along the total line-of-sight the infinitesimal $d\chi = cdt/R(t)$ contributions, one obtains the relationship between the comoving coordinate $\chi$ and $t_E$, or equivalently, using $1 + z = 1/a(t)$, with the redshift z :

$$\chi(z) = \frac{c}{R_0 H_0} \int_0^z \frac{dz}{E(z)}$$

The comoving coordinate is related to the proper distance $\ell$ through $\ell = R_0\chi$. The time corresponding to the proper distance is the look back time and is simply $D_P(z)/c$. It can be visualized by thinking of a measurement with a tape in any instant of the universe with expansion frozen. Practically it is not feasible since super luminous speed is not possible.

### Luminosity distance

The (bolometric) flux $F$ measured by the observer is related to the luminosity $\mathcal{L}$ of the source through :

$$D_L = \sqrt{\frac{\mathcal{L}}{4\pi F}} \tag{1.13}$$

which defines the luminosity distance $D_L$. Taking into account the surface of the sphere of radius $\chi$ around the source, and the drop of the photons energy due to their redshift, the luminosity distance reads :

$$D_L = R_0(1+z)\mathcal{S}_k(\chi(z)) \tag{1.14}$$

The dependency of the luminosity distance as a function of redshift, for $z < 5$, and the cosmological parameters $(\Omega_m, \Omega_\Lambda)$ is shown on figure 1.3, for different cosmological models : a flat and matter-only (Cold Dark Matter or CDM) universe, also known as the Einstein-de Sitter model, and a flat universe with matter and dark energy ($\Lambda$CDM) model. At high redshift, the deviation from the linear Hubble law $d \sim cz/H_0$ depends solely on the density parameters $\Omega_m$ and $\Omega_\Lambda$.

Standard candles are a class of object of supposed constant luminosity $\mathcal{L}$. They can be used to measure relative luminosity distances, their unknown luminosity $\mathcal{L}$ canceling out in the ratio : $F_1/F_2 = (D_2/D_1)^2$. Type Ia supernovae explosions have been used to this purpose, and permitted to demonstrate that the universe's expansion at present time is accelerating ([Per99], [Sch98]).

### Magnitudes

In practice we don't use the flux $F$ but the apparent magnitude $m$ defined as

$$m = -2.5 \times \log \frac{F}{F_0} \tag{1.15}$$

where $F_0$ is the flux of a reference source. The absolute magnitude $M$ is the magnitude of a source of same luminosity but situated at a distance of 10pc from the observer. It is thus possible to rewrite the relation 1.15 as:

$$m = M + 5 \times \log\left(\frac{d_L(z)}{10\text{pc}}\right) = M + \mu(z) \tag{1.16}$$

defining the distance modulus $\mu(z)$ as :

$$\mu = 5 \times \log\left(\frac{d}{1\text{Mpc}}\right) + 25 \tag{1.17}$$

### Angular Diameter Distance

The object of our interest has now a transverse physical size $d\ell$, and is observed subtending an angle $\delta\theta$. The angular diameter distance $D_A$ is then:

$$D_A = \frac{d\ell}{\delta\theta} \tag{1.18}$$

At emission $t = t_E$, the proper size of the object is $d\ell = r_E R(t_E)\delta\theta$, so that $D_A = \mathcal{S}_k(\chi)R_0/(1+z)$ and we can write :

$$D_A = \frac{D_L}{(1+z)^2} \tag{1.19}$$

Because of the relation 1.19, there is no point in attempting to separately determine both angular and luminosity distances if the redshift is accurately known, unless there is a need to check the Robertson-Walker metric or the origins of the cosmological redshift.

At low redshifts, we retrieve the Hubble law, and the angular diameter distance, the luminosity distance and the proper distance are equal :

$$D_P \simeq D_A \simeq D_L \simeq \frac{cz}{H_0} \quad \text{for} \quad z \ll 0.1$$

The dependency of the angular distance as a function of redshift, for $z < 5$, and the cosmological parameters $(\Omega_m, \Omega_\Lambda)$ is shown on figure 1.3.

The relative angular diameter distance between two objects at redshifts $z_1$ and $z_2$ is important in gravitational lensing analysis. It is not a straightforward subtraction between two angular distances, and is given by ([Hog99]):

$$D_{A12} = \frac{1}{1+z_2}\left[D_{m2}\sqrt{1+\frac{D_{m1}^2}{R_0^2}} - D_{m1}\sqrt{1+\frac{D_{m2}^2}{R_0^2}}\right], \quad D_m(z) = R_0\,\mathcal{S}(\chi(z)) \tag{1.20}$$

where $D_{m1}$ and $D_{m2}$ stand for the corresponding distance $D_m$ for the redshifts $z_1$ and $z_2$. This formula is only applicable for $\Omega_k \geq 0$.

A standard ruler is an object which physical scale $d\ell$ is known. Patterns in the distribution of distant galaxies provides such a ruler (see section 1.6.2). In this case, one can observe the pattern in angular coordinates ($\delta\theta = d\ell/D_A$) as well as in radial – in fact – redshift coordinates: $cdz = d\ell\, H(z)$. This requires to introduce the comoving volume distance :

$$D_V = \left[(1+z)^2 D_A^2(z)\frac{cz}{H(z)}\right]^{1/3} \tag{1.21}$$

## 1.5 An inhomogeneous universe

Although appearing homogeneous on large scale (>100 Mpc), the universe is clumpy on small scales (<50 Mpc) with voids and clusters of galaxies. We will concern ourselves in this section on the evolution of the matter density fluctuations and the formation on structures larger than galaxies. The formation of bound structures will be described in section 3.2.1.

The matter density $\rho(\vec{r}, t)$ is now a statistical field depending on the comoving coordinate $\vec{r}$ as well as time. Averaging the density over a large volume $V$ yields the mean energy density $\rho(t)$. Another point of view would be to average, at a fixed $(\vec{r}, t)$, the random field $\rho(\vec{r}, t)$ over many realizations.

The density contrast is defined as :

$$\delta(\vec{r}, t) = \frac{\rho(\vec{r}, t) - \rho(t)}{\rho(t)}$$

Large scale structures will grow through gravitational instability, starting from $\delta \ll 1$ up to $\delta \sim 1$ where they will leave the linear regime, collapse onto themselves, and in effect detach from the Hubble expansion (see section 3.2.1).

#### The power spectrum

To study the universe clumpiness as a function of the structures scale, we define the Fourier transform[5] of the density contrast $\delta$, omitting the time dependency :

$$\delta_{\vec{k}} = \frac{1}{V}\int \delta(\vec{r})e^{i\vec{k}.\vec{r}}d^3r$$

The comoving wavenumber $\vec{k}$ is associated to the comoving length through $r = 2\pi/k$. If $\delta(\vec{r})$ is isotropic, we can define without loss of information the power spectrum :

$$P(k) = \langle|\delta(\vec{k})|^2\rangle$$

[5] following here the conventions from [Pea99].

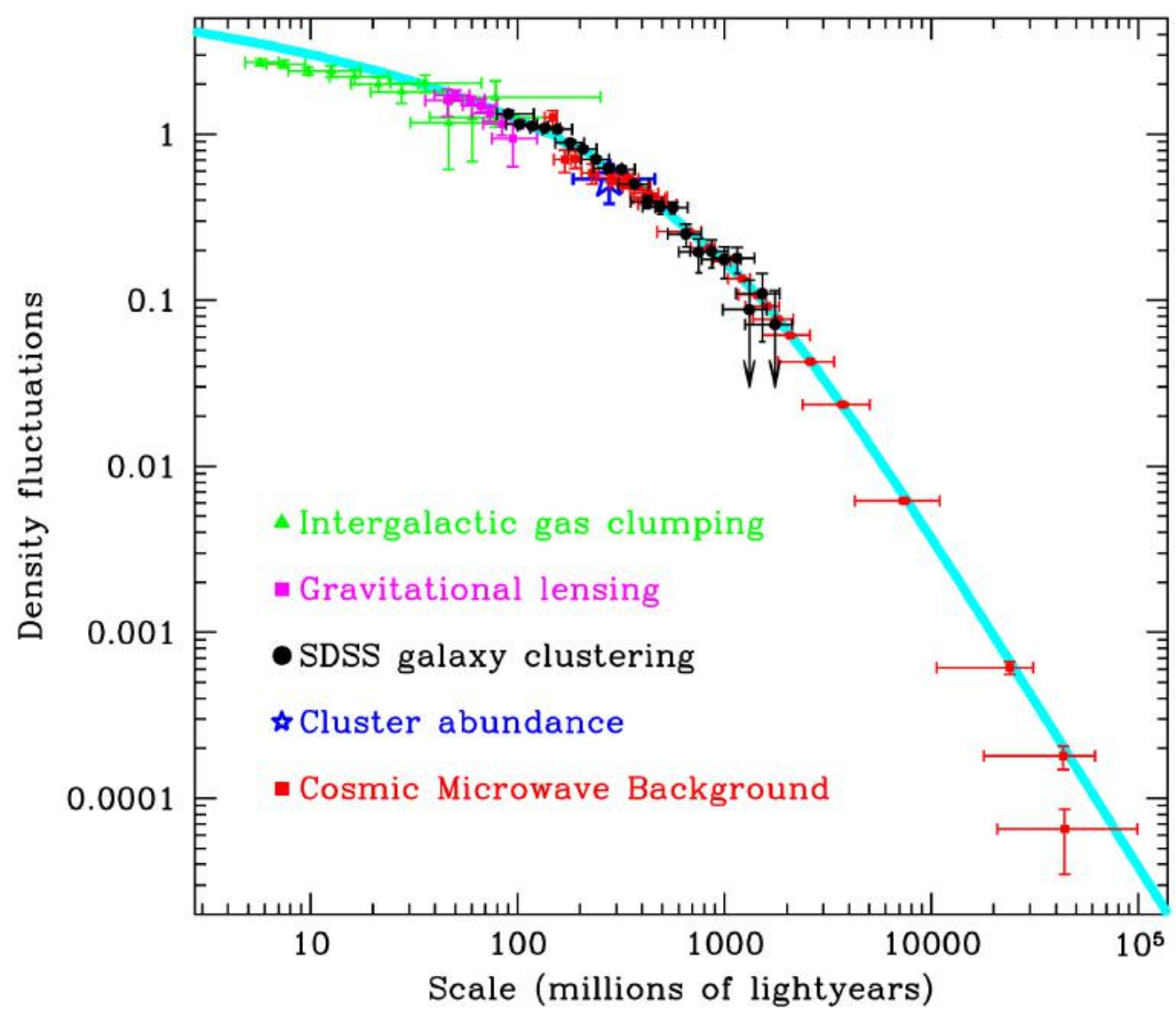

**Figure 1.4**: SDSS galaxies, CMB (WMAP), cluster, lensing and Ly-$\alpha$ forest constraints on the dimensionless power spectrum $\Delta(k)$ as a function of the structures scale $1/k$. Credits: [Teg04]

The average is again made over many realizations. Besides isotropy, we may also make the assumption that the statistical field $\delta_{\vec{k}}$ is gaussian, so that its characteristics are entirely characterized by its power spectrum. In particular, most inflationary scenarios predicts that inflation seeded at early epochs density fluctuations as a homogeneous, isotropic and gaussian field, via amplified quantum fluctuations from which structures formed at later times.

The power spectrum is related to the variance of the total mass $M_R$ in a sphere of comoving radius $R$, measuring the clumpiness at scale $R$ or equivalently $k \sim 1/R$. The mean mass is $\langle M_R \rangle = \frac{4\pi}{3}\rho_0 \times R^3$, and its variance reads :

$$\left\langle \left( \frac{M_R - \langle M_R \rangle}{\langle M_R \rangle} \right)^2 \right\rangle \sim \Delta_k^2$$

in which $\Delta_k$ is the dimensionless power spectrum per $\log k$ :

$$\Delta(k)^2 \equiv \frac{V}{(2\pi)^3} 4\pi k^3 P(k)$$

**The power spectrum law**

The power spectrum is often taken as a power law :

$$P(k) \propto k^n$$

In this case, the variance of $M_R$ becomes $\delta M/M \propto R^{-(3+n)/2}$. The spectral index $n$ governs the balance between the large and small scale power. The special case $n_s = 1$ is the Harrison-Zel'dovich-Peebles scale invariant spectrum ([Har70], [Eis98], [Zel70]), and corresponds to scale invariant fluctuations of the gravitational potential.

In the linear regime, the matter power spectrum is related to its primordial form $P_0(k) = A_S k^{n_S}$ through the transfer function $T(k) \propto \delta_k(z=0)/\delta_k(z=z_{\text{primordial}})$ : $P(k)_{lin} = P_0(k)T^2(k)$. During the radiation era, due to the microphysics at work, only density pertubations at scales larger than the Hubble radius could grow as $\delta \propto a^2$. When the matter domination era occurred, cold (i.e. non-relativistic particles) dark matter fluctuations could grow as $\delta \propto a$, whether they were super-Hubble or not. The caracteristical scale involved is the Hubble radius at radiation-matter equality, which translates into a comoving number $k_0 \sim 0.1h\mathrm{Mpc}^{-1} \times (\Omega_m h)/(0.3 \times 0.7))$. As a consequence, in the linear regime, starting out e.g. with a scale invariant shape after inflation, the power spectrum for cold dark matter is $\propto k$ at large scale, and bent over ($\propto k^{-3}$) for wavenumbers $k \gtrsim k_0$. The observations corroborate the CDM model (see figure 1.4) and gives a measurement of $\Omega_m h$. In the CDM model, the smaller scales structures merge to form the larger ones (bottom-up formation) as is confirmed by the observation of the earliest galaxies at $z \sim 10$.

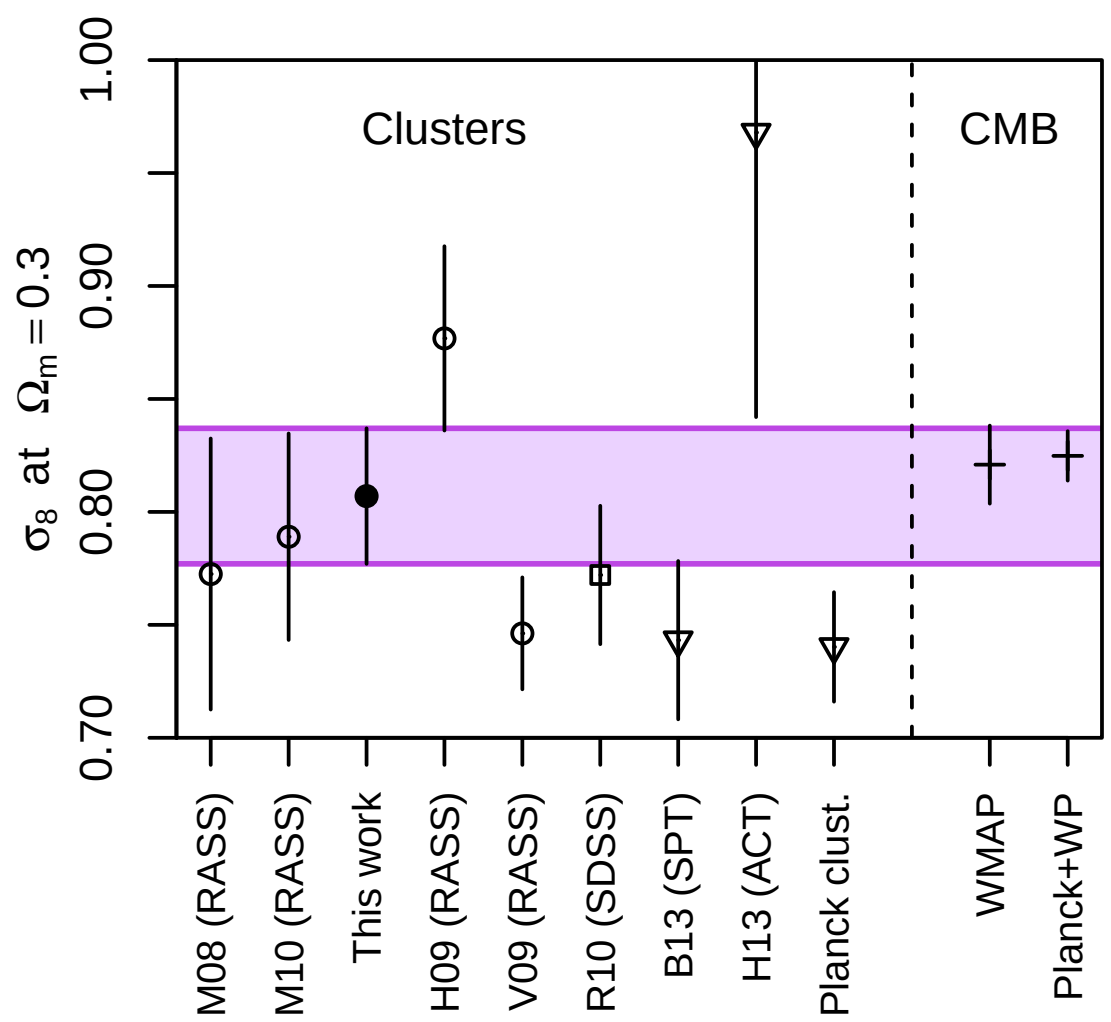

**Figure 1.5**: Measurements of $\sigma_8$ for different methods ($\Omega_m$ is set to its concordance value $\Omega_m = 0.3$) : results from clusters are shown by circles (X-ray surveys), squares (optical surveys) or triangles (SZsurveys), while crosses show CMB constraints. The shaded, horizontal band corresponds to the 68.3 per cent confidence interval for the result (filled circle) obtained by the project "Weighting the Giants". It is based on weak gravitational lensing measurements constraining the absolute mass scale of X-ray selected clusters detected in the ROSAT All-Sky Survey ([Man15]).

### The power spectrum normalization $\sigma_8$

The normalization of the power spectrum is customarily expressed in terms of $\sigma_8^2 = \langle(\delta M_R/M_R)^2\rangle$ for $R = 8h^{-1}$ Mpc, which corresponds to a cluster size. Observationnally, $\sigma_8^2 \sim 1$ so that below this scale, linear approximation is not applicable. The normalization of the power spectrum $\sigma_8$ can be estimated using several methods:

- measuring the CMB temperature anisotropies as the density perturbations induced the fluctuations in the CMB temperature on the sky.
- estimating the two point correlation function $\xi(r) = \langle\delta(\vec{x})\delta(\vec{x}+\vec{r})\rangle$ of galaxies. It is related to the probability to find a pair of galaxies at a distance $r$ : $d^2P(r) = n^2(1+\xi(r))dV_1\,dV_2$ ($n$ is their mean number per unit volume). It expresses the excess of probability over what is expected in the non-clustered case. The correlation function is the Fourier transform of the power spectrum, so that :

$$\xi(r) = \frac{V}{(2\pi)^3}\int P(k)\frac{\sin kr}{kr}4\pi k^2 dk \tag{1.22}$$

To translates galaxy counts into mass fluctuations, one relies on the assumption that they are related through $\delta_{n\,\text{galaxies}} = b\,\delta_{\text{CDM}}$ where $b$ is the bias parameter which can be complicated to estimate.

- counting clusters. Clusters counts estimated from optical or X-ray data or Sunyaev-Zeldovich(SZ) data ([Sad04], [Tab10]). The number density of clusters forming at a given epoch is related through the gravitational collapse mechanism to the matter density $\Omega_m$ and $\sigma_8$ values at that time.

Measurements of $\sigma_8$ for different methods are presented on figure 1.5, corresponding to a value of $\sigma_8 \simeq 0.8$.

## 1.6 Cosmological Probes

The parameters of the concordance ΛCDM cosmology model are the density parameters for matter, baryons and dark energy : $\Omega_m$, $\Omega_B$, $\Omega_{DE}$, the dark energy equation of state parameter $w$, the Hubble parameter $H_0$, the power spectrum spectral index $n$ and its normalization at e.g. $8h^{-1}$ Mpc. One can also add : the reionization optical depth $\tau$, which is related to the process that reionized the matter in the universe after the "dark ages", by the first galaxies and quasars at $z \sim 6-20$.

We will present below some of the cosmological probes used to estimate the cosmological parameters.

### 1.6.1 Cosmic Microwave Background

CMB is one of the major cosmological probe as the properties of our universe at the epoch of the photons emission, and along their path towards us, is imprinted in the CMB features.

Before recombination, the tightly coupled baryons and photons plasma was undergoing acoustic waves caused by two opposing pulls: the gravitational pull into the gravitational potential dark matter wells (dark matter producing almost all the gravitational force as this occurred during the matter domination era), and the outwards pull of radiation pressure. These acoustic oscillations are referred to as the Baryonic Acoustic Oscillations (BAO). After recombination, the universe became neutral and photons free-streamed with a mean free path larger than the Hubble distance, leaving the baryons free to fall in the dark matter gravitational wells. The angular size $\delta\theta$ of a temperature fluctuation in the CMB is related to its physical size $\ell$ on the last scattering surface, at an emission time $t_{\text{ls}} \sim t_{\text{rec}}$ corresponding to a redshift $z_{ls} \sim z_{rec} \sim 1100$, through : $\delta\theta = \ell/d_A(z \simeq 1100)$.

The CMB experiments measure the temperature anisotropies $\delta T/T$ on the celestial sphere. It is thus natural to write $\delta T/T$ as :

$$\frac{\delta T}{\overline{T}}(\theta,\phi) = \sum_{l=0}^{\infty}\sum_{m=-l}^{m=+l} a_{lm} Y_{lm}(\theta,\phi) \tag{1.23}$$

where $\theta$ and $\phi$ specify the directionality on the sky and $Y_{lm}$ are the spherical harmonic functions. The dipole $l = 1$ term corresponds to the Solar system peculiar velocity. The statistical properties of the temperature fluctuations are carried out by the angular spectrum :

$$\frac{\delta T^2}{\overline{T}^2} = \frac{l(l+1)}{2\pi} C_l, \qquad C_l = \left\langle |a_{lm}|^2 \right\rangle_m$$

The term $C_l$ is thus related to the temperature fluctuations at an angular scale $\delta\theta \sim \pi/l$.

The angular power spectrum as measured by Planck mission ([Tau04]) is presented on figure 1.6. The highest peak, at $l_S \sim 200$ or equivalently $\delta\theta_S \sim 1°$ corresponds to the scale of the sonic horizon at $t_{rec}$ i.e. the distance that a sound wave supported by the photon-baryon-electron plasma could travel between the Big-Bang and the recombination :

$$\ell_S = R(t_{rec})\chi_S = R(t_{rec}) \int_{z_{rec}}^{\infty} \frac{c_s dz}{H(z)} \tag{1.24}$$

where $c_s$ is the sound speed and depends on the baryon-to-photon ratio. The sonic horizon $\ell_S$ corresponds to the comoving distance $L = R(t_0)/R(t_{rec})\ell_S \sim 150$ Mpc. This physical scale is also observed closer to us, in the large scale structures (see section 1.6.2). The measurement of $\delta\theta_S = \ell_S/d_A(z)$ at $z \sim 1100$ is essentially sensitive to $H_0$ and to the curvature of the universe, i.e. $\Omega_T$, along, because of the sonic horizon size, with the baryonic content $\Omega_B h^2$, and the matter content $\Omega_m h^2$.

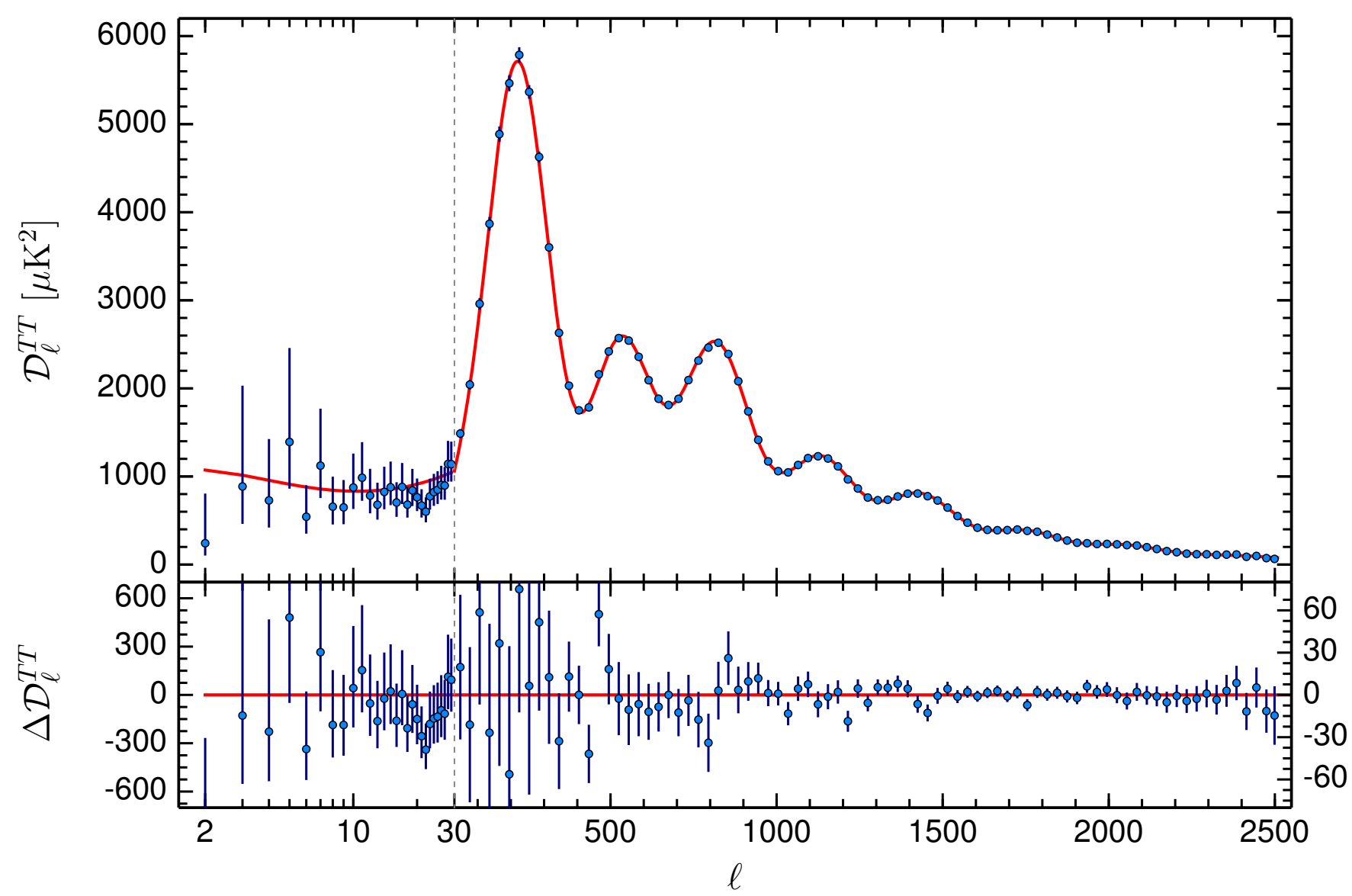

**Figure 1.6**: The CMB temperature power spectrum as measured by the Planck mission ([Ade15]). The position of the first peak, at $l \sim 220$ ($\delta\theta \sim 1°$) is related to the cosmological parameters describing the universe at the recombination epoch and the geometry and contents of the universe along the photons path towards us (curvature, dark energy). Superimposed is the best fit to the data, from which the cosmological parameters are inferred. The lower plot is the residual to the upper. The scatter at low $l$ is due to the cosmic variance, that is the lack of possibility to average over sufficiently numerous cells of size $\delta\theta$ on the sky. Image: [Ade15]

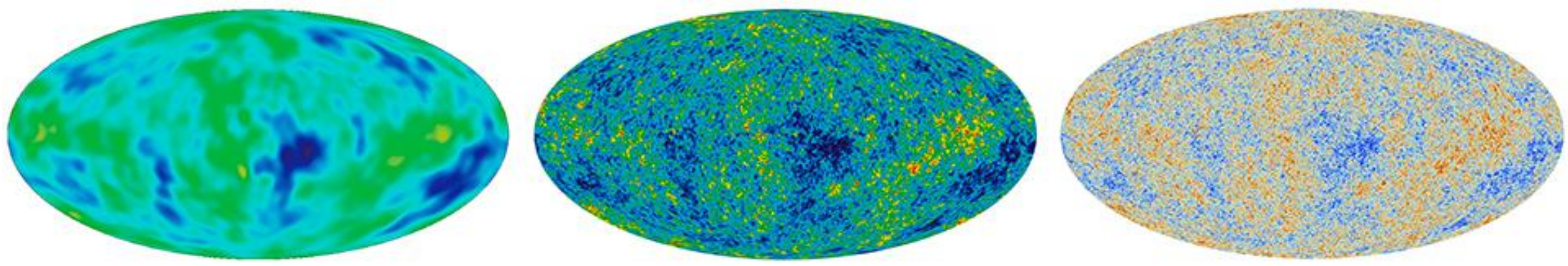

**Figure 1.7**: Sky map temperature anisotropies as measured by the successive COBE (Cosmic Background Explorer, launched in 1989), WMAP (Wilkinson Microwave Anisotropy Probe, launched in 2001) and Planck mission (launched in 2009).

CMB photons are polarized at the level of a few microkelvins, as Thomson scattering of radiation on electrons globally falling or escaping from a potential well generates linear polarization. The CMB polarization is described by the two fields $E$ and $B$, the $E$ modes are generated by the scalar density fluctuations, whereas primordial gravitational waves, foregrounds, and gravitational lensing on the photons line-of-sight generate both $E$ and $B$ modes. The detection of the primordial gravitational waves can only be confirmed from the detection of $B$ modes.

Succeeding to the Wilkinson Microwave Anisotropy Probe (WMAP) ([Pag03], [Ben03]) mission, itself a successor to COBE (Cosmic Background Explorer) mission, the ESA (European Space Agency) Planck satellite mission[6] ([The06], [Tau04]) was launched in 2009 and had 30 months of observation. The sky map of the temperature anisotropies is presented for these 3 successive missions in figure 1.7. The Planck mission has given indeed an impressive confirmation of the standard cosmological model. Planck TT, TE, and EE cross- and auto-spectra are extremely well fitted by a 6-parameters flat $\Lambda CDM$ model(Planck 2015 data, [Ade15]). Planck also provides a lensing potential map, which computation is based on the correlation induced at large scale (a few degrees) by the small (a few arcminutes) lensing deflections due to the large structures along the photon line-of-sight. The parameters describing the best fit model are as follows :

1. the density of baryonic dark matter : $\Omega_B h^2 = 0.02225 \pm 0.00016$
2. the density of cold dark matter: $\Omega_{\mathrm{CDM}} h^2 = 0.1198 \pm 0.0015$
3. the angle subtended by the sonic horizon at last scattering : $100\theta_S = 1.04077 \pm 0.00032$
4. the fraction of the CMB photons scattered by re-ionized matter along their path : $\tau = 0.079 \pm 0.017$
5. the amplitude of the initial density fluctuations power spectrum : $\log(10^{10} A_S) = 3.094 \pm 0.034$

[6] http://www.cosmos.esa.int/web/planck

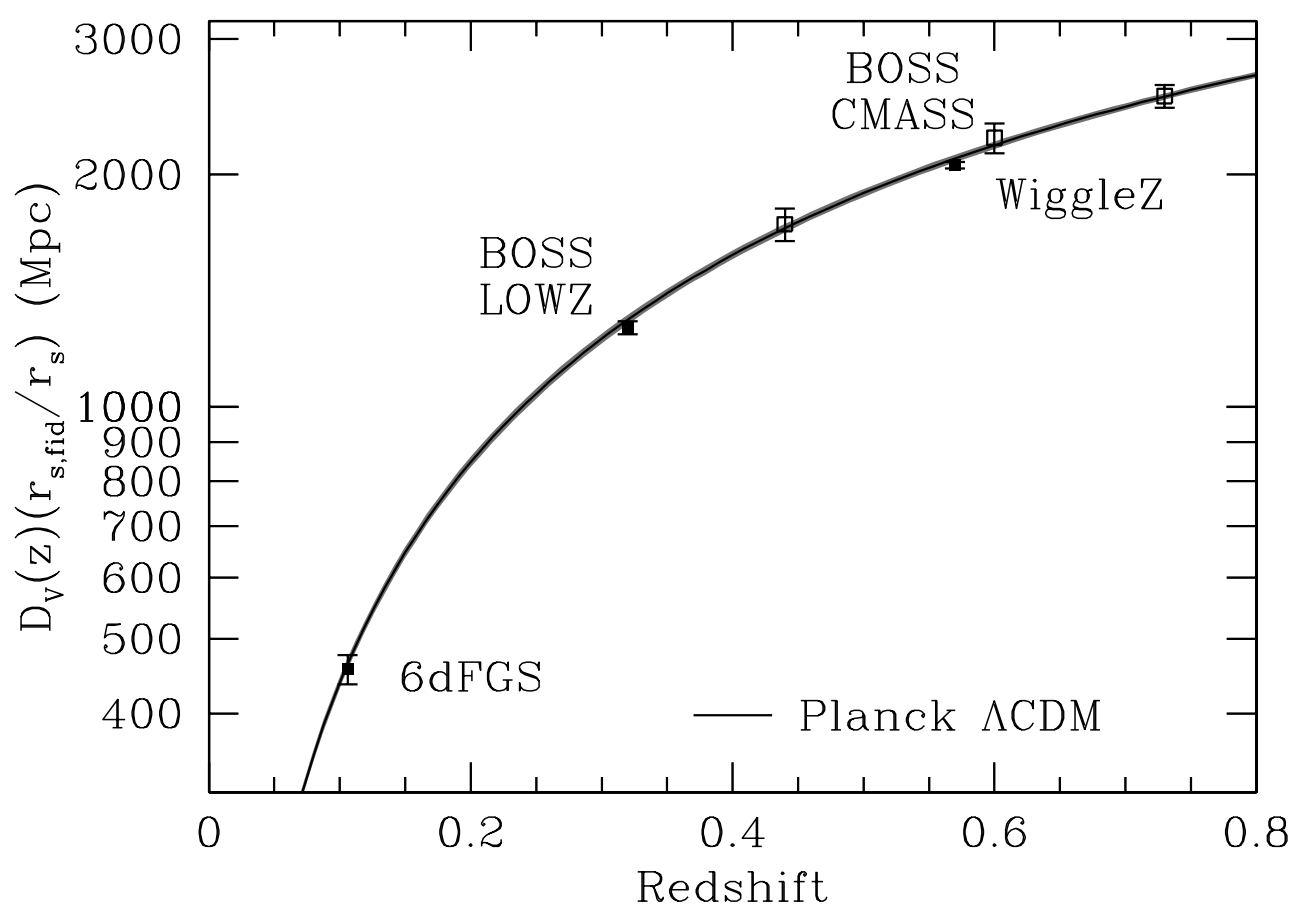

**Figure 1.8**: The distance $D_V$ as a function of redshift from BAO experiments : BOSS, on the CMASS and LOWZ SDSS galaxies sample, the 6dFGS survey ( [Jon09]) and the WiggleZ survey ([Bla11]). Superimposed is the flat ΛCDM model obtained using the CMB projects Planck and WMAP results ([Ade14]). There is a remarkable agreement between the different galaxy surveys and the Planck data. Image : [And14]

6. the slope of the initial density fluctuations power spectrum : $n_S = 0.9645 \pm 0.0049$

From this analysis are also inferred the value of the Hubble parameter $H_0 = 67.51 \pm 0.64 \text{km.s}^{-1}.\text{Mpc}^{-1}$ and $\Omega_m = 0.3121 \pm 0.0087$. Releasing the flat assumption, and adding observational constraints from the large scale structures and supernovae Ia projects, allows to estimate the spatial curvature $\Omega_k = 0.0008 \pm 0.004$. Together with data from large-scale structure (BAO) and type Ia supernovae, the gravitational lensing potential measurement constrains $\sigma_8 = 0.8159 \pm 0.0086$ at 68% c.l., in tension with clusters data.

### 1.6.2 Baryon Acoustic Oscillations

The Baryon Acoustic Oscillation (BAO) is an important cosmological probe for estimating the geometry of the universe and the dark energy content. As the acoustic waves ceased in the primordial baryons-electrons-plasma at recombination, their imprint, of characteristic size $\ell_S$, was left in the matter distribution. As a consequence, galaxies showed a slight preference to form, and thus be later on found at time $t$, at a separation distance $\sim L = a(t)/a_{\text{rec}}\ell_S$. This length scale $L_0 \sim 150$ Mpc is observed in the correlation function of galaxies, and is indeed a standard ruler that can be observed from $z \sim 0 - 3$ in structures to $z \sim 3200$ in the CMB.

Using BAO and comparing the sound horizon size of present with the one at recombination era permits to probe the expansion history of the universe and estimate the cosmological parameters, in particular $\Omega_m$. BAO measurements require the observation of millions of galaxies in a huge survey volume. Potential systematic errors arise from bias (i.e. the scale dependent difference between the galaxy and the dark matter clustering) and non linear gravitational evolution of structures.

The Sloan Digital Sky Surveys ([Lov02], [Ken94], [Yor00])[7] has observed since 2000 about 35% of the sky and has provided imagery for 500 million objects and spectroscopy for 3 million of them. It uses a 2.5 m optical telescope at Apache Point Observatory in New Mexico (USA) equipped with a 30 CCD's camera with a total of 120 Megapixels and multi-fiber spectrographs. It carries out astrophysical and cosmological surveys : SDSS-II (2005-2008) incorporates the Sloan Supernova

[7] http://www.sdss.org/

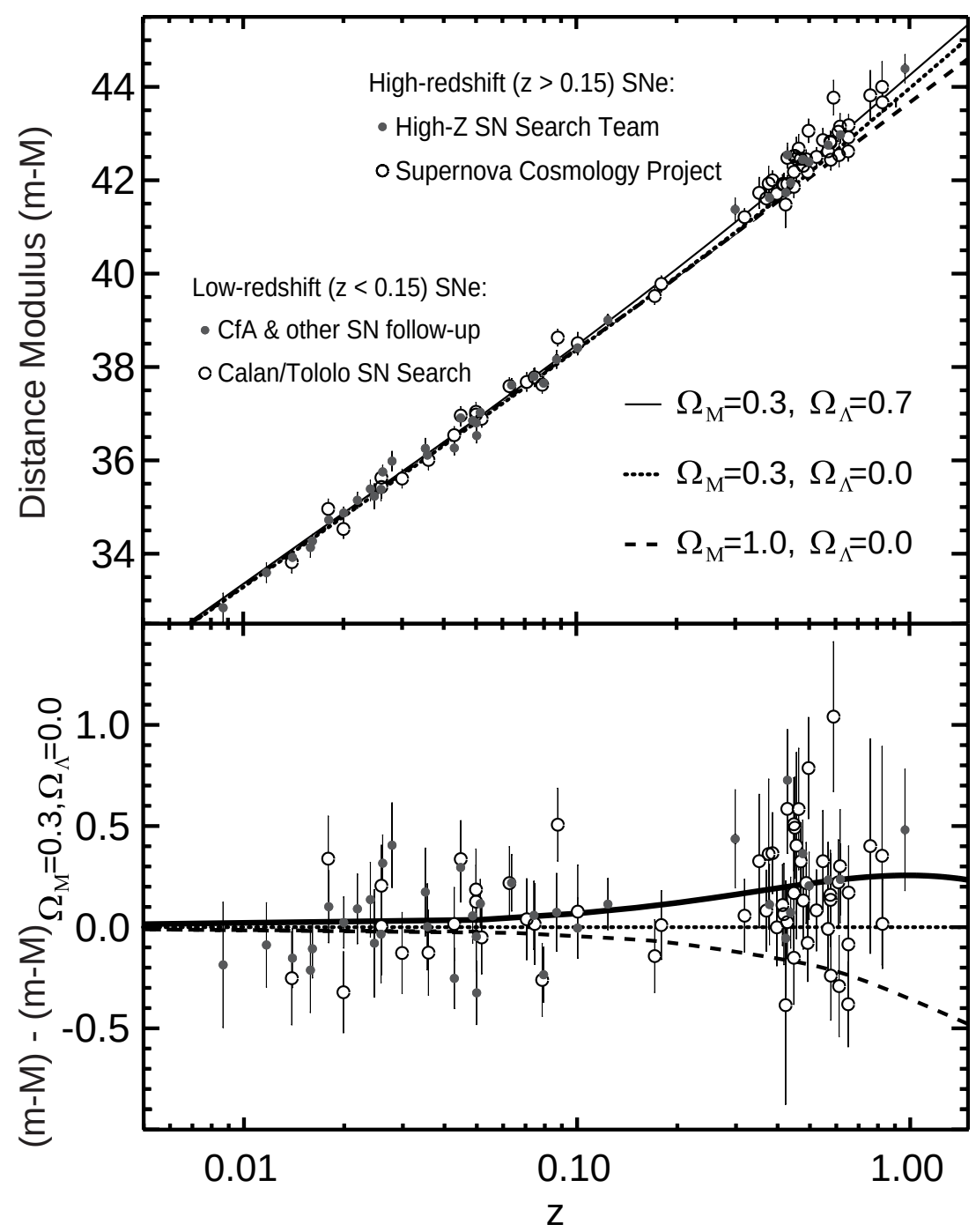

**Figure 1.9**: The original "discovery data", the Hubble diagram of SNe Ia compiled from the High-z Supernova team and the Supernova Cosmology Project's data. The bottom panel is the residual to the upper (distance modulus) corresponding to an open universe. The points can be seen to lie above the non-accelerating models (dashed and dotted lines). Image : [Fri08]

Survey which detected about 500 spectroscopically confirmed type Ia supernovae in the redshift range $z = 0.05 - 0.4$; SDSS-III included the Baryon Oscillation Spectroscopic Survey (BOSS) which mapped the spatial distribution of luminous red galaxies (LRGs) and quasars; SDSS-IV is scheduled to continue up to 2020 and includes the extended Baryon Oscillation Spectroscopic Survey (eBOSS).

Results from BOSS ([And14]) which observed $10000^2$ degrees of sky at a redshift range of [0.2,0.7] are presented on figure 1.8. Also presented on figure 1.8 are the measurement from the 6dFGS survey ([Jon09]) which observed $17000^2$ degrees, and the WiggleZ project ([Bla11]), with a redshift range of [0.4,1] and final data set covering a range of $800^2$ degrees of sky. The agreement with Planck data preferred ΛCDM model is remarkable.

### 1.6.3 Type Ia Supernovae

Type Ia supernovae (SNe Ia) are prime examples of standard candles with a very homogeneous intrinsic brightness. SNe Ia are the end points of stellar evolution when oxygen-carbon accreting white dwarfs reach the Chandrasekhar mass $M_{\rm Ch} \sim 1.4$ solar mass, the theoretical limit before electron degeneracy pressure is no longer sufficient to support against the inward gravitational pull. Their remarkable uniformity allows them to be used as distance indicators, and an efficient cosmological probe of the expansion history. Using SNe Ia observations, [Per97], [Rie98] built the Hubble diagram $D_L(z)$ up to $z \sim 0.5$ to demonstrate the presence of dark energy and the present time accelerated expansion of the universe (see figure 1.9).

SNe Ia can also be used to measure $H_0$ value. The challenge lies in the measurement of their absolute luminosity – or equivalently in the measurement of their distance independently of their redshift, using Cepheids distance to nearby galaxies, as did the project SH0ES (Supernova H0 for the Equation of State) with the HST Hubble Space telescope (HST). The analysis of SH0ES results taking into account the slight difference in luminosity for SNe Ia occurring in high or low star forming environment yielded $H_0 = 70.6 \pm 2.6\text{km.s}^{-1}.\text{Mpc}^{-1}$ within $\sim 1\sigma$ of the measurement based on the CMB power spectrum ([Rig15]).

More on supernova cosmology is being described in the next chapter where the Supernova Legacy Survey (SNLS) project is presented, along with the measurement of the dark energy equation of state parameter $w$. The joint analysis of the SNLS-3 years and the SDSS-II supernovae data, combined with CMB constraints, yields $w = -1.018 \pm 0.057$ for a flat universe (taking into account the statistic and the systematic errors), compatible with a cosmological constant ([Bet14]).

### 1.6.4 Gravitational Lensing

Gravitational lensing is an important cosmological probe of the matter distribution in the universe, and its overall geometry. The geometry of space time is dependent on the nature of the mass content of the universe hence the wavefronts of distant objects light traveling through any matter fluctuations during its flight will be distorted.

For most lines of sight in the universe, gravitational lensing occurs in the weak regime, and can be detected in the systematic tangential stretch of background sources around the lensing mass (see section 3.1.5). Cosmic shear is the shape distortion of many galaxies by the foreground large scale structures and is thus a statistical signal. The 2-point correlation function of the image distortions is directly related to the power spectrum of the density perturbations. More precisely, the shear angular power spectrum is related to the expansion history of the universe, the history of the growth of structures, and thus sensitive to the dark energy density. Main sources of systematics in weak lensing analysis originates from the incertitudes on the galaxy shape measurement (notably the point spread function – PSF – of the optical instrument) or on the source or lens populations (e.g. on their photometric redshift estimates).

Some recent weak lensing survey results includes:

- the Hubble Space Telescope Cosmic Evolution Survey (COSMOS), which combined space-based galaxy shape measurements with ground-based photometric redshifts to constrain cosmological parameters $\sigma_8$ and $\Omega_M$ ([Sch10]).

- the Canada- France-Hawaii Telescope Lensing Survey (CFHTLens) project ([Hey13]) which uses 154 square degrees of deep multi-colour data obtained by the CFHT Legacy Survey. They produced the largest contiguous maps of projected mass density obtained from gravitational lensing shear ([Van13]) and performed detailed cosmology analysis ([Fu14]).

- [Dem16] did exploratory works on optical/radio weak lensing analysis. They performed cross correlation weak lensing analysis with the VLA first survey and the SDSS, covering approximately $10,000$ square degrees. They probed shear power spectrum using cross power spectrum approach on large scales.

Some of the principal on-going and future lensing surveys are listed in table 1.2.

| Surveys | when | where | caracteristics | $n_{gal}$ | $f_{sky}$ arcmin$^2$ |
|---|---|---|---|---|---|
| Dark Energy Survey | now | CTIO, Chile | 4-m Blanco | 10 | 0.1 |
| Hyper Suprime Cam | now | Mauna Kea, Hawaii | 8.2-m Subaru | 20 | 0.048 |
| LSST | 2019 | Cerro Pachón, Chile | 8.4-m | 40 | 0.25 |
| EUCLID (ESA) | >2020 | space | 1.2-m, visible & NIR | 35 | 0.20 |
| WFIRST (NASA) | >2020 | space | 2.4-m, NIR | 45 | 0.05 |

**Table 1.2**: Major upcoming weak lensing surveys. $n_{gal}$ is the effective galaxy count, $f_{sky}$ is the fraction of the sky covered by the survey. Source : [Liu15]

### 1.6.5 Galaxy Clusters

Galaxies or galaxy clusters number counts as a function of their redshift provides another important cosmological probe that helps in constraining cosmological parameters, such as the density parameters and the equation of state parameter ($w$) or the density fluctuations power spectrum.

The number counting technique can be based on various effects and methods : the Sunyaev-Zeldovich(SZ) effects which distorts the CMB temperature, due to the Compton scattering of the photons of the intra-cluster hot baryonic gas ; the X-ray observations of this hot gas ; the weak gravitational lensing effect caused by the cluster dark matter halo ; the cluster galaxies velocity functions. The effectiveness of clusters as a probe depends on how robust is the association between the cluster observables with the cluster mass.

The number counts can be formalized as:

$$\frac{dN}{dzd\Omega} = \frac{dV}{dzd\Omega}(z) \int_{M_{min}} dM \frac{dn}{dM}(z, M) \tag{1.25}$$

where the left hand side gives the number of objects $dN$ observed in a given redshift slice of $dz$ at a solid angle interval of $d\Omega$. $dV$ is the comoving volume at redshift $z$, $n$ is the comoving number density at the corresponding redshift, and $M_{min}$ is the minimum mass of objects detected in that redshift slice. The knowledge of the comoving number density $n(z)$ is important and can be expressed with the Press-Schechter semi-analytic formalism:

$$\frac{dn}{dM}(z, M) = \sqrt{\frac{2}{\pi}} \frac{\rho_m}{M} \frac{\delta_c}{\sigma^2(M, z)} \frac{d\sigma(M, z)}{dM} \exp\left(-\frac{\delta_c^2}{2\sigma_M^2 D^2(z)}\right) \tag{1.26}$$

where $\rho_m$ is the present day matter density, $\delta_c \sim 1.68$ is the linear threshold overdensity collapse while $D(z)$ is the linear perturbation growth factor and $\sigma_M$ is the rms density fluctuations at mass scales of $M$ at present day ($z = 0$).

Various projects are observing clusters detected detected in X or CMB data and performing weak lensing studies, such as the Local Cluster Substructure Survey ([Oka10]) using high quality Subaru Suprime-Cam data or the Cluster Lensing And Supernova survey with Hubble (CLASH) which obtained accurate cluster mass profile measurements ([Ume14]).

Indeed, the main sources of systematics are the cluster mass-observable relations and the selection function, as the clusters selection method are based on redshift dependent observables. Number counts of galaxies and clusters provide a direct cosmological probe complementary to the CMB and the SNe Ia measurements, as shown on figure 1.10.

### 1.6.6 The Cosmic concordance

It is very remarkable that these various probes all favor the same cosmological model (see figure 1.10). Although very satisfactory, this situation nonetheless present us with challenging questions, such as the true nature of the dark energy, or the identity of the dark matter particle.

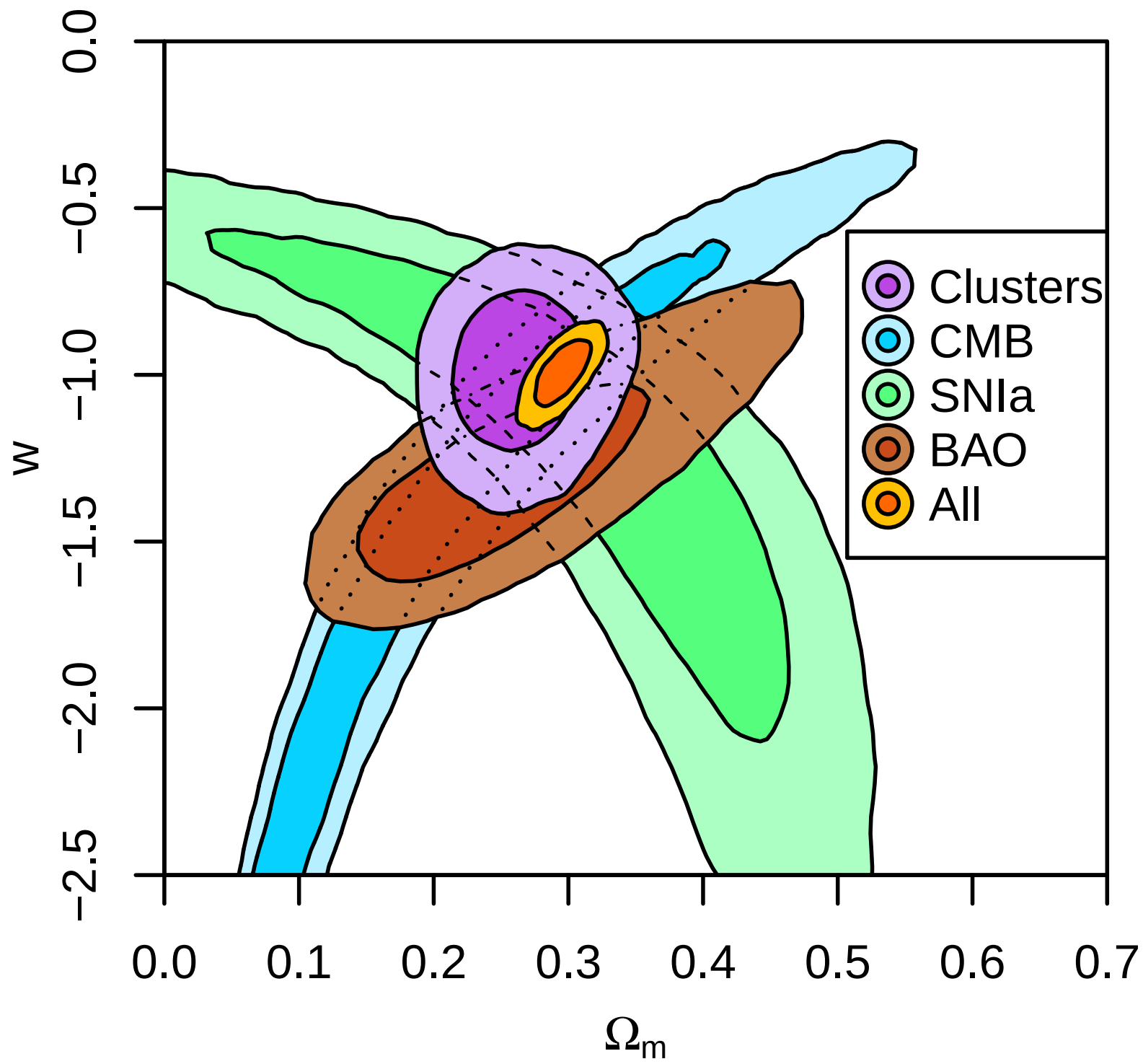

**Figure 1.10**: Constraints on constant $w$ dark energy models from the project Weighing the Giants, which uses weak gravitational lensing measurements to improve the measurements of the galaxy cluster mass function and its evolution, using X-ray selected clusters detected in the ROSAT All-Sky Survey. This measurement is compared with constraints form the CMB (WMAP, ACT and SPT), supernova (from the Union sample including SNLS-1 year data) and BAO (SDSS/BOSS, 6dF galaxy survey) data, and their combination. Dark and light shading respectively indicate the 68.3 and 95.4 per cent confidence regions, accounting for systematic uncertainties. Image : [Man15]

# Chapter 2

# Type Ia supernovae and the Supernova Legacy Survey

Type Ia supernovae are one of the most fascinating objects of the Universe. They produce the most violent explosions after the Big Bang, explosion of such magnitude (around $4 \times 10^9$ solar luminosity) that for a short while a supernova will outshine its entire host galaxy. The occurrences of supernovae are not very frequent at all per galaxy, around 2 to 3 per millennium. They are bright, uniform candles and stays so for a convenient period, about a month, during which they can be detected and observed. They are quasi-ideal calibratable standard candles, which makes them a very attractive and practical cosmological probe.

This chapter is structured in two sections. The first section presents the properties of type Ia supernovae and their use as a distance indicator. In the second section we shall describe the Supernova Legacy Survey experimental set-up and data analysis technique.

## 2.1 Type Ia supernovae as standard candles

### 2.1.1 Observations

Supernovae are categorized according to their spectral features near maximum brightness. Type II spectra exhibit hydrogen features, which are lacking in type I spectra. This category is furtherly separated into type Ia and Ib/c according to the presence or absence of silicon features (see figure 2.1).

Besides the clear Si II absorption lines in its spectra at around 4130 Å and 6150 Å, the type Ia spectral signature consists in a 'W' shaped absorption doublet at 5640 Å due to SII, and an absorption doublet at 3934 Å and 3968 Å due to Ca II (figure 2.2). All features exhibit a deep P-Cygni ( [Psk69], [Bra82], [Bra83]) profile characteristic of line formation in an expanding

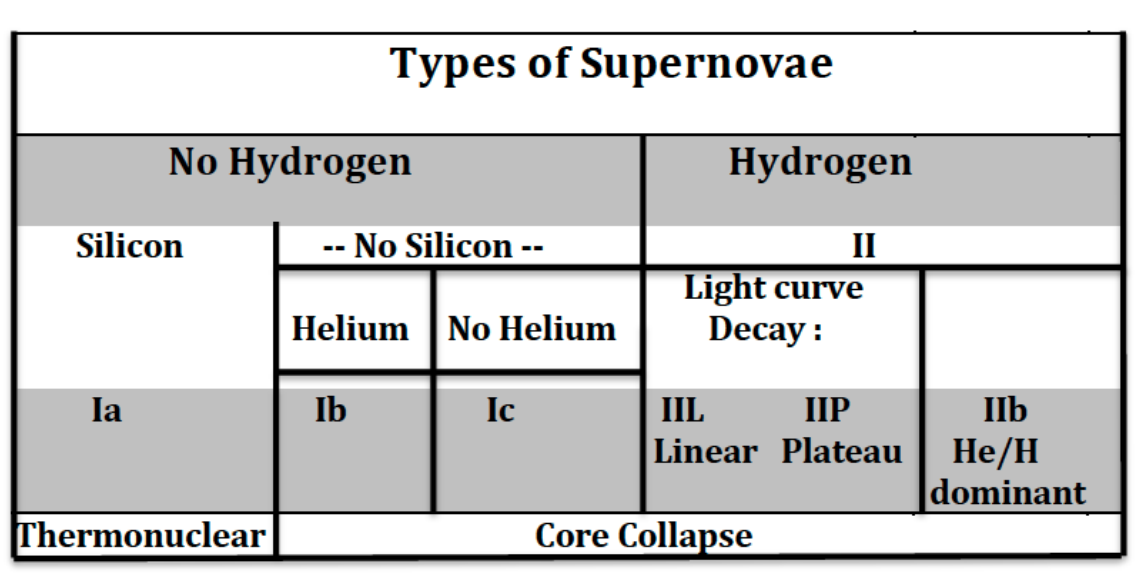

**Figure 2.1**: Supernovae categories

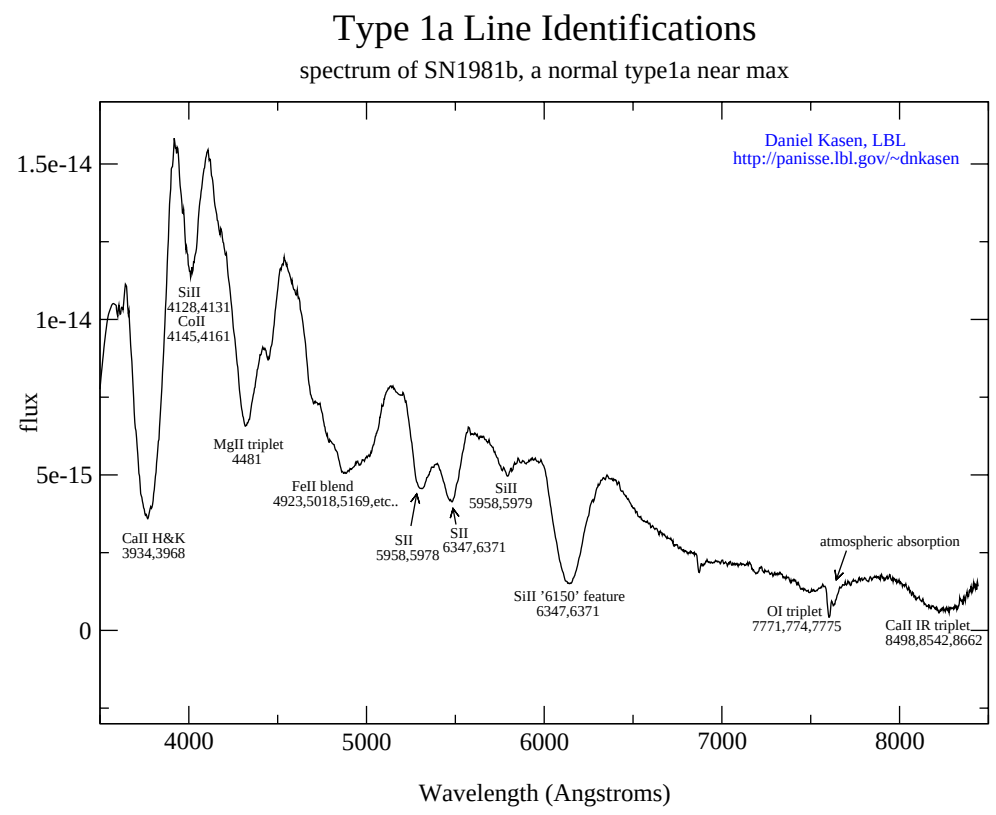

**Figure 2.2**: Example of a supernovae Ia spectrum near it's peak brightness. The clear dip in the spectra near the 6000Å indicates the presence of Si II which is the clear signature of a type Ia SN. Almost $3/4^{th}$ of the emitted light is in the optical range (3800-7500 Å). Image : Daniel Kasen, LBL

photosphere.

Type Ia supernovae (SNe Ia) form a homogeneous class in term of spectra and luminosity emission as a function of time (light curve). The peak brightness in the B band shows a dispersion of about 40% among "normal" SNe Ia. However, even if SNe light curves are quite similar, they are not identical. There exists correlations between the peak luminosity and some other observables : the luminosity variation time scale, as brighter SNe Ia dim slowlier – this is the so-called Phillips or *brighter-slower* empirical law ([Phi93]) ; the intrinsic color at peak – brighter SNe Ia are bluer (*brighter-bluer law*, [Tri99]). The use of SNe Ia as a distance indicator is based on these empirical relations (see section 2.1.3).

### 2.1.2 Theory

Type Ia supernovae (SN Ia) are believed to be the result of the thermonuclear disruption of a carbon-oxygen white dwarfs which reaches the Chandrasekhar-mass limit of stability ($M_{\rm Ch} \simeq 1.4 M_\odot$) by accreting matter from a companion ([Hoy60], [Whe73])[1].

This model explains the lack of H and He in SNe Ia spectra, their occurring in evolved parent galaxies, and their relative homogeneity, related to the existence of the Chandrasekhar limit. The nuclear energy arising from the explosion is entirely converted into kinetic energy and the luminous emission is powered by the decaying of the nickel $^{56}$Ni produced during the explosion. The reaction $^{56}\text{Ni} \longrightarrow {}^{56}\text{Co} \longrightarrow {}^{56}\text{Fe}$ sets the gradual dimming time-scale ([Col69]). The progenitor system identity (a single or a double WD), the hydrodynamic mechanism of the explosion (detonation or deflagration) are still the object of debate.

The mass of $^{56}$Ni produced is the primary determinant of the peak brightness of the SN Ia. Its variation ($M_{\rm Ni} \simeq 0.4 - 0.9 M_\odot$) could explain the SNe Ia diversity and the *brighter-bluer* and *brighter-slower* relations ([Hoe96]). A higher nickel mass implies a higher temperature and a modification of the opacity : e.g. a decreasing opacity with temperature in the blue B band means a slowlier dimming of the B magnitude ([Kas07]).

The SNe Ia physics is complicated at every step, from the physics of explosion to the processes

[1] The other supernovae type originate from the gravitational collapse of the iron core of massive stars $M > 8M_\odot$ ([Ibe83], [Woo02]) that cannot be supported by further exothermal thermonuclear reactions (core-collapse SNe).

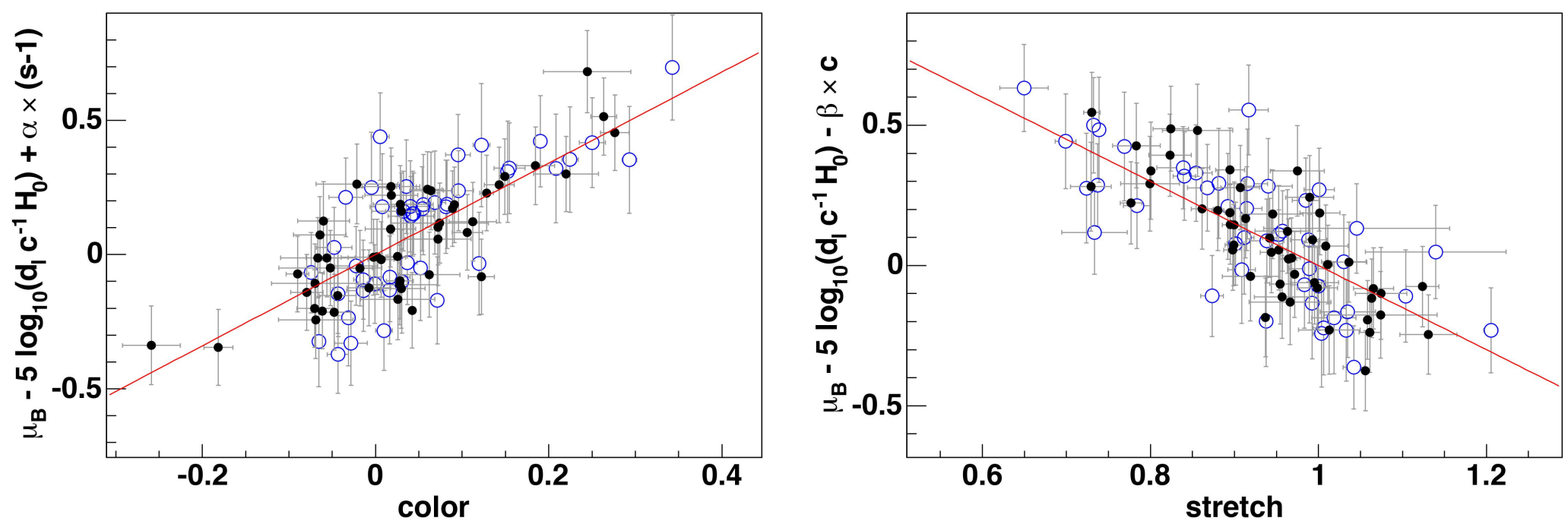

**Figure 2.3**: brighter-slower relation and the brighter-bluer relation : residual plots of $\mu_B - \mathcal{M}_B - m_B^\star$ (see eq. 2.1) showing the the relationships between the SN Ia luminosity and their light curve stretch factor $s$ or color $c$. The y-axis goes from brighter to fainter. Image [Ast06]

that convert the energy input into a visible emission through radiative transfer in the envelope, but it results nonetheless in a relatively homogeneous event. Model SN Ia light curves and spectra can fit the observations reasonably well ([Kas09]), but so far, their predictive power is limited: using SNe Ia as cosmological probes rests on empirical relations.

### 2.1.3 The distance indicator

In order to apply the *brighter-slower* correction, the light-curve characteristic time scale can be estimated by the stretch factor $s$ : a template light curve in the B (or V) band $F(t/s)$ is stretched to match the observations. Some other parametrization exists such as MLCS/MLCS2k2 ([Jha07]), $\Delta m15$ ( [Phi93]) . . . .

For the *brighter-bluer* relation, one defines $c$, the excess color at maximum as $c = (B - V)_{t_{max}} - \langle (B-V)_{max} \rangle$ (the average is taken over the SN Ia population).

The parameters $s$ and $c$ can be estimated by fitting a spectro-photometric model of the SN Ia emission $\phi(\lambda, t)$ such as the SALT2 model used in the SNLS data analysis ([Guy07]).

Taking into account the standardization relations, the distance modulus of a SN Ia is given by :

$$\mu = m_B^\star - \mathcal{M}_B + \alpha \times (s - 1) - \beta \times c \tag{2.1}$$

$m_B^\star$ is the peak apparent magnitude, estimated in the B band in the the supernova rest-frame (as opposed to the B band in the observer rest-frame). The stretch factor $s$, the color $c$ and the normalization $m_B^\star$ of the spectro-photometric model $\phi$ are estimated for each supernova, using its measured magnitudes.

The brighter-slower and brighter-bluer relation are parametrized by $\alpha$ and $\beta$. $\mathcal{M}_B$ combines the absolute magnitude of the standard $s = 1$, $c = 0$ supernova together with the Hubble parameter value $H_0$: $\mathcal{M}_B = M_B + 5\log_{10}(c/H_0/10\,\text{pc})$. These three parameters are fitted, along with the cosmological parameters, on the total supernova sample in the Hubble diagram $\mu(z)$ (see figure 2.3).

With this parametrization, there remain an intrinsic variation in the SN peak luminosity of $\sim 15\%$ which results in a distance measurement dispersion around 8% ([Ast06]).

Note that the $\beta \times c$ term is equivalent to an extinction correction $A_B = R_B \times E(V - B)$ calculated for an excess color $E(B-V) = (B-V)-(B-V)_0$, (the latter term $(B-V)_0$ standing for the color the object would have if it was not extincted) and a total to selective extinction ratio $R_B$. However, the estimation of the $\beta$ parameter with SNe Ia leads to an effective $R_B < 4.1$ ([Ost08]), so that the color corrections of SN Ia are incompatible with known dust properties.

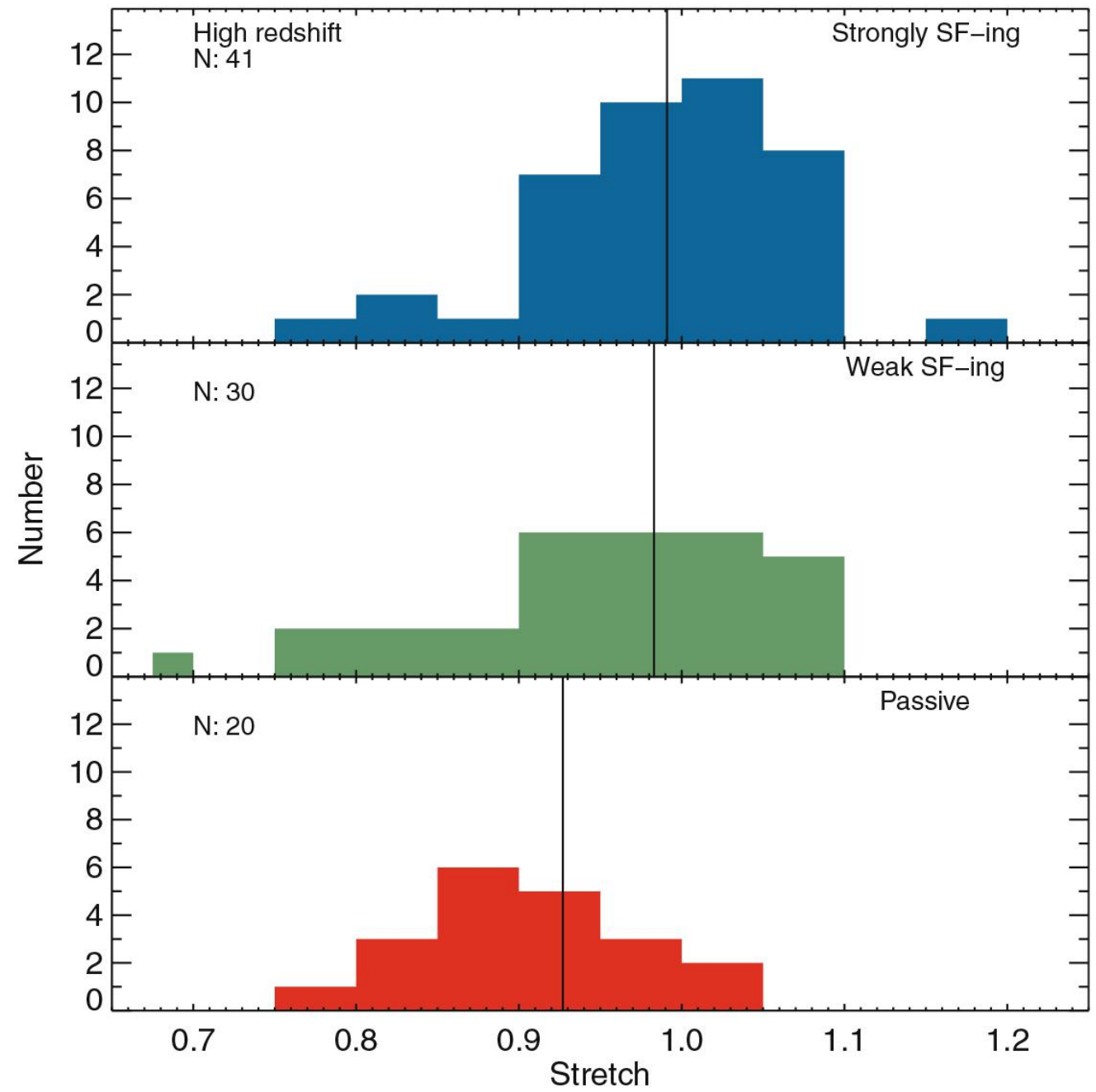

**Figure 2.4**: For for the supernovae of SNLS survey : light curve stretch parameter $s$ as a function of the host galaxy specific star formation rate, from active (top) to passive (bottom) galaxies. Vertical lines represent median stretch in the respective histograms. Image : [Sul06]

This difference could be explained by an intrinsic color variation of the SN Ia combined to dust extinction, and/or a specific extinction law due to the SN environment. Intrinsic and extrinsic color variations are not easily disentangled.

### 2.1.4 K-correction

In order to compare supernovae, their fluxes have to be expressed in the same way, and in particular using the same rest-frame band, e.g., as proposed above, the B band. A supernova is observed through fixed spectral bands, depending on the telescope set-up. Due to the redshift effect, the effective band-pass the supernova is seen through changes. K-corrections correct for this effect : according to [Ast12], K-correction is “transforming photometric data obtained with some filter into what would have been measured with some other filter”. If $S(\lambda)$ is the rest-frame spectrum of the supernova, observed at a distance $d$ and a redshift $z$, then the flux measured in the observed pass-band, e.g. the $r$ filter, is related to the flux as it would be observed in a rest-frame pass-band, e.g. the $B$ filter, by :

$$f_B = f_r \times \frac{\int S(\lambda) T_B(\lambda) d\lambda}{\int S(\lambda \times (1+z)) T_r(\lambda) d\lambda}$$

with $T$ being the filter transmission. To compute the rest-frame B band magnitude from the observed magnitudes thus requires a synthetic SED describing the supernova emission at $\lambda$ and $t$, a precise knowledge of the intervening effective pass-bands transmission for the instrument, together with its absolute calibration. As a consequence, supernova spectro-photometric modeling and instrument calibration are key-features in their cosmological exploitation.

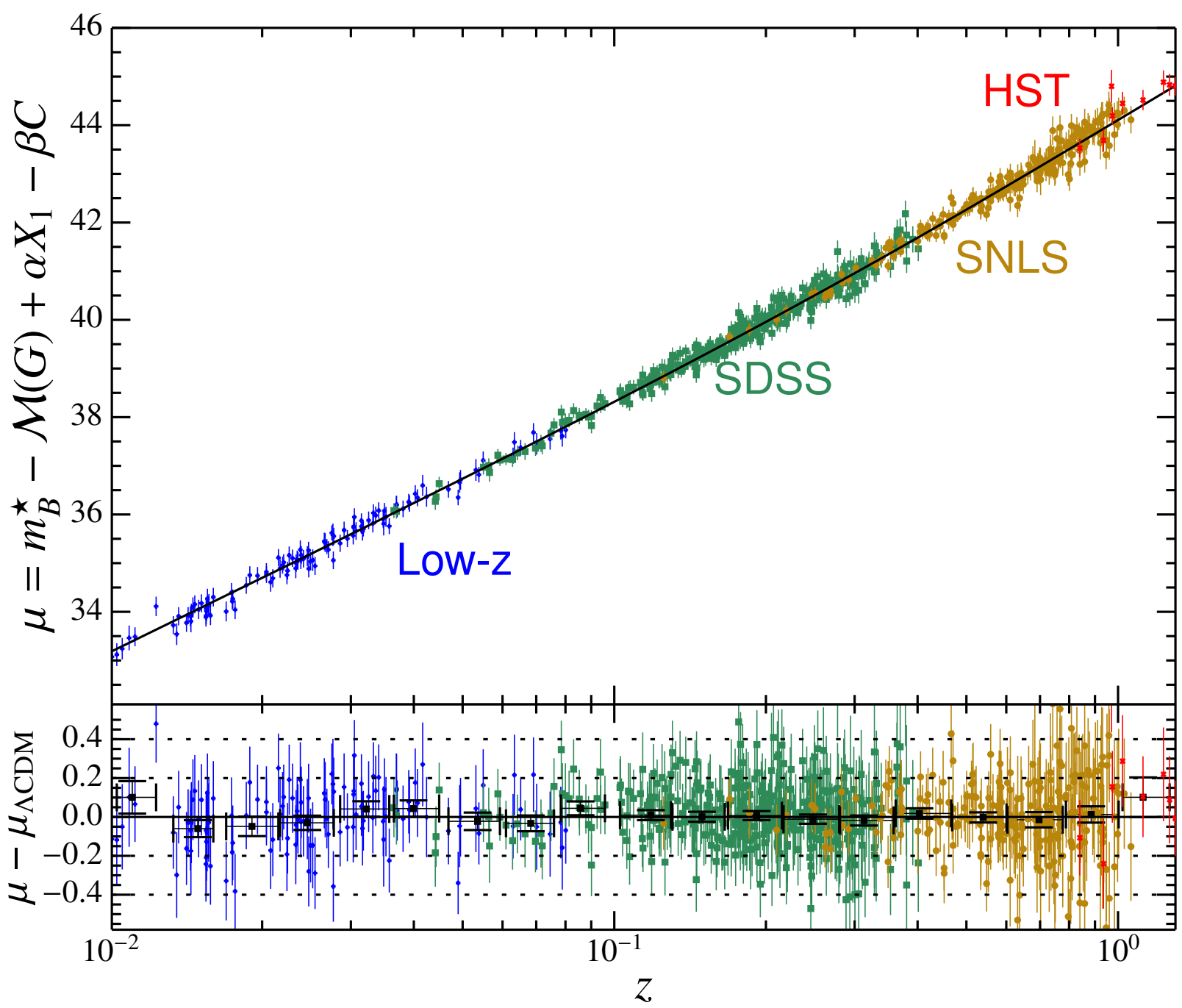

**Figure 2.5**: Hubble diagram from the joint sample of SNLS3, SDSS II, HST and low-z SNe (see §2.2.3). The bottom panel is the residual to ΛCDM cosmology fit. Image : Betoule [Bet14]

### 2.1.5 A third relation

SNe Ia properties correlate with the physical parameters describing its environment, such as it's host galaxy stellar mass, star formation rate, or metallicity. For instance, late type galaxies or galaxies with a higher specific star formation rate (sSFR) preferentially host SN Ia with higher stretch (e.g. [Ham00], [Gal05], [Gal08]). This property is illustrated on fig.2.4 for the Supernova Legacy Survey SNe Ia ([Sul06]). This also holds for the host galaxy stellar mass ([How09]) : massive galaxies host SNe Ia with lower stretch factor (which are then less luminous than the standard SN Ia with $s = 1, c = 0$). This effect could point out a relation existing between metallicity (higher in massive galaxies) and luminosity via the synthetized $^{56}$Ni mass ([Tim03]), although no conclusive evidence has been obtained ([Gal05], [How09]).

If the correlation between the SNe Ia light curve properties and their host galaxy stellar population are well established, no conclusive trends are seen between SNe Ia color and their host galaxy properties though.

Recently, it was demonstrated ([Kel10], [Sul10]) that the standard SN Ia was indeed about 10% more luminous in massive and passive galaxies. As a consequence, the *brighter-slower* and *brighter-bluer* relations are not sufficient to describe SN Ia variability. A *brighter-heavier* relation was thus introduced in recent cosmology analysis ([Con11], [Bet14]), to adjust the absolute magnitude parameter $\mathcal{M}_B$ in eq. 2.1 taking into account the SNe host galaxy properties. The SNe are assigned a host galaxy mass bin, which in turn assigns them an absolute magnitude : $\mathcal{M}_B$ (resp. $\mathcal{M}_B + \Delta\mathcal{M}_B$) for the SNe occurring in high (resp. low) mass galaxies.

Although SN Ia brightness are clearly linked to their host galaxy properties, the question of which parameter is the most pertinent (sSFR, mass, metallicity), and whether it should be local (e.g. sSFR at the SN location) or global (e.g. mass) ([Rig13]), is still in discussion.

### 2.1.6 Hubble diagram

Cosmological information are extracted from a Hubble diagram $\mu(z; \Omega_M, \Omega_{\rm DE}, w \ldots; \alpha, \beta \ldots)$ where nearby SNe ($0.01 < z < 0.1$) are compared to more distant ones ($z \sim 0.2 - 1$). The cosmological parameters $\Omega_M, \Omega_{\rm DE}, w \ldots$ are fitted along with the "nuisance" parameters $\alpha, \beta, \mathcal{M}_B, \Delta\mathcal{M}_B$. The joint Supernova Legacy Survey and SDSS Hubble diagram is shown on figure 2.5 and is presented in details in section 2.2.3. With $\sim 500$ supernovae, the statistical errors are of the order of the systematics, which must be accounted for and estimated thoroughly.

The main systematics sources are listed below,and their taking into account in the Supernova Legacy Survey will be presented in section 2.2.2.

- Modeling uncertainties : the systematic uncertainties arising from SN spectro-photometric modeling mostly depend on the assumption underlying the model.
- Photometric calibration : this step is the major source of systematics in the cosmology measurement. As explained in section 2.1.4, the comparison of the SNe Ia fluxes measured at different redshift, and in different band-pass, requires the calibration of the different instruments onto a common standard photometric system.
- Contamination from non SNe Ia : for high-z SNe, type Ia can sometimes be contaminated by type Ib and Ic.
- Supernova evolution : supernovae luminosity or standardization relations could be evolving with time, inducing this way a bias in the cosmological parameters estimation. These possibility is checked in cosmology analysis by comparing supernovae characteristics (e.g. spectroscopic features) between low-z and high-z SN, or testing for redshift dependence of the *brighter-bluer* or *brighter-slower* laws.
- Malmquist Bias : this selection bias affects any flux limited surveys.
- Peculiar velocities : this systematic uncertainty arises from the local velocity fields and affect the nearby supernovae samples in the Hubble diagram, even so they are required to be in the Hubble flow.
- Gravitational Lensing : gravitational lensing induces an additional scatter to the Hubble diagram (see section 3.3.1).

## 2.2 The Supernova Legacy Survey

The Supernova Legacy Survey has been one of the major projects till date using SNe as a probe to pursue the nature of dark energy and measure precisely its equation of state parameter $w$.

SNLS experimental set-up is described in section 2.2.1. Recent results, based on the joint data analysis of SNLS first 3 years and SDSS supernovae survey data are presented in section 2.2.3, along with the complete analysis of SNLS 5 years data which is still underway.

### 2.2.1 Experimental set-up

#### The Canada France Hawaii Telescope Survey (CFHTLS)

Canada and France joined a large fraction of their dark and grey telescope time from mid-2003 to early 2009 for a large project, the Canada-France-Hawaii Telescope Legacy Survey (CFHTLS). The CFHTLS uses the MegaCam camera ([Bou03]) set up on the 3.6-m Canada-France Hawaii telescope (CFHT). It consists of three surveys :

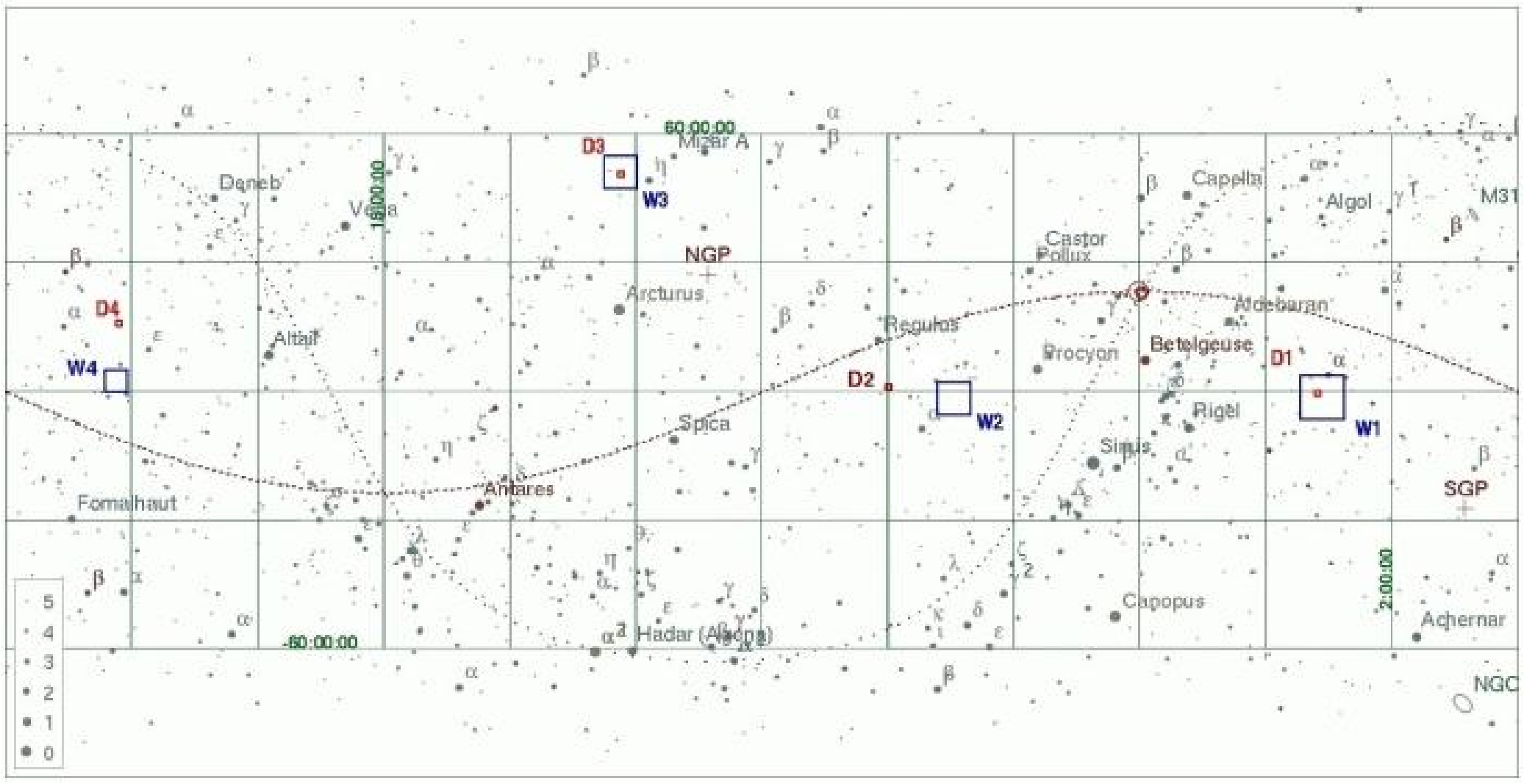

**Figure 2.6**: Positions of the Deep and the Wide fields on a full sky map. Image : CFHTLS website [Meg05]

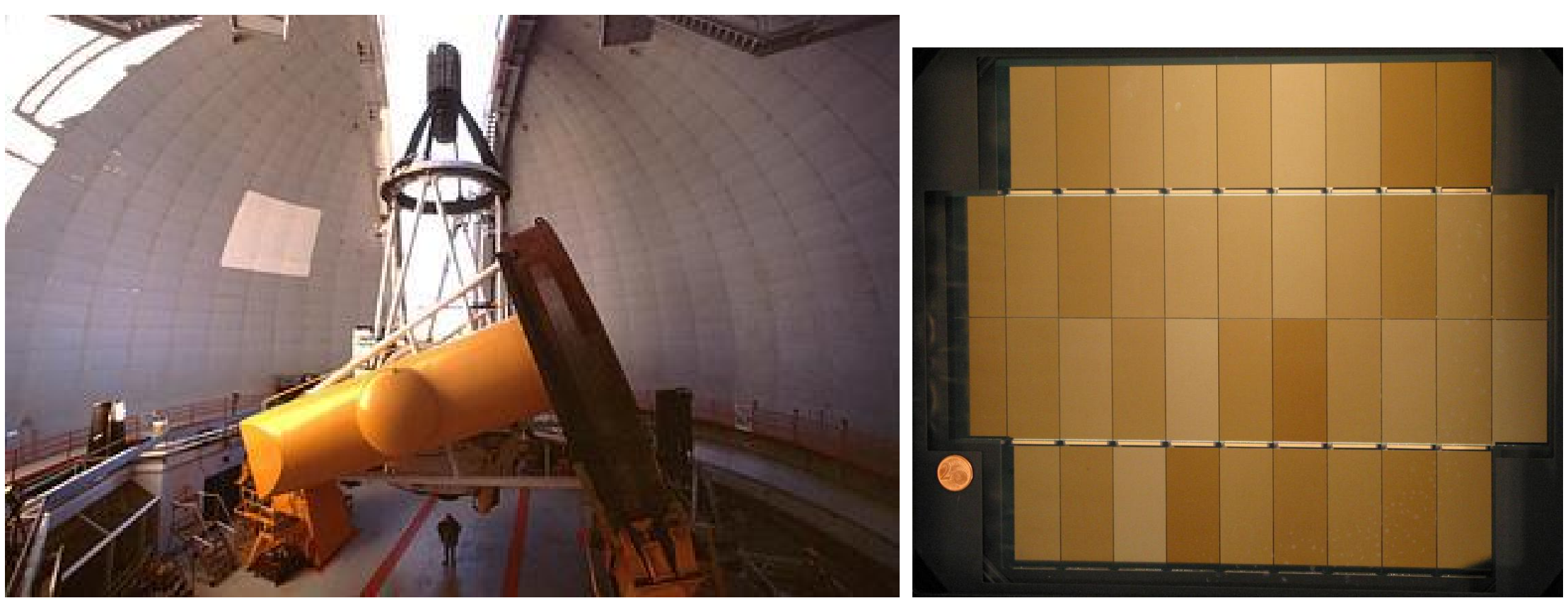

**Figure 2.7**: Left : The Canada-France-Hawaii Telescope. Right : The MegaCam CCD mosaic. Image : CFHTLS website [Meg05]

| Field | RA | Dec | E(B-V) | Other Observations | u | g | r | i | z |
|---|---|---|---|---|---|---|---|---|---|
| D1 | 02:26:00.00 | -04:30:00.0 | 0.027 | XMM Deep, VIMOS, SWIRE, GALEX | 33 | 33 | 66 | 132 | 66 |
| D2 | 10:00:28.60 | +02:12:21.0 | 0.018 | Cosmos/ACS, VIMOS, SIRTF, XMM | 33 | 33 | 66 | 132 | 66 |
| D3 | 14:19:28.01 | +52:40:41.0 | 0.010 | Groth strip, Deep2, ACS | 33 | 33 | 66 | 132 | 66 |
| D4 | 22:15:31.67 | -17:44:05.0 | 0.027 | XMM Deep | 33 | 33 | 66 | 132 | 66 |

**Table 2.1**: Summary of the four Deep fields. The last 5 columns show the amount of observing time in hours ensuring a deep coverage.

| | |
|---|---|
| CCD array | 4 rows × 9 columns |
| CCD's size | 2048 × 4612 pixels |
| Pixel size | 13.5 µ - 0.185" |
| Image size | 340 Megapixels |
| Required temperature | -120°C |
| Field of View | 0.96 deg × 0.94 deg |
| Readout Time | 35 seconds |
| Readout Noise | under 5 $e^-$ |
| Shutter Diameter | 1 meter |
| Minimum Exposure Time | 1 second |
| Filter Wheel | Provision for 8, 5 used |
| Filter Change Time | 2 min. |

**Table 2.2**: Summary of MegaCam specifications

- The Very Wide survey for the far ends of the solar system and our Galaxy survey. It covered 1300 square degree in the *gri* bands only.
- The Wide survey covers 155 square degree, spread over four patches : W1 to W4 in *ugriz*, reaching a limiting magnitude $i_{\mathrm{AB}} = 24.5$. The scientific goal is the study of the nearby universe and the large scale structures (lensing).
- The Deep survey was designed with four 1 square-degree fields D1 to D4 in *ugriz*.

The Wide and Deep fields are indicated on fig. 2.6 and the filter transmissions on fig.2.10. Post observational data processing took place at the Terapix (http://terapix.iap.fr/cplt/T0007/doc/T0007-doc.html) center. The Deep survey fields coverage are presented in table 2.1.

### MegaPrime

MegaPrime/MegaCam was the wide-field (1 square degree) optical imaging facility used at CFHT. The MegaCam camera specifications are summarized in table 2.2. The CFHT and MegaCam are presented on fig. 2.7.

### The Supernova Legacy Survey (SNLS)

SNLS is mainly a Canada-France collaboration with collaborators from the USA and the UK and some other European partners. SNLS used CFHTLS data for a 5 year period from 2003 to early 2009 with 450 nights of observation.

SNLS survey design has been specifically set up so as to control systematics ([Con11]):

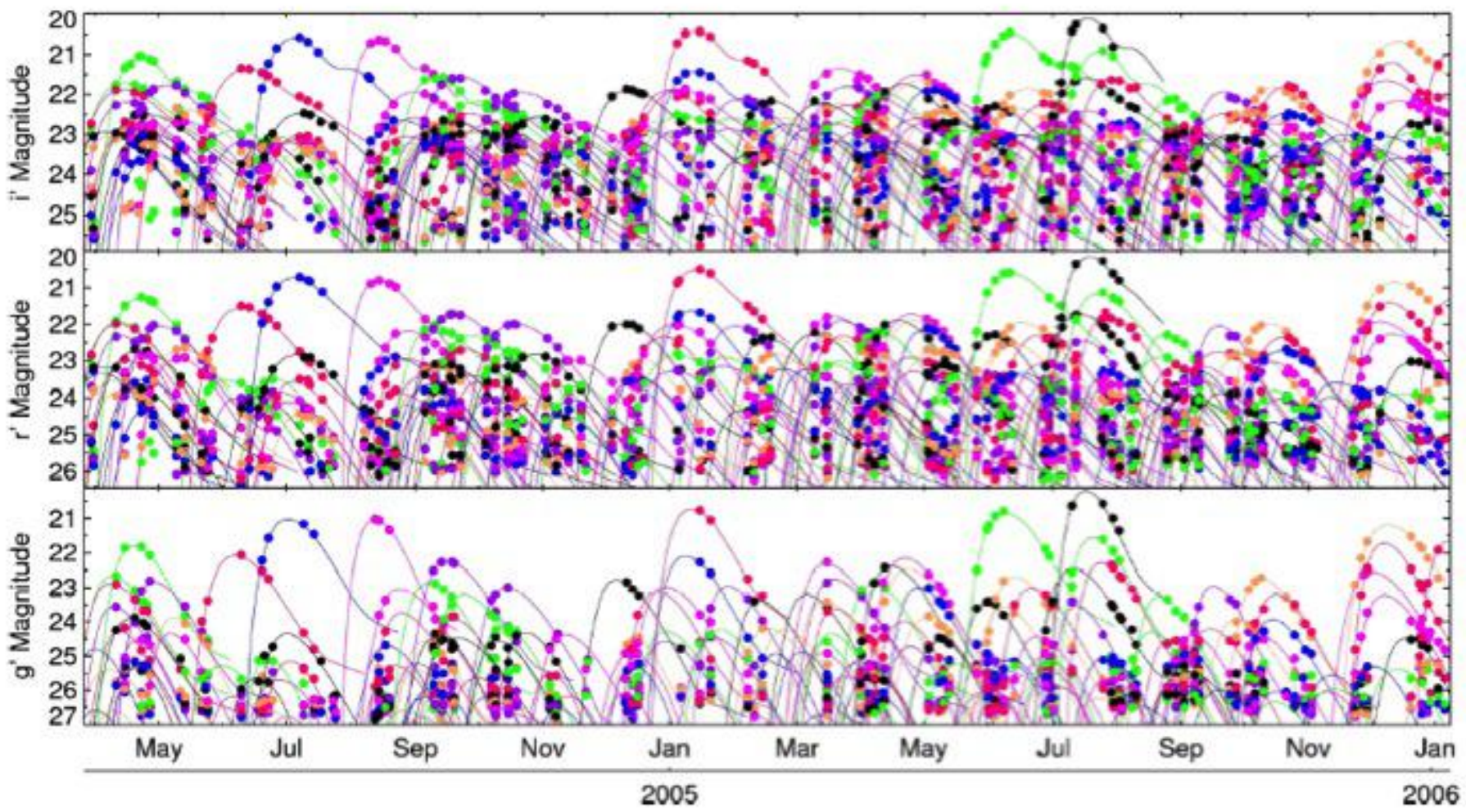

**Figure 2.8**: Rolling search : *gri* supernovae light curves as observed in spring 2004 to winter 2005. The newly detected SNe light-curves overlap with those of the already discovered SNe.

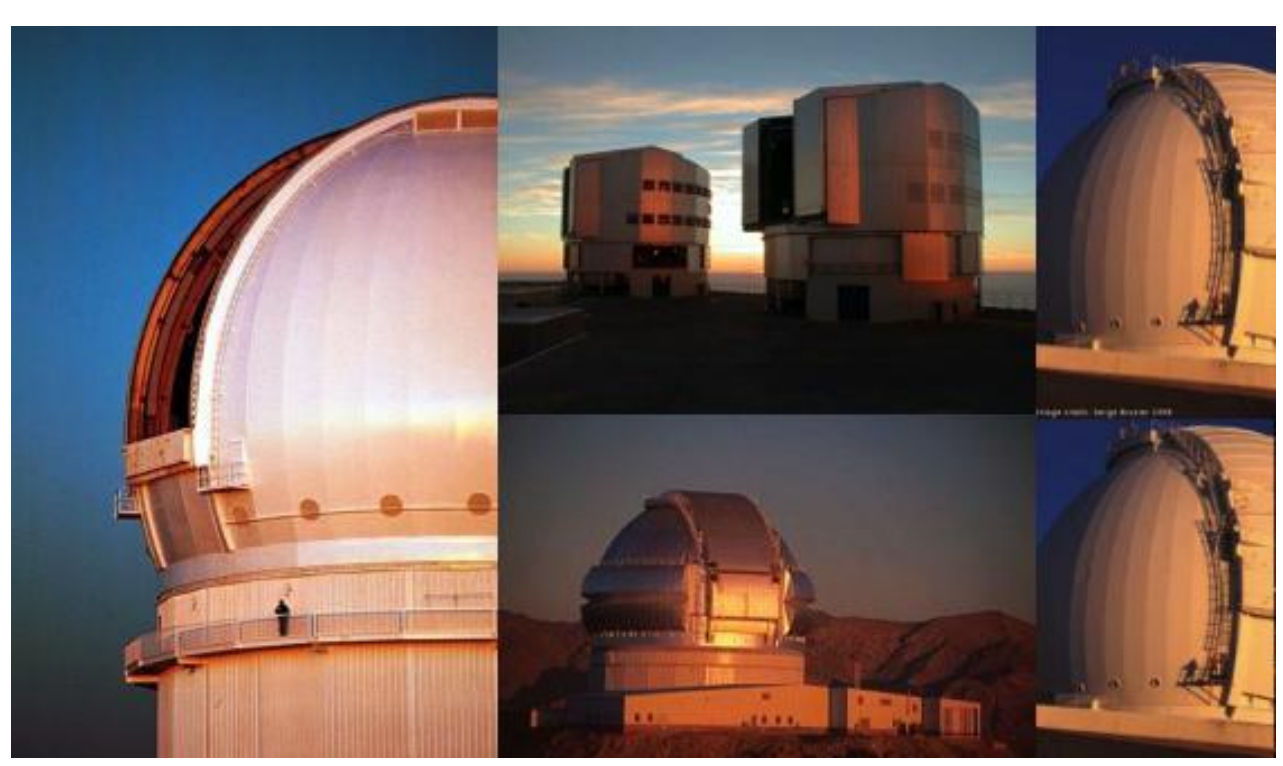

**Figure 2.9**: The spectroscopic follow up telescopes : UT1 Very Large Telescope, Chile (upper middle); Gemini-North (bottom middle) and Keck I and II telescopes (right and left sides), Mauna Kea, Hawaii. Image : CFHTLS

- SNe are both discovered and photometrically followed with one telescope, so that a special effort could be dedicated to the comprehension of the instrument calibration ([Reg09], [Bet13]). Within the survey, no intercalibrating between different follow-up telescope was needed.

- The survey four pass-bands *griz* permit to measure the SNe B rest-frame flux, as well as the B-V and U-B colors in a consistent way over the redshift range $z = 0.2 - 1$.

- The instrument design permit to observe SNe at a deep enough distance so that the Malmquist bias has a limited impact at $z \lesssim 0.6$ ([Per10]).

- SNLS is a rolling search, which consists in repeatedly observing the same part of the sky within a span of 2-3 days, thus continuously discovering new supernovae while monitoring the already discovered SNe (fig. 2.8). This method is in contrast to the old technique of observing the same field at different epochs spread over 2 months with the SN detection followed by scheduled photometric follow-up. This strategy ensures an adequate light-curve time sampling (every 3-4 days in the *gri* filters) as well as very good early-time coverage, which is crucial for the precise determination of each SN lightcurve parameters entering the cosmological fit.

- Each ∼500 SNe entering the cosmology analysis have been spectroscopically identified as a Ia, using data obtained on 8-m telescopes (fig. 2.9). About $\sim 200$ were identified at the Very Large Telescope (VLT), ESO, Chile, and $\sim 200$ at Gemini-North telescope (Hawaii). In the D3 field, $\sim 100$ were observed at the Keck Telescope (Hawaii). Spectra are published in [How05, Bro08, Ell08, Bal09]. The spectroscopic data also permitted to address the evolution question.

- Because of the survey duration, it is possible to construct very deep SN-free image stacks and get accurate colors of the host galaxies so as to study the relation between SN properties and host-galaxy environment to search for host-dependent systematic effects.

### 2.2.2 SNLS Data Analysis

We present here the main steps involved in the supernovae data processing for the SNLS 3-years (SNLS3) and five-years (SNLS5) data analysis.

The cosmology parameters are estimated through a a $\chi^2$ minimization between the distance indicator presented in section 2.1.3: $\mu_{\mathrm{SN}} = m_B^\star - \mathcal{M}_B + \alpha \times (s-1) - \beta \times c$, and the distance modulus $\mu(z_{\mathrm{SN}}; \Omega_M, w, \ldots)$ :

$$\chi^2 = \sum_i \frac{\left(\mu(z_{\mathrm{SN}i}; \Omega_M, w, \ldots) - \mu_{\mathrm{SN}i}\right)^2}{\sigma_i^2 + \sigma_{\mathrm{int}}^2} \tag{2.2}$$

For each supernova, the redshift $z$ and the parameters $m_B^\star$, $s$, and $c$ must be precisely measured. The cosmological parameters are then obtained from the fit process along with notably $\alpha$ and $\beta$, parametrizing the stretch-luminosity and color-luminosity relations. $\sigma_i$ accounts for the photometric uncertainty and $\sigma_{\mathrm{int}}$ is adjusted to account for the remaining intrinsic dispersion.

This diagonal expression for the $\chi^2$ is a simplification, as the covariance matrix $V$ is indeed not diagonal. It will be detailed below.

#### Detection

After a local pre-processing of the images performed on site using the Elixir pipeline, the $r$ and $i$ images are compared with reference earlier images to detect new transient events using an algorithm based on image subtraction. Two independent detection pipelines (one french and one

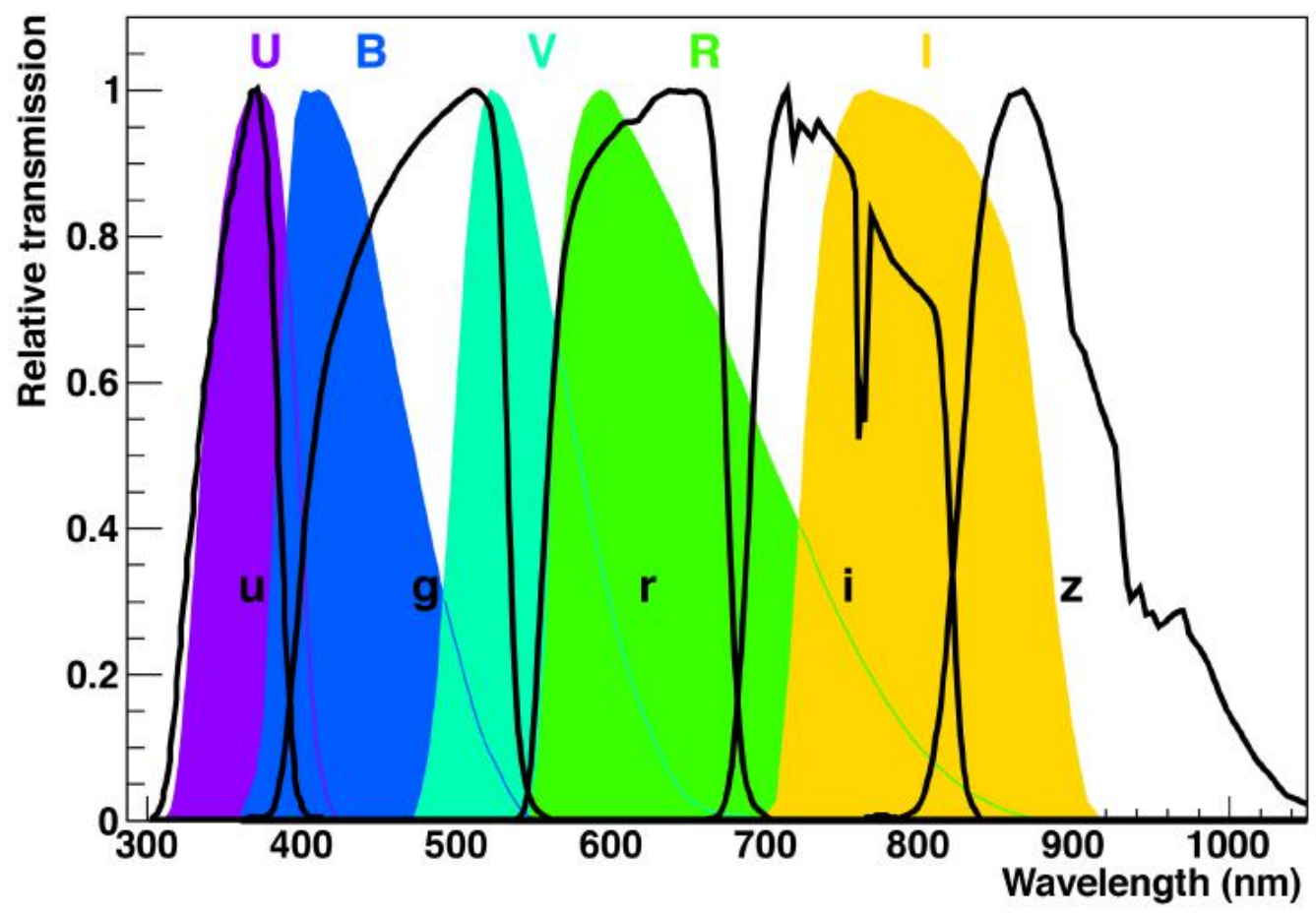

**Figure 2.10**: The transmission distribution as a function of the wavelength of the commonly used filter system: UBVRI and *ugriz*. Image : [Ast12]

Canadian), were performing the supernova detection. Although both pipelines uses a different algorithm (the Alard algorithm [Ala98] or a non-parametric approach), detections from the two different pipelines agree at a 90% level.

All identified transient objects cannot be forwarded for a spectroscopic follow up. Instead a primary photometric checking selects the most promising candidates.

The SNLS images were also submitted to a deferred processing independent of the SNLS real-time detection pipeline. This permits to complete the identification of the transients and perform a photometric analysis and identification of a "photometric" type Ia supernova sample. The photometric sample size is about 50% larger than the spectroscopically selected supernovae ([Baz11]).

### Photometry

The supernovae photometry is performed on the *griz* images. The supernova fluxes are estimated by fitting a point spread function (PSF) source, at the same sky position on the image series, on top of a a time-independent pixelized galaxy model. The local field (tertiaries) stars are measured through the same process (although setting the "galaxy model" to zero). What is measured is thus the ratio of the SN fluxes to those of the tertiary stars.

Two photometry algorithm, the "resampled simultaneous photometry" (RSP) and the "direct simultaneous photometry" (DSP) were implemented ([Ast13]). SNLS3 analysis is based on RSP : the images are first resampled to a common pixel grid to correct for the difference in telescope pointing, and fitted by a model combining a PSF centered at the SN position and a pixel grid (the underlying galaxy). The correlations introduced between neighboring pixels by resampling results in a sub-optimal estimation of the flux. SNLS5 analysis is based on DSP : to avoid the spatial correlations introduced by the images resampling, the DSP algorithm resamples instead the galaxy model onto the image pixel grid.

### Calibration

The purpose of photometric calibration is to relate the instrumental fluxes measured in the image pixels to the physical fluxes of the observed objects.

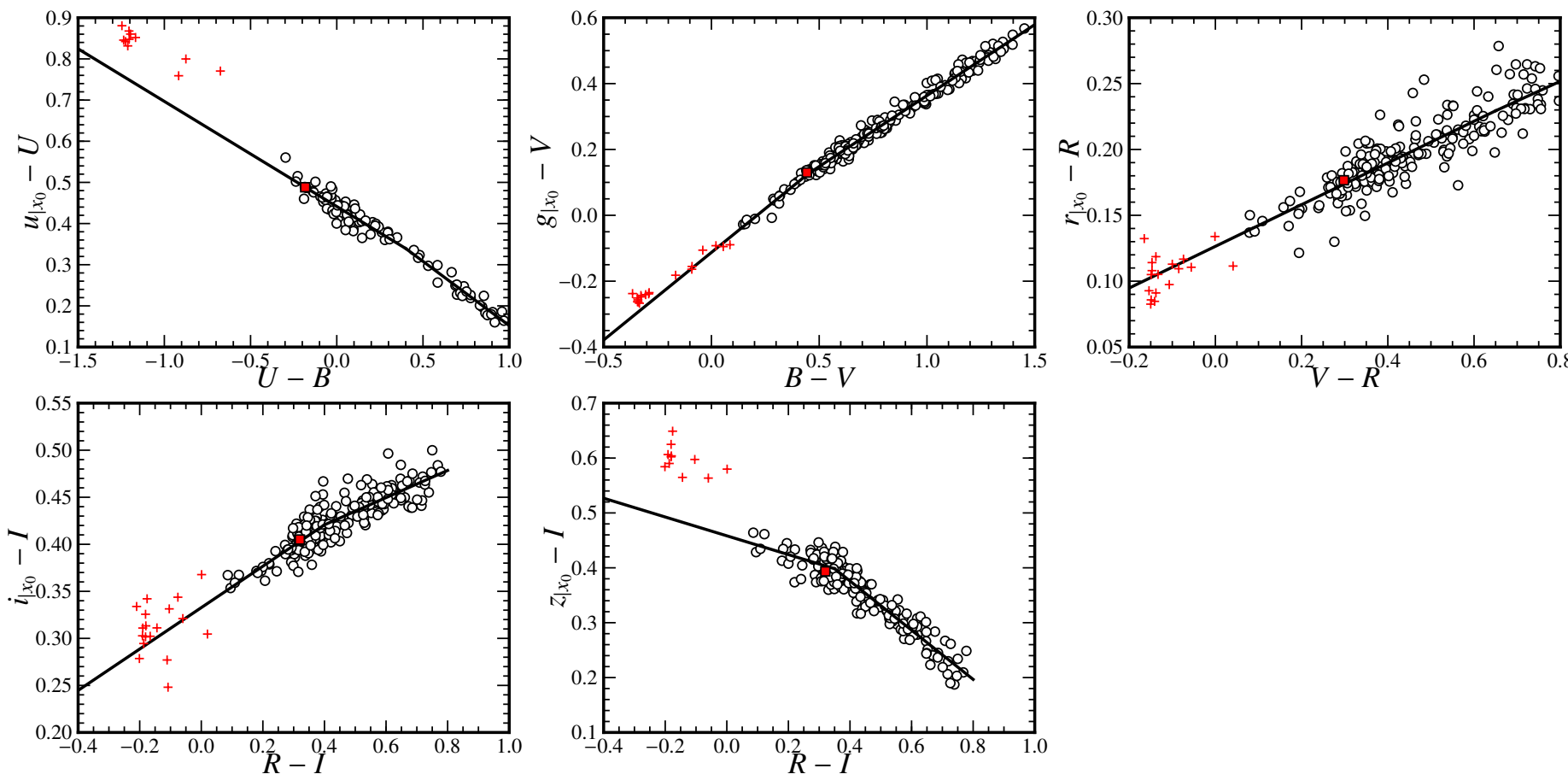

**Figure 2.11**: Color transformation between the Landolt system and the reference MegaCam system. Open black circles show the MegaCam and Landolt measurements of the Landolt secondary standards. The black line represents the average color transformation determined from the secondary stars measurements. The solid red square indicates the primary spectro-photometric standard star BD+17 4708 colors. Other primary standards from [Lan07] are displayed as red crosses. Image : [Bet13]

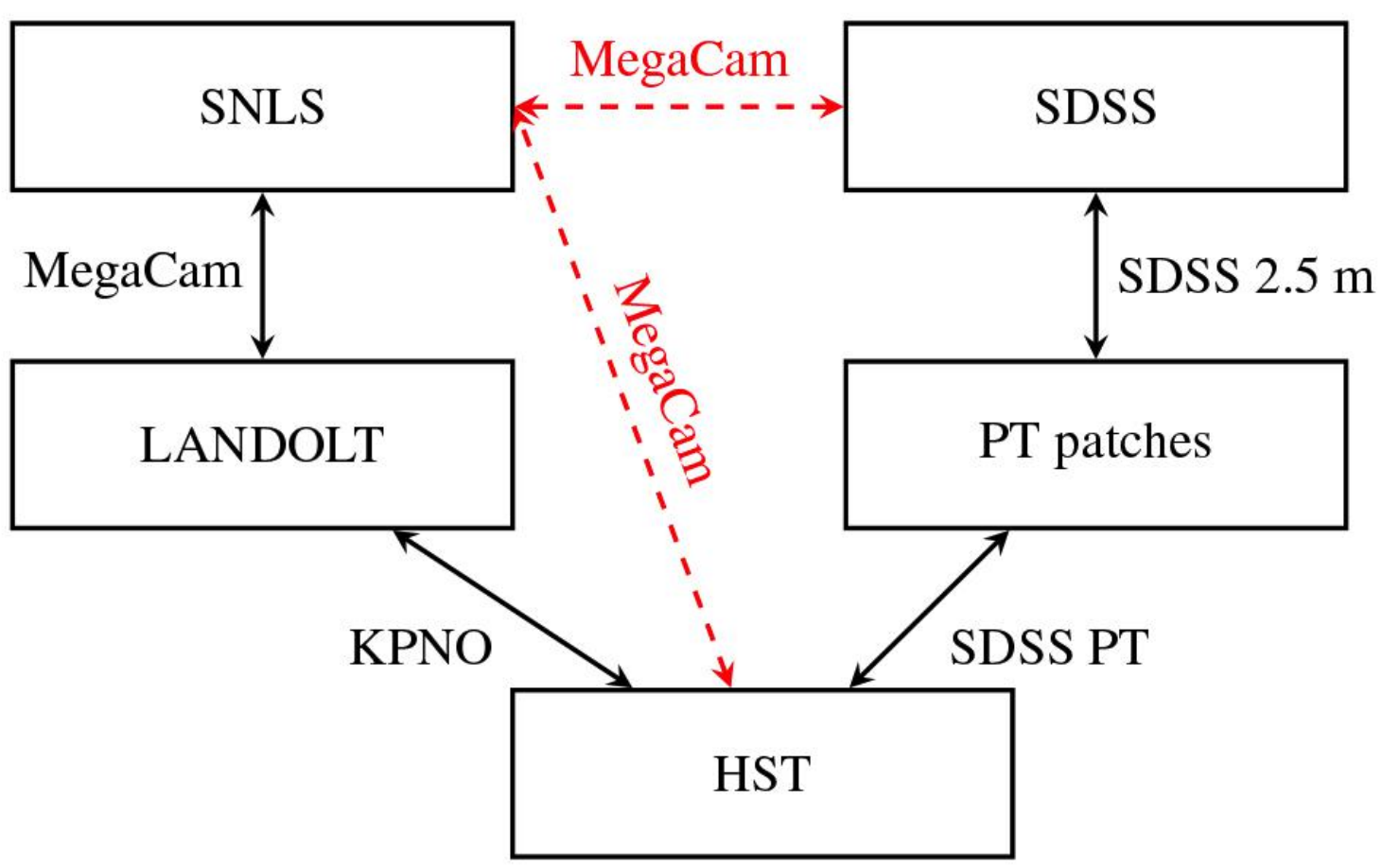

**Figure 2.12**: Flow chart of the calibration data process. The solid black line represents the SNLS3 calibration transfer scheme, while the red dotted line represents the new calibration path taken into account in JLA and SNLS5, and detailed in [Bet13]. Each box represents a set of standard stars established in that photometric system. The Instrument names indicate that both sets of stars on either side were measured using the same instrument hence making transfer of flux calibration data possible. Image : [Bet13]

The task is two-fold and thoroughly described in the two reference papers [Reg09] and [Bet13]. First, the effective instrument throughput variation in space and time must be characterized and homogenized to obtain a consistent data set. Secondly, calibration must provide a path to obtain from the homogenized flux measurement of an astrophysical source its physical flux or equivalently its calibrated broadband magnitudes $m$:

$$m_b \equiv -2.5 \log_{10} \left( \frac{\int \lambda T_b(\lambda) S(\lambda)}{\int \lambda T_b(\lambda) S_{\text{ref}}(\lambda)} \right)$$

$S$ is the spectral energy density (SED) of the object above the earth's atmosphere in units of $\text{erg}^{-1}\text{cm}^{-2}\text{Å}^{-1}$, $T_b$ is the instrument transmission in the filter $b$, and $S_{\text{ref}}$ is a reference spectrum.

The precise mapping of the spatial non-uniformities of the imager photometric response using dithered observations of dense stellar fields and the modeling of the effective pass-bands of the instrument as a function of the position on the focal plane are explained in [Reg09] and updated in [Bet13]. The effective pass-bands are obtained from laboratory or in situ transmission measurements of the CCD and filter pass-bands, combined with on site measurements of the mean atmospheric absorption.

The calibration process of the supernovae fluxes involved several steps and is described below.

The photometry of supernovae is made relative to the stars, referred to as tertiary standards, surrounding the supernovae in the science fields. This process delivers instrumental fluxes $\phi$ for the supernovae and the tertiary standards in consistent but arbitrary units.

To convert the instrumental flux $\phi_{SN}$ into a physical flux $S_{\text{SN}}$, one relies on the choice of a fundamental spectro-photometric standard star of known SED $S_S(\lambda)$, and defines the zero point $ZP_b$ in the chosen filter $b$ as :

$$\phi_b \, 10^{-0.4 \times ZP_b} \equiv \frac{\int \lambda T_B(\lambda) S_{SN}(\lambda}{\int \lambda T_B(\lambda) S_S(\lambda)} \tag{2.3}$$

In SNLS3 analysis, the chosen primary standard star was BD+17 4708, a HST spectro-photometric fundamental flux standard of precisely estimated SED [2]. The flux scale now relies on the theoretical modeling of the atmosphere of pure hydrogen WD observed by the HST, transferred to three solar analogs by HST observations with the STIS instrument ([Boh04, Boh10]). These 3 solar analogs were directly observed by MegaCam for the JLA calibration.

The $ZP$ estimation is performed following a path as short as possible. It formerly involved a secondary stars catalog, located in the science fields, of well known magnitudes, to assign magnitudes to the tertiary stars : the Landolt Catalog ([Lan92]) stars, with anchoring to the HST flux scale indirectly provided by [Lan07] observations of HST standards in the Landolt UBVRI photometric system, implying to deal with the transmission of this new set of filters. The BVRI system differs from MegaCam *griz* (see fig. 2.10), so that piece-wise linear color transformation laws had to be used between the two systems. The color transformation between the Landolt system and the reference MegaCam system are presented in figure 2.11. The use of solar analogs permits now to bypass this step.

The next step is to calibrate tertiary stars using the same photometry than the one performed on the calibration stars observed with MegaCam (aperture photometry), and the transformation laws from MegaCam physical fluxes to the chosen magnitude system. From there, the zero point for a given observation is estimated using:

$$ZP = \langle m_{\text{aper}} + 2.5 \log_{10} \phi_{\text{aper}} \rangle$$

[2] https://www.eso.org/sci/facilities/paranal/instruments/xshooter/tools/specphot_list.html

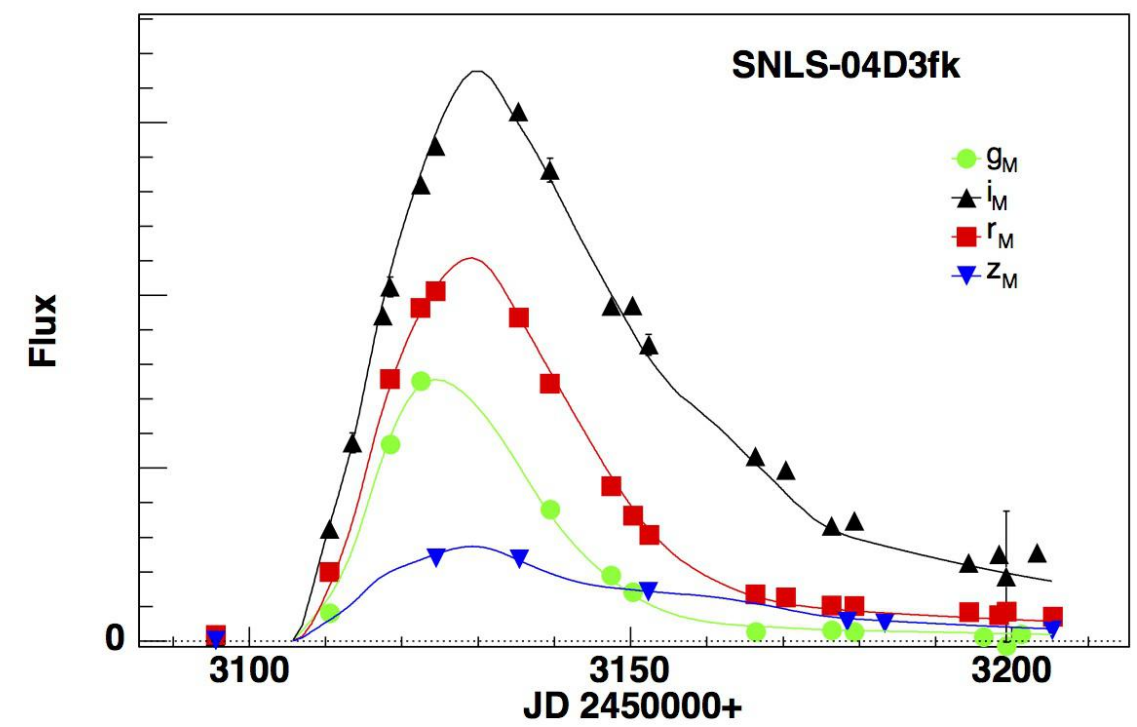

**Figure 2.13**: Typical SN Ia light curve in different bands. The points are observations done with the *griz* pass-band filters using the MegaCam (see §2.2) for the supernova 04D3fk, while the curves are the model fits from the SALT2 lightcurve fitting software.

where the mean is taken on the field tertiary stars.

The calibration of the tertiary stars is finally transferred to the SN using the ratio of their fluxes measured with the very same photometry, i.e. the PSF photometry.

For the SDSS survey, the calibration is performed by observing the HST spectro-photometric standard and the SDSS field with a dedicated photometric telescope (PT). The calibration process of both projects is presented on figure 2.12. The short-cut which permits to suppress the use of Landolt secondary stars for SNLS calibration is indicated.

### Spectro-photometric modeling: SALT2 and SIFTO

For each SN Ia, the peak flux $m_B^\star$ in the rest-frame B band, the color $c$ and the stretch factor $s$, are estimated by fitting the SN observed magnitudes with a spectro-photometric model (fig. 2.13). The "lightcurve fitters" used in SNLS analysis are SALT2 ([Guy07]), developed by the french team, and SIFTO ([Con08]), on the Canadian side.

SALT stands for "Spectral Adaptive Lightcurve Template". It aims at producing an average spectral sequence of SNe Ia with the addition of the fewest degrees of freedom for reciprocating the empirical observations of the SN Ia. The SN rest-frame flux is parametrized by three parameters, $X_0$, $X_1$, and $c$ :

$$\phi_{SNIa}(t,\lambda) = X_0 \times [M_0(t,\lambda) + X_1 M_1(t,\lambda)] \times e^{c \times CL(\lambda)} \tag{2.4}$$

where $t$ is the phase of the event ($t = 0$ corresponds to peak luminosity). $M_0(t,\lambda)$ is the average spectrum at a given phase $t$ and $X_0$ is the flux normalization in the rest-frame B-band, equivalent to $m_B^\star$. $M_1(t,\lambda)$ is the correction template to the average spectrum, it allows without forcing this behavior for the time-stretching of the light curve. $X_1$ is equivalent to the stretch parameter $s$, with roughly $X_1 \simeq s - 1$. $CL(\lambda)$ is the time dependent color law and $c$ is by construction the SN color as it corresponds to the difference in intensity of the rest frame B and the V band at peak luminosity.

$M_0(t,\lambda)$, $M_1(t,\lambda)$, $CL(\lambda)$ are computed during the model training based on spectro-photometric data from nearby supernovae, and also SNLS supernovae photometry. These latter were quite useful to sample the rest-frame U band. No additional assumptions is made on the wavelength dependency of the color law, and it does indeed differ from the Galactic dust absorption law $R_\lambda$ (e.g. [Car89]). The SALT2 model is thus at contrast with the MLCKS2 model ([Jha07]) which makes strong assumptions on CL wavelength dependency and $c$, and $\beta \equiv R_B$, possible values.

| Redshift | Number | Survey |
|---|---|---|
| $[0, 0.1]$ | 239 | SNLS |
| $[0.03, 0.4]$ | 374 | SDSS |
| $[0.1, 1.1]$ | 118 | Several Low-z samples |
| $[0.8, 1.3]$ | 9 | HST |

**Table 2.3**: Break up of the JLA SNe Ia sample

The SIFTO model uses a time-stretched template time series. The color law is implemented through linear relations relating the colors $c = B - V$, $U - B$ and the stretch $s$. The obtained results are similar to SALT2. SALT2 and SIFTO results are mitigated for the SNLS3 analysis, and SALT2 is used for SNLS5 analysis.

#### Supernovae host galaxies

The supernova host galaxy identification and photometry is performed on deep stacked images (see chapter 4). The host galaxy stellar mass is computed using PEGASE.2 ([Fio99]), so that each supernova can be assigned a host mass bin, below and above $M = 10^{10} M_{\odot}$. In the process of the cosmology fit, two different absolute magnitudes $\mathcal{M}_B$ and $\mathcal{M}_B + \Delta\mathcal{M}_B$ are fitted for two host-mass bins.

#### Systematic uncertainties

To include the identified systematics in the cosmology fit, eq. 2.2 can be generalized by replacing the diagonal contribution by a more general covariance matrix $V$ combining systematic and statistical errors. The $\chi^2$ equation then reads:

$$\chi^2 = \Delta\vec{\mu}^T V \Delta\vec{\mu} \tag{2.5}$$

Constraints form other cosmological probes can also be taken into account by adding supplementary terms in the $\chi^2$ expression.

The intrinsic dispersion $\sigma_{\text{int}}$ is fitted separately on each sample completing the SNLS supernovae in the Hubble diagram : nearby supernovae from nearby searches and intermediate redshift supernovae from the SDSS-II survey. The gravitational lensing supplementary dispersion is accounted for by introducing $\sigma_{\text{lensing}} \simeq 0.055 \times z$.

As all SNe share the same photometric model, the statistical part of the covariance matrix $V_{\text{stat}}$ dealing with $m^{\star}_B, c, s$ measurements is non-diagonal. The systematic part $V_{\text{syst}}$ comprises the calibration uncertainties, which affect the photometric measurement of the supernovae, but also the spectro-photometric model trained with these data. This is to furtherly reduced the contribution of calibration to the systematic uncertainties, and be able to fully exploit the data from SNLS and SDSS survey, that [Bet13] established a common calibration at a level of 0.4% in $gri$.

### 2.2.3 Measuring the dark energy equation : JLA and SNLS5

The first set of distance measurements data was released from SNLS $1^{st}$ year's observation data [Ast06]. With 71 high redshift SNe Ia, together with the constraints from the SDSS baryon acoustic oscillations ([Eis05]), the equation of state parameter $w$ was measured at a precision of ∼10%.

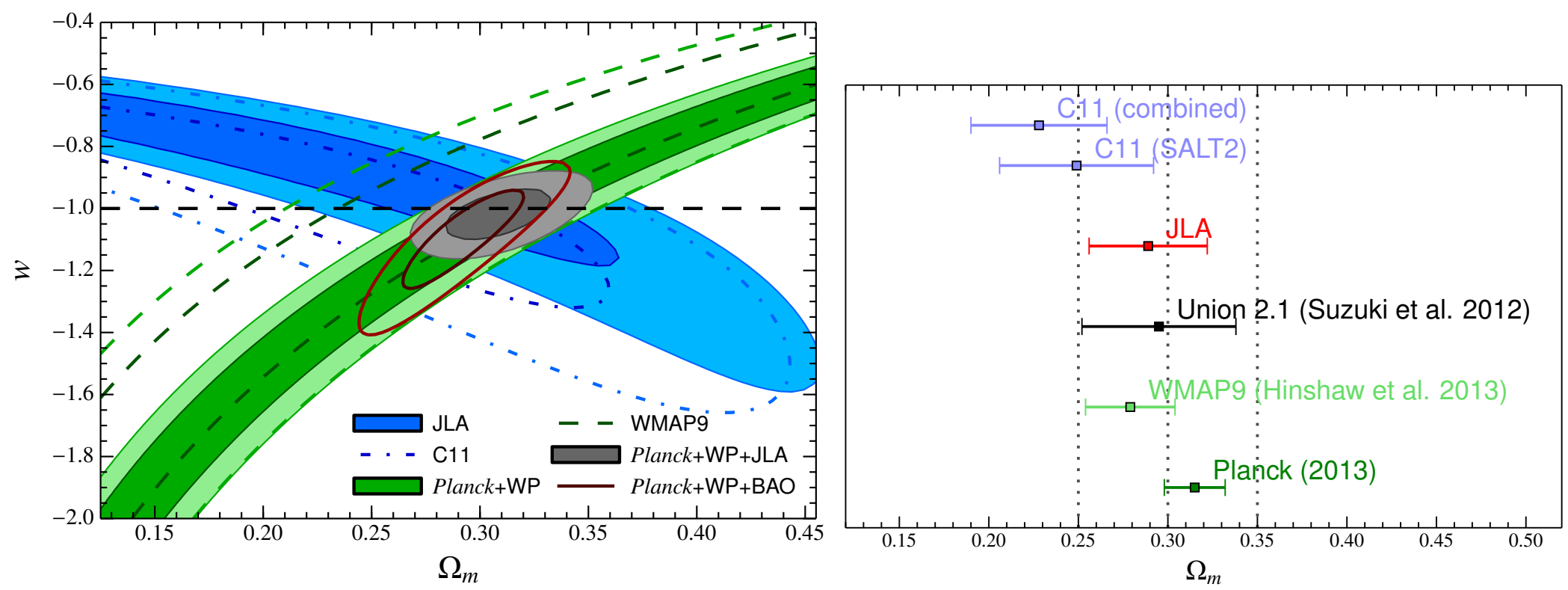

**Figure 2.14**: Left: Phase plot of dark energy equation of state parameter $w$ vs matter density $\Omega_m$. Right: Comparison of various measurements of $\Omega_m$ from different collaborations. The JLA and the Planck measurements are in good agreement with Planck (2013) measurements. They are also in good agreement with the final Planck (2015) data release [Ade15]. Image : [Bet14]

### SNLS3 (2011)

The SNLS 3 years data analysis ([Con11]) confirmed this measurement. SNLS3 consisted of 242 SNe Ia at $0.3 < z < 1.1$. As described in the previous section, two independent analysis were carried out by the Canadian and the french counterpart for the SN photometry, photometric calibration and light curve fitting. Precise photometric calibration and improved SNe Ia light curve modeling were done, models were trained using the SNLS data. Inclusion of a host mass term$\Delta\mathcal{M}_B$ was done and the related systematics were added to the cosmological fits.

### The Joint Light Curve Analysis (2014)

An intermediate update has been presented with the Joint Light Curve Analysis (JLA) which was a joint SNLS3 and SDSS-II data analysis ([Bet14]). The JLA sample consists of 740 spectroscopically confirmed type Ia SNe with high quality light curves. The sample comprised low redshift SNe Ia with $z < 0.1$ ; mid-range SNe Ia from the SDSS sample at $0.05 < z < 0.4$ ; higher redshift SNLS SNe Ia at $z < 1$ ; a few SNe observed at z $\sim$ 1 with the HST. The inclusion of the SDSS sample (374 SN Ia) was fully exploited thanks to the inter-calibration of the SNLS and the SDSS data (fig. 2.5).

The findings of the JLA are consistent with the CMB [Pla14] observations (e.g. matter density parameter $\Omega_m$ and dark energy equation of state parameter $w$). When combining SNe constraints with the the Planck measurement [Pla14] of the CMB temperature fluctuations and the WMAP measurement of the large-scale fluctuations of the CMB polarization [Ben13], the precision obtained on $w$ parameter is now less than 6% (see fig. 2.14).

The previous disagreement on $\Omega_m$ at a $\sim 2\sigma$ level reported by [Con11] was greatly reduced with the improvement in the calibration accuracy using the joint recalibration methods (see fig. 2.14).

### SNLS5 (2014-)

The SNLS 5 years are based on $\sim$ 400 spectroscopically confirmed and $\sim$ 300 photometrically confirmed SNe Ia covering the full period of observation between 2003-2008. The SNLS5 analysis is described in [EH14]. With respect to JLA analysis, SNLS5 is now relying on the DSP photometry algorithm, and is using a fully re-trained SALT2 model.

We will be using the still cosmologically blinded[3] SNLS5 sample analysis in these thesis. The SNe are selected as follows. To ensure that the supernovae parameters lie within the bulk of the distribution, the condition $-3 < X_1 < 3$ and $-0.3 < c < 0.3$ are set. Poorly sampled supernovae are excluded by requiring that it be observed before $t_{\max}$, and that $\sigma(t_{\max}) < 2$ and $\sigma(X_1) < 1$. A cut on the light-curve $\chi^2$ fit is also applied. In addition, all light curves that enter the Hubble diagram have been visually inspected. This resulted in the exclusion of a few SDSS supernovae. The contributions of all sample included in SNLS5 analysis are presented in table 2.4.

| mean redshift | Number | Survey |
|---|---|---|
| 0.34 | 960 | ALL |
| 0.02 - 0.04 | 212 | Several Low-z samples |
| 0.2 | 351 | SDSS |
| 0.62 | 389 | SNLS |
| 1. | 8 | HST |

**Table 2.4**: Break up of the SNLS5 SNe Ia sample

[3]The values of the supernovae fluxes have been artificially changed so as to alter the cosmology they describe.

# Chapter 3

# Gravitational Lensing

Gravitational lensing is a consequence of Einstein's General Relativity[1]: the gravitational field of a massive object causes light rays passing close to that object to be bent. As the gravitational deviation of light depends solely on the lens mass distribution, this phenomenon permits to study the dark matter distribution and growth history, from haloes to large scale structures, as well as probing the universe overall geometry and expansion history.

This chapter is essentially based on [Nar96], [Men15], and [Har16b] notes.

## 3.1 Basics of gravitational lensing

We will present here the basics of gravitational lensing.
We will be working under three assumptions :

1. the lens mass only weakly perturbs the FRW metric describing the homogeneous universe : the lens mass Newtonian potential $\Phi$ is assumed to be small, $\Phi \ll c^2$. As a consequence, it is possible to consider that the individual lenses are embedded in a locally flat (Minkowskian) space-time. For a galaxy cluster where galaxies peculiar velocities are about $v \sim 10^3$km.s$^{-1}$, the condition is met, with $\Phi/c^2 \sim 10^{-5}$. Under this assumption, the field equations of General Relativity can be linearised.

2. The lens velocity, and the velocity of its constituents, are slow ($v \ll c$), which is again the case for galaxies and clusters.

3. Thin lens approximation: the lens physical size $L$ is much smaller than the relative distances between the source, the lens and the observer. This is again the case in this thesis, with distances to galaxies and galaxy clusters of interest being larger that their typical sizes : $L \ll c/H_0$. The path of light propagation can then be broadly broken up into 3 parts. From the source to the lens, and the lens to the observer, light travels through an unperturbed space-time described by the Friedman metric. In the lens vicinity, described as locally Minkowskian, the ray, corresponding to an impact parameter $\vec{\xi}$, can be approximated as a straight line and its change of direction $\vec{\hat{\alpha}}$ can be considered to be instantaneous.

   Under this condition, the lens mass is assumed to lie in a plane distribution (see figure 3.1) and is described by it's surface density:

$$\Sigma(\vec{\xi}) = \int \rho(\vec{\xi}, z)\, dz \tag{3.1}$$

   where $\vec{\xi}$ is a two dimensional vector in the lens plane and $z$ is the line-of-sight coordinate.

[1]Newton pointed out about the possibility of deflection of light rays. Later on Eddington calculated it to be roughly twice the deflection of Newtonian formalism, the missing factor of 2 from Newtonian calculations was due to the time dilation factor which is present in G.R. See for example: [Ric09] or [Sch92].

### 3.1.1 The deflection angle

The deflection angle is calculated by the integral along the line of sight of the potential gradient along the normal perpendicular to the path of light propagation :

$$\widehat{\vec{\alpha}} = \frac{2}{c^2} \int \overrightarrow{\nabla_\perp} \Phi dl \tag{3.2}$$

Because of assumption (1), on expects the deflection angle to be small. So that the gravitational potential along the deflected trajectory can be approximated by the potential along the undeflected trajectory (Born approximation).

For a point mass $M$ and an impact parameter $\vec{\xi}$, the deflection reads:

$$\widehat{\vec{\alpha}} = \frac{4GM}{c^2 \xi} \frac{\vec{\xi}}{\xi} \tag{3.3}$$

The Schwarzchild radius:

$$R_S = \frac{2GM}{c^2} \tag{3.4}$$

sets the scale at which assumption (1) is valid: $\xi \gg R_S$. For $M = 10^{11} M_\odot$, $R_S \simeq 10^{-2}$ pc.

In the case of the thin lens approximation relation, the deflection angle is obtained by summing all the contributions from all the mass elements:

$$\overrightarrow{\widehat{\alpha}}(\vec{\xi}) = \frac{4G}{c^2} \int \frac{(\vec{\xi} - \vec{\xi'})\Sigma(\vec{\xi'})}{\left|\vec{\xi} - \vec{\xi'}\right|^2} d^2\vec{\xi'} \tag{3.5}$$

The integral is over $\vec{\xi'}$ in the lens mass distribution plane, and the origin of the coordinates system from which $\vec{\xi}$ and $\vec{\xi'}$ vectors are taken can be chosen anywhere in this plane. For an axially symmetric lens, the light rays, deflected and undeflected, lay in the same plane than the source-lens-observer and the deflection angle becomes one-dimensional (see figure 3.1):

$$\alpha(\xi) = \frac{4GM(\xi)}{c^2 \xi} \tag{3.6}$$

$\xi$ is now the distance from the center of the 2-dimensional mass profile. The deflection is determined by the mass enclosed within the radius $\xi$, $M(\xi)$ :

$$M(\xi) = 2\pi \int_0^\xi \Sigma(\xi')\xi' \, d\xi' \tag{3.7}$$

where, for the integration in the lens plane, the lens symmetry center is chosen as origin of the coordinates system.

### 3.1.2 The lens equation

The lensing situation is described on figure 3.1. The reference axis originates from the observer O and is orthogonal to the lens mass plane. The lens equation relates : the angle $\vec{\beta}$ between the source direction S and the reference axis; the deflection angle $\vec{\hat{\alpha}}$; the angular separation $\vec{\theta}$ between the source image and the reference axis; the distance $D_S$ between the source plane and the observer; the distance $D_L$ between the lens plane and the observer; the distance $D_{LS}$ between the source plane and the lens plane. The derivation is obtained explicitly on figure 3.2 and yields:

$$\vec{\beta} D_S = \vec{\theta} D_S - \vec{\hat{\alpha}} D_{LS} \tag{3.8}$$

The distances are angular distances as defined in 1.18 so that $\vec{\eta} = \vec{\beta} D_S$, $\vec{\xi} = \vec{\theta} D_L$. Note that in general, $D_L + D_{LS} \neq D_S$. However, in a flat universe, $D_S(1 + z_s) = D_L(1 + z_L) + D_{LS}(1 + z_s)$.

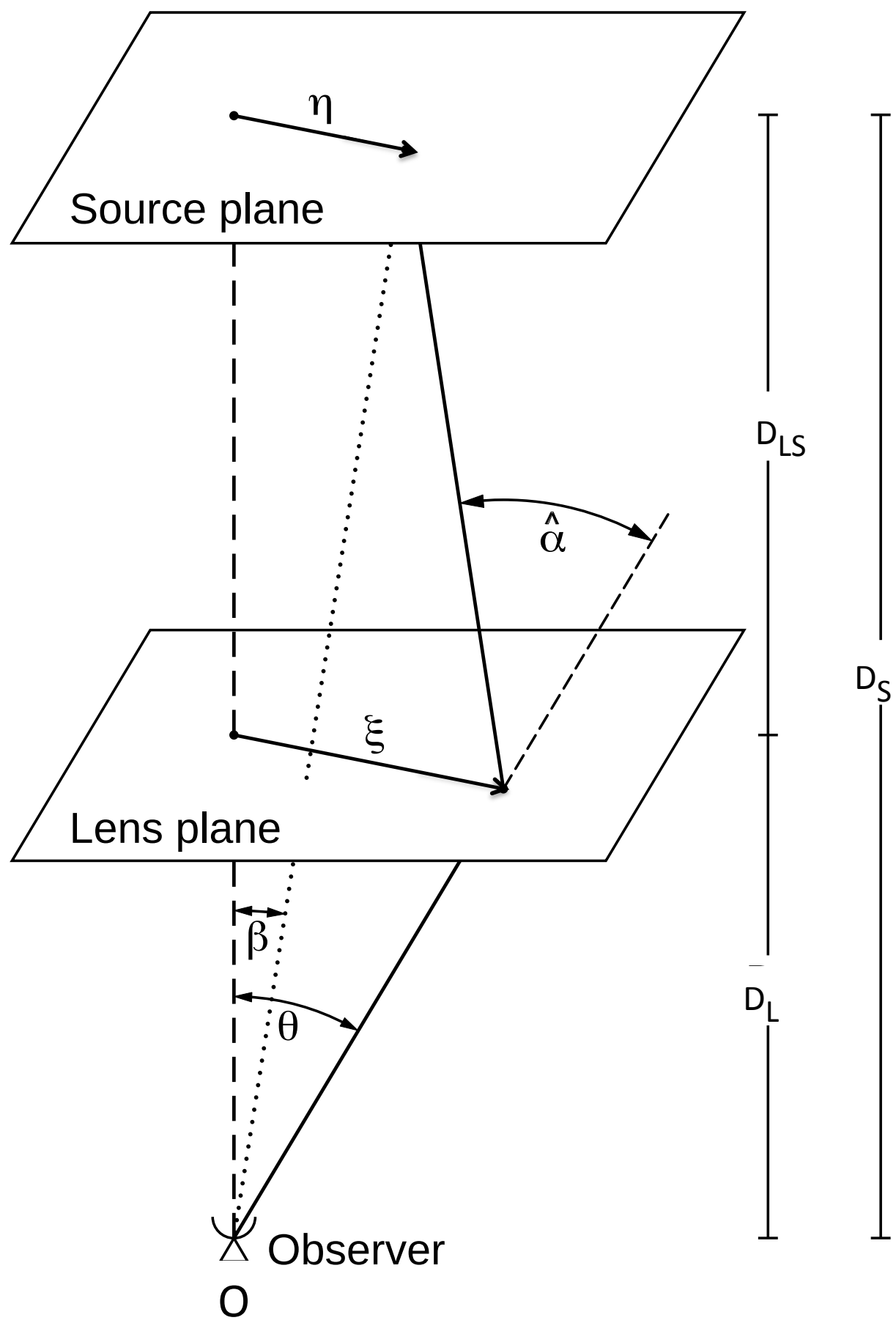

**Figure 3.1**: Schematic diagram of gravitational lensing. The reference axis originates from the observer and is orthogonal to the lens mass plane. The position of the lens mass distribution symmetry center L, should it have any, is not specified : it lies somewhere in the lens plane. The light ray originates from a point S which position is set by $\vec{\eta}$ in the source plane and travels unperturbed until it hits the lens plane, at the position given by $\vec{\xi}$. It then instantaneously changes direction (set by $\widehat{\alpha}$). In the special case where the lens is spherically symmetric with respect to a symmetry center $L$, the lens plane can be taken to be orthogonal to the direction $(LO)$. The reference axis can be taken as the optical axis $(LO)$, and $\xi$ is then the impact parameter with respect to $L$. This is not possible when considering multiple lenses. Image: adapted from [Bar01]

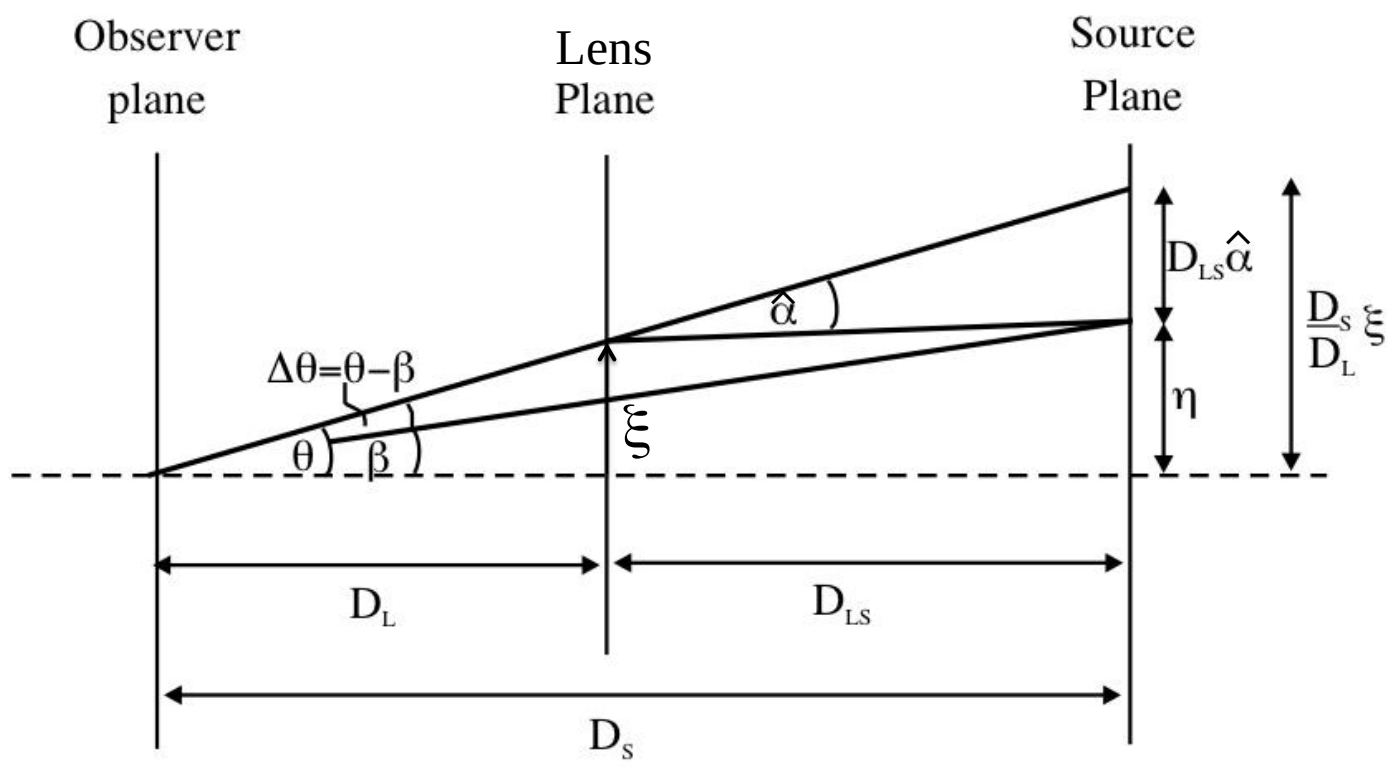

**Figure 3.2**: Derivation of the lensing equation. The rationale is done between the source plane and the lens plane using Thales theorem: $(D_L/D_S)\xi = \eta + D_{LS}\widehat{\alpha}$. Image: adapted from [Ama15]

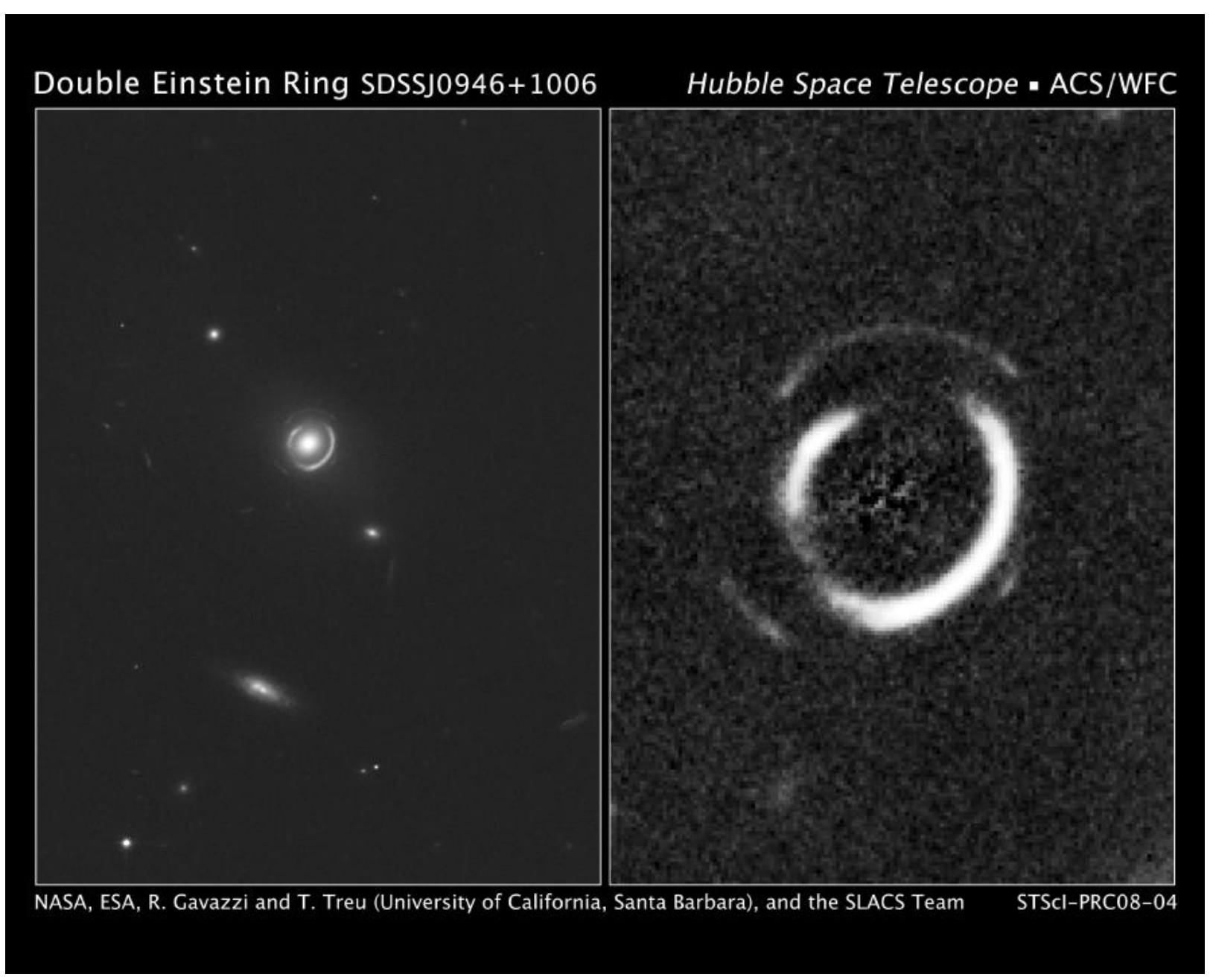

**Figure 3.3**: Einstein Ring Image. The point-like source and the point mass lens are nearly perfectly aligned with the observer. Credit : NASA/HST

Defining the reduced deflection angle: $\vec{\alpha}(\vec{\theta}) \equiv (D_{LS}/D_S)\,\vec{\hat{\alpha}}$, the lens equation can be written as:

$$\vec{\beta} = \vec{\theta} - \frac{D_{LS}}{D_S}\,\vec{\hat{\alpha}} = \vec{\theta} - \vec{\alpha}(\vec{\theta}) \quad \text{(lens equation)} \tag{3.9}$$

This equation in $\vec{\theta}$ may be non-linear and have multiple solutions, so that a given single point source at $\vec{\beta}$ has multiple images $\vec{\theta}$.

### 3.1.3 The Einstein radius and the critical surface density

In the special case of a point mass $M$ lying exactly behind the lens ($\beta = 0$), the image is a ring (see figure 3.3) which angular size is the Einstein radius:

$$\theta_E \equiv \sqrt{\frac{4GM}{c^2}\frac{D_{LS}}{D_S D_L}} \tag{3.10}$$

The Einstein radius sets the characteristic angular scales of lensed images : when multiple images are produced, they are separated by $\sim 2\theta_E$. For a typical galaxy mass and effective distance $D = D_L D_S / D_{LS}$, it is of order :

$$\theta_E \simeq 1" \left(\frac{M}{10^{11} M_\odot}\right)^{1/2} \left(\frac{D}{1\,\mathrm{Gpc}}\right)^{-1/2}$$

$\Sigma_{\rm cr}$ is the critical density[2]:

$$\Sigma_{\rm cr} \equiv \frac{c^2}{4\pi G}\frac{D_S}{D_L D_{LS}} \tag{3.11}$$

Note that the critical density is purely a geometric quantity, and corresponds to the mean surface mass density inside the Einstein radius. Numerically, defining the characteristic distance $D' = (D_L D_{LS})/D_S$:

$$\Sigma_{\rm cr} \simeq 0.35\,\mathrm{g.cm^{-2}} \left(\frac{D'}{1\,\mathrm{Gpc}}\right)$$

The critical density or equivalently the Einstein radius gives roughly the boundary beyond which multiple imaging can occur, and distinguishes between the strong and the weak lensing regime.

### 3.1.4 Magnification

Because of Liouville theorem[3], gravitational lensing preserves surface brightness but changes the apparent solid angle of the source. The flux of a source of surface brightness $I_\nu$ and solid angle area $\delta\Omega$ is $F_\nu = I_\nu\,\delta\Omega$. The lensing alters the solid angle area to $\delta\Omega_{\rm lensed}$ , so that the lensed flux $F_{\nu,\,\rm lensed} = I_\nu\,\delta\Omega_{\rm lensed}$ is modified by the magnification factor:

$$\mu \equiv \frac{\delta\Omega_{\rm lensed}}{\delta\Omega} \tag{3.12}$$

The image distortion can be described by the lens mapping Jacobian matrix $\mathcal{A}$:

$$\mathcal{A} \equiv \frac{\partial\vec{\beta}}{\partial\vec{\theta}} = \mathbb{1} - \frac{\partial\vec{\alpha}}{\partial\vec{\theta}} \tag{3.13}$$

so that:

$$\mu \equiv \frac{1}{\det\mathcal{A}} \tag{3.14}$$

[2] For a lens with constant surface density $\Sigma = \Sigma_{cr}$, $\beta = 0$ for any $\theta$, so that the lens focuses perfectly.

[3] See e.g. [Ser10] for a demonstration.

It is useful to define the effective lensing potential:

$$\psi(\vec{\theta}) = \frac{D_{LS}}{D_L D_S}\frac{2}{c^2}\int \Phi(D_L\vec{\theta}, z)dz$$

Its gradient yields the deflection angle $\vec{\alpha}$:

$$\vec{\alpha}(\vec{\theta}) = \nabla_{\vec{\theta}}\psi$$

and its Laplacian is proportional to the surface-mass density at $\vec{\xi} = D_L\vec{\theta}$ (using Poisson equation for $\Phi$):

$$\nabla^2_{\vec{\theta}}\psi = 2\frac{\Sigma(D_L\vec{\theta})}{\Sigma_{\rm cr}} \equiv 2\,\kappa(\vec{\theta}) \tag{3.15}$$

and introducing the convergence $\kappa$. We may then write the Jacobian matrix $\mathcal{A}$ in term of the effective lensing potential Hessian :

$$\mathcal{A} = \mathbb{1} - \frac{\partial^2\psi(\vec{\theta})}{\partial^2\vec{\theta}} = \begin{pmatrix} 1-\kappa-\gamma_1 & -\gamma_2 \\ -\gamma_2 & 1-\kappa+\gamma_1 \end{pmatrix} \tag{3.16}$$

where $\kappa = \psi_{11} + \psi_{22}$ according to the definition 3.15, and $\gamma_1 = (\psi_{11} - \psi_{22})/2$, $\gamma_2 = \psi_{12}$ defines the shear, with $\widetilde{\gamma} = \gamma_1 + i\gamma_2 = \gamma e^{2i\phi}$. In the matrix $\mathcal{A}$, the convergence $(1-\kappa)\mathbb{1}$ term produces an isotropic image expansion or contraction, while the shear rotation matrix stretches the intrinsic shape of the source. For an initially circular source of apparent size $r$, the image is elliptical with a an ellipticity $(a-b)/(a+b) = \gamma/(1-\kappa) = g$: $g$ is the reduced shear. With these notations, the magnification $\mu$ now reads :

$$\mu = [(1-\kappa)^2 - \gamma^2]^{-1} \tag{3.17}$$

In the weak lensing regime where $\kappa \ll 1$ and $\gamma \ll 1$, the magnification is simply related to the convergence by:

$$\mu \simeq 1 + 2\kappa \simeq 1 + 2\frac{\Sigma(\vec{\xi})}{\Sigma_{\rm cr}} \qquad \text{(weak regime)}$$

### 3.1.5 Types of Lensing

One usually distinguishes three regimes : strong lensing, of which microlensing is a special case, and weak lensing. The distinction between these regimes depends on the steepness of the gravitational potential along the light ray path, and the source-lens-observer geometry.

**Strong Lensing:** strong lensing produces multiple images or highly distorted images (such as arcs) and occurs when the lens system is critical, i.e if $\kappa \geq 1$ (or the source is within an Einstein radius off the optical axis). As a consequence it is very sensitive to the inner lens density profile and its substructure and can be used to compute a detailed model of the mass distribution of the lens.

Microlensing can be regarded as a special case of strong lensing. It was noted by Einstein in 1936 ([Ein36]) that a star could gravitationnaly lens another one that would appear inside its Einstein radius. Given the number of (lensing) stars per unit area of the sky, the probability of observing such an event would be of 1:1 million of observed (source) stars. As the typical image separation distance is $\Delta\theta \approx 2\theta_E \sim 1$ mas, far below the observations limiting resolution, only the magnification of the source star can be observed, providing the lens is moving transversely to the line of sight. The background star is progressively magnified as the lens passes in front of it and enters gradually the Einstein radius, exhibiting a characteristic light curve, symmetric in time, and achromatic. Microlensing is suited for looking for planets ([Gou06]) or the once candidate for Byronic dark matter, the massive halo compact objects (MACHOS), from $10^{-3}$ to $10^3 M_\odot$

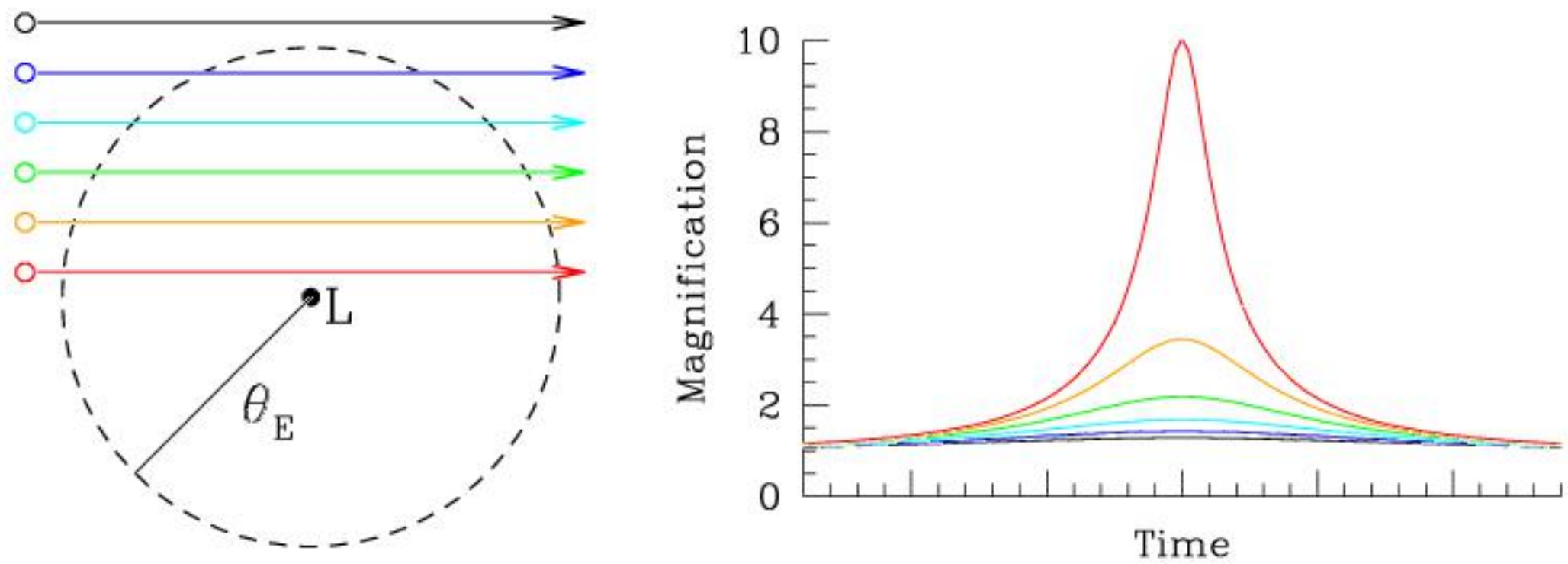

**Figure 3.4**: Microlensing light curves for a point-like source and a point-mass lens (right). The curves are color coded to the corresponding minimum impact parameters (left) ($y_0 = 0.1$ (top, red), 0.3,...1.1 (bottom, black)) given by their trajectories across the Einstein ring ($\theta_E$). Image : Professor Penny D Sackett & [Koc06]

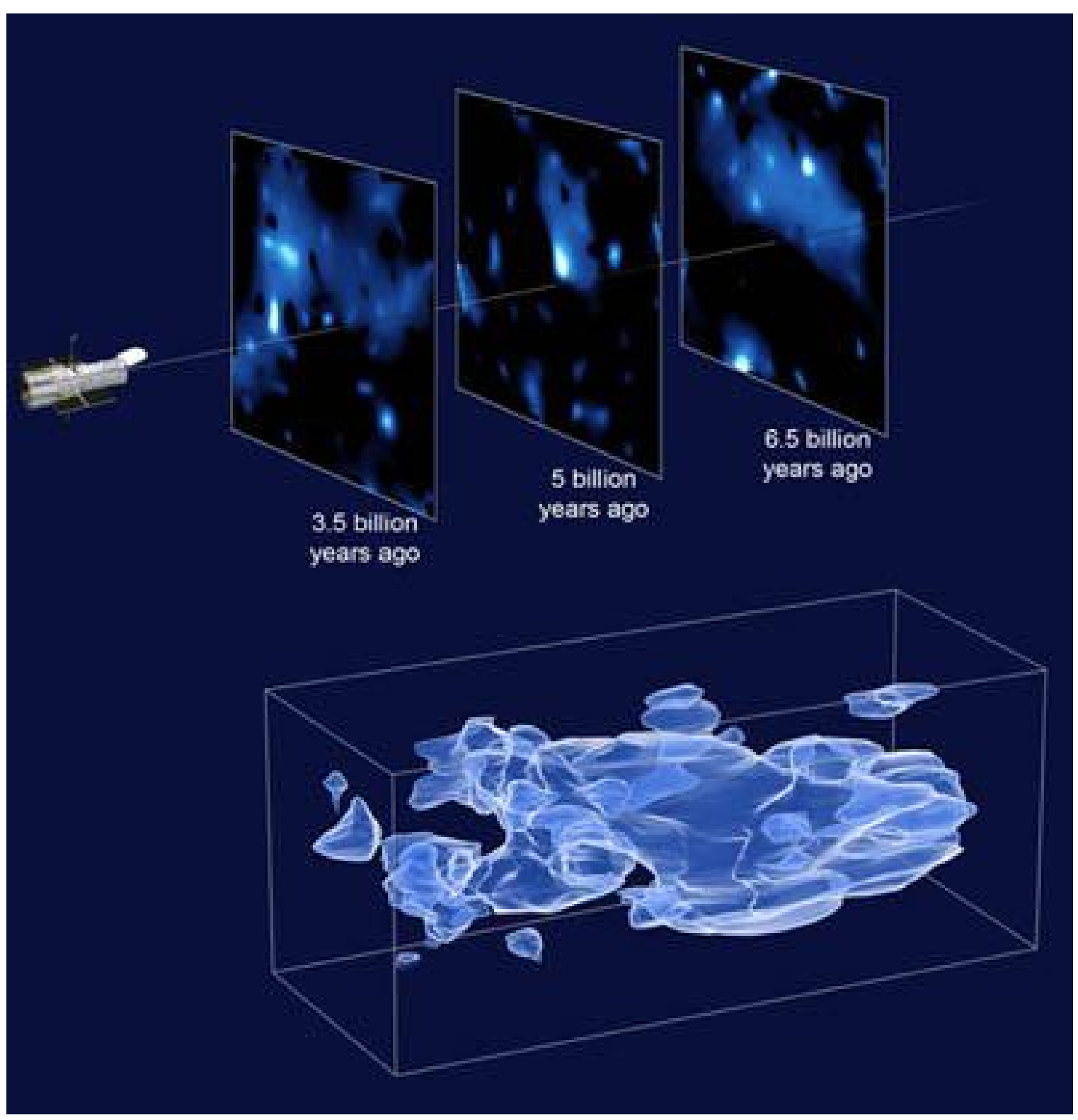

**Figure 3.5**: Dark matter distribution mapped with Hubble Space Telescope's Cosmic Evolution Survey (COSMOS), obtained with a tomography technique in 3 bins of redshift at $z \sim 0.3, 0.5, 0.7$. The observed shear field is converted into a convergence map (using Kaiser-Squire inversion technique [Kai93]), which is proportional to the two dimensional, projected mass ([Mas07]). Image credit: NASA

([Alc96], [Tis07]).

**Weak Lensing:** the most frequent occurrence of gravitational lensing is in the weak regime, when $\kappa \ll 1$ and $\gamma \ll 1$. The lens is not strong enough to form multiple images or arcs. The source image can still be nonetheless distorted: stretched (shear) and magnified (convergence). Galaxy clusters distort this way the background galaxies shapes, and distant galaxies can also be lensed by galaxies closer to us (galaxy-galaxy lensing).

The galaxies shape distortion is minute: typical image ellipticities induced by weak lensing are of order a few per cent. Moreover, as galaxies are not intrinsically circular, the weak-lensing distortions cannot be inferred from one individual galaxy: several of them need to be averaged, assuming that their intrinsic ellipticities $\widetilde{\varepsilon}$ are randomly oriented and would thus average to zero in absence of lensing. For these reasons, it is only by studying a large number of sources statistically, that information about the foreground lenses can be inferred (see figure 3.5).

The average (over a finite area of the sky) ellipticity induced by lensing is:

$$\langle \widetilde{\varepsilon} \rangle = \left\langle \frac{\widetilde{\gamma}}{1-\kappa} \right\rangle$$

Because in the limit of the weak lensing, the convergence $\kappa(\vec{\theta})$ can be inferred from the shear $\widetilde{\gamma}(\vec{\theta})$, this relation permits to iteratively recover $\kappa(\vec{\theta})$ and therefore $\Sigma(\vec{\theta})$.

### 3.1.6 Multiple lens plane method

The dark matter distribution in the Universe is clumpy, so that gravitational lensing by successive local density inhomogeneities along the line of sight have to be considered.

Under the assumption that the light rays originating from the source are undergoing small deflections due to the gravity of the intervening clumps, the inhomogeneities can be taken into account by dividing the space into thin layers, assuming that each lens belong to one lens plane. The plane are perpendicular to the light bundle. The position of the lensing planes can be distributed according to theoretical prescriptions or to N-body simulation, so as to obtain a statistical description. But they can also be set realistically, according to the positions taken in a galaxy and cluster catalog, to study a real lensing situation.

In backward ray-tracing methods, light rays are propagated from the observer to the source with deflections only occurring in successive lens planes (see e.g. [Sch92]). The lensing effect for the corresponding lens plane is then calculated and the linear summation of all the lens planes gives the net lensing signal.

The multiple-lens situation is described on figure 3.6 for 2 lenses, $L_1$ and $L_2$ ($L_1$ is the closest to the observer). Starting from the source $S$, one can use the same rationale presented in figure 3.2 (see [Sch92], chap. 9). So that, generalizing to $N$ lenses:

$$\vec{\eta} = \vec{\xi}_1 \frac{D_S}{D_1} - \sum_k \widehat{\vec{\alpha}}_k(\vec{\xi}_k) D_{kS}$$

Equivalently, setting $\vec{\xi}_1 = D_1 \vec{\theta}$ and $\vec{\xi}_k = D_k \vec{\theta}_k$, and defining $\vec{\alpha}_k$ as previously:

$$\vec{\beta} = \vec{\theta} - \sum_k \widehat{\vec{\alpha}}_k(D_k \vec{\theta}_k) \frac{D_{kS}}{D_S} = \vec{\theta} - \sum_k \vec{\alpha}_k(\vec{\theta}_k) \qquad (3.18)$$

The Jacobian matrix $\mathcal{A}$ of the mapping is then:

$$\mathcal{A} = \frac{\partial \vec{\beta}}{\partial \vec{\theta}} = \mathbb{1} - \sum_k \frac{\partial \vec{\alpha}_k}{\partial \vec{\theta}_k} \frac{\partial \vec{\theta}_k}{\partial \vec{\theta}} \qquad (3.19)$$

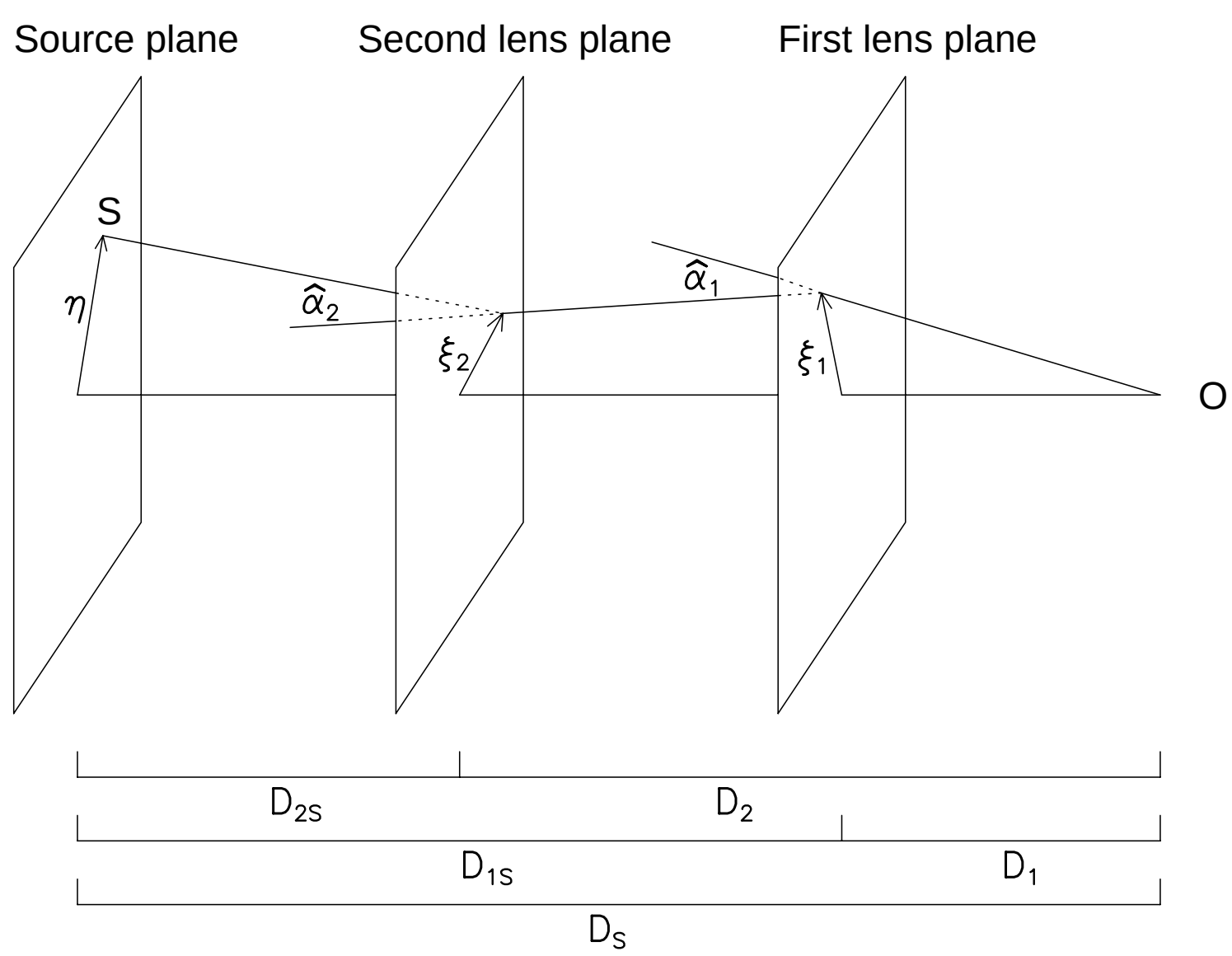

**Figure 3.6**: Schematic diagram of the observer's backward ray tracing in the multiple-lens-plane approximation. A light ray experiences a deflection $\widehat{\alpha}_k$ only when passing through each lens plane ($L_k$), located at a distance $D_k$ from the observer and $D_{kS}$ from the source. Image: [Pre98]

To obtain equations that are as similar as possible to the single lens-plane equations we set:

$$\kappa_k(\vec{\xi}_k) = \frac{4\pi D_i D_{kS}}{D_S}\,\Sigma_k(\vec{\xi}_k)$$

In the weak lensing case, where $\kappa_k \ll 1$ and $\gamma_k \ll 1$, it can be shown (see again [Sch92], chap. 9) that at first order:

$$\mu \simeq 1 + 2\sum_k \kappa_k \qquad \text{(weak approximation)}$$

## 3.2 Lens models

Weak gravitational lensing, whether of type Ia supernovae or galaxies, probes the distribution of matter along the line of sight in the form of dark matter haloes around galaxies or clusters. We shall present in this section two popular halo models, along with the scaling relations used for their parametrization. But first, we introduce some concepts and analytic results from the theory of non-linear structure formation which underlay the use of haloes as cosmological probes.

### 3.2.1 Non-linear structure formation

The linear growth of density perturbations through gravitational attraction was described in section 1.5 where we introduced the matter density contrast:

$$\delta = \frac{\rho(\vec{r},t) - \rho(t)}{\rho(t)}$$

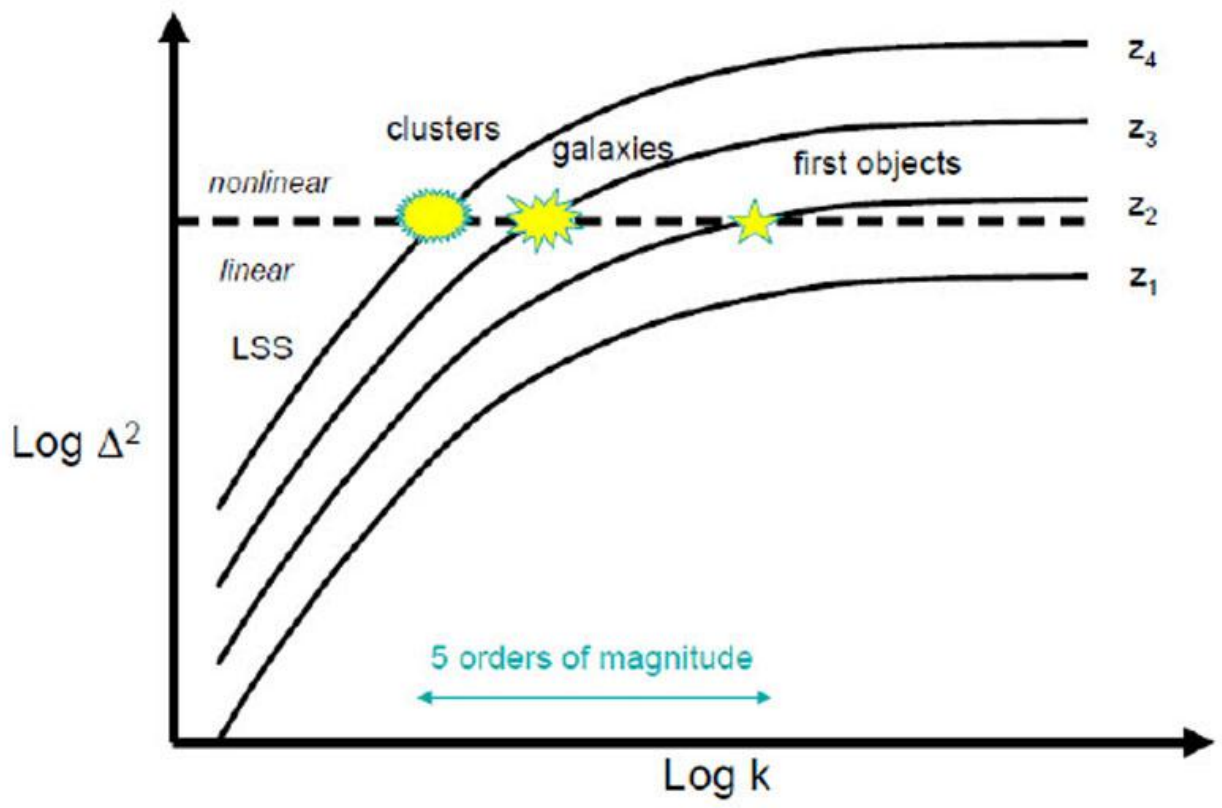

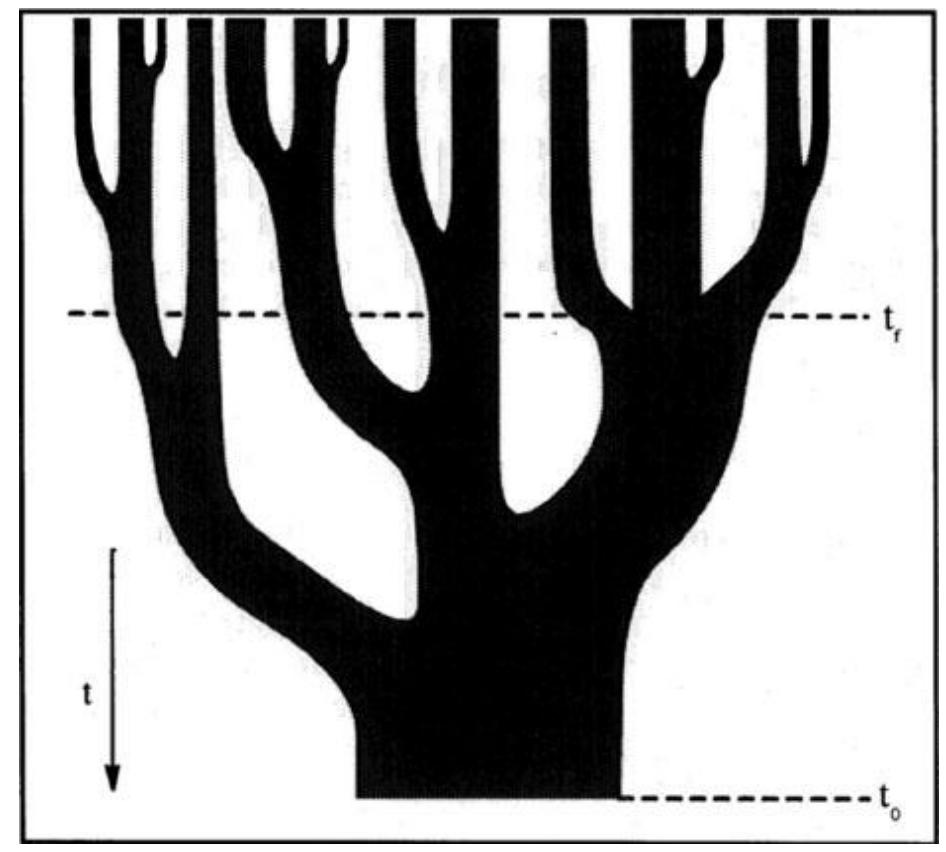

**Figure 3.7**: Left: growth of matter fluctuations in the linear regime. Non linearity is defined where $\Delta^2(k) = 1$. All scales $k$ reaching this threshold collapsed into bound objects, in a bottom-up fashion (smaller scales first). Credits: [Nor10]. Right: Schematic diagram of the hierarchical growth. Smaller halos merge into bigger halo as the time increases from top to bottom. The horizontal slice on the tree at any instant gives the halo distribution at the corresponding time. Credits: [Lac93].

Let's recall the evolution of a density fluctuation (with $\delta \ll 1$) in the linear regime: it grows as $\propto a^2(t)$ (if of super-Hubble size) before radiation-matter equality, and as $\propto a(t)$ during matter domination (whatever its size). Once $\Lambda$ starts to dominate, the universe ceases to behave as an Einstein-de Sitter universe, and the linear growth factor is written as $\delta \propto ag(a)$. Most of the evolution of the density perturbation for structure formation has thus occurred during the matter dominated era, in a bottom-up fashion, reaching a non-linearity threshold when $\delta > 1$ (see figure 3.7).

Using a simple non-linear analytic model, it is possible to describe the spherical collapse of a localized density perturbation, much smaller than the horizon (see [Ric09] for a detailed study). Considering an overcritical "bubble", embedded in an overall critical Einstein-de Sitter CDM, $\Omega = 1$ universe, it can be shown that, when the initially slightly denser bubble will have reached the state of virialization, where dissipative physics turn the kinetic energy of collapse into random motions and end up satisfying the virial theorem $T_{\rm vir} = -V_{\rm vir}/2$, its density contrast will be of the order of :

$$\text{at } t_{\rm vir} : \quad \delta_c + 1 \simeq 178$$

In a nutshell, over densities at virialization are approximately 100-200 times that of the background cosmic mean density at that epoch. This simple model is consistent with results from N-body simulations. Note using the linear theory would have yield an underestimated value for the density contrast $\delta_c^{\rm lin} \simeq 2$. The behavior of $\delta_{\rm lin} - \delta$ is shown on figure 3.8 for an EdS and a F$\Lambda$CDM cosmology.

These quantities play an important role in defining key parameters in dark matter halo models, such as the "virial" radius $r_{\rm vir}$ and the related "virial" mass $M_{\rm vir}$ through:

$$M_{\rm vir} = \frac{4\pi}{3}\,\Delta_{\rm vir}\,\rho(z)\,r_{\rm vir}^3 \tag{3.20}$$

The relation above may be simple, but the variety of choices for $\Delta$ and $\rho$ can create confusion. Following the simple model, the classic choice $\Delta = 200$ defines $M_{200}$ and $r_{200}$:

$$M_{200} \equiv \frac{4\pi}{3}\,\Delta_{200}\,\rho_{\rm cr}(z)\,r_{200}^3, \qquad \rho_{\rm cr}(z) = \frac{3H^2(z)}{8\pi G}, \qquad \Delta_{200} \equiv 200 \tag{V1}$$

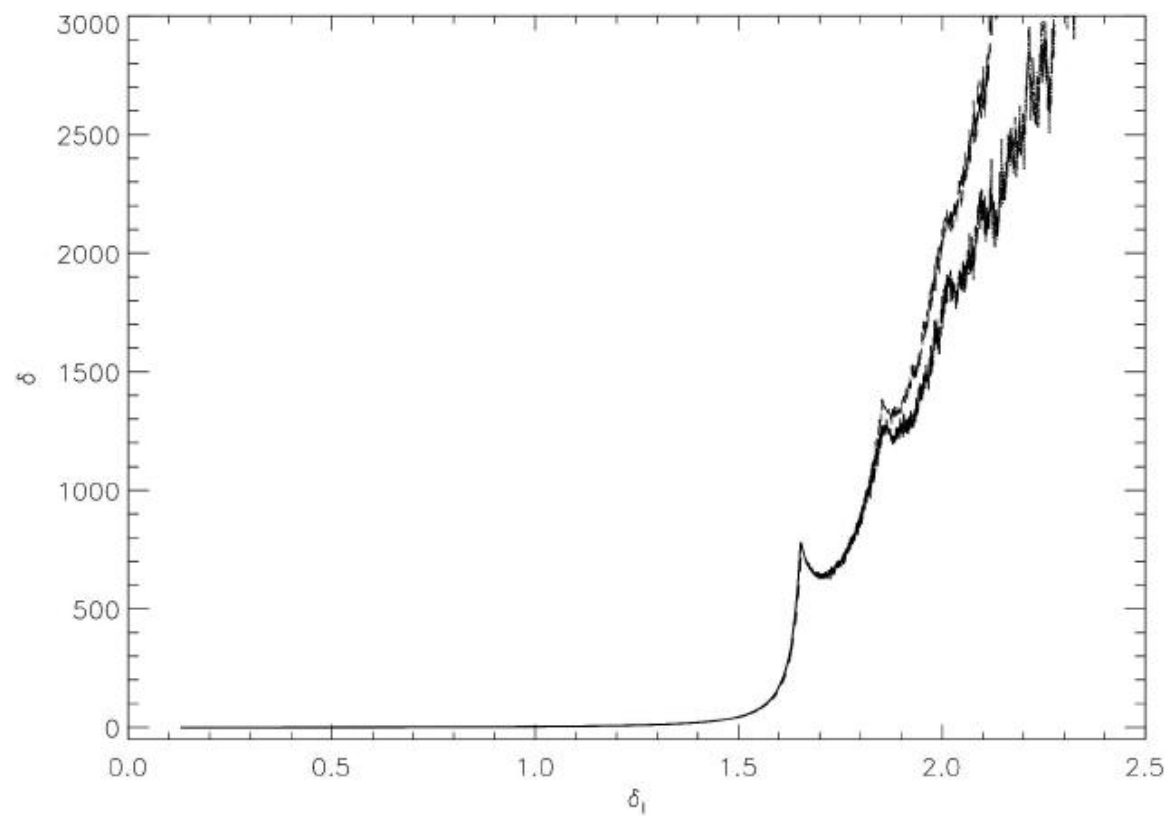

**Figure 3.8**: The difference $\delta - \delta_{\rm lin}$ as a function of $\delta_{\rm lin}$ for two different cosmologies. From down to top, the curves correspond to the Einstein-deSitter case and to the $\Lambda \neq 0$ cosmology (a virial mass of $M_{\rm vir} = 3 \times 10^{12} h^{-1} M_\odot$ was used in all the cases). The departure from the linear case, leading to the structure collapse and virialization, is clearly seen after $\delta_{\rm lin} \sim 1.5$. Credits: [SC07]

One can also use a fitted formula to the numerical computation corresponding to the spherical collapse model in the frame of e.g. a flat $(\Omega_m, \Omega_\Lambda)$ universe. [Bry98] propose:

$$M_{\rm vir} \equiv \frac{4\pi}{3}\, \Delta_c(z)\, \rho_{\rm cr}(z)\, r_{\rm vir}^3 \tag{V2}$$

with:

$$\Delta_c(z) \equiv 18\pi^2 + 82x - 39x^2, \quad \text{with} \quad x = \Omega(z) - 1 \tag{3.21}$$

introducing the ratio $\Omega(z)$ of the matter density to the critical density:

$$\Omega(z) \equiv \frac{\rho_m(z)}{\rho_{\rm cr}(z)} = \frac{\Omega_m(1+z)^3}{E^2(z)}, \qquad E^2(z) \equiv \frac{H^2}{H_0^2} = \Omega_m(1+z)^3 + \Omega_\Lambda$$

Note that $18\pi^2 \simeq 178$ is the factor we already encountered. For a $\Lambda$CDM model, with $\Omega_m = 0.3$, $\Delta_c(z=0) \simeq 100$.

Once a convention is chosen, there is a one-to-one correspondence between $M_{\rm vir}$ and $r_{\rm vir}$.

### 3.2.2 Halo Models

We describe here two popular halo models characteristics, along with their lensing properties.

#### The Singular Isothermal Sphere (SIS) Model

Spiral galaxies are observed to have flat rotation curves out to a large radii ([Rub78], [Beg87], see figure 3.9), suggesting the existence of a spherical halo embedding the galaxy, with a profile density $\rho \propto r^{-2}$ (e.g. [Ost73]). Stellar dynamics of elliptical galaxies (e.g., [Rix97]) are also consistent with a constant dispersion velocity $\sigma_v$.

The Singular Isothermal Sphere (SIS) profile is derived by assuming that the halo matter content behaves as an isothermal self-gravitating, equilibrium sphere of collisionless particles. Its density profiles:

$$\rho(r) = \frac{\sigma_v^2}{2\pi G r^2} \tag{3.22}$$

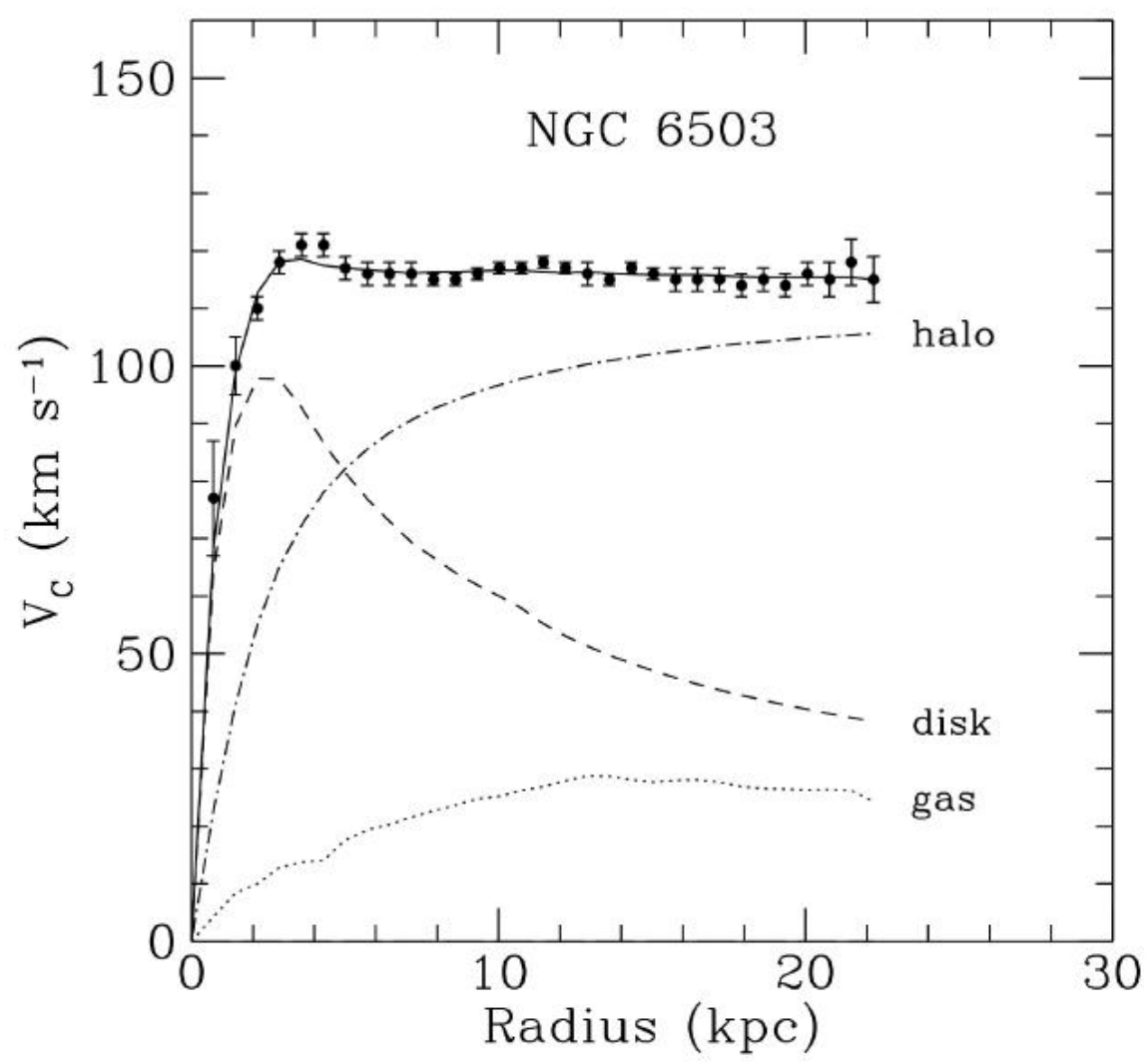

**Figure 3.9**: The rotational curve measured in the 21-cm line of neutral hydrogen for spiral galaxy NGC6503. The solid line shows the dark halo fitted to the data. The profile is an isothermal sphere with a finite core radius $\rho \propto (r_c^2 + r^2)^{-1}$. Also shown are the visible (dashed), gas (dotted) and dark halo (dash-dot) components curves. Image : [Beg91]

is parametrized solely by the constant dispersion velocity $\sigma_v$. The velocity dispersion in a SIS model ranges in $100 - 200$km.s$^{-1}$ ([Bri13]). It is singular at $r = 0$[4]. The mass $M(r)$ within a given radius $r$ is proportional to $r$:

$$M(r) = (2\sigma_v^2/G)\, r \tag{3.23}$$

insuring a flat rotation curve, as:

$$v_{\rm rot}^2 = \frac{GM(r)}{r} = 2\sigma_v^2$$

$M(r)$ is divergent at large $r$, therefore in practice, the model is truncated at some radius, for example at the virial radius $r_{\rm vir}$ defined as $M_{\rm vir} = (2\sigma_v^2/G) r_{\rm vir}$. Using the relation between $M_{\rm vir}$ and $r_{\rm vir}$ (depending on the chosen convention, e.g. eqns. V1 and V2), one can compute the virial radius $r_{\rm vir}(\sigma_v)$ or the virial mass $M_{\rm vir}(\sigma_v)$ :

$$r_{\rm vir} = \frac{2}{\sqrt{\Delta}} \frac{\sigma}{H_0 E(z)} \tag{3.24}$$

Some related lensing quantities for the SIS model are ([Gun05]):

- the (constant) deflection angle:

$$\widehat{\alpha} = 4\pi \frac{\sigma_v^2}{c^2} = 1.15" \left( \frac{\sigma_v}{200\,\text{km.s}^{-1}} \right)^2$$

- the Einstein Radius and the lens equation:

$$\theta_E = 4\pi \left(\frac{\sigma_v}{c}\right)^2 \frac{D_{LS}}{D_S}, \qquad \beta = \theta - \theta_E \frac{\theta}{|\theta|}$$

  $\theta = \theta_E$ corresponds to the tangential critical curve of the SIS. For a source at a distance at $z = 0.8$ and a lens with $\sigma_v = 200\,\text{km.s}^{-1}$ at $z = 0.4$, $\theta_E \simeq 0.5"$.

[4] This can be overcome by introducing a core radius $r_c$ so that $\rho \propto (r_c^2 + r^2)^{-1}$, see figure 3.9.

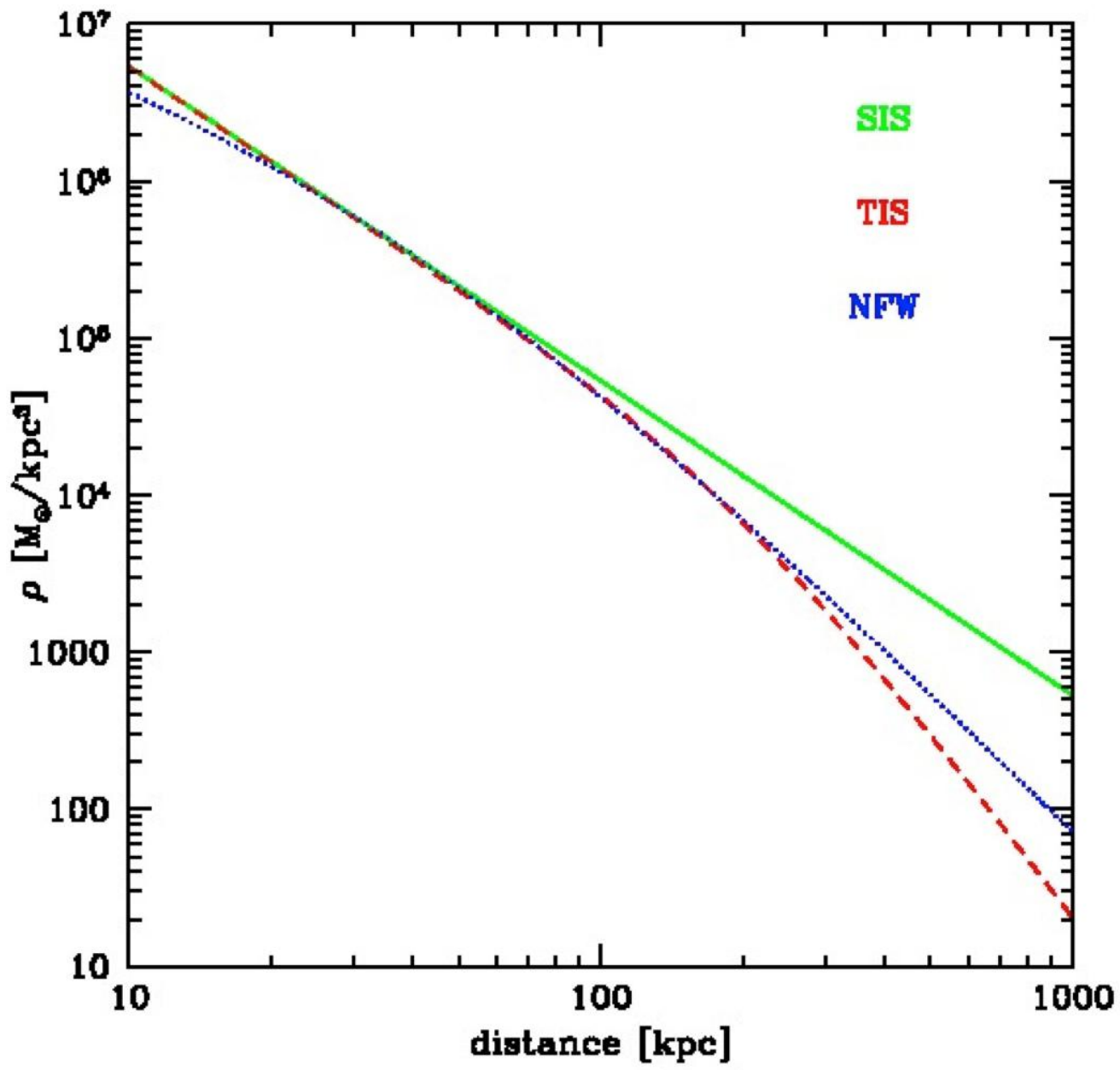

**Figure 3.10**: Comparison of density profiles of halo models: in green the singular isothermal sphere(SIS) profile, in blue the universal(NFW) profile and in red the truncated isothermal model (TIS) with $\rho_{\rm TIS}(r) = (\sigma_v^2)/(2\pi G r^2)\ s^2/(s^2+r^2)$. Image : Stella Seitz

- the surface mass density:

$$\Sigma(\xi) = \int_{-\infty}^{\infty} \rho\left(\sqrt{\xi^2+z^2}\right)\, dz = \frac{\sigma_v^2}{2G}\frac{1}{\xi}$$

  which can be rewritten in term of $x = \xi/\xi_0$, choosing the length scale $\xi_0$ in the lens plane:

$$\xi_0 = 4\pi\frac{\sigma_v^2}{c^2}\frac{D_{LS}D_L}{D_S} \tag{3.25}$$

$$\Sigma(x) = \frac{\Sigma_{\rm cr}}{2x}$$

  For a source at a distance at $z = 0.8$ and a lens at $z = 0.5$, $\xi_0 \simeq 2.75$ kpc.
- The convergence:

$$\kappa = \frac{1}{2x}$$

- The magnification

$$\mu(x) = \frac{|x|}{|x|-1}$$

### The Navarro-Frenk-White (NFW) model

In general, structure formation is too complicated to be analytically derived: it is necessary to turn to numerical simulations, that can now include, besides the gravitational interaction of dark matter "macroscopic" particles, hydrodynamic and radiative process. On the basis of N-body simulation results, Navarro, Frenk and White (NFW, [Nav95, Nav96, Nav97]) proposed a "universal" halo profile, consistent with observational data from the scales of dwarf galaxies to galaxy clusters, over a wide range of masses:

$$\rho(r) = \frac{\rho_s r_s^3}{r(r + r_s)^2} \tag{3.26}$$

where the parameter are $r_s$ the scale radius and and $\rho_s$ the amplitude of the density profile. Another choice is to chose the parameters $r_s$ and $\delta_c$ the relative amplitude of the density profile, defined as:

$$\rho_s = \delta_c \, \rho_{cr}(z)$$

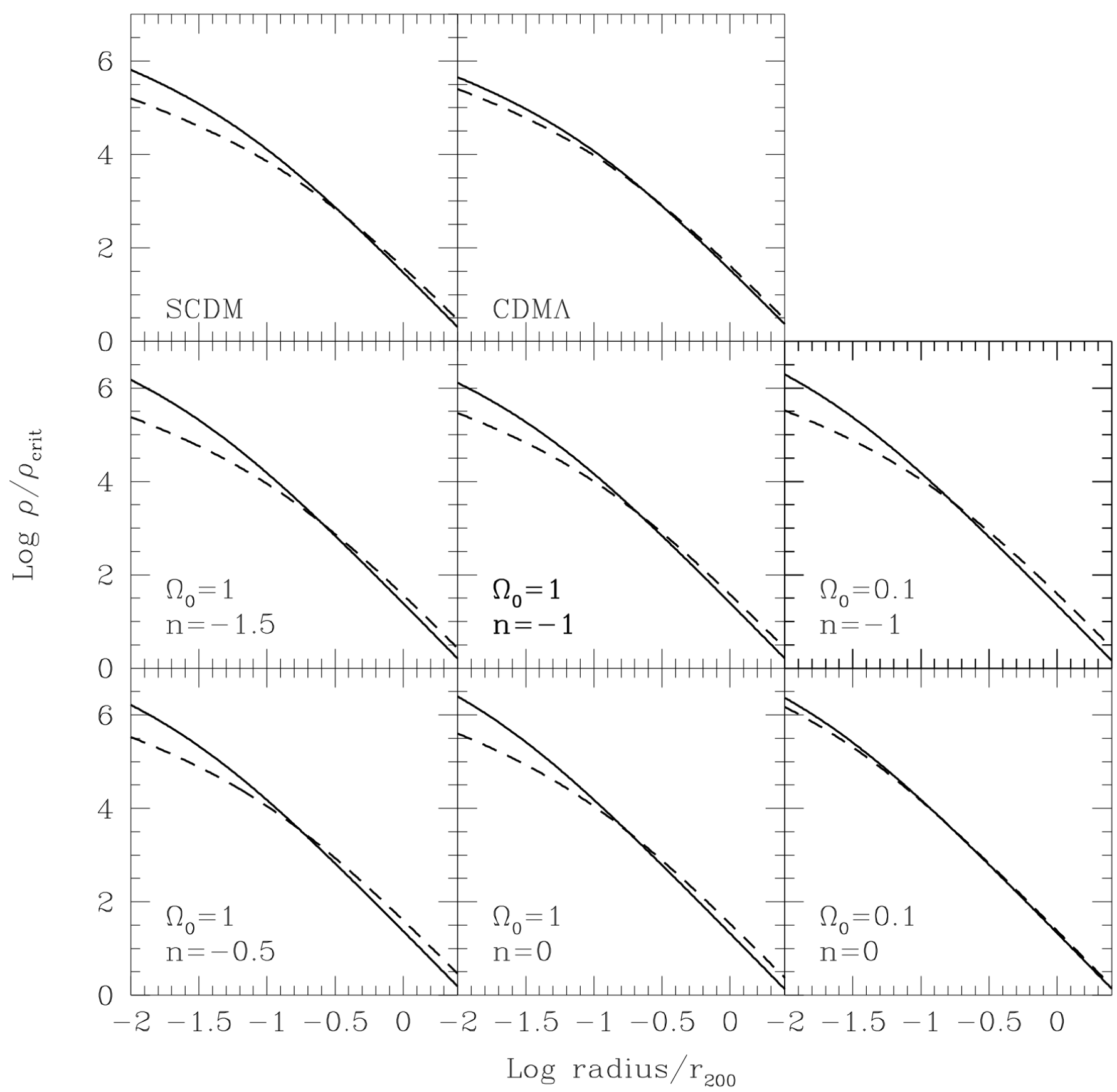

**Figure 3.11**: NFW density profile fit to N-body simulations for eight different cosmologies: the best fitting profile is shown for the low mass (solid) and the high mass (dash) halos. SCDM is the EdS model, $\Omega_0$ is the matter density parameter and $n_s$ is the spectral index of the initial density fluctuation power spectrum $P(k) \propto k^{n_S}$ seeding the simulation. The radius is scaled to $r_{200}$ and the density to $\rho_{crit}(z=0)$. Lower mass halos are more concentrated near the center than higher mass halos. As a consequence their concentration parameter is higher. Image : [Nav97]

Although singular at $r = 0$, this profile does not lead to a singularity in mass, being $\propto r^{-1}$, and thus shallower than an isothermal profile. At $r \gg r_s$, the profile is $\propto r^{-3}$, so steeper than an isothermal profile, and the integrated mass rises logarithmically.

The parametrization of the NFW model is worth discussing in details. We shall see that it can be parametrized by $M_{\rm vir}$ and the concentration parameter $c$:

$$c \equiv \frac{r_{\rm vir}}{r_s}$$

A large value for $c$ implies that the halo is highly concentrated in the inner regions. The halo virial mass $M_{\rm vir}$, enclosed within the virial radius $r_{\rm vir}$, is computed as:

$$M_{\rm vir} = \int_0^{r_{\rm vir}} \rho(r) 4\pi r^2 dr = 4\pi \rho_S \, r_s^3 \, f\left(\frac{r_{\rm vir}}{r_s}\right)$$

with $f(x)$ defined as:

$$f(x) = \int_0^x \frac{du}{u(u+1)^2} = \ln(1+x) - \frac{x}{1+x}$$

So that the NFW profile can be written as:

$$\rho(r) = \frac{M_{\rm vir}}{4\pi f(c)} \frac{1}{r(r + r_{\rm vir}/c)^2} \tag{3.27}$$

and is now parametrized with $M_{\rm vir}$ and the concentration parameter $c$. Recall that there is a one-to-one relation between $r_{\rm vir}$ and $M_{\rm vir}$ once a convention is chosen (eqn. V1 or V2). It is worth noting that with a chosen "virial convention" as e.g. V2, the relative amplitude of the density profile $\delta_c$ defined by $\rho_s = \delta_c \rho_{\rm cr}(z)$ is related simply to the concentration parameter $c$ with:

$$\delta_c = \frac{\Delta_c}{3} \frac{c^3}{f(c)}$$

Simulations show that the concentration index $c$ is related to the halo (virial) mass and redshift ([Nav96], [Bul01]): small halos are significantly denser than large halos as a result of the fact that small, low-mass halos formed at higher collapse redshifts when the density of the universe was higher. Observations also show that dark matter halos represent a one-parameter family with self similar density profiles ([Bur97]) and that the NFW concentration parameter $c$ decreases with halo mass, from around 10 for galactic halos to 4 for cluster halos ([Man08]). This dependency can be described with a power law:

$$c(z, M) = A \left(\frac{M}{M^\star}\right)^B (1+z)^C \tag{3.28}$$

where the pivot $M^\star$ is the median halo mass. A value $C = -1$ can be obtained from analytic considerations ([Bul01]). There is an intrinsic scatter $\sigma(\log_{10} c) \sim 0.1 - 0.2$ about the median $c(M)$ relation ([Nav97, Duf08]). Different estimations of law 3.28 parameters are listed in table 3.1.

| concentration parameter $c(z,M) = A\,(M/M^{\star})^{B}\,(1+z)^{C}$ | | | | |
|---|---|---|---|---|
| Author | $hM^{\star}/M_{\odot}$ | A | B | $C$ |
| Seljak, 2000 ([Sel00]) | $3.42\,10^{12}$ | 10 | $-0.2$ | -1 |
| Neto et al., 2007 ([Net07]) | $10^{14}$ | 4.67 | $-0.11$ | $(z=0)$ |
| A. Duffy et al., 2008 ([Duf08]) | $2.\,10^{12}$ | 7.85 | $-0.081$ | -0.71 |
| A. Maccio et al., 2008 ([Mac08]) | $10^{12}$ | 6.76 | $-0.098$ | -1 |

**Table 3.1**: Parametrization of the concentration parameter law $c(M,z) = A(M/M^{\star})^{B}(1+z)^{C}$. The values are referring to $c_{\rm vir}$, except for [Net07] which gives $c_{200}$.

Note that $c$ depends on the chosen virial definition, with $c_{200}$ and $c_{\rm vir}$ usually referring respectively to convention V1 and V2. The conversion between different conventions for $c$ are given in [Joh07].

Having dealt with the NFW parametrization, we can now present the related lensing quantities ([Gun05]), taking $x = r/r_s$:

- The surface mass density:

$$\Sigma(x) = 2\rho_s r_s F(x) \tag{3.29}$$

  where $F(x)$ reads:

$$F(x) = \begin{cases} \frac{1}{x^2-1} - \frac{2}{(x^2-1)^{3/2}} \operatorname{arctan}\sqrt{\frac{x-1}{x+1}} & (x>1) \\ \frac{1}{x^2-1} + \frac{2}{(1-x^2)^{3/2}} \operatorname{artanh}\sqrt{\frac{1-x}{x+1}} & (x<1) \end{cases}$$

- The convergence:

$$\kappa(x) = 2\kappa_s F(x), \quad \kappa_s = \frac{\rho_s r_s}{\Sigma_{cr}}, \quad \kappa(1) = \frac{2}{3}\kappa_s \tag{3.30}$$

### 3.2.3 Mass scaling relations

The halo parameters describing the mass distribution of a galaxy are inferred from observational quantities through scaling relations, relating the total mass, or the density profile kinematics properties, to the galaxy luminosity.

The singular isothermal sphere (resp. NFW) halo model presented in section 3.2.2 can be solely parametrized by the dispersion velocity $\sigma$ (resp. the virial mass $M$). The scaling relations thus usually takes the form:

$$\frac{\sigma}{\sigma^{\star}} = \left(\frac{L}{L^{\star}}\right)^{\eta_\sigma} \tag{3.31}$$

$$\frac{M}{M^{\star}} = \left(\frac{L}{L^{\star}}\right)^{\eta_M} \tag{3.32}$$

We shall present here the scaling relations derived from galaxy-galaxy gravitational lensing studies, together with the empirical Tully-Fisher and Faber-Jackson relations for galaxies.

#### The Tully-Fisher relation

The Tully-Fisher relation ([Tul77]) states that the maximum rotational velocity $V_{\rm max}$ in spiral galaxies are observed to be closely related to their luminosity $L$ (figure 3.12):

$$L \propto V_{\rm max}^{\alpha}$$

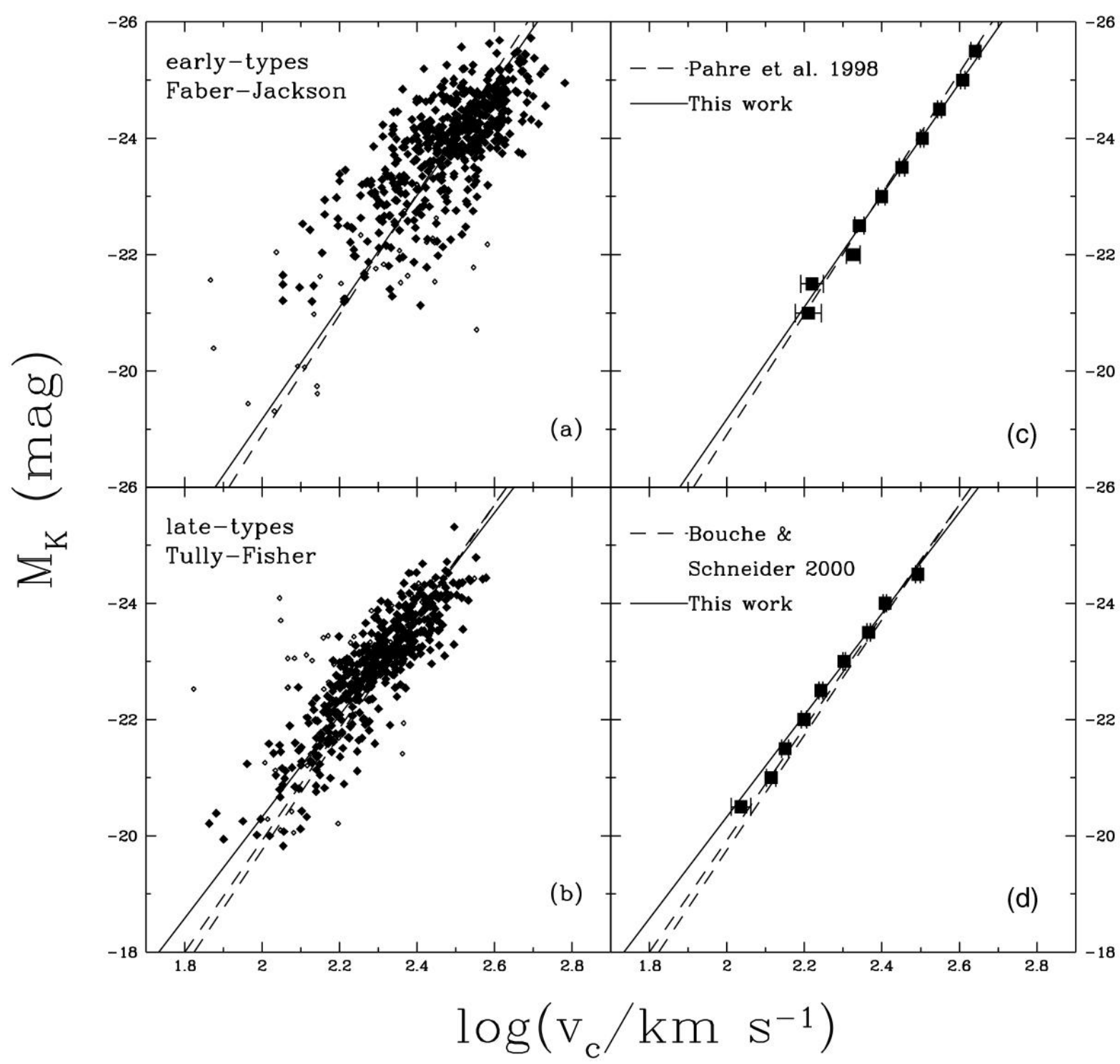

**Figure 3.12**: Faber Jackson relations (top) and Tully Fisher relation (bottom) for 2MASS galaxies in $K$ band. The left panel shows individual galaxies while the right panel shows the mean relations. Note the larger scatter of the Faber Jackson plots than in the Tully Fisher. Image: [Koc06].

In the frame of the SIS halo model, the rotational velocity can be converted into an equivalent isotropic velocity dispersion $\sigma$ using $V_{\max} = \sqrt{2}\sigma$. We shall present here the relation established by [Boe04], which is used in this thesis and in some previous supernovae gravitational studies ([Gun06], [Jon08], [Kro10]). Based on the measurement of spatially resolved rotation curves for 77 FORS Deep Field ([Nol04], [Hei03]) spiral galaxies with a redshift between $z = 0.1$ et $z = 1$ using the Very Large Telescope in Multi Object Spectroscopy, they estimate a T-F relation that takes into account an evolution in redshift. It thus differ from the local relation established by [Pie92]:

$$M_B = -7.48 \times \log V_{\max} - 3.52$$

corresponding to an index $\eta_\sigma \simeq 0.33$, to which they add a redshift dependent correction:

$$\Delta M_B = -(1.22 \pm 0.56) \times z - (0.09 \pm 0.24)$$

so that it is equivalent to:

$$\log V_{\max} = -0.134 \times (M_B + 1.22 \times z + 3.61) \tag{3.33}$$

As a consequence, for the same mass, at higher redshift, a galaxy is more luminous in the B-band, owing to its younger star population. For a magnitude $M_B = 21$, the velocity dispersion is $\sigma \simeq 150\,\mathrm{km.s^{-1}}$. The observed scatter in the relation established by [Pie92] is:

$$\sigma_{M_B} = 0.41$$

which results in:

$$\sigma(\log V_{\max}) = 0.055 \tag{3.34}$$

This scatter arises from the observational errors, a variation from the mass function of stars and an intrinsic contribution ([Eis96]). This result is corroborated by [Fer09] study, based on 612 galaxies from the DEEP-2 spectroscopic survey at a redshift between $z = 0.2$ and $z = 1.2$.

### The Faber-Jackson relation

The Faber-Jackson ([Fab76]) relation expresses the loose correlation between the luminosity $L$ and the velocity dispersion $\sigma$ observed in the center of early type galaxies (figure 3.12):

$$L \propto \sigma^\gamma \tag{3.35}$$

The observed aperture-corrected central velocity dispersion is a good estimate for the velocity dispersion of the dark matter halo velocity dispersion when modeled as a spherical isothermal sphere ([Koc94]). The F-J law is a projection of the tighter 3-dimension relation between the effective radius $R_e$, the surface brightness $I \propto L/R_e^2$ and the velocity dispersion $\sigma$, known as the Fundamental Plane: $R_e \propto \sigma^{1.4} I^{-0.85}$ ([deV82, Djo87]) or equivalently the $D_n - \sigma$ relation ([Dre87]).

We will make use in this thesis of ([Mit05], eq. 33) measurement, following again previous work from [Gun06] and [Kro10]. Using approximately 30,000 SDSS surveyed elliptical galaxies at a redshift between $z = 0.01$ and $z = 0.3$ with a spectroscopic measured velocity dispersion, they obtain:

$$\langle \log(\sigma) \rangle = 2.2 - 0.091 \times (M_r + 20.79 + 0.85 \times z) \tag{3.36}$$

corresponding to a Faber-Jackson index $\gamma = 4.4$ or $\eta_\sigma = 0.275$. For $M_r \simeq -22.34$ (corresponding roughly to $M_B \simeq -21$[5]), the velocity dispersion is $\sigma \simeq 220\,\mathrm{km.s^{-1}}$.

The observed scatter about this relation is taken from [She03]:

$$\sigma_{\log\sigma} = 0.079 \times [1 + 0.17 \times (M_r + 21.025 + 0.85 \times z)] \tag{3.37}$$

[5]Assuming, following [Gun06], a mean galactic AB color $M_r - M_B = -1.2$ and a conversion from AB to Vega system $B_{\mathrm{AB}} = B_{\mathrm{Vega}} - 0.12$.

### Galaxy-Galaxy lensing estimation

Galaxy-galaxy lensing is a powerful tool to study the mean radial profile of a ensemble of galaxies, acting as gravitational lenses on a high number density of background galaxies. From the shear of the background sources, one directly measured $\Sigma(R)-\langle\Sigma(R)\rangle$ for the lensing galaxies as a function of the distance R to the galaxy centers. This stacking or averaging technique yields the mean properties of the sample. It is thus possible, using simple halo profile models, to establish scaling relations describing simultaneously galaxy haloes of various luminosity. Over the last 15 years, high-quality galaxies imaging data together with redshift information (either spectroscopic or photometric redshifts) has become available, leading to progress in the use of galaxy-galaxy lensing, thanks to major surveys such as the COMBO-17 survey (e.g. [Kle04]), the Red-Sequence Cluster Survey (e.g. [Hoe04], [van11]), the Sloan Digital Sky Survey (e.g. [Man06]), the Canada-France-Hawaii Telescope (CFHT) Survey or the Dark Energy Survey ([Cla16]).

We present here results obtained by [Bri13] with data from the CFHT Legacy Wide Survey, based on a high statistics of 12 million objects observed in *ugriz*, spawning a surface of 124 square degree. Selecting a sample of lenses and sources with $0.05 < z_L < 1$ (with a mean redshift $\langle z_L\rangle \simeq 0.35$) and $0.05 < z_S < 2$, they derive scaling relations for the halo properties, separating the lenses into blue and red lenses according to their color, at $B-V = 0.7$.

For a SIS model, they obtain, for a fiducial luminosity $L_r^\star = 1.6\,10^{10}\,h^{-2}\,L_{r\odot}$, corresponding to $M_{r\,V} = 21.7$ in the Vega system [6]:

$$\begin{aligned}\sigma_{\rm red} &= 162 \pm 2\,{\rm km.s}^{-1}\left(\frac{L}{L_r^\star}\right)^{0.24\pm0.03}\\ \sigma_{\rm blue} &= 115 \pm 3\,{\rm km.s}^{-1}\left(\frac{L}{L_r^\star}\right)^{0.23\pm0.03}\end{aligned}$$

The precision refer to the fitting procedure, and the scatter around the fitted law is indeed larger, of the order of $30\,{\rm km.s}^{-1}$. To take into account that the galaxy luminosity evolves with look-back time and thus redshift, they also considered an evolving luminosity according to $L \propto (1+z)$. The fitted amplitude is moderately increased ($\sigma^\star_{\rm red} = 173 \pm 2\,{\rm km.s}^{-1}$, $\sigma^\star_{\rm blue} = 123 \pm 3\,{\rm km.s}^{-1}$) in comparison to the values without evolution, but the scaling relation index $\eta_\sigma \simeq 0.25$ remains almost unaffected. The generic relation with evolution then reads:

$$\frac{\sigma}{\sigma^\star} \sim \left(\frac{L}{L_r^\star}\right)^{0.25} \times (1+z)^{-0.25}$$

A comparison of [Boe04], [Mit05] and [Bri13] results are presented in table 3.2, for a reference redshift $z = 0.45$ representative of the lenses redshift for the lensing analysis in this thesis. The galaxy-galaxy lensing results are completely compatible with the spectroscopic measurements.

[6] Taking $h = 0.72$, a solar magnitude $M_{r\,AB\,\odot} = 4.68$ and a conversion from AB to Vega system $r_{\rm AB} = r_{\rm Vega} + 0.125$.

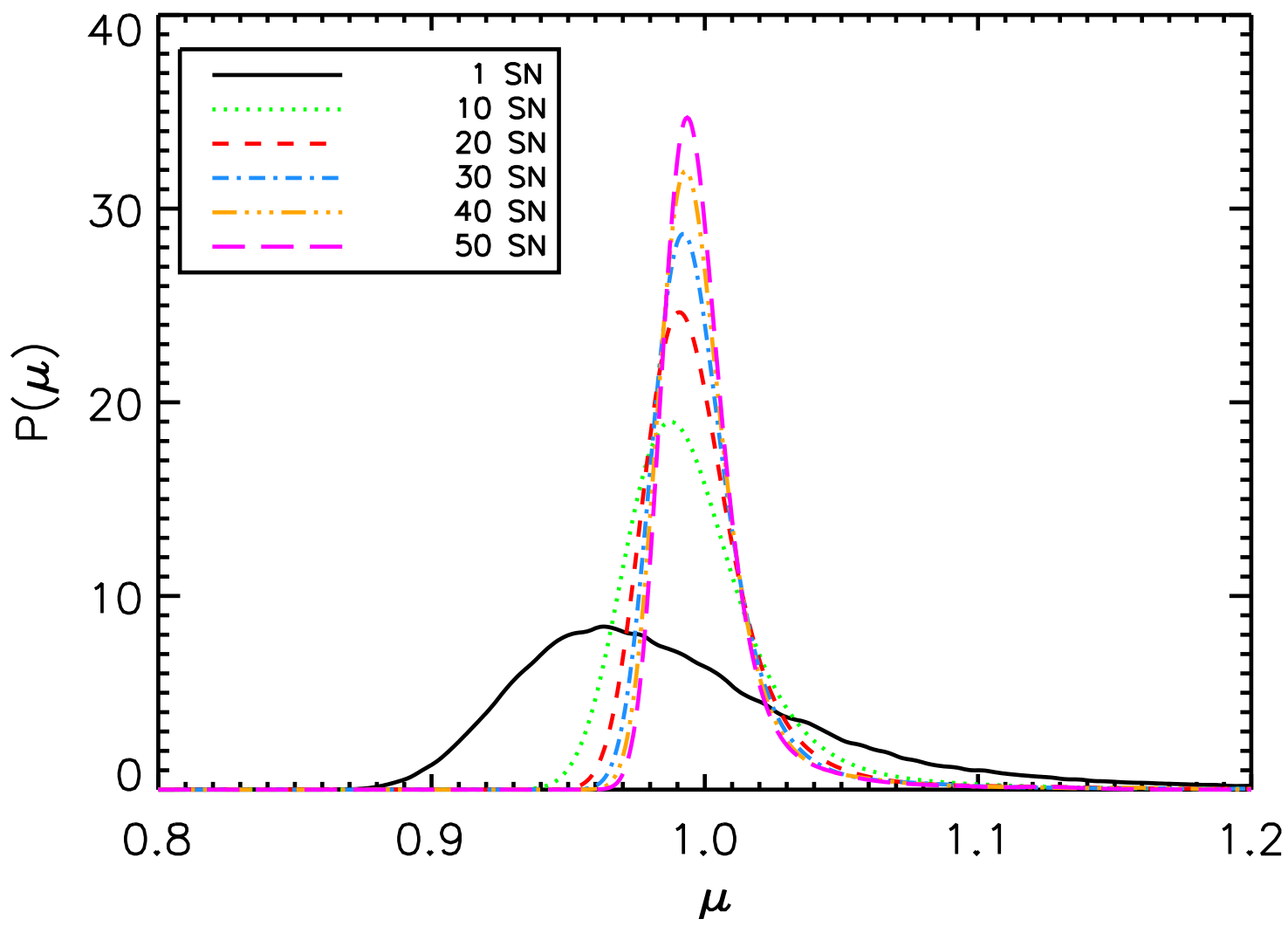

**Figure 3.13**: Lensing magnification distributions for a perfect standard candles sample at $z = 1.5$ in a ΛCDM cosmology. Because of flux conservation the mean magnification value is $\mu = 1$. As more sources are observed, the distribution approaches a Gaussian and eventually converges on a $\delta$-function. Image: [Hol05].

| Author | $L^\star$ | redshift | scaling index $\eta_\sigma$ | $\sigma^\star$ (km.s$^{-1}$) |
|---|---|---|---|---|
| [Boe04] (TF) | $10^{10}\,h^{-2}\,L_{B\odot}$ | $z = 0.45$ | 0.33 | $101 \pm 11$ |
| [Mit05] (FJ) | $10^{10}\,h^{-2}\,L_{B\odot}$ | $z = 0.45$ | 0.275 | $172 \pm 30$ |
| [Bri13] | $1.6\,10^{10}\,h^{-2}\,L_{r\odot}$ | | | |
| red sample | – | no evolution | $0.24 \pm 0.03$ | $162 \pm 2$ |
| blue sample | – | no evolution | $0.23 \pm 0.03$ | $115 \pm 3$ |
| red sample | – | $z = 0.45$ | $0.25 \pm 0.03$ | $157 \pm 2$ |
| blue sample | – | $z = 0.45$ | $0.24 \pm 0.03$ | $112 \pm 3$ |

**Table 3.2**: A comparison of velocity dispersion values $\sigma^\star$ for the fiducial luminosity $L_r^\star = 10^{10}\,h^{-2}\,L_{r\odot}$, roughly equivalent to $L_B^\star = 1.6\,10^{10}\,h^{-2}\,L_{B\odot}$, using the Vega solar magnitude $M_{B\odot} = 5.47$, and, following [Gun06], a mean galactic AB color $M_r - M_B = -1.2$ .

## 3.3 Gravitational Lensing and Supernovae Ia

The impact of gravitational lensing by matter inhomogeneity on standard-candle use as distance indicators for cosmology has been considered since the 90's. It was noted along that standard candles could also serve as a probe of the dark matter distribution along their line of sight. [Smi14] can be consulted for a recent thorough review on this subject.

### 3.3.1 Lensing impact on the Hubble diagram

The gravitational impact of lensing on the Hubble diagram is to increase the scatter of the distance measurement around the value predicted in a homogeneous universe model ([Fri96], [Hol05], [Gun06], [Sar08]). Flux conservation guarantees that the mean flux of the SN population does not change, so that the mean magnification stays at the unlensed magnification value $\mu = 1$ (see section XXX-normalization-XXX). The resulting distribution ([Wan99, Wan02, Hol05, Sar08]) is however non-gaussian, with the majority of the supernovae being de-magnified, and rarely substantially magnified (see fig. 3.13). The resulting scatter increases with the redshift, but statistically diminish $\propto 1/\sqrt{N}$ with the number of standard candles involved.

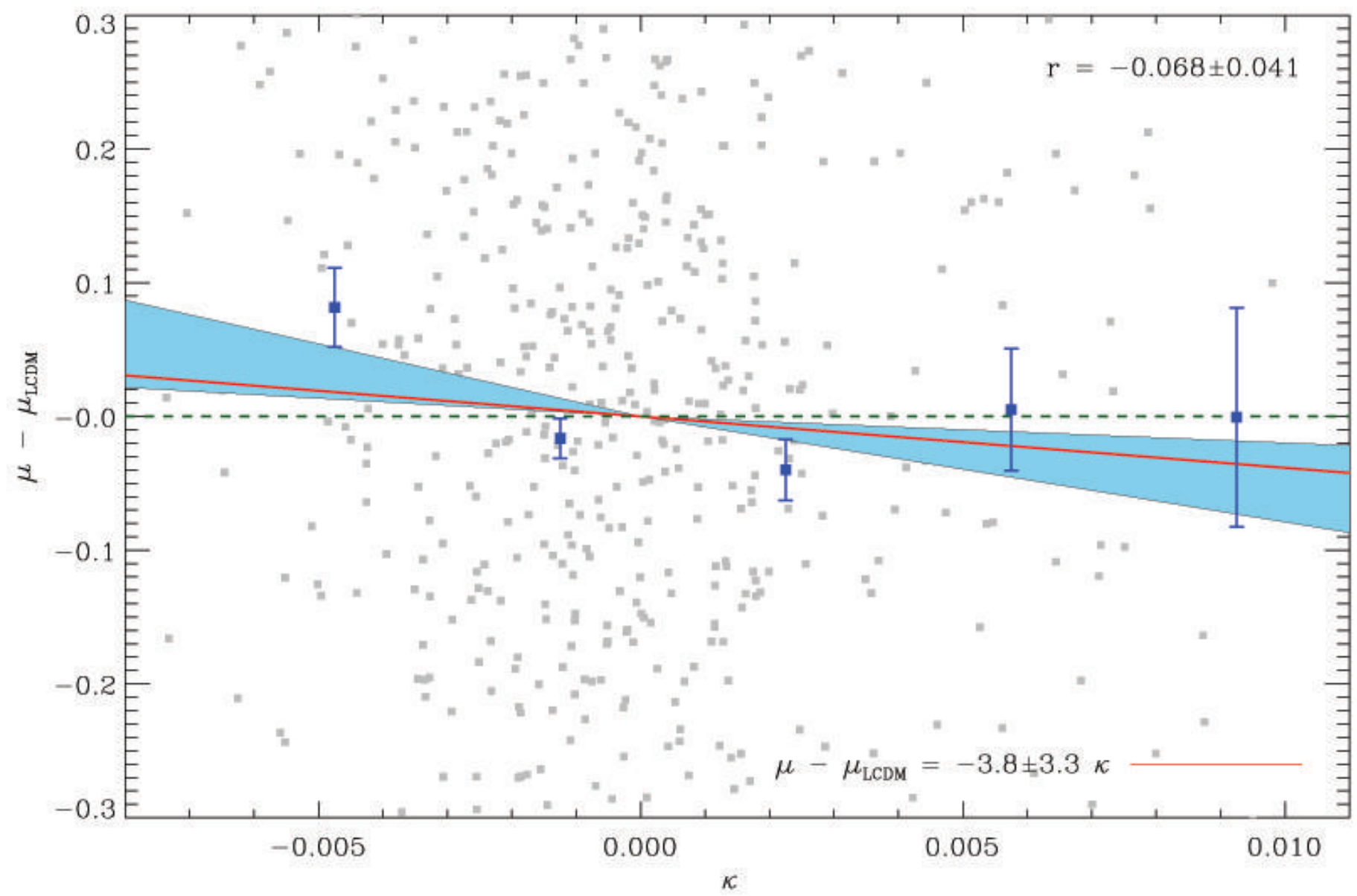

**Figure 3.14**: The residual to the Hubble diagram for 608 Sne from the SDSS II-BOSS sample vs the corresponding convergence $\kappa$. The red line is the line of best fit with a correlation coefficient $\rho = -0.068 \pm 0.41$. The anticipated range of correlation is in blue shade. Image : [Smi14]

This effect has to be considered in cosmology analysis, in the SNe selection (discarding outliers in the Hubble diagram can bias the distribution), and in the statistical treatment of the data. It will not significantly bias the cosmological parameters ([Sar08], [Jon08]) but the resulting scatter must be accounted for ([Gun06]), although within SNLS sample's redshift range, it won't have a significant influence. In SNLS3, a supplementary statistical uncertainty $\sigma \simeq 0.05 \times z$ was included in the cosmology fit computation.

The magnification distribution $p(\mu)$ and especially its skewness is directly related to the large scale structure distribution along the line-of-sight, which will make it a probe for cosmology ([Met99], [Dod06]), attainable for future survey such as LSST.

### 3.3.2 Supernovae lensing signal detection

Supernovae magnification by galaxies along the line of sight is a probe for dark-matter distribution on the haloes scale and its detection has been elusive still then.

The lensing signal is detected by cross-correlating the SN residual $r$ to the cosmological fit in the Hubble diagram:

$$r = \mu_{SN} - \mu_L(z, \text{cosmology}) \tag{3.38}$$

to the SN expected magnification computed by taking into account the matter distribution on the line of sight. We are of course considering here weak lensing, although interestingly, in 2014, strong lensing of a supernova was detected for the first time ([Kel15]) in the MACS J1149.6+2223 cluster with four images of the exploding star arrayed as an Einstein cross.

In the weak lensing domain, introductory work were done to detect the lensing signal by [Wil04], who used 50 high-z supernovae from the SCP and High-Z projects, and [Mén05] and [Wan05] with [Rie04] SN sample. It lead to contradictory results : [Wil04] found that brighter SNe preferentially lied behind denser foreground galaxies with a statistical significance $> 99\%$, a result not supported by [Mén05] and [Wil04] owns. At higher redshift ($0.4 < z < 0.8$), [Jon07] finally detected a non-zero correlation at 90% confidence with 24 SNe from the GOODS sample.

Regarding SNLS supernovae, [Kro10] estimated a correlation of $\rho \sim 0.2$ at $\sim 2.4\sigma$ on 171 SNe from the SNLS3 sample. Independently and with the same data, [Jon10] detected a signal at 92% confidence, constraining the dark matter halo parameters used to assign the galaxies mass.

Recently, [Smi14] with SDSS-II and BOSS survey's combined data set of 608 SNe obtained a $1.7\sigma$ significance of a lensing signal detection (fig(3.14).

The aim of this thesis is to extend [Kro10] study to the SNLS5 sample. The data analysis and signal computing will be presented in chapters 4 and 5.

# Chapter 4

# Line-of-sight modeling and magnification computation

In this chapter we will be presenting the main steps involved in the line-of-sight modeling that lead to the computation of the supernovae magnification.

Following the same road-map as [Kro10] (hereafter K10), the entire analysis chain for the lensing signal detection is outlined below. It consists of :

1. build deep galaxy catalogs (section 4.1).

2. clean catalogers from stars, host galaxies, and polluted areas by e.g. bright stars (section 4.1.5).

3. compute accurate photometric redshifts along with galaxies absolute luminosity and rest-frame colors (section 4.2).

4. assign each galaxy a halo mass or velocity dispersion using the scaling relations presented in section 3.2.3.

5. select the supernovae in the SNLS5 supernovae sample presented in section 2.2.3. The supernova scoring is detailed in section 4.3.

6. compute the supernovae magnification. The algorithms are presented in section 4.4.

7. compute the magnification normalization using randomly selected line-of-sights (section 4.5).

   The last steps below are presented in chapter 5.

8. compute the lensing signal by estimating the correlation between the residual to the cosmology fit in the Hubble diagram and the expected supernova magnification. The lensing signal consists in detecting a positive correlation.

Compared with K10, each involved step (but step 4.) benefits from a re-analysis making use of SNLS 5-years data set, or/and new methods implementation.

## 4.1 Galaxy catalogs building

We present here the key steps involved in the galaxy catalog construction, realized prior to this thesis.

This part is based on [Har12] and [Roman, *in. prep.*].

The observed luminosity of the galaxies (B-band absolute luminosity) situated in close proximity to the line of sight of a supernova are converted to mass using scaling relations, such as the Tully-Fisher and Faber Jackson law, selected according to the galaxy rest-frame color U-V. The galaxy photometric redshift is required to compute the lensing galaxy distance. For these reasons, a deep multi-band galaxy catalog is essential for the lensing analysis.

Making use of the complete SNLS 5-years data set, deep *ugriz* stacked image frames were built for each field, on which multi-band photometry was then performed.

### 4.1.1 Image selection

The CFHT-LS *ugriz* observations of the Deep fields are pre-processed through the Elixir pipeline to correct for the instrumental effects that affect the pixels flux. It includes flat-fielding and fringe-correction in $i$ and $z$ data.

The individual CCD images are then processed through the SNLS pipeline ([Ast13]). A weight map is generated that incorporates bad pixels information. The CCD image catalog is built using SExtractor ([Ber96]). A smooth sky-background map is subtracted. Each image is calibrated by performing a large diameter ($D = 15 \times \sigma_{\text{seeing}}$) aperture photometry on the tertiary stars catalog from [Reg09].

The *ugriz* 36-CCDs frames entering the stacking step are selected using cuts on : the photometric zero point, the mean seeing (FWHM<1.1" for *griz* bands, 1.3" in $u$), the sky variance, the mean star shapes, the number of saturated stars. About 60% of the best quality images are kept in *griz*, which corresponds to 300 to 400 36-CCDs frames.

### 4.1.2 Image stacking

Each field is observed during a “season” of 6 consecutive months. Depending of the field, there are 5 to 6 observing seasons available. The selected 36-CCDs frames are combined on a per-season basis to construct 1-square-degree *griz* “per-season” deep stacked images.

The individual CCD images, rescaled to a common photometric zero-point ZP=30, are co-added using SWARP V2.17.1 package[1] using the median-filter option, which permits to reject satellites and cosmic rays. A total weight map is also produced.

To avoid supernova light contamination, the supernova host photometry is performed on the “excluded-season” stacked images, obtained by co-adding all seasons but the supernova season stacks.

The deep catalog of the field galaxies, of-interest for the line-of-sight study, is built by using, for the sources detection and the photometry, the “all-seasons” stacked images, combining all the seasons stacks.

### 4.1.3 Photometry

Getting a very clean galaxy catalog is necessary for obtaining good photometric redshifts. Proper identification of all objects, especially bright stars and SN host galaxies (see following sections) is important.

The detection of the sources and the related photometry are performed using SExtractor ([Ber96]) in the dual-image mode, with a detection threshold level of 2.5-$\sigma$ for 3 contiguous pixels. The detection is realized on the $i$-frame and the photometry on the *ugriz* frames.

The magnitudes are estimated within SExtractor Kron ([Kro80]) elliptical aperture magnitudes (AUTO-magnitudes) providing a measurement of the galaxies total flux.

The summary of SNLS four fields catalogs is presented in table 4.1.

[1] http://terapix.iap.fr/soft/swarp/

| Field | effective surface square degrees | Number of galaxies | Number of galaxies per arc minutes squared | Limiting magnitude i-band (S/N>9) |
|---|---|---|---|---|
| D1 | 0.78 | 133476 | 48 | 24.8 |
| D2 | 0.8 | 129112 | 45 | 24.7 |
| D3 | 0.87 | 153520 | 49 | 24.9 |
| D4 | 0.77 | 117591 | 42 | 24.7 |

**Table 4.1**: Summary of SNLS four galaxy field catalogs

At the limiting magnitude $i \simeq 24.8$, it corresponds to a magnitude uncertainty of 0.12 mag.

### 4.1.4 Stars identification

Stars identification is important as they must be excluded from the galaxy catalogs, and bright star light contamination interferes with the detection and the photometry of galaxies in their vicinity.

Star identification is done on the $r$-band images.

#### Star cluster method

The star identification method relies on the 2-D profile intensity (PSF) characteristics of the stars and is described in [Ast13]. The Gaussian weighted first and second moments are iteratively calculated for each sources:

$$M_g = 2 \times \frac{\Sigma_{pixels}(x_i - x_c)(x_i - x_c)^T W_g(x_i) I_i}{\Sigma_{pixels} W_g(x_i) I_i} \tag{4.1}$$

$$W_g(x_i) = exp\left[-\frac{1}{2}(x_i - x_c)^T M_g^{-1}(x_i - x_c)\right] \tag{4.2}$$

where $x_i$ are the pixels coordinates, $x_c$ the Gaussian weighted centroid, and $I_i$ the sky subtracted image value at the pixel $i$.

The method is iterative. First a robust stars subset is identified, using the fact that stars tend to cluster in the $\sqrt{M_{xx}} - \sqrt{M_{yy}}$ plane because their characteristic sizes follow the PSF shape. Then a mean star shape model $M_{uu,\text{model}}$ ($u = x$ or $u = y$) is computed on this stars subset. Finally, using the model, the quantity $\Delta s_{uu} = \sqrt{M_{uu}} - \sqrt{M_{uu,\text{model}}}$ is estimated for every sources, and is used to refine the stars identification, as $\Delta s_{uu} \simeq 0$ for star-shaped sources.

This method can be implemented either on the collection of individual CCD images or the deep stacked frames. The method is illustrated on figure 4.1.

#### Identification on the collection of individual CCD images

All MegaCam single CCD images taken at different epochs undergo a pre-processing in the SNLS pipeline. The star identification process is based on some outputs of this pre-processing, particularly the image catalog of sources, for which the Gaussian-weighted second moments $M_{uu}$ are computed, and the image mean star shape $M_{uu,\text{model}}$ model.

We can then, for a given source, collect on every single CCD images it appears on its measurement of $\Delta s_{uu}$. If the source is a star, all its shape measurements will cluster around $(0, 0)$ in the $\Delta s_x - \Delta s_y$ plane. The source is selected as a star by requiring that its shape measurements cluster within a 2-$\sigma$ ellipse from the origin.

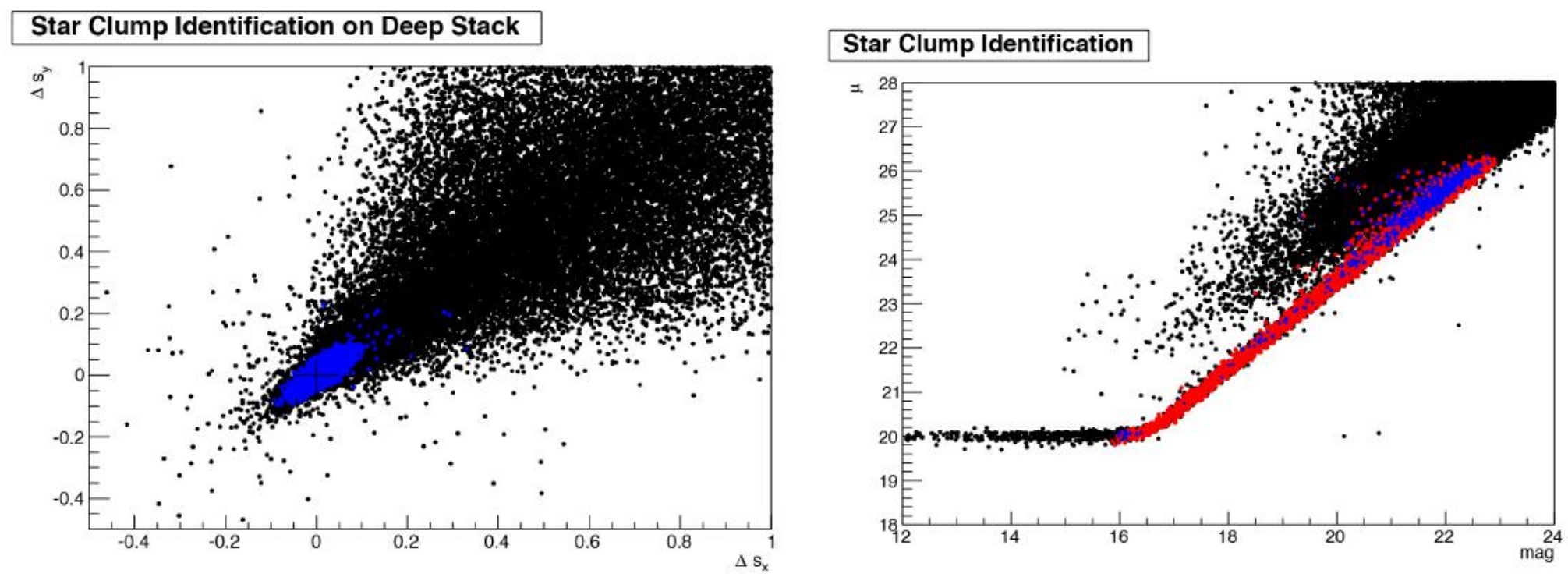

**Figure 4.1**: Right: Selection of the stars on the deep stacked frame, using their second order moments, which form a clump in the $\Delta s_x - \Delta s_y$ plane. The stars are selected in the blue 5-$\sigma$ ellipse centered at (0,0). Left: in the surface brightness ($\mu$) - magnitude plane, the stars selected using the “star clump” method. In red are the stars identified on the deep stack, in blue are the supplementary stars identified using individual CCDs. The bright stars ($r \leq 16$) are selected by neither methods.

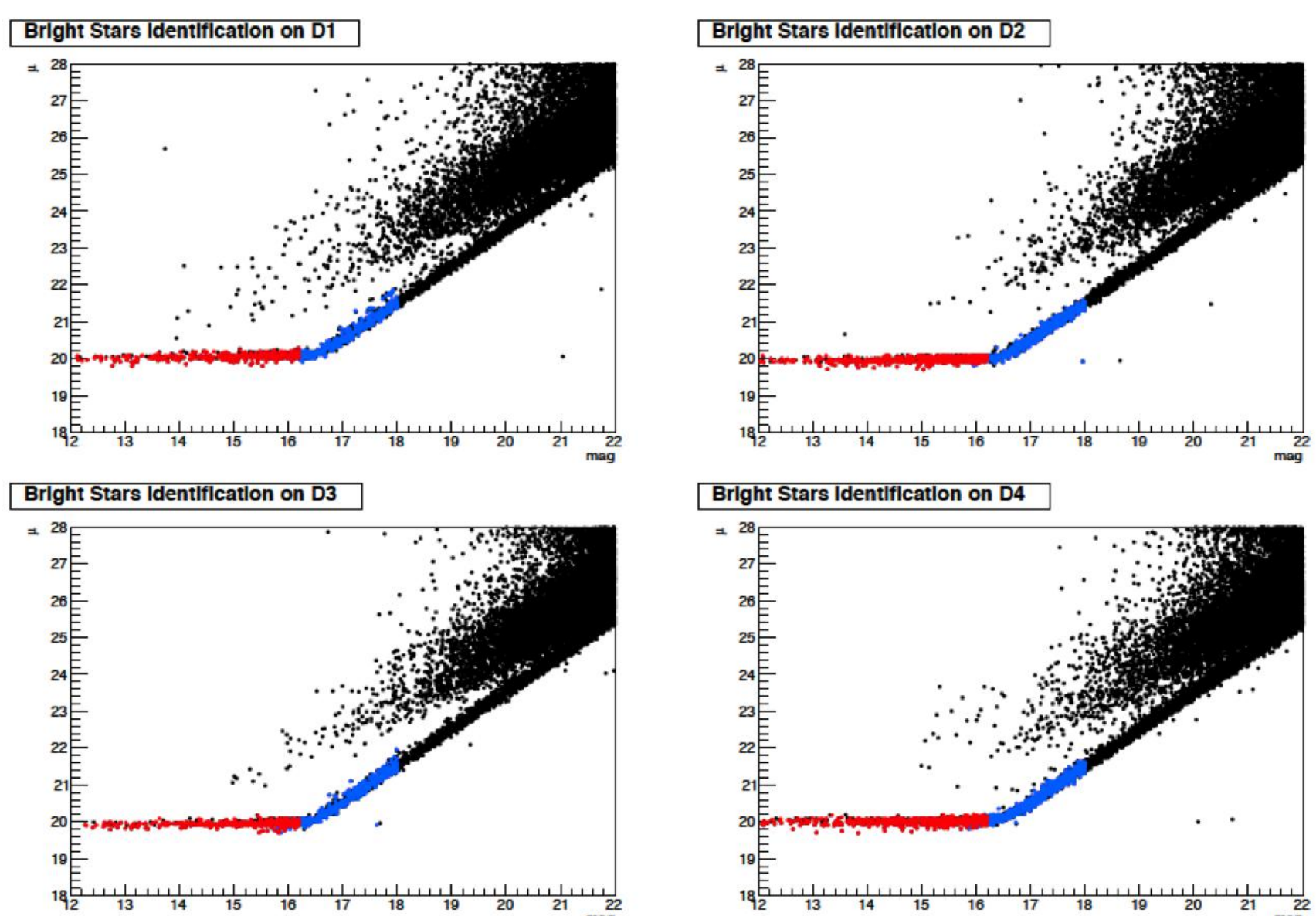

**Figure 4.2**: Stars catalog in the $\mu - m$ plane for the four fields. In blue: the stars selected by the star-clump method. In red: the bright stars selected on their magnitude and surface brightness.

#### Identification on the deep stacked frames

It is also possible to estimate the gaussian weighted second moments $M_{uu}$ for each sources throughout the $r$ 1-square-degree deep stacked frame. To take into account the PSF variation across the focal plane, the star shape model is computed on $9 \times 9$ sub-images, and interpolated by a $4^{th}$ order polynomial in $x$ and $y$ to obtain a star shape model valid across the whole frame.

Using this model, $\Delta s_u$ can be computed for each source detected on the stacked frame. In the $\Delta s_x - \Delta s_y$ plane, stars lies in a clump centered on the origin $(0, 0)$, and are selected within a 5-$\sigma$ ellipse (figure 4.1).

#### Star catalog completion

The star catalog produced by these two methods are identical within 90%. The result of both selection methods are presented on figure 4.1 in the magnitude ($r$) - surface brightness ($\mu$) plane, where the surface brightness is defined as the logarithm of the source peak value : $\mu = -3.5 \log_{10}(\text{peak}) + ZP$. It shows that bright stars ($r \sim 16$) are identified by neither of the two methods.

Indeed, bright stars are often saturated and one can't rely on their shape parameters for their identification. They can be however identified easily in the magnitude - surface brightness plane to complete the star catalog. For this we select: a) all sources with mag$= r \leq 17$ for which $(\mu - \text{mag}) < 4$; b) all sources brighter than $r \leq 16.2$ for which the peak value is above $\geq 80\%$ of the saturation level value [2] on the frame.

The bright stars selection for the four fields is shown on figure 4.2.

### 4.1.5 Catalog masking

Field edges and areas polluted by bright stars light makes it necessary to mask out some regions in the catalogs because they impaired the photometry accuracy.

The presence of a bright star is causing many defects : a halo (of a roughly constant size) around the star position, which are a consequence of the internal reflections in the optics ; "bleeding" pixels problem which arises when the electrons from the saturated pixel in the CCD sensor overflow to the surrounding pixels ; diffraction spikes due to the support rods of the camera. These spikes sizes depend on the star flux.

All the location and shape of those areas have to be listed, so that catalogs can be cleaned from objects in problematic regions.

#### Camera Edges

Removal of the 1-square-degree field edges are done straight away to each fields, excluding bout 8-9% of the field surface.

#### Stars haloes

Haloes positions and sizes had been thoroughly listed for K10 analysis, and masked out by a disk of large radius, positioned according to the star center, shifted according to the star location on the camera field, to take into account the halo spatial shift due to optic features. We have included this list in the SNLS5 mask.

[2] The saturation level is the minimum of the saturation levels of all the CCDs entering the stacking process. Its value ranges in the $r$ frame from 7000 to 8000 on the stack normalized to a $ZP = 30$.

#### CARS polygons ([Erb09])

To complete the haloes masking and take into account stars spikes, the Deep fields image masks from the CFHTLS-Archive-Research Survey (CARS, [Erb09]) were used.

They consist of a polygon list built using a semi-automatic procedure. First, SExtractor is run on a deep co-added image $r$ stack, with a fixed background set to zero and a very low detection threshold of $0.6\sigma$. Areas of significant over-densities and strong gradients in the object density distribution are identified and their polygonal boundaries are computed. Secondly, bright stars positions are identified, and cross-checked with published star catalogs (such as the standard stars catalog GSC-1 [Las90], GSC-2.3.2 [Mor01], [Las08] and USNO-A2 [Mon98]) and the extent of their central light halo and their cross-shaped diffraction spikes is modeled as a function of their apparent $r$ magnitude. The very bright stars haloes is modeled in a similar way as in K10, including their radial offset towards the camera center. Camera edges and satellite tracks are also included in the CARS mask as rectangles but we do not take them into account : the edges location is specific to the stacking procedure, and satellites are removed by construction from our deep stacked frames because of our using a median-filter co-adding algorithm.

A example of the CARS mask is presented on figure 4.3.

#### Bright stars

We make use of the bright stars identification realized in section 4.1.4 to complete the mask by vetoing disk areas around bright stars, which radii are set according to the stars magnitude $r$ : for $15 < r < 17$, $R = 75$ pixels, for $14 < r < 15$, $R = 100$ pixels, and for $r < 14$,$R = 150$ pixels.

An example of the masking procedure is presented on figure 4.3. The catalog masking reduce the field effective area by about 20% (see table 4.1).

## 4.2 Photometric redshifts

The galaxies photometric redshifts are estimated by using a template matching algorithm onto their broadband photometry, based on a set of synthetic spectra that are trained on spectroscopic data available on the D3 field. The photometric redshift procedure method is described in K10 and has not been modified for this analysis. The training was re-performed on the SNLS5 stacks photometry.

This work was done prior to this thesis. This part is based on [Kro10], and [Har12].

### 4.2.1 Spectral template sequence

The photometric redshift algorithm is based on an initial set of galaxies spectra which are used to construct a continuous spectral template sequence. The continuous spectral template sequence $F(a_\star, \lambda)$ is parametrized using a single parameter which is equivalent to the mean age of the stellar population $a_\star$, closely related to the galaxies rest-frame color : early type galaxies (elliptical) are redder than the late-type (spiral) galaxies which are bluer.

The spectra $F$ were calculated using the galaxy evolution model PEGASE.2 ([Fio99], specifying an initial zero-metallicity and a galaxy "exponential" law for the star-formation rate : $SFR \propto (t/\tau) \times \exp(-t/\tau)$ and a galaxy age also depending on $\tau$, so that the mean age of the stellar population ranges in 50 Myr to 13 Gyr, permitting to reproduce the observed galaxy colors.

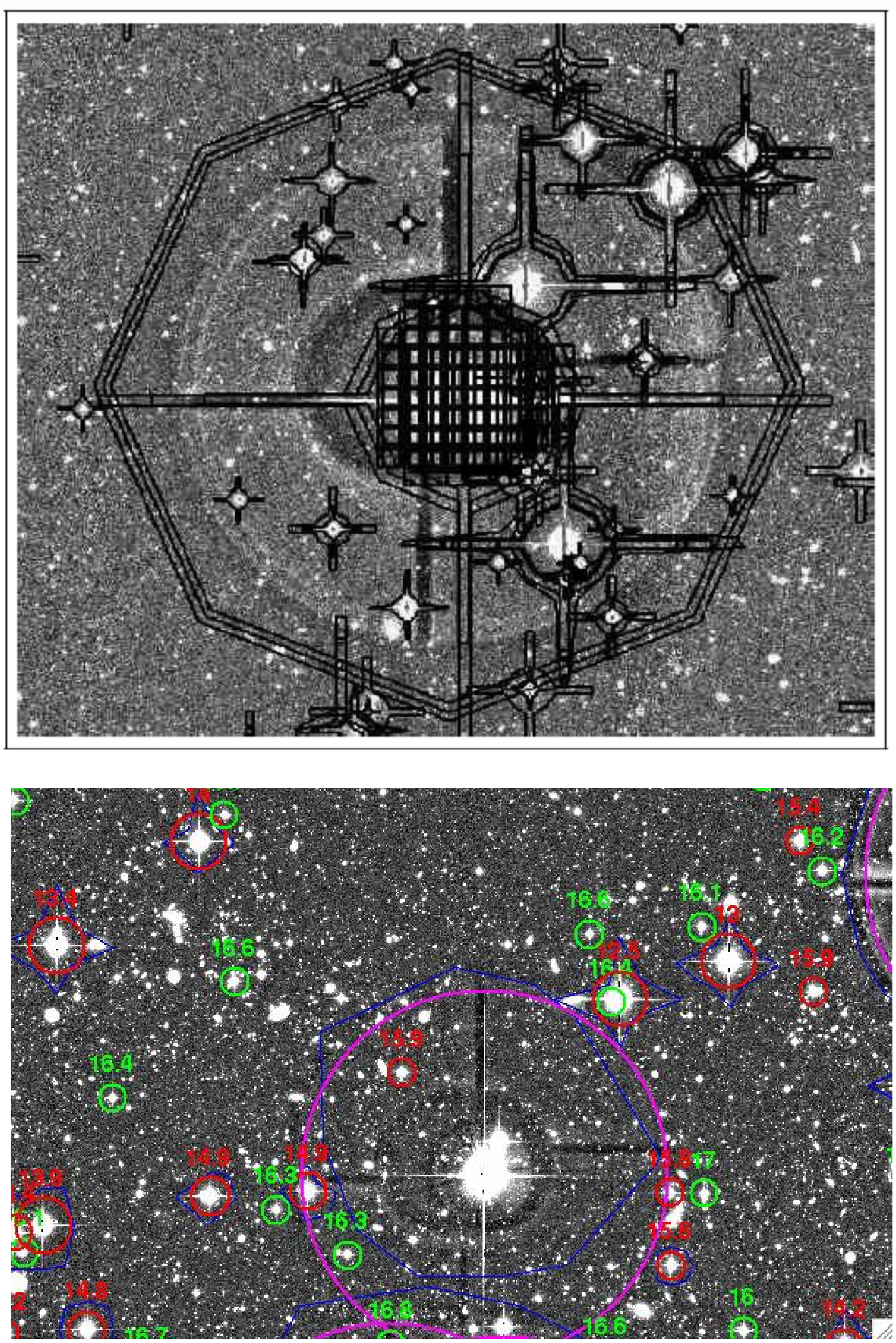

**Figure 4.3**: Upper : CARS mask for star haloes and diffraction spikes. Credit: [Erb09]. Lower : Final implementation of the masking procedure. In blue the polygons from CARS mask ([Erb09]). In magenta K10 bright stars haloes mask. In green and red circles, the bright stars mask, the radius is set according to the star $r$ magnitude.

### 4.2.2 Spectral template sequence training

The spectral template has to be optimized to described the data better. For this, the template are trained using the magnitudes of $\sim 6000$ galaxies in the D3 field catalog, with known spectroscopic redshift from the DEEP-2 spectroscopic survey ([Dav03, Dav07]).

The training is an iterative process. It consists of fitting for $a_*$ on the galaxy magnitudes through a $\chi^2$-minimization :

$$\chi^2(z, a_\star) = \sum_{\text{bands}\, b} \left( \frac{m_b(\text{obs.}) - m_b(z, a_\star)}{\sigma_b} \right)^2 \tag{4.3}$$

involving the magnitude residuals, i.e. the difference between the galaxy observed magnitudes $m_b$(obs. and the synthetic model magnitudes $m_b(z, a_\star)$ computed by redshifting and integrating $F(a_\star, \lambda)$. A third-order spline correction $f(a_\star, \lambda)$ to $F(a_\star, \lambda)$ is then computed, again by $\chi^2$-minimization, to minimize the magnitude residuals to the fitted model. A set of magnitude offsets is also computed in this iterative process (4.2).

| | u | g | r | i | z |
|---|---|---|---|---|---|
| Δm | -0.0033 | 0.0006 | 0 | -0.0141 | -0.0040 |

**Table 4.2**: Magnitude offsets which were computed during the training process.

The trained spectral $\mathcal{F}$:

$$\mathcal{F}(a_\star, \lambda) = F(a_\star, \lambda) \times f(a_\star, \lambda) \tag{4.4}$$

is presented on figure 4.4. The correction $f$ can be as high as 30% of the initial template value, alleviating the importance of the initial template set choice.

### 4.2.3 Photometric redshift accuracy

Once the photometric redshift spectral templates have been trained on SNLS5 photometry of Deep-2 galaxies, all four fields galaxies catalogs are processed, yielding a photometric redshift estimation for each galaxy.

To control the accuracy of the obtained photometric redshift, spectroscopic data available on D1, D2, D4 fields are used. D1 is centered in the Deep area of the VVDS field[Le 04], D2 is in the Cosmic Evolution Survey (COSMOS, [Sco07]) area (HST/ACS), and some field galaxies were targeted at the same time as the SN primary target during observations with FORS-2 in multi-slit mode in field D4.

The resolution of the photometric redshift is checked by computing the redshift residuals $\Delta z = z_{\text{photo}} - z_{\text{spectro}}$ and $D_z = \Delta z/(1 + z_{\text{spectro}})$, as well as the catastrophic error rate for $|D_z| > 0.15$, at a limiting magnitude $i = 24$.

The resolution is presented in table 4.3. It is comparable to the one published by [Ilb06] and [Ilb09] (dispersion on $\Delta_z$ of 0.03, catastrophic error rate of 3.7%).

Note that all spectroscopic redshifts, when available, supersede the photometric estimation in the final galaxy catalog.

### 4.2.4 Galaxy classification

The scaling relations of Tully-Fisher and Faber Jackson used for the luminosity-mass conversion of the galaxies require to classify galaxies in elliptical or spiral. Since morphological classification was not possible for the galaxies using the SNLS data, a color based classification system was used following K10, as elliptical galaxies are redder while spiral galaxies are bluer.

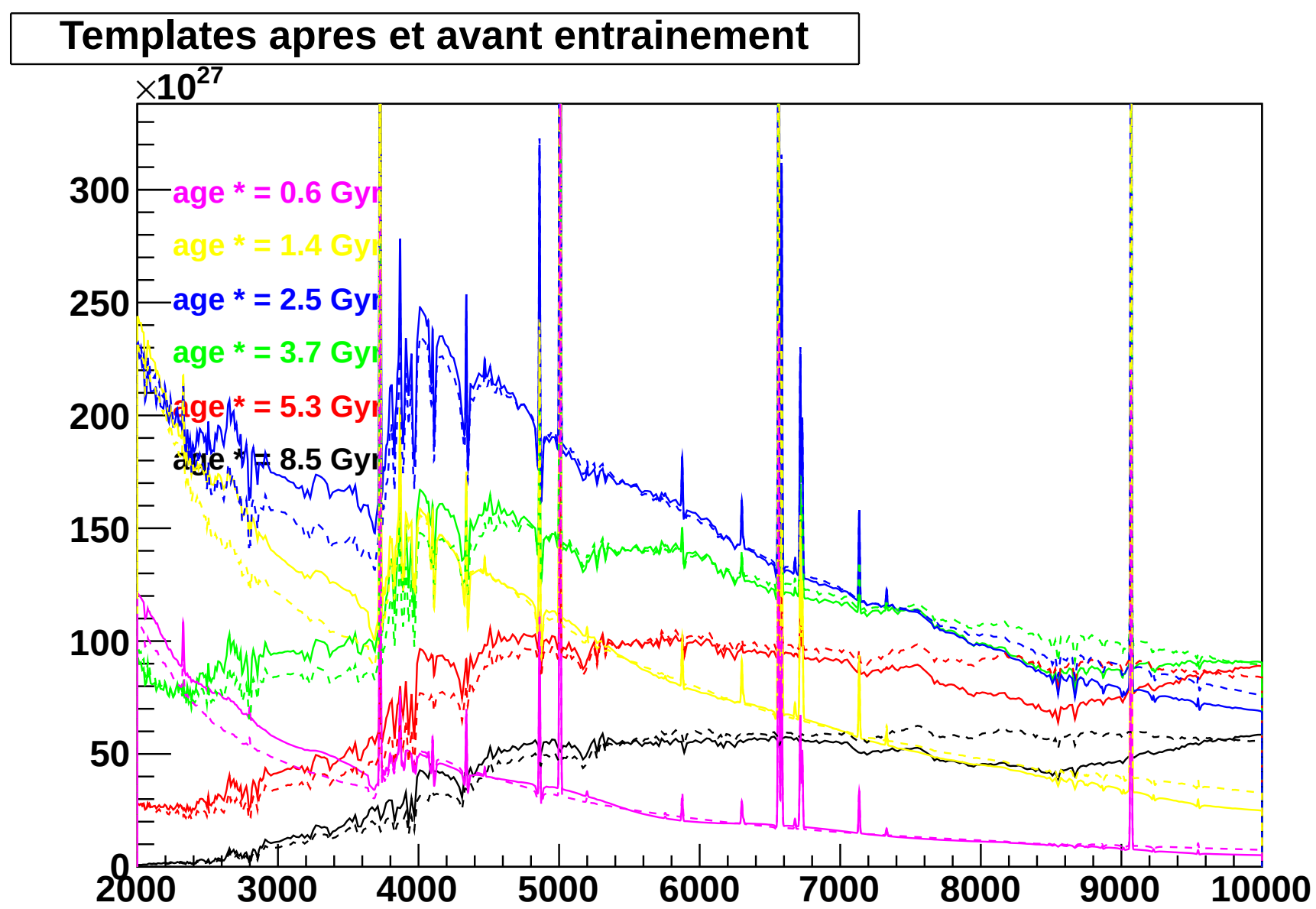

**Figure 4.4**: The trained spectral templates SED for different values of the mean stellar age $a_*$. The flux is in erg.Å$^{-1}$.s$^{-1}$.$M_\odot{}^{-1}$.

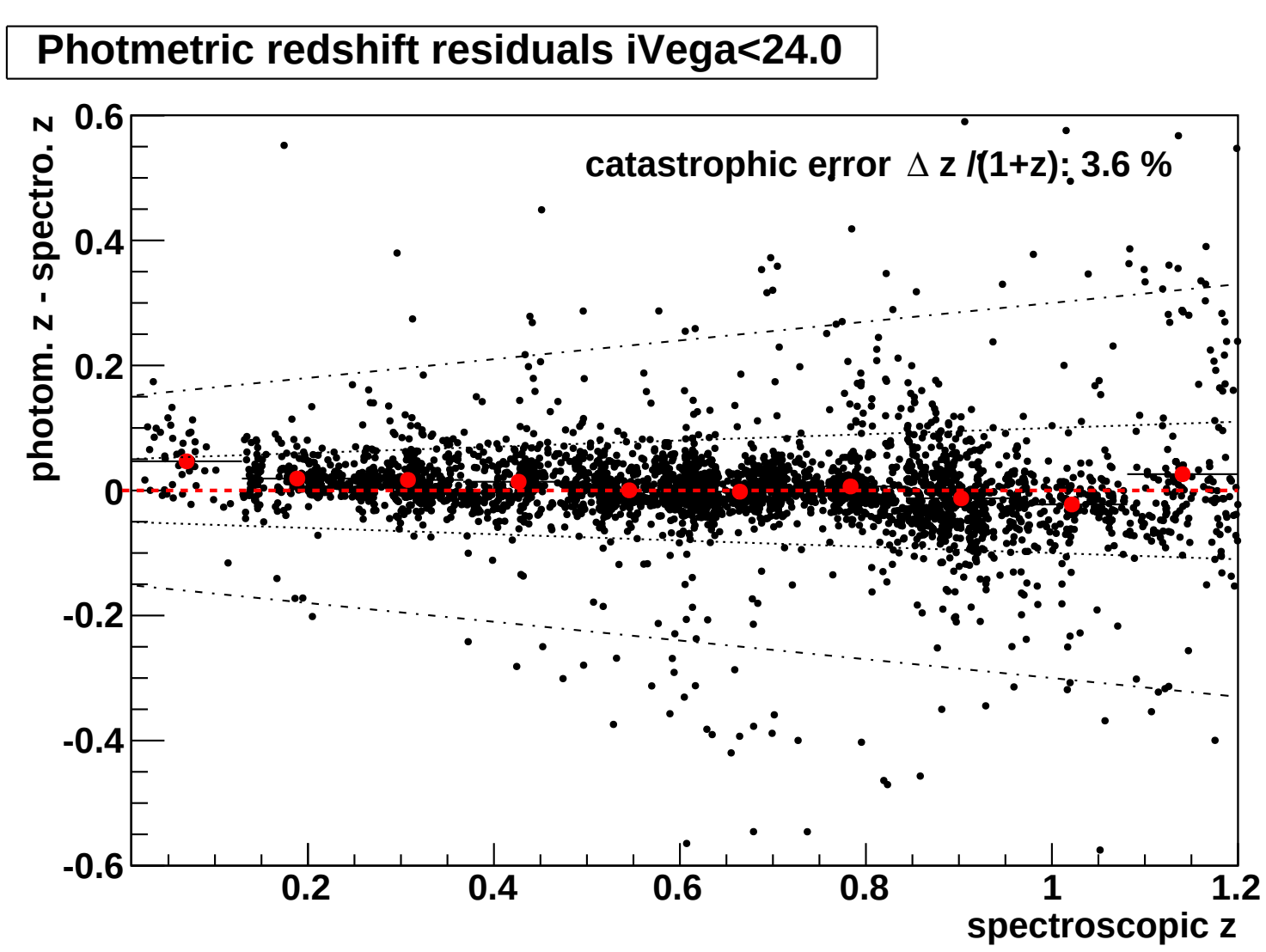

**Figure 4.5**: Redshift residual as a function of the spectroscopic redshift obtained from the VVDS data set [Le 04]. No redshift dependent systematic bias is seen.

| Field | N gal. | i lim. | % cata. | $\langle\Delta z\rangle$ | RMS($\Delta z$) | $\Delta z/(1+z)$ | RMS($\Delta z/(1+z)$) |
|---|---|---|---|---|---|---|---|
| D1 | 3691 | 24 | 3.62% | 0.0035 | 0.0551 | 0.0033 | 0.0323 |
| D2 | 5537 | 24 | 2.85% | 0.0076 | 0.0474 | 0.0062 | 0.0295 |
| D3 | 6006 | 24 | 3.17% | 0.0091 | 0.0544 | 0.0064 | 0.0311 |
| D4 | 200 | 24 | 3.06% | 0.0169 | 0.0339 | 0.0119 | 0.0253 |

**Table 4.3**: The resolution of the photometric redshift computation in the four fields for galaxy with a Vega magnitude $i < 24$. The catastrophic error rate at $\Delta z/(1+z) > 0.15$ is also indicated.

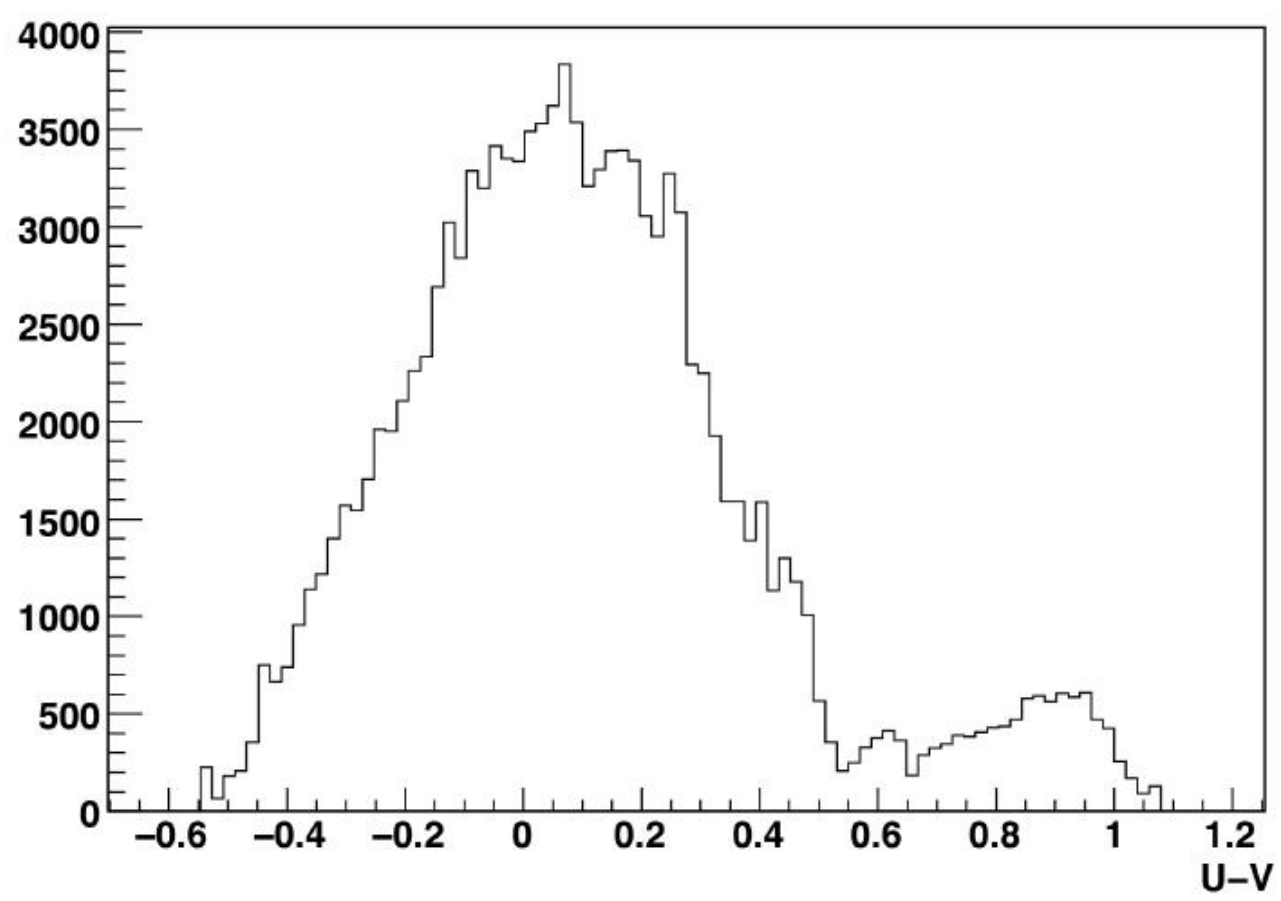

**Figure 4.6**: Distribution of the galaxies rest-frame U-V, which is used to separate the galaxy types into red and blue : below (resp. above) $U - V = 0.54$ lay the bluer spirals (resp. redder elliptical) galaxies. Image: K10

For this computation, the rest frame color (U-V) is estimated with the best-fit template obtained in the photometric redshift computation. The distribution of galaxies $U - V$ rest-frame color permits to separate clearly two populations of red and blue galaxies at $U - V = 0.54$ (see figure 4.6).

## 4.3 The SNLS5 supernovae sample

We will consider here the 439 SNe Ia identified as such in the SNLS5 data base. Note that the cosmology analysis will apply stringent cuts on their light-curve quality and parameters, which will reduce this sample to 389 SNe.

In SNLS3, out of 233 SNe available, only 171 were used for the analysis, and the visual inspection of the sample was important. For the SNLS5 sample was developed a more systematic scoring method.

The D2 field additional HST images were used for a supplementary visual check.

### 4.3.1 SN host galaxies identification

To compute a supernova magnification, it is necessary to exclude its host galaxy from the line-of-sight galaxy selection.

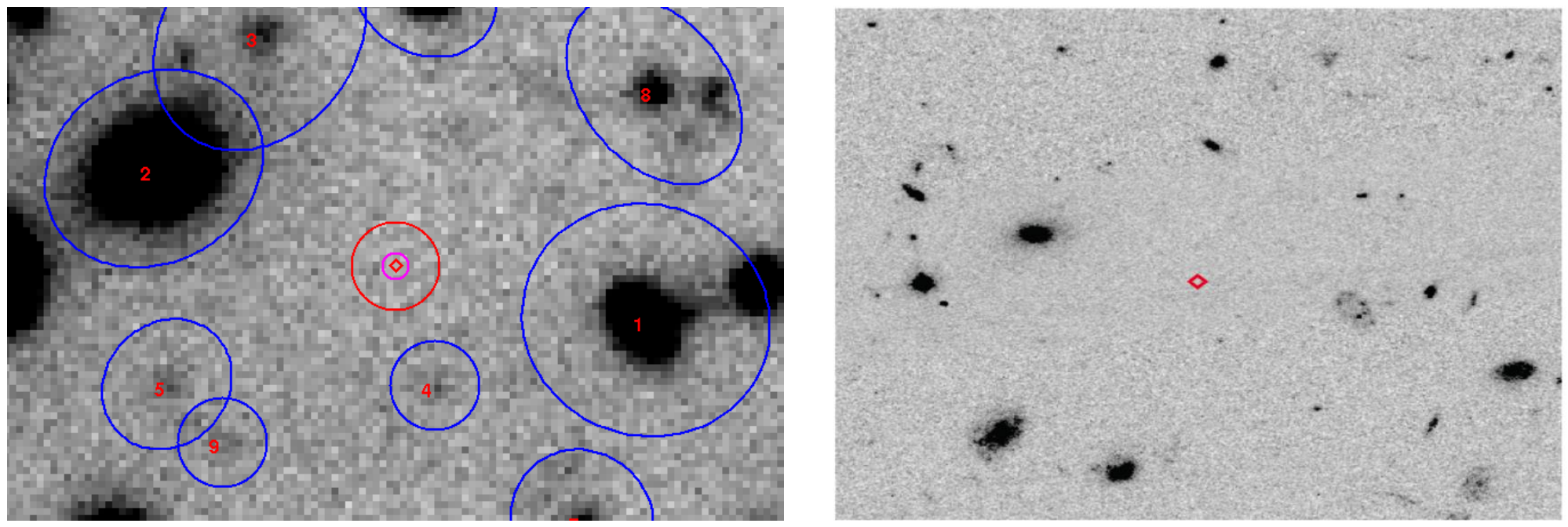

**Figure 4.7**: SN 04D2iu at $z = 0.7$ with no identified host. Left : deep stacked $i$ image used for the host identification. Right : HST image of the same location.

The supernova host identification procedure was primarily set so as to obtain stringent criteria, insuring the accuracy of the derived host galaxy characteristics, such as its stellar mass, that are taken into account in the cosmology fit. As a consequence, a dubious identification of the supernova galaxy implied that it was problematic and tagged as having "no host information". For the lensing analysis, we will pay further attention to these problematic cases, which sometimes correspond to interesting potential lensing situation. These criteria were defined on the SNLS3 supernovae sample prior to this thesis work.

Firstly, the supernova host identification is based on the normalized elliptical distance $d$, so that $d < 1$ corresponds to the elliptical aperture, used to estimate the galaxy AUTO-magnitude by the SExtractor software. The equation of the $d = 1$ ellipse defining the photometry aperture reads (in a coordinate system centered on the galaxy position) :

$$ax^2 + by^2 + cxy = d^2$$

where the ellipse is computed using the galaxy second-order moments, scaled with the KRON factor defined in SExtractor to take into account the galaxy light profile. The supernova host galaxy is identified as the galaxy within which the supernova lays at the closest normalized elliptical distance $d$. When the closest galaxy lays at $d > 1.8$, the supernova is declared has having "no host". Having the *no-host* label results in the SN to be assigned in the lowest host stellar mass bin for the cosmology analysis. An example is presented on figure 4.7 where the D2 field HST image is shown along with the deep stacked $i$ image used for the host identification.

Supplementary checks are performed so as to detect problematic cases :

- SNe for which the identified host may be polluted by a nearby bright star are rejected.
- We also require that the closest galaxy photometric redshift be consistent with the supernova spectroscopic redshift : $\Delta z/(1 + z) < 0.15$.
- In the case of many galaxies laying close to the supernovae within $d < 1.8$, the galaxy with the closest photometric redshift is selected as the host. If more than one of these close galaxies meet the criteria $\Delta z/(1 + z) < 0.15$, then the supernova is declared has having a "dubious host information" and tagged as such.

When the SN host identification is problematic, the SN distance modulus computed for the cosmology analysis is assigned a supplementary error, as the SN cannot be assigned any host stellar mass bin.

These criteria ensuring a consistent cosmology analysis results in $\sim 6\%$ of the SNe being declared as having no host galaxy detected on the deep image stacks, and $\sim 87\%$ as having a detected host galaxy. The remaining SNe are either dubious or problematic cases. This classification, indicated in table 4.4, is furtherly refined for the lensing analysis.

| Host galaxy identification | N SN | % |
|---|---|---|
| All | 439 | 100 |
| problematic (all) | 30 | 6.8 |
| polluted photom. | 10 | 2.3 |
| inconsistent photo.-z | 17 | 3.9 |
| dubious host id. | 9 | 2.1 |
| no id. host | 27 | 6.2 |

**Table 4.4**: Summary of SNLS5 SNe Ia host galaxy situation classification. Note that 6 SNe classified as having a dubious host identification are also classified as having a redshift inconsistent with the closest galaxy photometric redshift.

### 4.3.2 Supernovae scoring for the lensing analysis

SNe classified for the cosmology analysis as not problematic are kept for the lensing analysis, and set to a score = 1. The 27 SNe in a *no-host* situation are assigned a score = 4 and are also kept for the lensing analysis.

The *polluted* photometry situation are not kept and assigned a score = 2. The check for the presence of bright stars will moreover be furtherly performed in a thorough automatic way that is presented in section 4.3.3.

The *dubious* cases classification are re-examined, as some may indicate an interesting lensing situation, such as the presence of a foreground galaxy lying close to the line-of-sight. If the closest galaxy photometric redshift $z_{\rm gal}$ is confirmed by [Cou09] published estimation, the following cases are distinguished :

- if $z_{\rm gal} < z_{\rm SN}$: the galaxy is situated in-front-of the SN, so that the SN score is set to score=3 and the SN is kept. Examples of this situation are shown on figure 4.8.

- If $z_{\rm gal} > z_{\rm SN}$, then the galaxy is behind the SN and will not be included in the line-of-sight modeling anyway. The score is set to score= 1.

- if the closest galaxies lies within $z_{\rm gal} \simeq z_{\rm SN}$, they will not be included in the line-of-sight modeling anyway either. The score is set to score= 1.

As a result of these further checking, all *dubious* cases were kept with a score= 3 or = 1. This classification, presented in table 4.5, selects 425 SNe, which line-of-sight will be furtherly checked for imagery problems, such as the bright-star light pollution.

### 4.3.3 Line-of-sight selection and masking

To compute a SN magnification, the involved galaxies are selected on their radial distance $R$ to the SN line-of-sight. In principle all galaxies would have a lensing effect on a given SN, but a reduced sample of galaxies is chosen at a maximum radial distance $R_{\rm LOS}$. To set $R_{\rm LOS}$ value, the total magnification as a function of the angular radius $R$ has been studied in K10 analysis, concluding that a radius cut off at :

$$R_{\rm LOS} = 60 \text{ arcsec.}$$

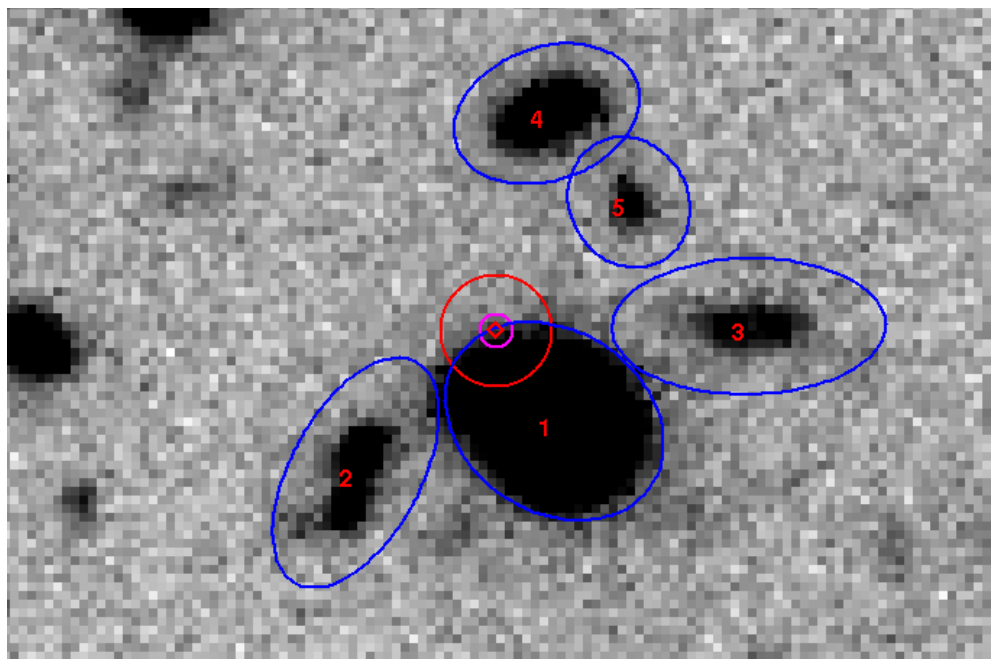

(a) SN 04D3cp at $z = 0.83$, the closest galaxy is at $z = 0.29$.

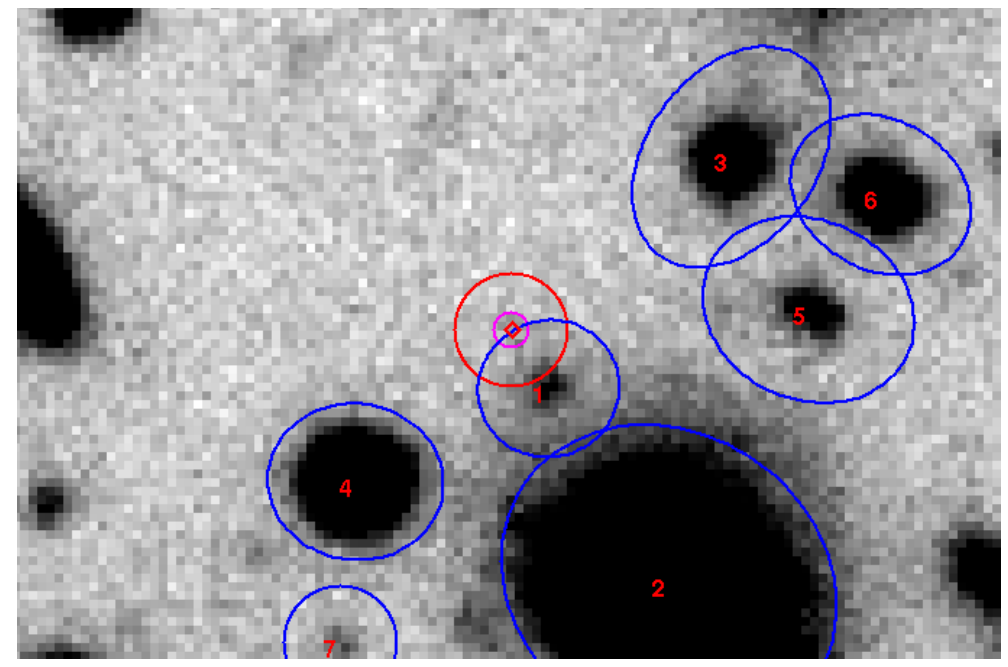

(b) SN 07D3bt at $z = 0.91$, the closest galaxies are at $z = 0.32$ and $z = 0.35$.

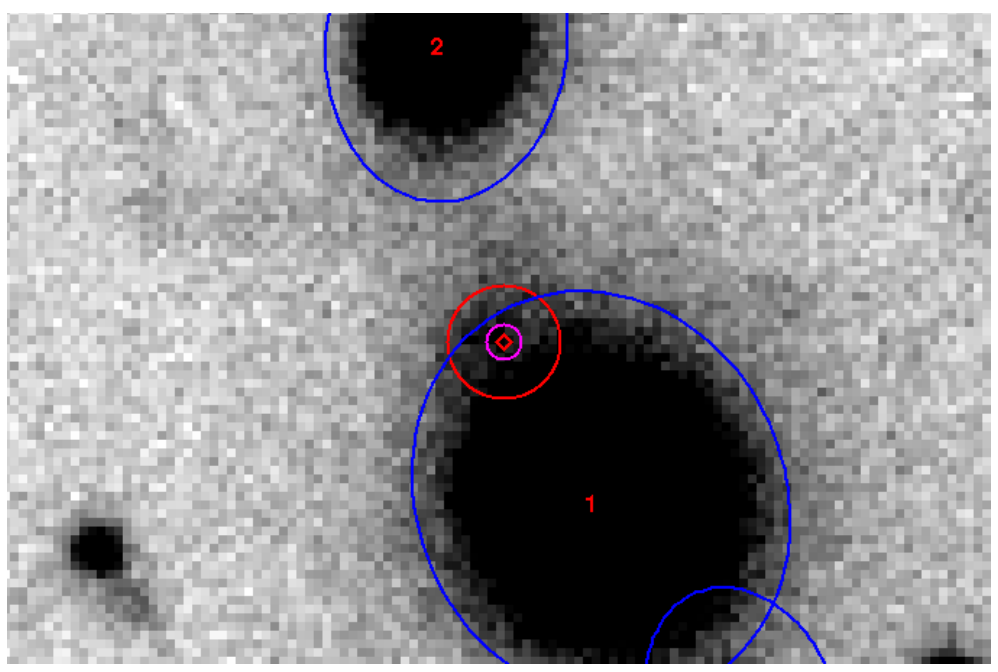

(c) SN 05D3km at $z = 0.96$, the closest galaxy is at $z = 0.43$.

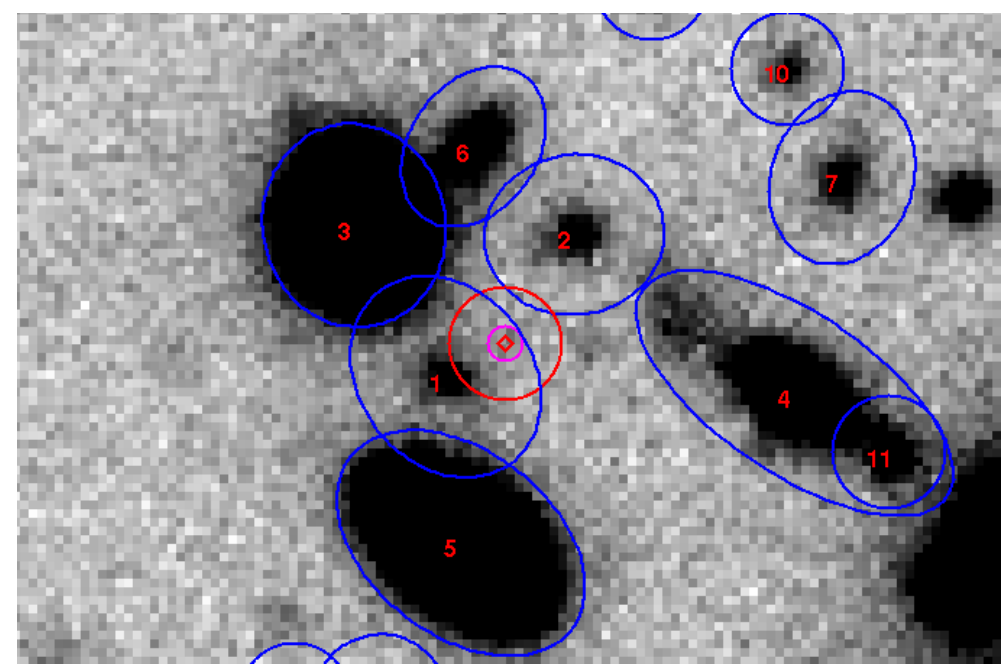

(d) SN 04D4bq at $z = 0.55$, the closest galaxies are at $z \sim 0.35$.

**Figure 4.8**: Examples of SNe for which the closest galaxy is a foreground galaxy on the line-of-sight (score=3).

| Selection for lensing analysis | N SN | % |
|---|---|---|
| All | 439 | 100 |
| selected (all) | 425 | 96.8 |
| closest gal. is a foreground gal. | 6 | 1.4 |
| no host gal. is identified | 27 | 6.2 |
| problematic | 14 | 3.2 |

**Table 4.5**: Summary of SNLS5 SNe Ia lensing situation classification. The selected SNe line-of-sights will then be furtherly checked for imagery problems, such as the bright-star light pollution.

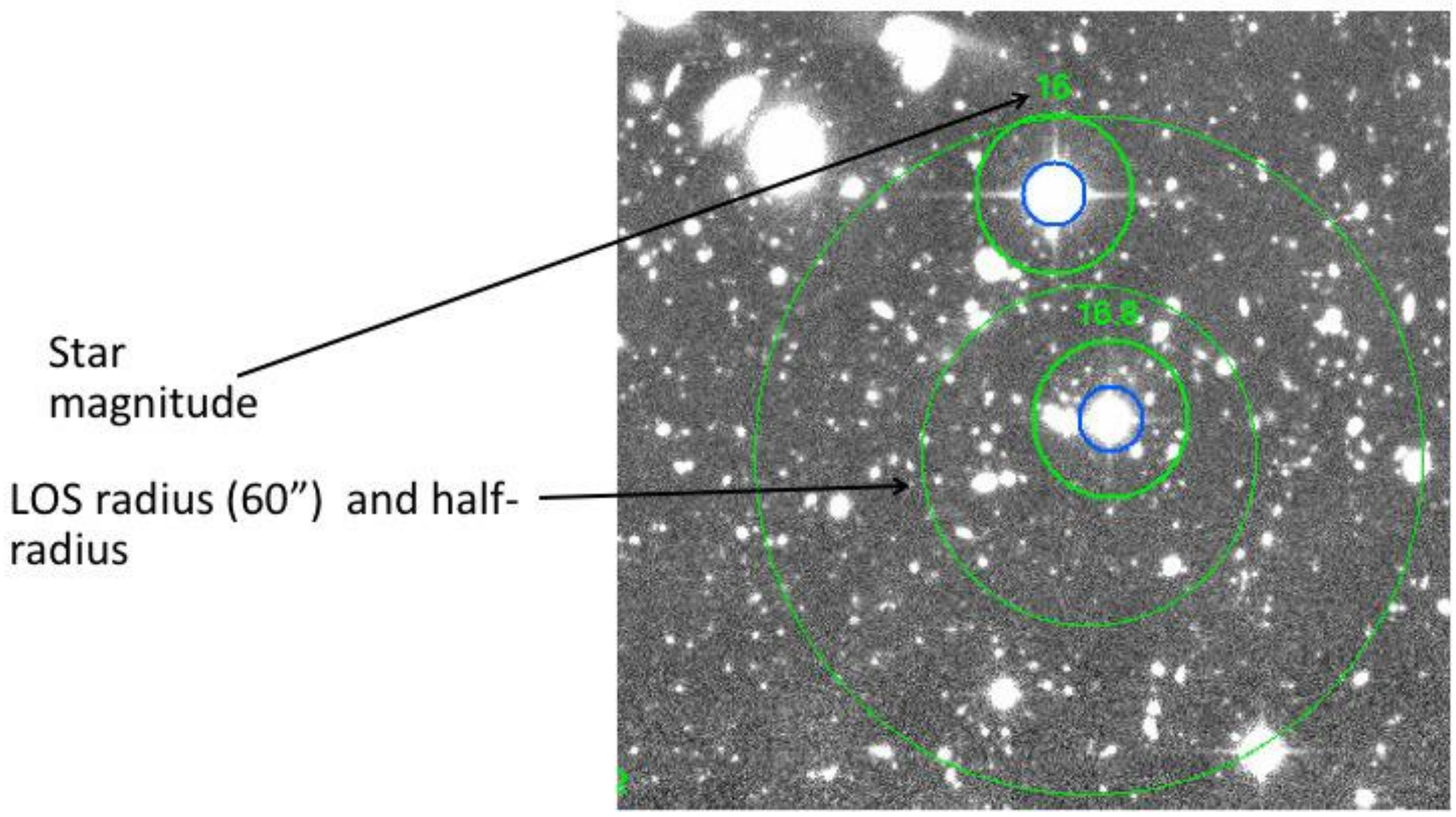

**Figure 4.9**: Line of sight inner disk and outer annulus at $R_{\mathrm{LOS}} = 60$ arcsec and $R_{\frac{1}{2}\mathrm{LOS}} = R_{\mathrm{LOS}}/2$, defined for the fractional contamination estimation.

resulted in a corresponding error of less than 1% loss on the total magnification. For a given SN, we will then take into account all galaxies situated at a radial distance less than $R_{\mathrm{LOS}}$, and of course in front of the SN.

The line-of-sight defined this way can include bright stars or polluted area on the image. It is thus necessary to identify and reject them. For this, we have implemented an algorithm aiming at producing in a systematic way a cleaner line of sight. The technique relies on the use of both the mask defined in section 4.1.5, that includes the camera edges, very bright stars haloes, and bright star luminous disks and spikes, and the bright stars list that is entering the mask construction.

For a given SN we define two zones in the LOS disk of radius $R_{\mathrm{LOS}}$ (see fig. 4.9) : an inner disk of radius $R_{\frac{1}{2}\mathrm{LOS}} = R_{\mathrm{LOS}}/2$, completed by an outer annulus between radius $R_{\frac{1}{2}\mathrm{LOS}}$ and $R_{\mathrm{LOS}}$. Pixels in these two zones that are flagged in the mask are vetoed, and a cut is apply on the number of vetoed pixels. For the D2 and D4 (resp. D1 and D3) fields, the fraction of vetoed pixels must be less than 20% (resp. 30%) in the inner disk and less than 15% (resp. 20%) in the outer annulus. The differences in these two sets of cuts is essentially due the intrinsic size of the [Erb09] polygons.

Furtherly, a supplementary cut is applied, based on the distance from the bright stars to the center of the line-of-sight. All SNe LOS containing a bright star (magnitude $r < 17$) within the inner disk, extended to 1/4th of the bright star luminous disk masking radius, is rejected.

The masking cuts were set to obtain a compromise between limiting the rejection rate and obtaining a cleaner sample. In K10 analysis, about 30% of the sample was semi-automatically (automatic masking and visual inspection) rejected, due to bright star luminous contamination or various problems. With the criteria defined above, about $\sim 40\%$ of the SNe are automatically rejected, permitting to obtain a clean sample for this first analysis of the SNLS5 data.

## 4.4 Magnification computation

We describe here the hypothesis and input on which the SNLS5 SNe magnification computation is based.

### 4.4.1 Weak lensing approximation

Lensing magnification values in K10 analysis were calculated using the publicly available Fortran software called Q-LET (quick lensing estimation tool) by Christopher Gunnarsson [Gun05], [Gun04]. Q-LET uses the multiple plane method described in section 3.1.6. It traces the light-rays recursively from the image plane to the source plane, computing the multiple deflections along the light path. K10 used the truncated Singular Isothermal Sphere model and specified the filled beam approximation i.e. that the universe is homogeneously filled with matter as described by the cosmological parameters $h_0 = 0.7$, $\Omega_m = 0.27$, $\Omega_\Lambda = 0.73$, with superimposed dark matter haloes of galaxies. The obtained magnification is always greater than 1., which implies to furtherly apply a normalization factor, as discussed in section 4.5.

Because most line-of-sight are passing away of the intervening haloes, we made use in this analysis of the weak approximation, as suggested in [Jon10]. As the convergence $\kappa \ll 1$, the magnification can be approximated by :

$$\mu = \frac{1}{\left((1-\kappa)^2 - |\gamma|^2\right)} \simeq 1 + 2\kappa \tag{4.5}$$

The contributions of all the haloes met along the line of sight can be cumulatively added to obtain the total magnification :

$$\mu \simeq 1 + 2\sum_k \kappa_k \qquad \text{(weak approximation)}$$

[Jon10] used the same SNLS SNe Ia sample as K10 to to investigate the properties of dark matter haloes of galaxies. They advocated that the weak lensing approximation has only a 5% margin of deviation from the ray tracing algorithm for magnification computation.

In what follows, we will compare the ray tracing computations from Q-LET and the approximated computation used for this analysis.

### 4.4.2 Galaxy halo models

For this thesis, we developed a dedicated C++ code for the computation of the convergence of the intervening haloes on the line of sight, which permitted to compute each supernova magnification in a cumulative way. We have set the cosmological parameters to $h_0 = 0.7$, $\Omega_m = 0.27$, $\Omega_\Lambda = 0.73$. All the magnification values are given with respect to a homogeneous Universe and the distances are computed with the filled beam approximation. The impact parameter $\xi$ is computed from the radial distance of the given galaxy (lens) center to the supernova (source).

The halo galaxy model is parametrized within the SIS model frame. The velocity dispersion $\sigma$ of each galaxy is computed from their absolute magnitude $M_B$, using the Tully-Fisher ([Boe04]) and Faber-Jackson ([Mit05]) redshift dependent scaling laws $\sigma(M_B, z)$ as given in eq. 3.33 and 3.36. The color cut $U - V = 0.54$ permits to separates galaxies between spiral and elliptical-likes. The value of the velocity dispersion is important as it gives the lensing strength of the corresponding halo.

From there, the virial radius $r_{200}$ and mass $M_{200}$ are estimated from the velocity dispersion $\sigma$ using eq. 3.24 :

$$r_{\text{vir}} = \frac{2}{\sqrt{\Delta}} \frac{\sigma}{H_0 E(z)}$$

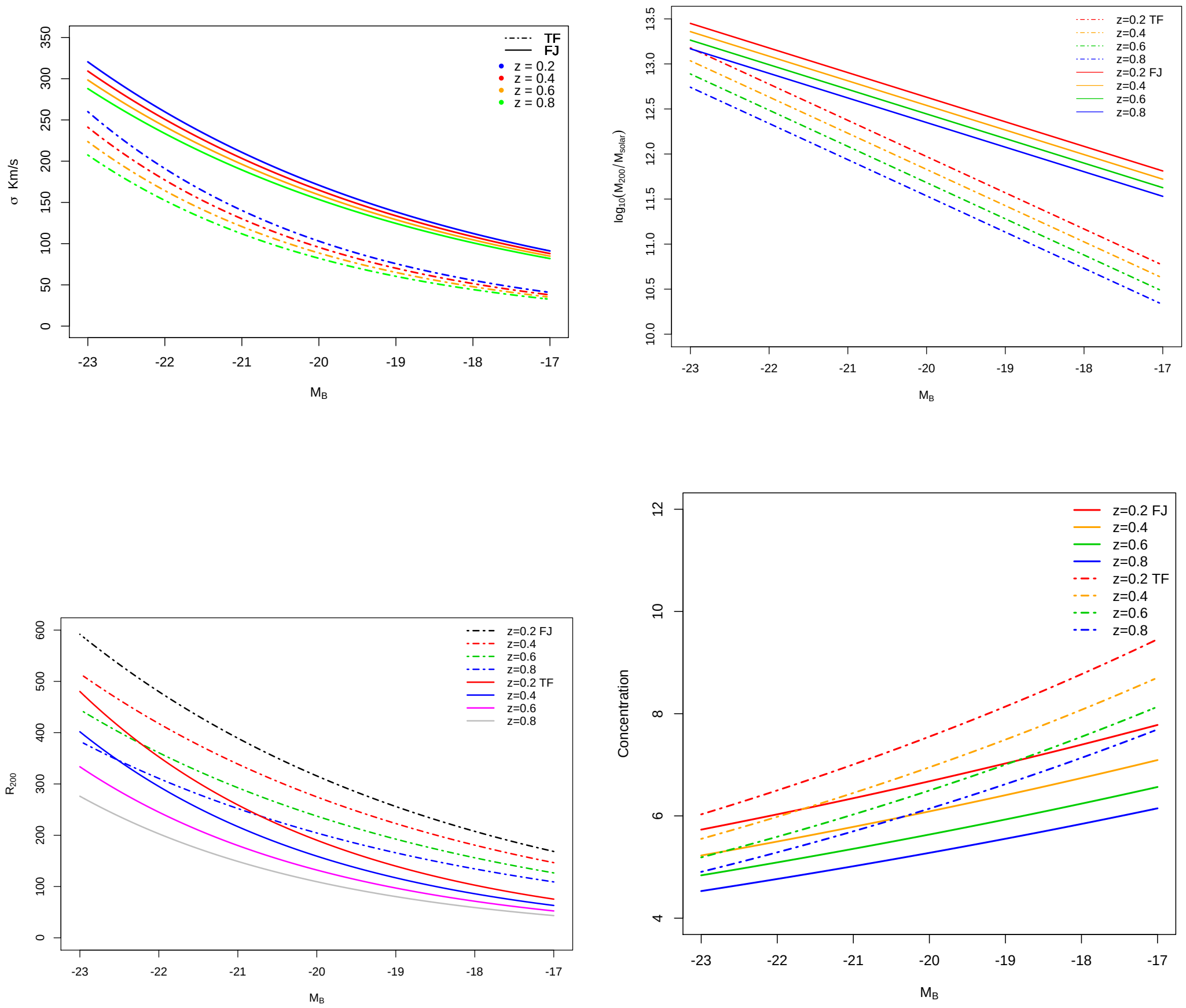

**Figure 4.10**: Parametrization of the halo models : the SIS velocity dispersion is modeled with the TF and FJ relations from [Boe04] and [Mit05]. The velocity dispersion is then converted into a virial mass $M_{200}$ and radius $r_{200}$ values. For the NFW concentration parameter $c$, we used the published [Duf08] $c(M_{200}, z)$ law, converted into a $c(M_B, z)$ law.

and the relation for a SIS halo mass (cf. eq. 3.23) : $M(r) = (2\sigma_v^2/G)\,r$. The scaling laws for the velocity dispersion $\sigma(M_B, z)$ and the resulting relations for the virial mass $M_{200}(M_B, z)$ are presented in fig. 4.10.

### Truncated SIS halo model

We implemented for the analysis the truncated SIS model, with a truncation radius set at the virial radius value. Following the formalism of section 3.2.2, we define $x = \xi/\xi_0$ and $x_t = r_{200}/\xi_0$, with the length scale $\xi_0$ as defined in eq. 3.25.

The truncated SIS convergence reads :

$$\kappa(x) = \begin{cases} \frac{1}{\pi x} \arctan \frac{\sqrt{x_t^2 - x^2}}{x} & (x \leq x_t) \\ 0 & (x > x_t) \end{cases}$$

### Truncated NFW halo model

We also implemented for this analysis the truncated NFW model. We used $M_{200}$(SIS) computed from the velocity dispersion in the frame of the SIS model, as described above. We did not use the NFW profile to convert the velocity dispersion into the virial mass because of the complexity of this relation: the velocity dispersion depends on radius for a NFW profile. However, [Gun04] checked that when using an input velocity of $\sigma = 170$ km/s to get the parameters for the NFW and then computing the corresponding NFW velocity dispersions, the obtained values vary roughly between 5% to −15% around 170 km/s between radii $0.1r_{200}$ to $2r_{200}$. Using $M_{200}$(SIS) to parametrize the NFW model provides moreover the possibility to compare easily and consistently the two models.

For the concentration parameter $c$, we relied on [Duf08] law $c(M_{200}, z)$ as presented in eq. 3.28, which translates into a $c(M_B, z)$ relation when using the TF and FJ $\sigma(M_B)$ scaling laws. The law is presented on figure 4.10.

We set again the truncation radius at the virial radius value. Defining now $x = \xi/r_S$ and $x_t = r_{200}/r_S = c$, the convergence is given by :

$$\kappa(x) = 2\kappa_s F(x), \quad \kappa_s = \frac{\rho_s r_s}{\Sigma_{cr}} = \frac{M_{200} c^2}{4\pi r_{200}^2 f(c)} \times \frac{1}{\Sigma_{cr}} \tag{4.6}$$

where $f(c) = \ln(1+c) - c/(1+c)$ as previously defined. The function $F(x)$ is provided by eq. (27) in [Tak03]:

$$F(x) = \begin{cases} -\frac{\sqrt{x_t^2 - x^2}}{(1-x^2)(1+x_t)} + \frac{1}{(x^2-1)^{3/2}} \operatorname{arcosh} \frac{x^2 + x_t}{x(1+x_t)} & (x < 1 < x_t) \\ \frac{\sqrt{x_t^2 - 1}}{3(1+x_t)} \frac{x_t + 2}{x_t + 1} & (x = 1 < x_t) \\ -\frac{\sqrt{x_t^2 - x^2}}{(1-x^2)(1+x_t)} - \frac{1}{(1-x^2)^{3/2}} \arccos \frac{x^2 + x_t}{x(1+x_t)} & (1 < x \leq x_t) \end{cases}$$

Note that this expression differs from the relation proposed in [Gun05] for $x < 1$. Both relations from [Gun05] and [Tak03] are strictly equivalent for $x > 1$ though.

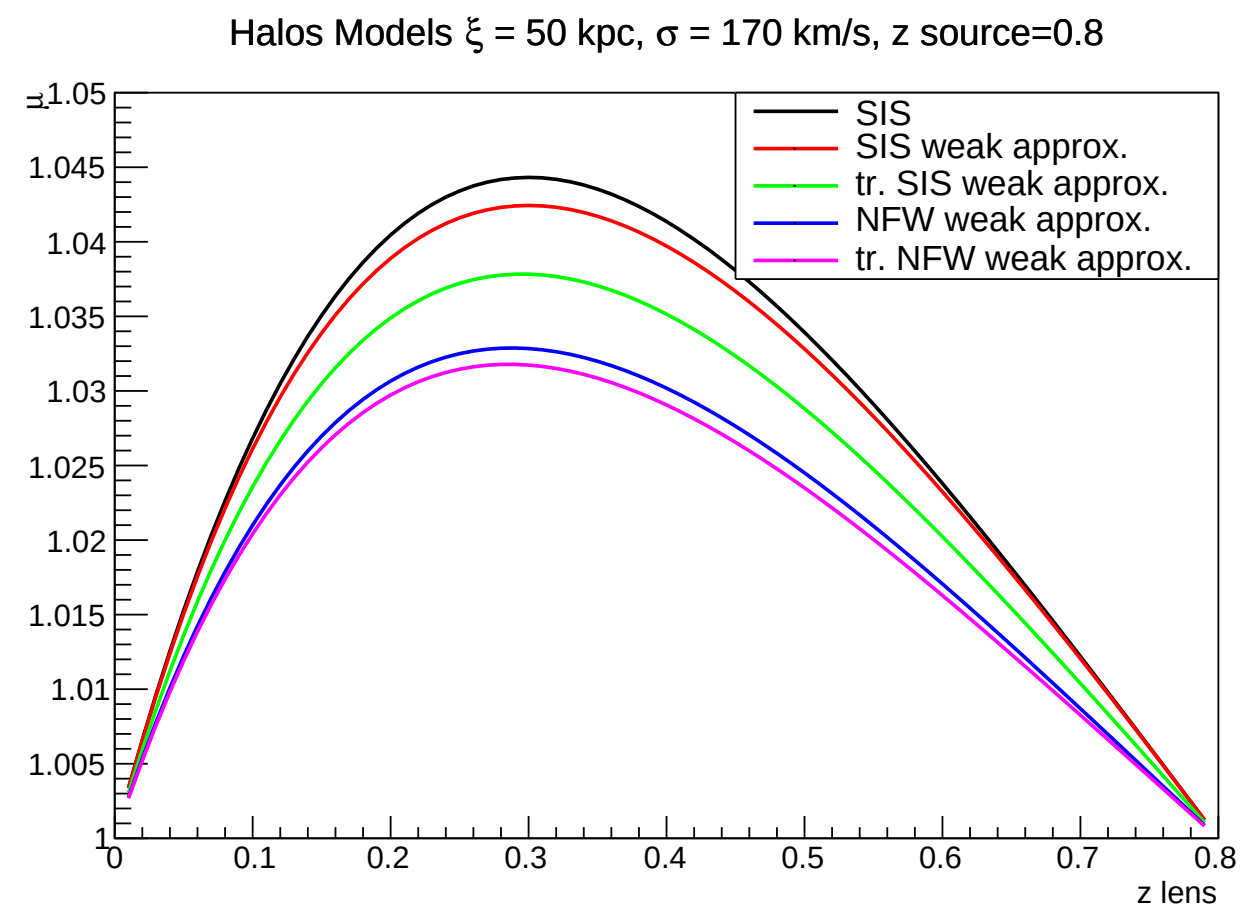

**Figure 4.11**: Magnification computation for a typical lensing situation and different models.

### Comparison of the magnification computation models

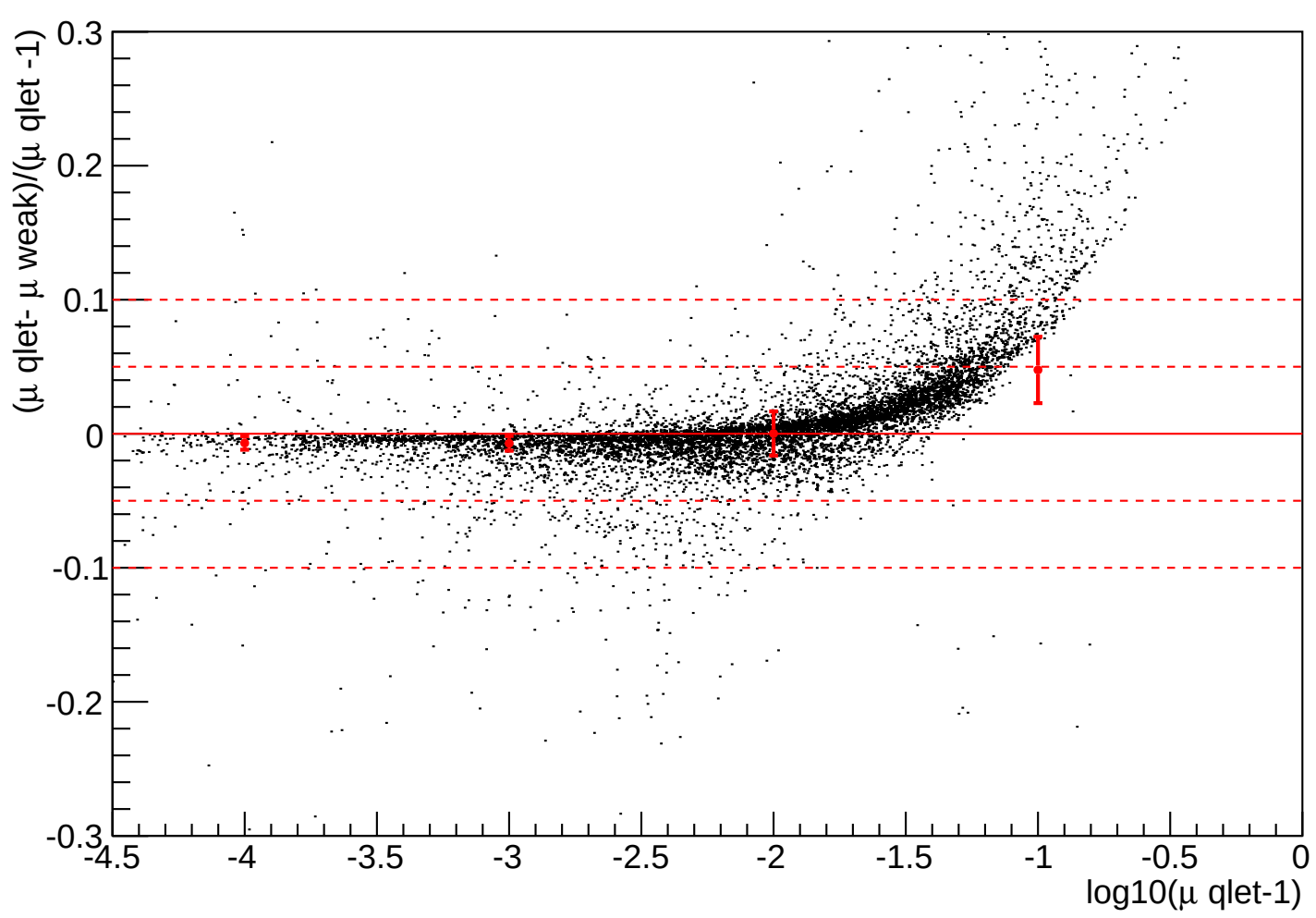

**Figure 4.12**: Comparison between ray tracing (Q-LET) and the weak lensing approximation adopted for this work for a truncated SIS model.

Different computation for a typical case are presented on figure 4.11. The impact parameter is set at $\xi = 50$ kpc, and the source is at $z_S = 0.8$. The lens velocity dispersion is set at $\sigma = 170$ km/s, and the lens redshift varies from 0 to $z_S$. The corresponding virial mass, virial radius and concentration parameters, depending on the lens redshift lay within these intervals : $M_{200} \sim 3 - 4\,10^{12}\,M_{\odot}$, $r_{200} \sim 200 - 300$ kpc, $c \sim 5 - 7.5$. The magnification $\mu$ is computed with the exact formula for a SIS model, and within the weak approx $\mu \simeq 1 + 2\kappa$ for the SIS, the truncated SIS, the NFW and the truncated NFW models. The lensing effect is maximum for a lens at $z \sim 0.3$.

We checked for the validity of the weak lensing approximation by comparing, for a truncated SIS model, the values computed with Q-LET algorithm and with our code based on the weak approximation. The result is presented on figure 4.12. Both values are within 5% until the magnification starts to get bigger than $1+10^{-1.5}$, and start to deviate when $\mu > 1+10^{-1}$, which marks the gradual limit towards a stronger lensing regime.

## 4.5 Magnification normalization

The magnification computed as explained in previous section needs to be normalized, to insure that $\langle\mu\rangle = 1$ at all redshifts.

Following [Sch92] (but see also [Bab91]), the magnification is given by :

$$\mu = \frac{F}{F_0} \tag{4.7}$$

where $F$ is the flux in the inhomogeneous clumpy universe and $F_0$ is the flux in the homogeneous universe, related to the source luminosity $L$ through :

$$F_0 = \frac{1}{4\pi}\frac{L}{D_L^2(z)} \tag{4.8}$$

The distance $D_L(z)$ is the luminosity distance as defined in eq. 1.14 for a smooth Friedmann-Lemaitre model, fitted to the inhomogeneous universe ([Sch92]). Because of flux conservation, the average of the flux $\langle F\rangle_z$, obtained by "moving" the source of luminosity $L$ over the sphere corresponding to a fixed redshift $z$, is equal to $F_0$. This implies that the average magnification is $\langle\mu\rangle_z = 1$.

When computing the magnification, we superimposed haloes on a universe already homogeneously filled with matter, obtaining systematically a magnification $\mu > 1$. As a consequence, it is necessary to compute $\langle\mu\rangle_z > 1$ to normalize the computed magnification per redshift bin to 1.

Following K10, we thus carry out for each field a simulation, randomly positioning line-of-sight pointing towards fictitious sources, distributed over 12 redshift bins from $z = 0.1$ to $z = 1.2$ in steps of $\Delta z = 0.1$. We simulate 1000 sources per redshift bin, a choice justified by a required precision of $10^{-3}$ on the computed magnification mean. The magnification distribution for the redshift bin $z = 1$ is shown on 4.14. Note that the distribution of $\log_{10}(\mu-1)$ is roughly gaussian, with a compatible RMS for the two considered model, truncated SIS and NFW in the weak approximation.

For each field, we compute the mean magnification per redshift bin, and fit the 12 points by a 3rd-order polynomial to obtain a normalization $N(z)$ valid across the redshift interval $0 < z < 1.2$. The normalization function $N(z)$ is presented on figure 4.13 for the four fields and the three models : truncated SIS with ray tracing (Q-LET) and weak approximation, truncated NFW with the weak approximation.

The normalized magnification for the $z = 1$ bin is presented on 4.14. The distributions peaks slightly below 1 (i.e. for SNe slightly de-magnified), keeping its skewed tail towards magnification values greater than 1.

In the next chapters we will analyze the results obtained from the magnification computation pipeline, including their significance, their dependence on the halo model choice.

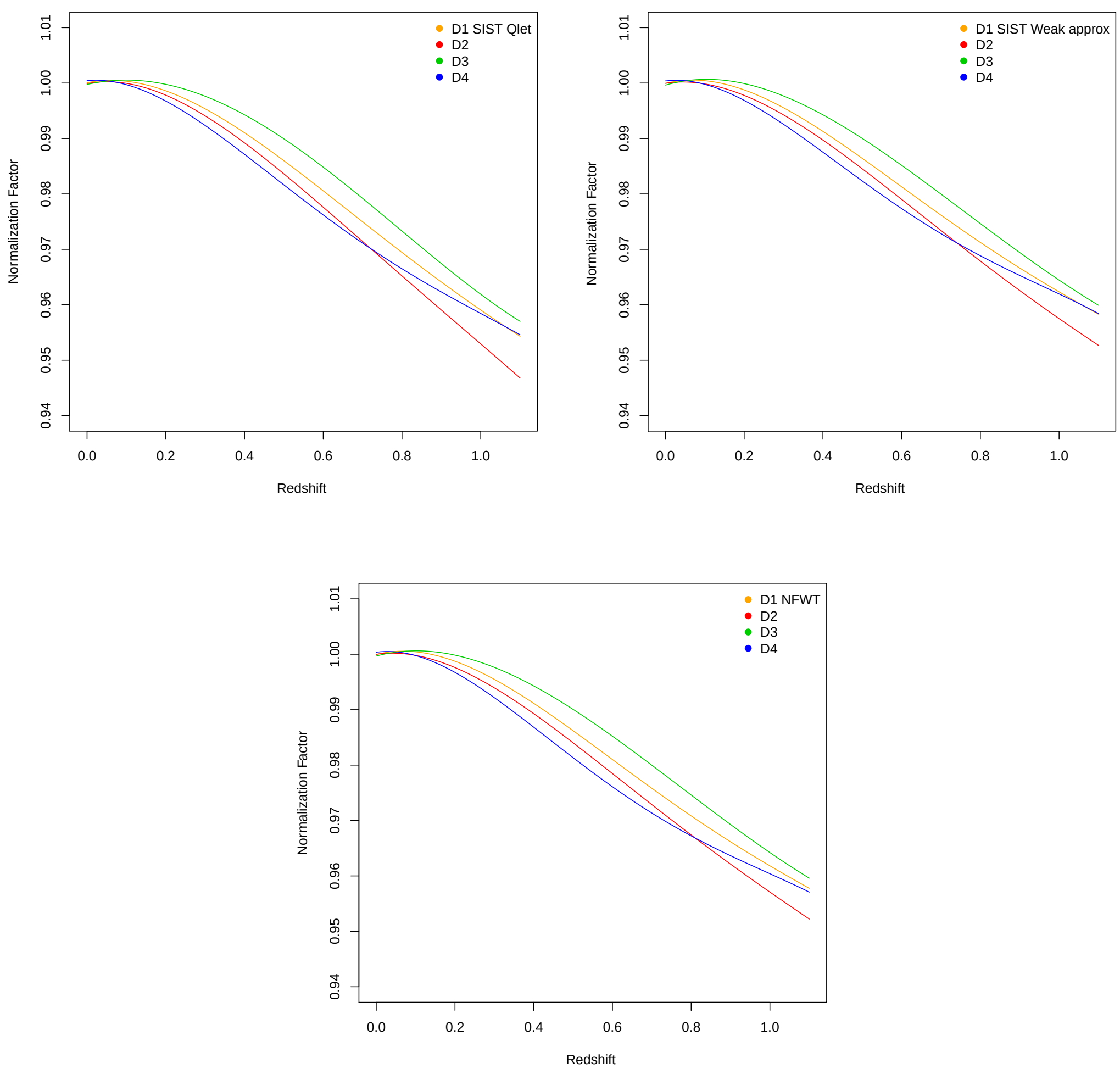

**Figure 4.13**: Normalization factor computed for the four fields, for the truncated SIS model (Q-LET and weak approximation), and within the weak approximation for the truncated NFW model.

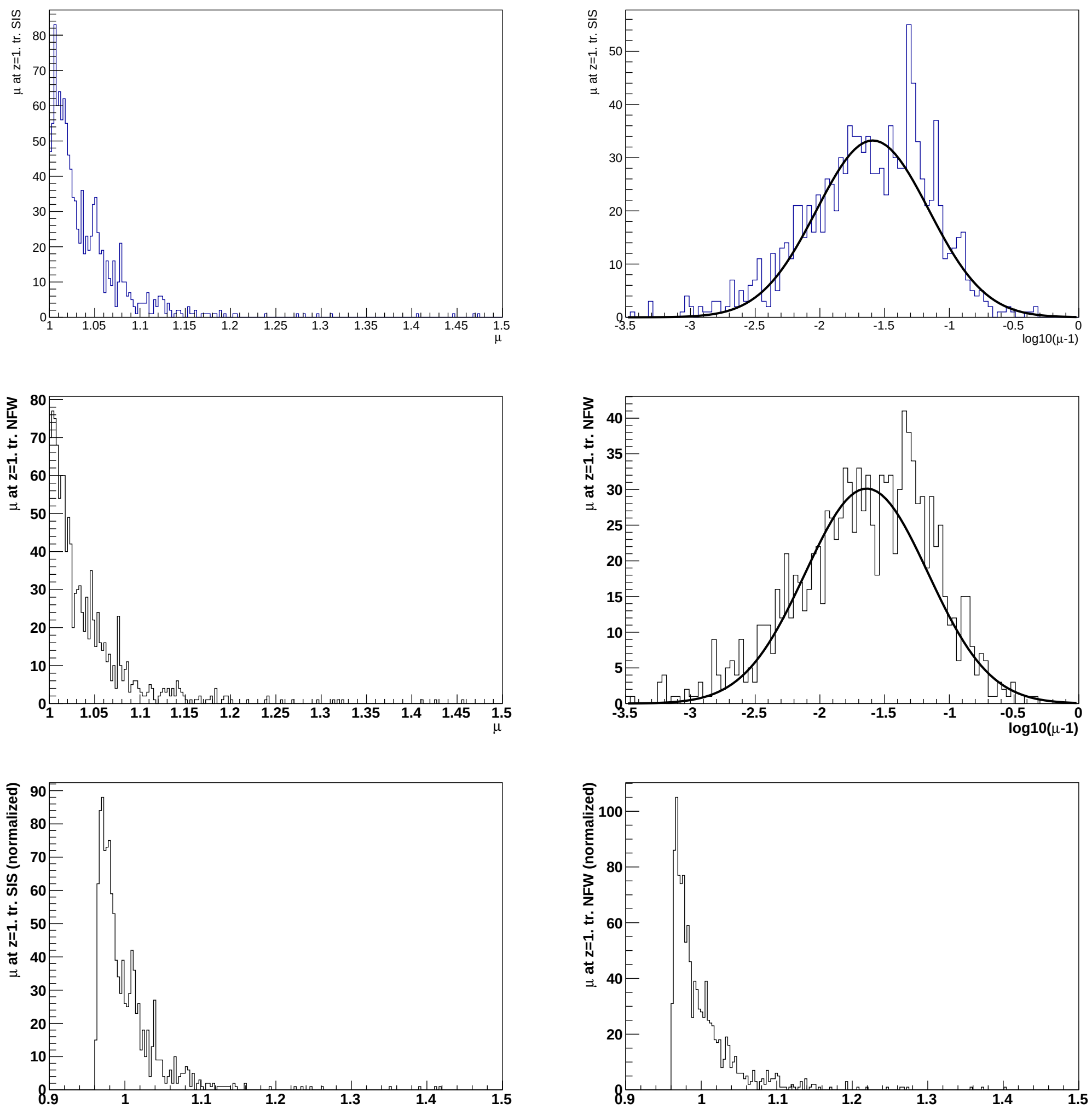

**Figure 4.14**: Upper left : for the D1 field , distribution of the magnification computed for 1000 simulated SNe at $z = 1.$ and the truncated SIS (weak approximation). The mean magnification for this redshift bin is 1.03957. Upper right : distribution of the logarithm of the un-normalized magnification : $\log_{10}(\mu - 1)$, which follows approximately a gaussian distribution, with a RMS of 0.43. Middle: same with the NFW truncated SIS model (weak approximation). The RMS of the gaussian distribution is 0.47. Lower: Normalized magnification distribution for the two models.

# Chapter 5

# The lensing signal detection

## 5.1 The lensing signal detection and prospects

We review here the lensing detection signal method and prospects as described in K10.

### 5.1.1 The lensing signal computation

The lensing signal detection is based on the computation of the correlation coefficient between the Hubble residuals and the magnification computed as described in chapter 4 :

$$\rho = \frac{\mathrm{cov}(\mathrm{magnification}, \mathrm{residual})}{\sqrt{\mathrm{var}(\mathrm{magnification})\,\mathrm{var}(\mathrm{residual})}} \tag{5.1}$$

The magnification term refers to :

$$\mu_m = -2.5 \log_{10}(\mu)$$

The Hubble residual is defined by the difference between the SN distance modulus $\mu_{SN}$ from eq. 2.1 and its predicted value $\mu_L$ by the cosmological model fitted on the Hubble diagram, given the SN redshift :

$$r = \mu_{SN} - \mu_L(z; \Omega_m, w)$$

The weighted covariance $\mathrm{cov}(x, y)$ is defined as :

$$\mathrm{cov}(x, y) = \frac{\sum w_i x_i y_i}{\sum w_i} - \overline{x}\,\overline{y}$$

where the weighted means $\overline{x}$ and $\overline{y}$ are computed as :

$$\overline{x} = \frac{\sum w_i x_i}{\sum w_i}$$

It was shown by K10 that weighting with the inverse of the Hubble residual variance, $w = 1/\sigma^2(r)$ is maximizing the signal detectability. Indeed, K10 showed that the magnification error is increasing with the magnification value :

$$\sigma(\mu_m) = 0.008 - 0.17 \times \mu_m$$

The principal contribution to the magnification error is the dispersion on the TF and FJ scaling laws used to parametrized the galaxy haloes. As a consequence, weighting with both the Hubble residual and the magnification variance is sub-optimal, as it lowers the weight of the most magnified events that carry the signal.

Based on the SNLS3 sample of 171 SNe selected for the lensing analysis, K10 obtained the value : $\rho = 0.18$.

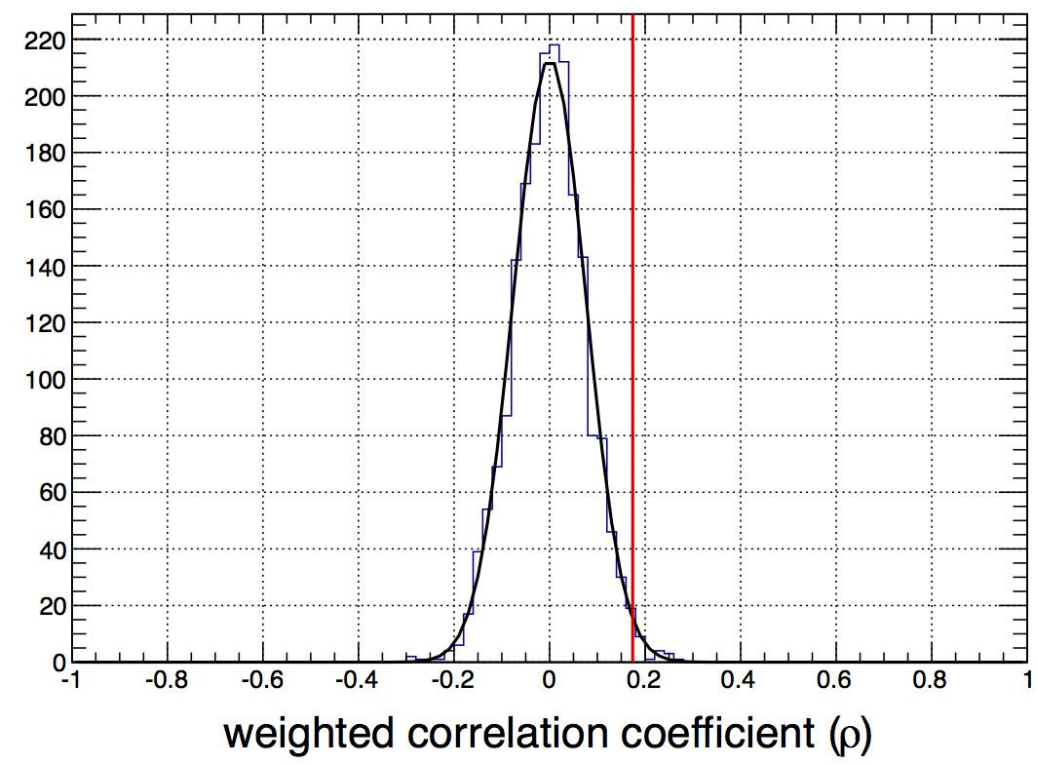

**Figure 5.1**: Distributions of the weighted correlations coefficient. $\rho = 0.18$ at 99% C.L. detection. Image : [Kro10]

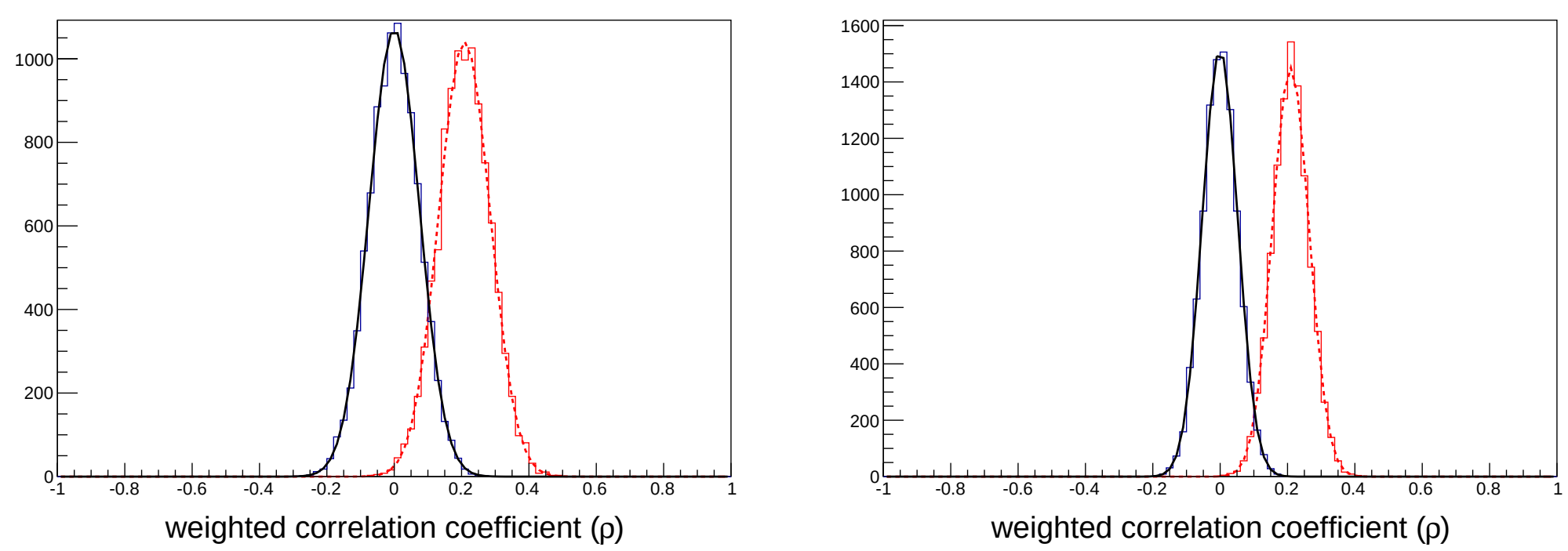

**Figure 5.2**: Left: Correlation coefficient distribution for 10 000 sample mimicking SNLS5 data, in the un-correlated case, where the SNe residuals are centered on 0, and in the correlated case, where the SN residual is supposed to equate the SN magnification. Each simulated sample contains $N = 225$ SNe. Right: same with a double statistics $N = 450$. Note the accordingly lower RMS of the distribution..

### 5.1.2 Detected signal significance

The signal significance is assessed by randomly associating Hubble residuals and expected magnifications of the real sample, obtaining this way a "shuffled" sample. The method for evaluating the result is to compute the chance of obtaining, for the $N = 10000$ "shuffled" samples, a correlation coefficient higher than the actual value obtained on the true sample, $\rho = 0.18$.

The distribution of the correlation coefficient for these shuffled sample is of course centered on 0, and gives the significance of the true sample correlation coefficient measurement (see fig. 5.1) : 2.3-$\sigma$, meaning that 99% of the shuffled samples have a $\rho$ value lower than $\rho = 0.18$.

### 5.1.3 Signal detection prospects

We use here the simulation code developed by K10 to assess the lensing signal detection prospects in SNLS5. The simulated SNe redshift interval can be chosen so as to be representative of the experiment. For the SNLS sample, it is a gaussian centered on $z = 0.65$ and with a RMS of 0.2, and $N = 225$ SNe where drawn for each sample.

For each individual SN, the un-normalized magnification value $\mu' > 1$ is drawn assuming a gaussian distribution for $\log(\mu' - 1)$ (see fig. 4.14) which mean and RMS depend on the SN redshift. The magnification value $\mu' > 1$ is subsequently normalized as described in section 4.5 to obtain the "true" magnification value $\mu$. The SN residual value $r$ is then drawn from a gaussian centered either on 0 – to obtain an un-correlated sample – or $\mu_m$ – for a correlated sample – and with a dispersion $\sigma_r = 0.16$.

| JLA sample | K10 sample + K10 selection | selection (mask + score, sec. 4.3.3 ) | N | $\rho$ | n-$\sigma$ |
|---|---|---|---|---|---|
| | $\surd$ | | 170 | 0.184 | 2.40 |
| $\surd$ | $\surd$ | | 155 | 0.206 | 2.48 |
| $\surd$ | $\surd$ | $\surd$ | 110 | 0.276 | 2.78 |
| $\surd$ | | $\surd$ | 133 | 0.252 | 2.84 |

**Table 5.1**: The different SNLS3 SNe samples selected by requiring : that the SN is included in JLA sample ; that the SN was selected in K10 sample ; that the SN is selected using the masking and scoring criteria as described in section 4.3.3. Note that on the first line, for K10 complete sample, we used the magnification computed in this thesis, and the residuals as published in K10 and not the JLA residuals.

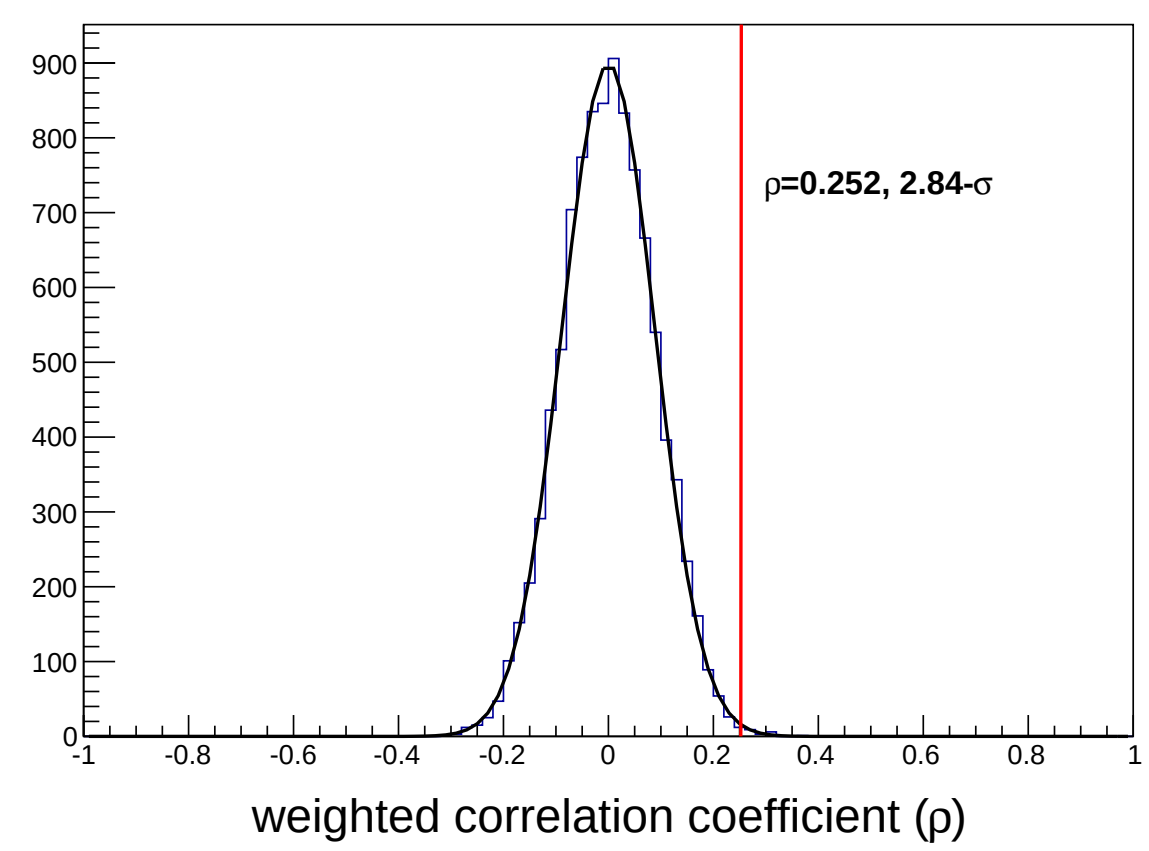

**Figure 5.3**: Measurement of the correlation on 133 SNe selected in the JLA sample. The histogram of the correlation coefficient computed on the "shuffled" samples. The measured value $\rho = 0.252$ is indicated.

The correlation coefficient is computed for each 10 000 sample. For the un-correlated samples, the distribution of the correlation coefficients is of course centered on 0, with a dispersion $\sigma_\rho \simeq 0.075$ (see fig. 5.2). The distribution of the correlation coefficients for the correlated samples peaks on a slightly positive value, $\overline{\rho}_{\text{corr.}} \simeq 0.21$ for a similar dispersion. So that about 42% of the correlated samples obtain a correlation coefficient $\rho > 3\sigma_\rho$ corresponding to a lensing signal detection at 3-$\sigma$.

The prospect depends on the number of SNe in the sample, and on the redshift range : with a double statistics of $N = 450$ SNe, there is a 80% chance of lensing signal detection at 3-$\sigma$.

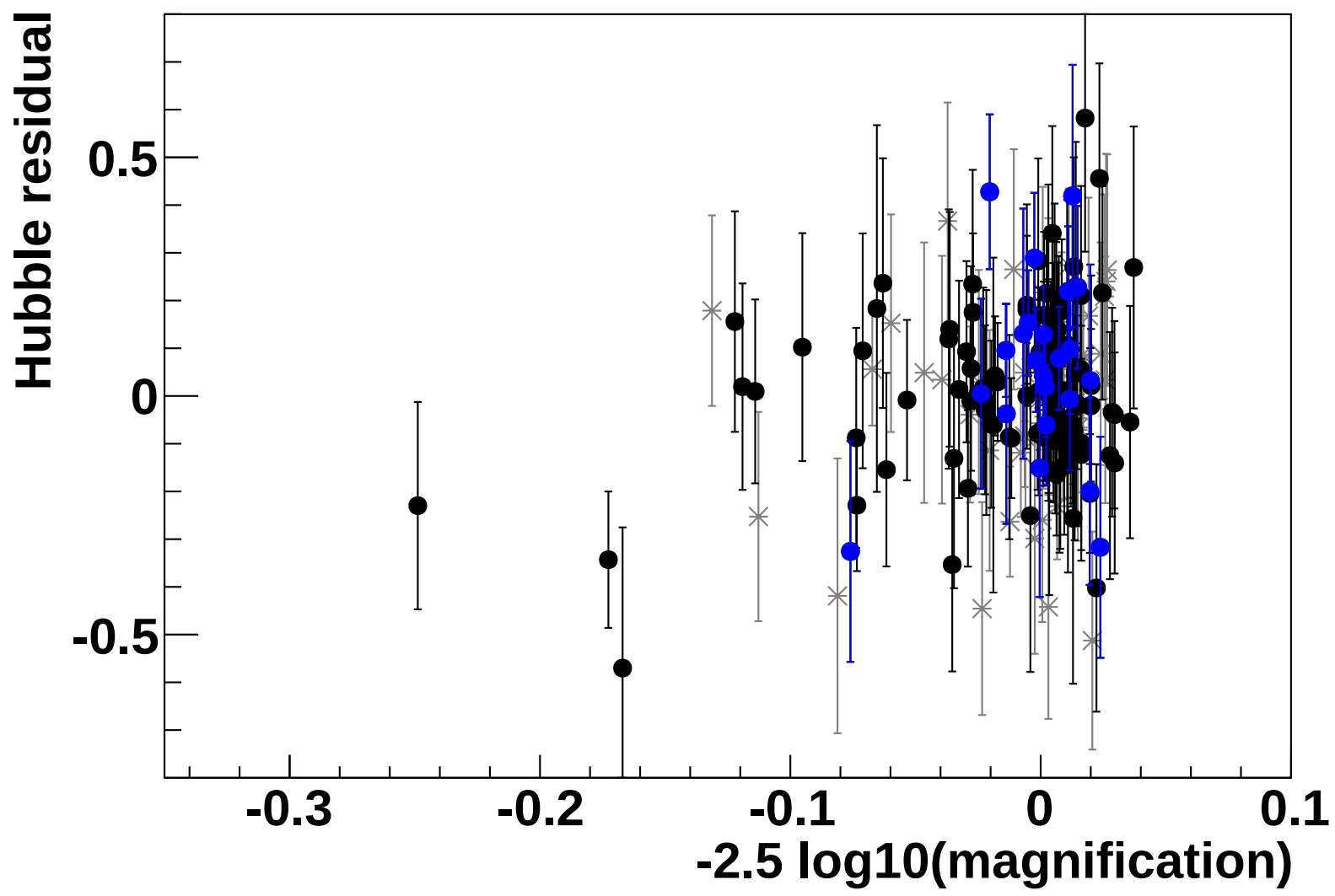

**Figure 5.4**: Measurement of the correlation on 133 SNe selected in the JLA sample. The dots indicate the 133 SNe. In blue, the 23 SNe that were not included in K10 sample. Stars in grey: the 45 SNe from K10 sample that were rejected by the masking process described in this thesis.

## 5.2 SNLS5 lensing signal detection

### 5.2.1 The SNLS3 sample

We first checked our newly computed SNe magnifications against K10 previous results, restricting thus ourselves to SNLS3 data. We will be using the cosmology residuals from the JLA analysis. Note that K10 SNe are not all included in the JLA sample and vice-versa, as the criteria for the selection of SNe Ia "good for cosmology" were slightly different for these two analysis : only 155 of the 170 K10 SNe are included in the JLA cosmology sample.

We use here the magnification computed using the weak approximation and the truncated SIS model.

The masking and scoring steps select 133 SNe out of the 239 SNe in the JLA cosmology sample, which correspond to a rejection of 45% of the SNe, within 1-$\sigma$ of the announced rejection rate of 40%.

Of the 133 JLA SNe selected by our masking and scoring procedure, 110 are included in K10 sample. Compared to K10 selection, we reject 45 SNe, with one third being on the edge of the camera, which were not masked out in K10 ; two-third because of bright star light contamination. On the reverse, we kept 23 supplementary SNe : 8 SNe were masked out in K10 selection, plus 15 SNe were not included in the preliminary SNLS3 cosmology sample used for K10.

The samples summary and their associated correlation coefficients are presented on table 5.1. Note that rejecting 45 SNe from K10 sample helps to strengthen the signal, which does not change when the supplementary 23 SNe are added. As a conclusion, the result from K10 is confirmed by this new analysis. The results for the 133 SNe selected in JLA sample is shown on figures 5.3 and 5.4. A correlation coefficient $\rho_{\mathrm{JLA;133}} = 0.252$ is measured, corresponding to a detection level at 2.8-$\sigma$.

| SNLS5 sample | in JLA sample | Mask+Score | N | $\rho$ | n-$\sigma$ |
|---|---|---|---|---|---|
| $\surd$ | $\surd$ | $\surd$ | 128 | 0.228 | 2.47 |
| color cut | | $\surd$ | 225 | 0.177 | 2.51 |
| color cut | excluded | | 97 | 0.098 | 0.92 |
| color cut, D1 & D2 | | $\surd$ | 109 | 0.122 | 1.19 |
| color cut, D3 & D4 | | $\surd$ | 116 | 0.218 | 2.23 |
| color cut, $\mu > 0.99$ | | $\surd$ | 164 | 0.254 | 3.04 |

**Table 5.2**: The different SNLS5 SNe samples selected using the masking and scoring criteria as described in section 4.3.3 and by requiring : that the SN is included in SNLS5 sample, with an additional color cut $|c| < 0.25$ eventually and the exclusion of 03D4gl ; that the SN is included in JLA sample. We used the residuals from SNLS5 analysis.

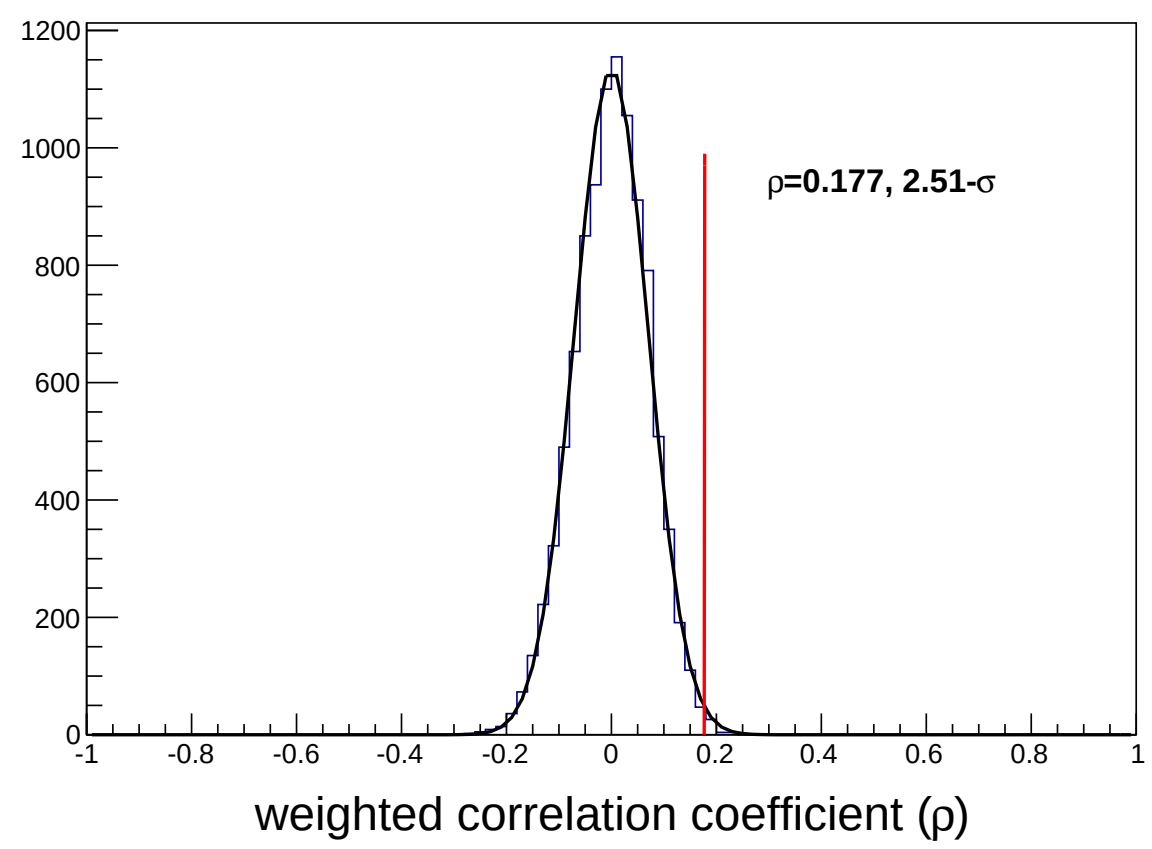

**Figure 5.5**: Measurement of the correlation on 225 SNe selected in the SNLS5 sample. The histogram of the correlation coefficient computed on the "shuffled" samples. The measured value $\rho = 0.177$ is indicated.

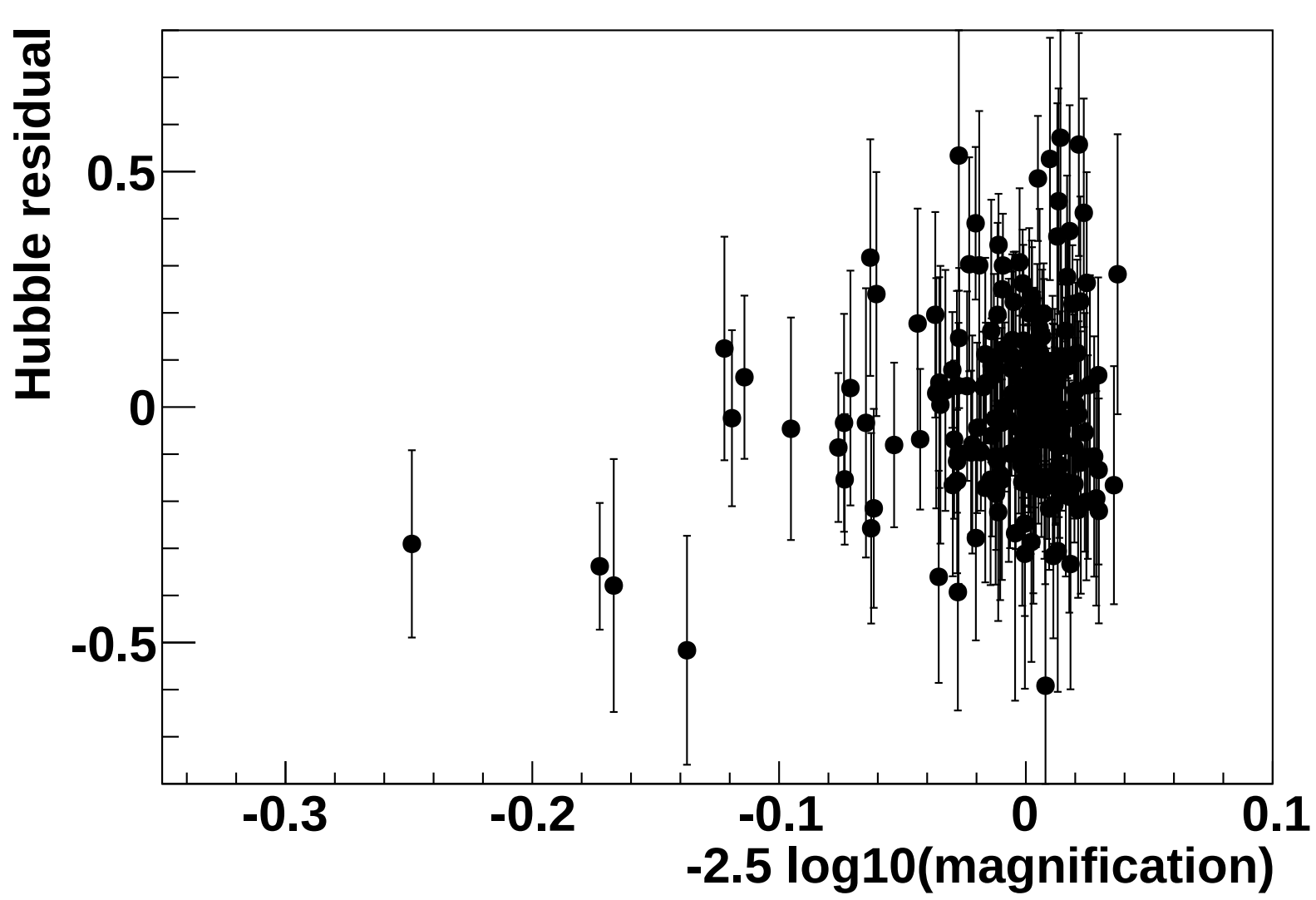

**Figure 5.6**: Measurement of the correlation on 225 SNe selected in the SNLS5 sample.

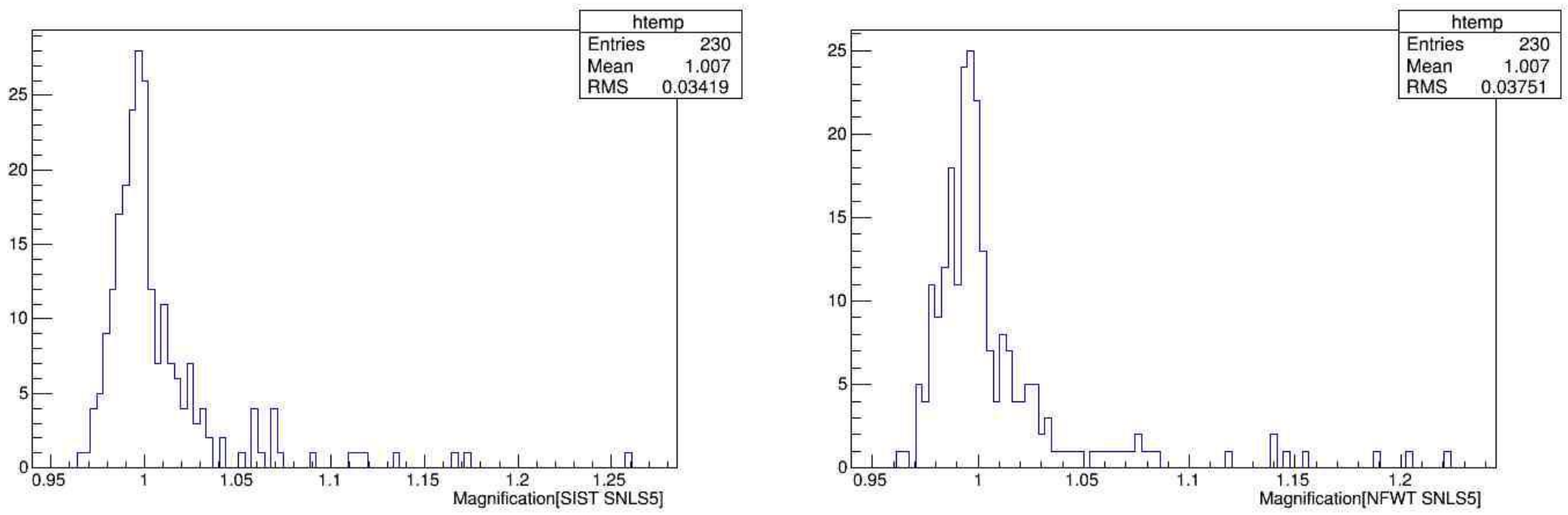

**Figure 5.7**: Normalized magnification of SNLS5 SNe for the truncated SIS and the truncated NFW halo model.

### 5.2.2 The SNLS5 sample : a preliminary analysis

The SNLS5 cosmology analysis is not finished yet. The first consequence is that the sample of "good for cosmology" SNe is not finalized yet. The second consequence is that the analysis is still blinded, so that we do not use the true residuals. For these reasons, the analysis presented below is still preliminary.

We add a supplementary cut on color to SNLS5 sample so as to further select "good" SNe Ia, demanding that $|c| < 0.25$. Following [Bet14], we also eliminate 03D4gl which has no post-max data. We obtain this way a sample of 384 SNe.

The masking and scoring steps select 225 SNe (60%) out of the 384 SNe in the SNLS5 cosmology sample. This 225 SNe sample can be divided in 128 SNe that were in the JLA sample used in the previous section, and 97 SNe supplementary SNe.

We obtain a correlation coefficient $\rho_{\mathrm{SNLS5;225}} = 0.177$ which corresponds to a detection level at 2.51-$\sigma$. The result is shown on figures 5.5 and 5.6. We can notice that the signal is carried out by the most magnified SNe : by requiring that the magnification $\mu > 0.99$, we select 164 SNe out of 225 (a fraction of 75%), and obtain a correlation coefficient $\rho = 0.254$.

Note that without a supplementary cut on SNe quality, we obtain, with now 230 SNe, including 3 very red and one very blue SNe plus 03D4gl, a correlation coefficient of $\rho_{\mathrm{SNLS5;230}} = 0.138$ which corresponds to a detection level at 1.97-$\sigma$. This points the necessity to work with a SN sample with a strict selection on color and, in general, quality.

The correlation measured on the 97 SNe added to the previous JLA sample is weak : $\rho_{\mathrm{SNLS5;97}} = 0.098$, corresponding to a 0.92-$\sigma$ level. To assess the significance of this performance, we randomly selected N=2000 samples of 97 SNe into the N=225 selected SNe : 25% of the selected samples yield a correlation coefficient lower than $\rho_{\mathrm{SNLS5;97}}$. Their is thus a reasonable chance when including 97 SNe that they perform as poorly, regarding their ability at producing a lensing detection signal.

For comparison, we can also divide our sample in two halves, according to the fields numbers. The results are summarized in table 5.2.

### 5.2.3 Magnification computation method

We have computed the magnification using the truncated SIS and the truncated NFW model. The distribution of the magnification for both models is presented on fig. 5.7. The distribution is indeed skewed, peaking slightly below $\mu = 1$, with a mean value close to 1. The magnification of the SNe as a function of redshift is indicated on figure 5.8. Except for 04D2kr, the results obtained using both models are quite similar. Indeed, the magnification of 04D2kr, computed

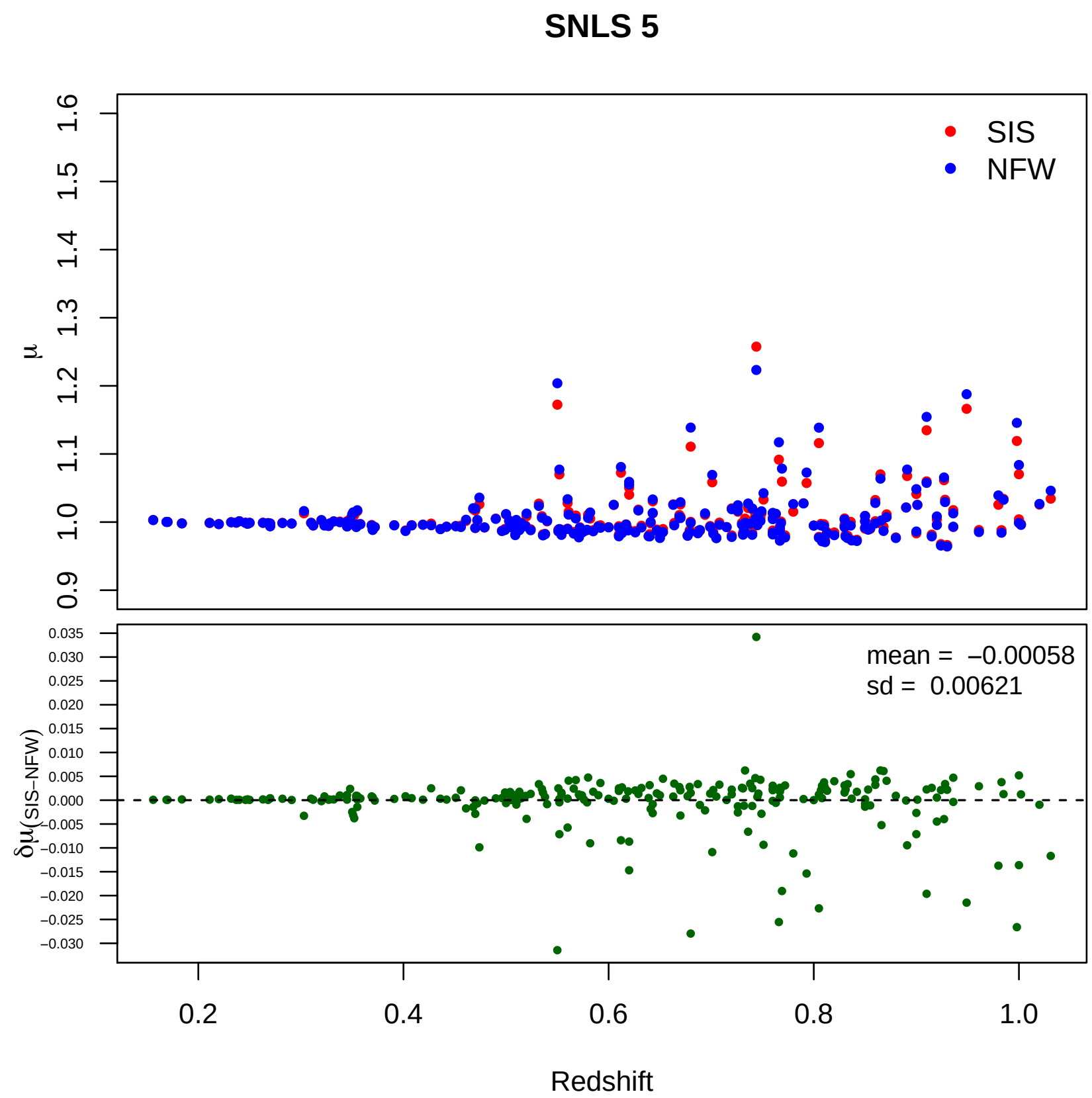

**Figure 5.8**: Normalized magnification of SNLS5 SNe for the truncated SIS and the truncated NFW halo model as a function of redshift. Except for 04D2kr which is not in the weak regime, the obtained values are very similar.

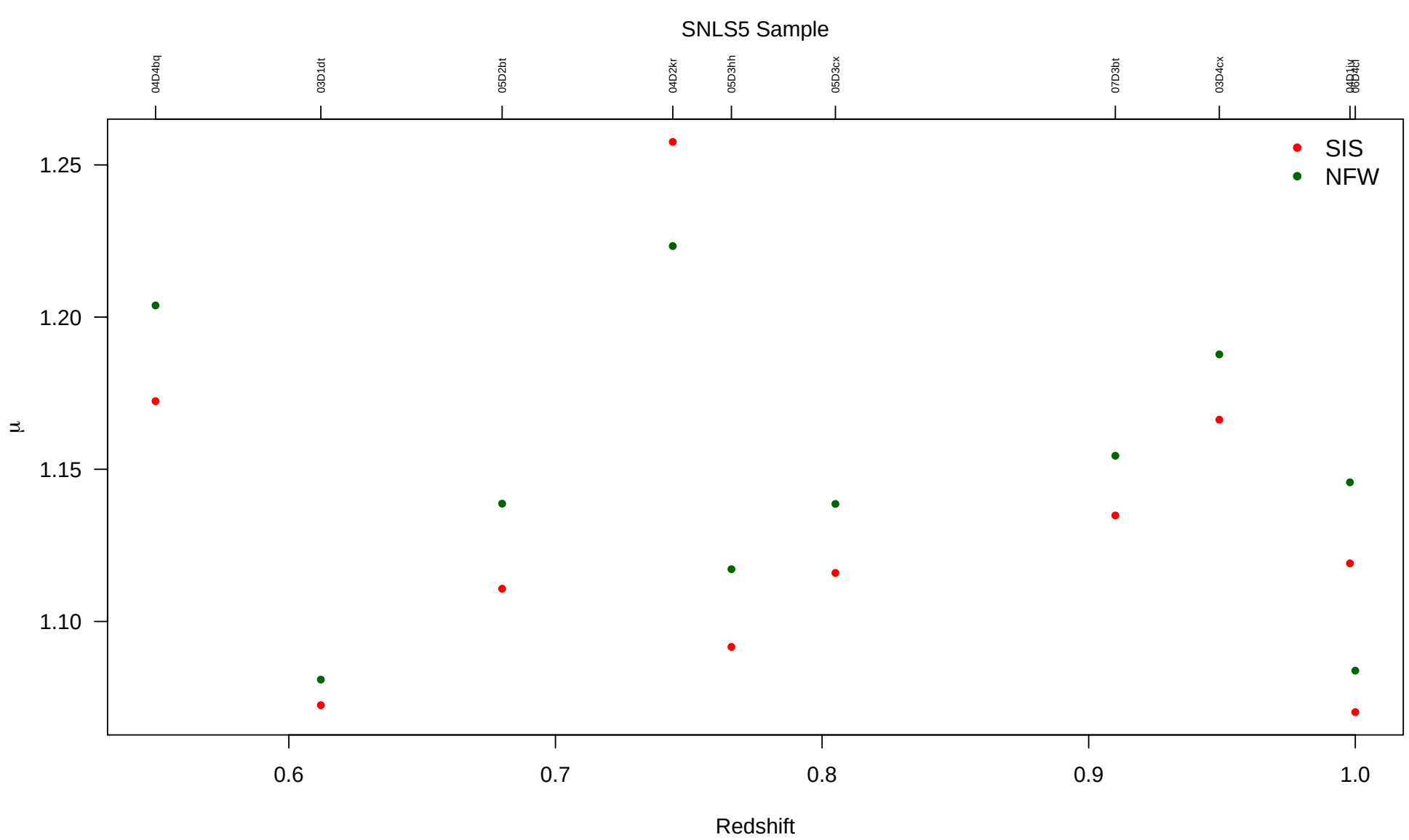

**Figure 5.9**: The 10 most magnified SNe from the SNLS5 sample

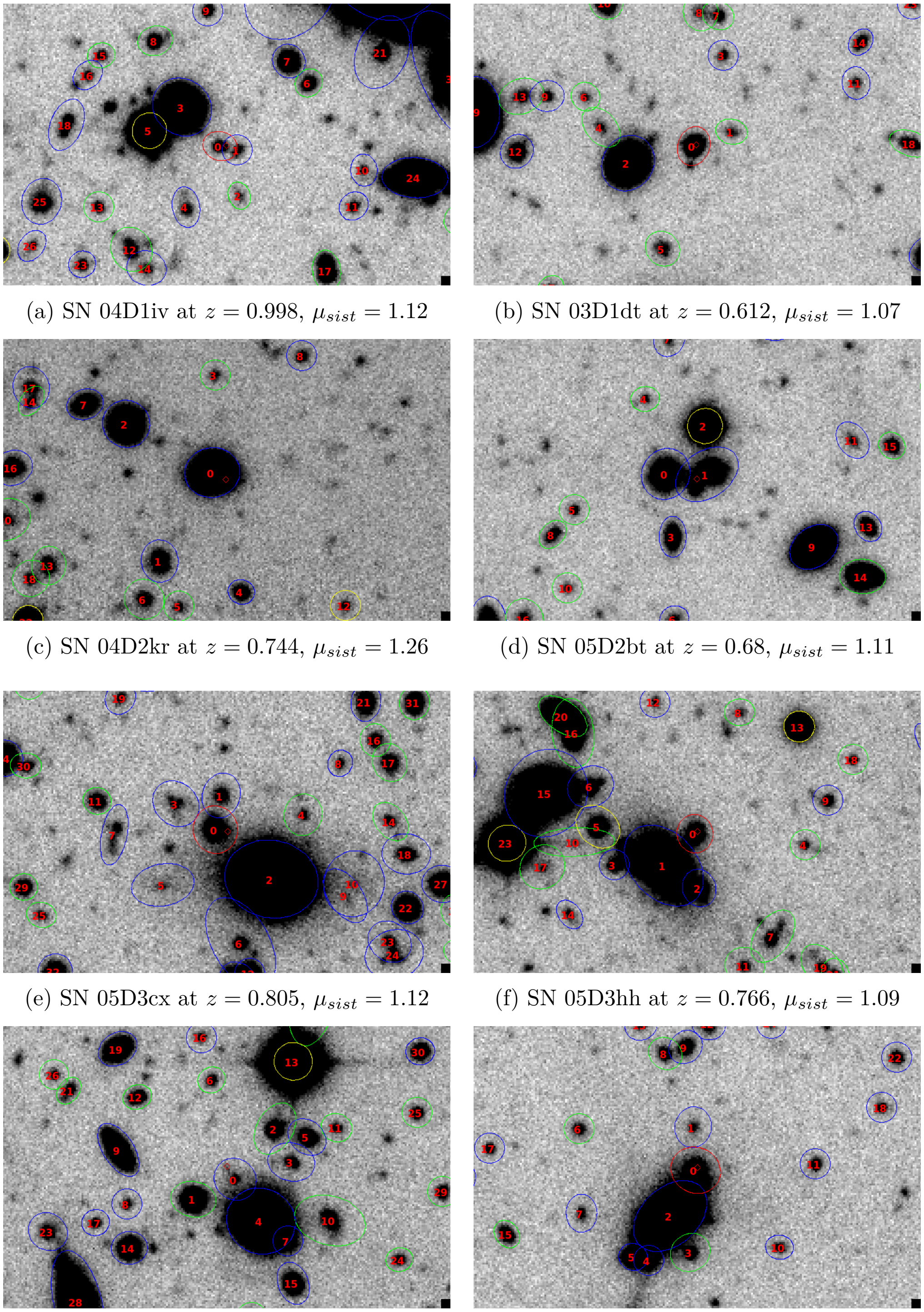

(a) SN 04D1iv at $z = 0.998$, $\mu_{sist} = 1.12$

(b) SN 03D1dt at $z = 0.612$, $\mu_{sist} = 1.07$

(c) SN 04D2kr at $z = 0.744$, $\mu_{sist} = 1.26$

(d) SN 05D2bt at $z = 0.68$, $\mu_{sist} = 1.11$

(e) SN 05D3cx at $z = 0.805$, $\mu_{sist} = 1.12$

(f) SN 05D3hh at $z = 0.766$, $\mu_{sist} = 1.09$

(g) SN 07D3bt at $z = 0.91$, $\mu_{sist} = 1.13$

(h) SN 03D4cx at $z = 0.949$, $\mu_{sist} = 1.16$

**Figure 5.10**: Ten most magnified SNe (grouped according to their Field). The SN position is indicated by a red diamond. A blue ellipse indicates a foreground galaxies, a green one a background galaxy, and the red one the host galaxy. A yellow circle indicates a star.

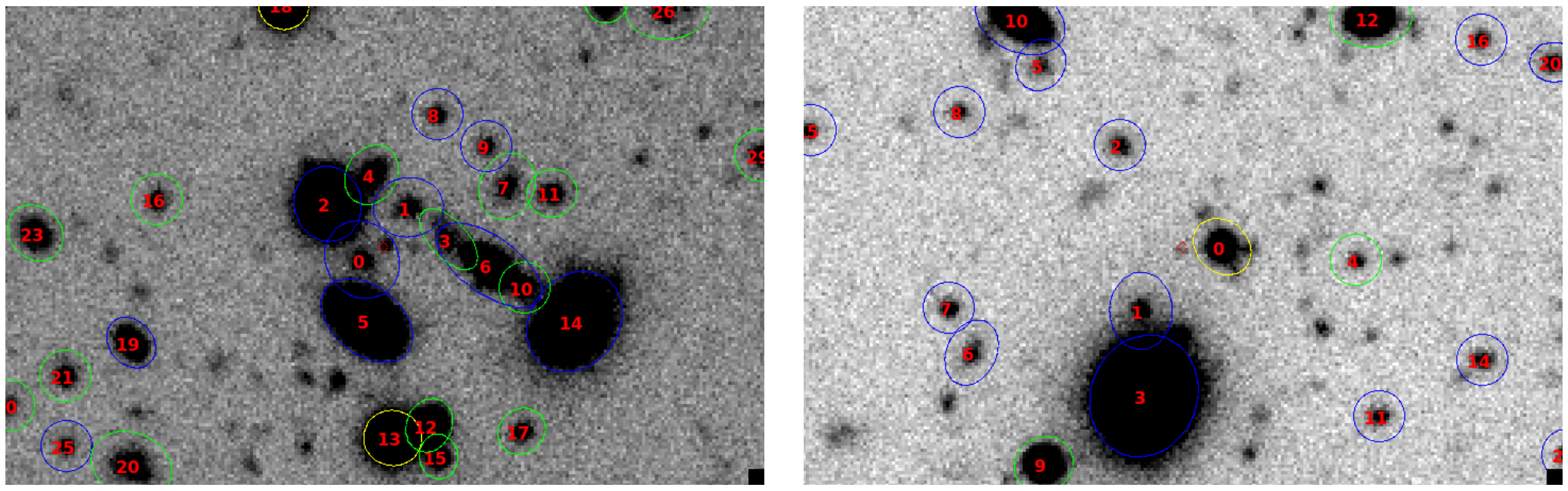

(a) SN 04D4bq at $z = 0.55$, $\mu_{sist} = 1.17$ (b) SN 06D4cl at $z = 1.0$, $\mu_{sist} = 1.07$

**Figure 5.11**: The ten most magnified SNe (grouped according to their Field), cont. from fig. 5.10.

using Q-LET ray-tracing algorithm, is $\mu = 1.37$, which cannot be considered to be in the weak approximation regime.

The correlation coefficient computed with the truncated NFW model, $\rho_{\rm NFW} = 0.176$, does not differ much from the values obtained with the truncated SIS ($\rho_{\rm SIS} = 0.177$). As previously noted and advocated by [Jon06] and [Jon10], the ray tracing algorithm gives similar results within $\sim$5% to the weak approximation computation (except for 04D2kr) : $\rho_{\rm Q-LET} = 0.184$.

This permits to conclude that the result is quite robust with regards to the choice of the halo parametrization. The ten most magnified SNe are presented on figure 5.9. Their visual is presented on figure 5.10 to 5.11.

## 5.3 Conclusion and discussion

We conclude that with a new analysis pipeline, we estimate the lensing signal on the preliminary SNLS5 cosmology sample and obtain a consistent result with the one obtained on the JLA SNLS3 sample : the measured correlation coefficient is $\rho = 0.177$ corresponding to a detection at 2.5-$\sigma$. The necessity to work with a SN sample selected with strict quality cuts, and also a "clean" line of sight, was demonstrated.

We would like to point the future steps to complete this analysis.

- Error analysis : the magnification error computation was left aside, as it was already performed in K10. It includes the propagation of the photometric errors on the galaxies luminosity and photo-z computation, but also the observed dispersion on the scaling laws : TF and FJ (see eq. 3.34 and 3.37). In the case of the NFW model, the scaling law for the concentration parameter also exhibits a large dispersion.
- Masking : the masking procedure rejected a large fraction (40%)of the SNe sample. It would be interesting to consider a masking procedure relying on a different algorithm : the computation of the SN magnification with and without the masked areas, and the comparison of the difference to the error on the magnification, to reject or keep the SN line-of-sight.
- Photometric sample : including the SNLS5 photometric SNe sample is a major step to complete the SNLS5 analysis. There will be about 300 photometric SNe Ia. Using a photometric sample prefigures the analysis that will be done within the LSST project which will provide thousands of SNe.

# Conclusion

We will conclude the thesis with the current chapter, here we will focus on presenting the advantage of the lensing methodology used in this thesis and suggest a future roadmap. Using the luminosity properties of SNe Ia, cosmological distance measurements have been better constrained in the last 15 years. It also gave impetus to the study of dark energy. In this thesis we worked on the effect of the line of sight dark matter mass distribution on the SNe Ia brightness distributions or typically the effect of gravitational magnification on SN Ia by the foreground galaxies modelled with dark matter halo models. This has been studied in the weak lensing regime. The completion of gravitational lensing study of SNe Ia with 5 years full SNLS data is second in series to the primary lensing analysis started with the three years SNLS lensing data by [Kro10].

Presence of dark matter mass in the line of sight distribution, induces dispersion in the SNe brightness distribution. $\kappa$, the convergence parameter in the lensing theory gives a measure of this isotropic brightness variations (responsible for inducing magnification). Thus it was magnification computation that was the primary step in the lensing analysis. Leaving out the shear component (which was justified given the small angular size of the SNe), our lensing magnification computation expression was

$$\mu = 1 + 2\kappa \tag{5.2}$$

This is evidently the first order approximated expression of the convergence component. The use of the weak lensing approximation formula has been justified by [Jon10]. We rechecked this approach and our results were found to match the predicted 5% deviation in value within the weak lensing regime as was suggested by [Jon10].

The presence of lensing signal can be seen as a source of contamination to the original SN brightness distribution, with a reduction in the brightness mode. Consequently this induces a dispersion in the observed Hubble diagram. Making use of this fact, we computed the correlation between the magnification computed and the residuals to the Hubble diagram. Detection of a positive correlation would give us confidence on the prediction of presence/detection of weak lensing signal. We obtained $\rho = 0.177$ at $2.53\sigma$ from 225 SNe Ia at $z \lesssim 1$. However it is to be noted that this is not the final computation as the final sample selection of the SNLS 5 SNe is still not set. In addition there is scope to increase the sample size in future, with the use of additional $\sim 300$ photometric SNe candidates.

Following the analysis presented in [Kro10], we present in this thesis the comprehensive pipeline for the computation and signal detection of SNe brightness magnification via weak lensing mechanism. This technique could provide a complementary technique to more widely used galaxy-galaxy lensing techniques.

The SNLS 5 lensing analysis was modified from its predecessor with the aim to make it more automatic, as required for future big data analysis. Detection and removal of contaminations in the lensing analysis (masking and scoring) had been upgraded to suit the upcoming bigger surveys. Also the weak lensing analysis code that we developed for the SNLS5 is not based on the rigorous ray-tracing algorithm unlike it's predecessor : QLET and thus saves from program complexity and time. We also made provisions for two halo models in it to increase the acceptability of our code. Together with all these upgrades and modifications we hope that with large scale upcoming surveys such as e.g. the LSST, EUCLID, JWST. There will be more scope for narrowing down on the constraints of weak lensing detection signals from the SNe Ia.

# Appendix A

# SNLS5 SNe Ia magnifications

| No. | redshift | $\mu_{SIS}$ | $\mu_{NFW}$ | name |
|---|---|---|---|---|
| 1 | 0.408 | 0.996 | 0.995 | 03D1ar |
| 2 | 0.504 | 0.996 | 0.994 | 03D1au |
| 3 | 0.582 | 1.010 | 1.010 | 03D1aw |
| 4 | 0.496 | 0.987 | 0.987 | 03D1ax |
| 5 | 0.865 | 1.070 | 1.060 | 03D1bk |
| 6 | 0.679 | 0.988 | 0.985 | 03D1co |
| 7 | 0.612 | 1.070 | 1.080 | 03D1dt |
| 8 | 0.868 | 0.993 | 0.987 | 03D1ew |
| 9 | 0.332 | 1.000 | 1.000 | 03D1fc |
| 10 | 0.800 | 0.995 | 0.995 | 03D1fq |
| 11 | 0.211 | 0.999 | 0.999 | 04D1dc |
| 12 | 0.768 | 1.000 | 0.998 | 04D1de |
| 13 | 0.369 | 0.996 | 0.995 | 04D1hd |
| 14 | 0.560 | 0.990 | 0.990 | 04D1hx |
| 15 | 0.850 | 1.000 | 1.000 | 04D1hy |
| 16 | 0.998 | 1.120 | 1.150 | 04D1iv |
| 17 | 0.585 | 0.988 | 0.987 | 04D1kj |
| 18 | 0.590 | 0.994 | 0.993 | 04D1oh |
| 19 | 0.915 | 0.982 | 0.979 | 04D1ow |
| 20 | 0.515 | 0.999 | 0.998 | 04D1pg |
| 21 | 0.639 | 0.980 | 0.979 | 04D1pu |
| 22 | 0.767 | 0.973 | 0.973 | 04D1qd |
| 23 | 0.436 | 0.990 | 0.990 | 04D1rh |
| 24 | 0.985 | 1.030 | 1.030 | 04D1rx |
| 25 | 0.702 | 0.985 | 0.983 | 04D1si |
| 26 | 0.663 | 1.030 | 1.030 | 04D1sk |
| 27 | 0.842 | 0.974 | 0.972 | 05D1az |
| 28 | 0.632 | 0.995 | 0.992 | 05D1cb |
| 29 | 0.830 | 0.996 | 0.993 | 05D1cl |
| 30 | 0.580 | 1.010 | 1.010 | 05D1dx |
| 31 | 0.312 | 0.995 | 0.995 | 05D1ej |
| 32 | 0.860 | 1.030 | 1.030 | 05D1er |
| 33 | 0.763 | 1.010 | 1.010 | 05D1if |
| 34 | 0.490 | 1.010 | 1.000 | 05D1ix |

*Continued on next page*

Table A.1 – *Continued from previous page*

| No. | redshift | $\mu_{SIS}$ | $\mu_{NFW}$ | name |
|---|---|---|---|---|
| 35 | 0.248 | 0.998 | 0.998 | 05D1iy |
| 36 | 0.760 | 1.010 | 1.010 | 06D1bg |
| 37 | 0.620 | 1.040 | 1.050 | 06D1bo |
| 38 | 0.820 | 0.985 | 0.981 | 06D1bt |
| 39 | 0.619 | 0.990 | 0.988 | 06D1cm |
| 40 | 0.860 | 1.000 | 0.998 | 06D1cx |
| 41 | 0.767 | 0.993 | 0.991 | 06D1dc |
| 42 | 0.240 | 1.000 | 1.000 | 06D1du |
| 43 | 0.350 | 1.010 | 1.010 | 06D1fd |
| 44 | 0.524 | 0.989 | 0.988 | 06D1fx |
| 45 | 0.980 | 1.030 | 1.040 | 06D1gl |
| 46 | 0.345 | 0.994 | 0.994 | 06D1hf |
| 47 | 0.327 | 0.994 | 0.994 | 06D1hj |
| 48 | 0.641 | 0.998 | 1.000 | 06D1jf |
| 49 | 0.561 | 1.020 | 1.010 | 06D1kf |
| 50 | 0.320 | 1.000 | 1.000 | 06D1kg |
| 51 | 0.328 | 0.998 | 0.998 | 07D1ab |
| 52 | 0.617 | 0.997 | 0.996 | 07D1bs |
| 53 | 0.626 | 0.987 | 0.985 | 07D1bu |
| 54 | 0.853 | 0.991 | 0.989 | 07D1cc |
| 55 | 0.500 | 1.010 | 1.010 | 07D1cf |
| 56 | 0.705 | 0.977 | 0.976 | 07D1cl |
| 57 | 0.730 | 0.999 | 0.996 | 07D1cx |
| 58 | 0.540 | 1.000 | 1.000 | 08D1aa |
| 59 | 0.836 | 1.000 | 0.995 | 04D2al |
| 60 | 0.220 | 0.997 | 0.997 | 04D2bt |
| 61 | 0.568 | 1.010 | 1.010 | 04D2cw |
| 62 | 0.732 | 0.987 | 0.988 | 04D2gp |
| 63 | 0.740 | 0.984 | 0.981 | 04D2ja |
| 64 | 0.744 | 1.260 | 1.220 | 04D2kr |
| 65 | 0.513 | 0.990 | 0.988 | 04D2mj |
| 66 | 0.323 | 0.996 | 0.995 | 05D2ab |
| 67 | 0.479 | 0.992 | 0.992 | 05D2ac |
| 68 | 0.184 | 0.998 | 0.998 | 05D2ah |
| 69 | 0.920 | 1.000 | 1.010 | 05D2ay |
| 70 | 0.680 | 1.110 | 1.140 | 05D2bt |
| 71 | 0.891 | 1.070 | 1.080 | 05D2by |
| 72 | 0.427 | 0.998 | 0.995 | 05D2cb |
| 73 | 0.574 | 0.986 | 0.985 | 05D2dt |
| 74 | 0.510 | 1.000 | 1.000 | 05D2dy |
| 75 | 0.733 | 1.010 | 0.999 | 05D2fq |
| 76 | 0.348 | 1.000 | 1.000 | 05D2ie |
| 77 | 0.354 | 0.993 | 0.993 | 05D2mp |
| 78 | 0.924 | 0.968 | 0.966 | 05D2ob |
| 79 | 0.310 | 0.999 | 0.999 | 06D2ag |
| 80 | 0.499 | 0.990 | 0.988 | 06D2bk |

*Continued on next page*

Table A.1 – *Continued from previous page*

| No. | redshift | $\mu_{SIS}$ | $\mu_{NFW}$ | name |
|---|---|---|---|---|
| 81 | 1.000 | 1.000 | 0.999 | 06D2cb |
| 82 | 0.532 | 1.030 | 1.020 | 06D2cc |
| 83 | 0.552 | 0.989 | 0.990 | 06D2ck |
| 84 | 0.345 | 1.000 | 1.000 | 06D2ff |
| 85 | 0.560 | 1.030 | 1.030 | 06D2hm |
| 86 | 0.850 | 1.010 | 1.010 | 06D2iz |
| 87 | 0.726 | 1.020 | 1.020 | 06D2ja |
| 88 | 0.900 | 0.983 | 0.986 | 06D2ji |
| 89 | 0.600 | 0.993 | 0.992 | 06D2js |
| 90 | 0.927 | 1.060 | 1.070 | 06D2ju |
| 91 | 0.900 | 1.040 | 1.050 | 06D2jw |
| 92 | 0.780 | 1.020 | 1.030 | 07D2ah |
| 93 | 0.793 | 1.060 | 1.070 | 07D2be |
| 94 | 0.551 | 0.989 | 0.986 | 07D2bi |
| 95 | 0.535 | 1.010 | 1.010 | 07D2bq |
| 96 | 0.694 | 1.010 | 1.010 | 07D2cb |
| 97 | 0.749 | 1.010 | 1.020 | 07D2cl |
| 98 | 0.738 | 1.000 | 0.998 | 07D2co |
| 99 | 0.746 | 1.010 | 1.010 | 07D2cq |
| 100 | 0.538 | 0.983 | 0.982 | 07D2du |
| 101 | 0.270 | 0.998 | 0.998 | 07D2ec |
| 102 | 0.871 | 1.010 | 1.010 | 07D2fv |
| 103 | 0.720 | 0.980 | 0.978 | 07D2fy |
| 104 | 0.354 | 0.996 | 0.995 | 07D2kc |
| 105 | 0.731 | 0.984 | 0.981 | 07D2kh |
| 106 | 0.554 | 0.989 | 0.987 | 08D2ad |
| 107 | 0.689 | 0.987 | 0.988 | 08D2ag |
| 108 | 0.474 | 1.030 | 1.040 | 08D2ch |
| 109 | 0.831 | 0.981 | 0.979 | 08D2cl |
| 110 | 0.355 | 0.997 | 0.996 | 08D2dr |
| 111 | 0.650 | 0.978 | 0.977 | 08D2dz |
| 112 | 0.833 | 0.980 | 0.977 | 08D2id |
| 113 | 0.291 | 0.998 | 0.998 | 03D3ba |
| 114 | 0.355 | 1.020 | 1.020 | 03D3bl |
| 115 | 0.715 | 0.993 | 0.993 | 03D3bn |
| 116 | 0.156 | 1.000 | 1.000 | 04D3bf |
| 117 | 0.620 | 1.050 | 1.060 | 04D3co |
| 118 | 0.643 | 1.010 | 1.010 | 04D3cy |
| 119 | 1.002 | 0.997 | 0.996 | 04D3dd |
| 120 | 0.470 | 0.991 | 0.991 | 04D3df |
| 121 | 0.610 | 0.994 | 0.992 | 04D3do |
| 122 | 0.263 | 0.999 | 0.999 | 04D3ez |
| 123 | 0.358 | 0.997 | 0.997 | 04D3fk |
| 124 | 0.910 | 1.060 | 1.060 | 04D3gx |
| 125 | 0.552 | 1.070 | 1.080 | 04D3hn |
| 126 | 0.983 | 0.988 | 0.984 | 04D3lp |

*Continued on next page*

Table A.1 – *Continued from previous page*

| No. | redshift | $\mu_{SIS}$ | $\mu_{NFW}$ | name |
|---|---|---|---|---|
| 127 | 0.813 | 0.985 | 0.983 | 04D3mk |
| 128 | 0.810 | 0.997 | 0.994 | 04D3ny |
| 129 | 0.643 | 1.030 | 1.030 | 05D3ax |
| 130 | 0.419 | 0.996 | 0.996 | 05D3cf |
| 131 | 0.890 | 1.020 | 1.020 | 05D3cq |
| 132 | 0.805 | 1.120 | 1.140 | 05D3cx |
| 133 | 0.805 | 0.978 | 0.977 | 05D3ha |
| 134 | 0.766 | 1.090 | 1.120 | 05D3hh |
| 135 | 0.338 | 1.000 | 1.000 | 05D3hq |
| 136 | 0.664 | 0.999 | 0.995 | 05D3hs |
| 137 | 0.901 | 1.030 | 1.030 | 05D3ht |
| 138 | 0.745 | 0.996 | 0.995 | 05D3jb |
| 139 | 0.736 | 1.020 | 1.030 | 05D3jk |
| 140 | 0.579 | 0.988 | 0.989 | 05D3jq |
| 141 | 0.850 | 0.990 | 0.992 | 05D3kp |
| 142 | 0.936 | 0.993 | 0.993 | 05D3la |
| 143 | 0.647 | 0.989 | 0.988 | 05D3lb |
| 144 | 0.461 | 1.000 | 1.000 | 05D3lc |
| 145 | 0.670 | 1.030 | 1.030 | 05D3mh |
| 146 | 0.760 | 0.988 | 0.985 | 05D3mn |
| 147 | 0.246 | 0.999 | 0.999 | 05D3mq |
| 148 | 0.470 | 1.020 | 1.020 | 05D3mx |
| 149 | 0.169 | 1.000 | 1.000 | 05D3ne |
| 150 | 0.232 | 1.000 | 1.000 | 06D3cn |
| 151 | 0.442 | 0.993 | 0.993 | 06D3df |
| 152 | 0.726 | 1.020 | 1.020 | 06D3do |
| 153 | 0.282 | 0.999 | 0.999 | 06D3dt |
| 154 | 0.519 | 0.995 | 0.994 | 06D3el |
| 155 | 0.576 | 0.987 | 0.987 | 06D3et |
| 156 | 0.268 | 0.999 | 0.998 | 06D3fp |
| 157 | 0.720 | 1.020 | 1.020 | 06D3gh |
| 158 | 0.250 | 0.999 | 0.999 | 06D3gn |
| 159 | 0.760 | 1.010 | 1.000 | 06D3gx |
| 160 | 0.237 | 0.999 | 0.999 | 07D3ae |
| 161 | 0.355 | 0.997 | 0.996 | 07D3af |
| 162 | 0.451 | 0.994 | 0.994 | 07D3ap |
| 163 | 0.920 | 0.996 | 0.996 | 07D3bo |
| 164 | 0.910 | 1.130 | 1.150 | 07D3bt |
| 165 | 0.708 | 0.999 | 0.996 | 07D3cc |
| 166 | 0.807 | 0.997 | 0.995 | 07D3cp |
| 167 | 0.808 | 0.976 | 0.976 | 07D3cs |
| 168 | 0.512 | 0.993 | 0.993 | 07D3cu |
| 169 | 0.837 | 0.973 | 0.973 | 07D3da |
| 170 | 1.020 | 1.030 | 1.030 | 07D3do |
| 171 | 0.740 | 1.020 | 1.020 | 07D3ey |
| 172 | 0.830 | 1.010 | 1.000 | 07D3gm |

*Continued on next page*

Table A.1 – *Continued from previous page*

| No. | redshift | $\mu_{SIS}$ | $\mu_{NFW}$ | name |
|---|---|---|---|---|
| 173 | 0.669 | 1.010 | 1.010 | 07D3gt |
| 174 | 0.391 | 0.995 | 0.995 | 07D3gw |
| 175 | 0.670 | 1.010 | 1.010 | 07D3hl |
| 176 | 0.572 | 0.992 | 0.991 | 07D3hu |
| 177 | 0.351 | 1.010 | 1.010 | 07D3hv |
| 178 | 0.748 | 1.010 | 1.000 | 07D3hw |
| 179 | 0.505 | 0.992 | 0.992 | 07D3hz |
| 180 | 0.680 | 1.000 | 0.999 | 07D3ib |
| 181 | 0.928 | 1.030 | 1.030 | 08D3dx |
| 182 | 0.170 | 1.000 | 1.000 | 08D3gb |
| 183 | 0.352 | 1.010 | 1.010 | 08D3gf |
| 184 | 0.767 | 0.990 | 0.987 | 08D3gu |
| 185 | 0.468 | 1.020 | 1.020 | 03D4au |
| 186 | 0.270 | 0.994 | 0.994 | 03D4cj |
| 187 | 0.949 | 1.170 | 1.190 | 03D4cx |
| 188 | 0.610 | 0.982 | 0.979 | 03D4dy |
| 189 | 0.592 | 0.995 | 0.992 | 03D4gg |
| 190 | 0.571 | 0.979 | 0.978 | 03D4gl |
| 191 | 0.613 | 0.986 | 0.984 | 04D4an |
| 192 | 0.880 | 0.978 | 0.977 | 04D4bk |
| 193 | 0.550 | 1.170 | 1.200 | 04D4bq |
| 194 | 0.811 | 0.973 | 0.971 | 04D4dm |
| 195 | 1.031 | 1.030 | 1.050 | 04D4dw |
| 196 | 0.629 | 1.020 | 1.020 | 04D4fx |
| 197 | 0.936 | 1.020 | 1.010 | 04D4hf |
| 198 | 0.699 | 0.994 | 0.993 | 04D4ib |
| 199 | 0.687 | 0.987 | 0.984 | 04D4ic |
| 200 | 0.866 | 0.997 | 1.000 | 04D4ii |
| 201 | 0.751 | 1.030 | 1.040 | 04D4im |
| 202 | 0.472 | 1.000 | 1.000 | 04D4ju |
| 203 | 0.961 | 0.988 | 0.985 | 04D4jw |
| 204 | 0.930 | 0.966 | 0.964 | 04D4jy |
| 205 | 0.640 | 0.982 | 0.979 | 05D4ag |
| 206 | 0.509 | 0.982 | 0.981 | 05D4av |
| 207 | 0.701 | 1.060 | 1.070 | 05D4bj |
| 208 | 0.372 | 0.993 | 0.993 | 05D4bm |
| 209 | 0.790 | 1.030 | 1.030 | 05D4cs |
| 210 | 0.855 | 0.990 | 0.991 | 05D4dw |
| 211 | 0.810 | 0.979 | 0.975 | 05D4dy |
| 212 | 0.605 | 1.030 | 1.030 | 05D4ef |
| 213 | 0.536 | 0.982 | 0.980 | 05D4ek |
| 214 | 0.402 | 0.988 | 0.987 | 05D4ff |
| 215 | 0.808 | 0.975 | 0.972 | 05D4gw |
| 216 | 0.552 | 0.989 | 0.989 | 06D4bo |
| 217 | 1.000 | 1.070 | 1.080 | 06D4cl |
| 218 | 0.303 | 1.010 | 1.020 | 06D4dh |

*Continued on next page*

Table A.1 – *Continued from previous page*

| No. | redshift | $\mu_{SIS}$ | $\mu_{NFW}$ | name |
|---|---|---|---|---|
| 219 | 0.769 | 1.060 | 1.080 | 06D4dr |
| 220 | 0.677 | 0.981 | 0.980 | 06D4fc |
| 221 | 0.566 | 0.986 | 0.983 | 06D4jh |
| 222 | 0.760 | 0.984 | 0.982 | 06D4jt |
| 223 | 0.456 | 0.995 | 0.993 | 07D4cy |
| 224 | 0.743 | 1.010 | 1.000 | 07D4dp |
| 225 | 0.554 | 0.983 | 0.981 | 07D4dq |
| 226 | 0.772 | 0.981 | 0.977 | 07D4dr |
| 227 | 0.653 | 0.990 | 0.985 | 07D4ec |
| 228 | 0.520 | 1.010 | 1.010 | 07D4ed |
| 229 | 0.370 | 0.989 | 0.988 | 07D4ei |
| 230 | 0.503 | 1.000 | 1.000 | 07D4fl |

**Table A.1**: List of magnifications for all selected SNe from SNLS5 sample computed with two halo models

# Bibliography

[Ade14] P. A. R. Ade, P. Collaboration, N. Aghanim, et al. (2014). *Planck 2013 results. xvi. cosmological parameters.* Astronomy & Astrophysics, 571:A16. http://arxiv.org/abs/1303.5076, arXiv:1303.5076. v, vi, 8, 18

[Ade15] P. A. R. Ade, P. Collaboration, N. Aghanim, et al. (2015). *Planck 2015 results. xiii. cosmological parameters.* ArXiv e-prints. http://arxiv.org/abs/1502.01589, arXiv:1502.01589. v, vii, 5, 16, 17, 38

[Ala98] C. Alard and R. H. Lupton (1998). *A Method for Optimal Image Subtraction.* Astrophysical Journal, 503:325–331. astro-ph/9712287. 33

[Alc96] C. Alcock, R. A. Allsman, D. Alves, et al. (1996). *The MACHO Project: Limits on Planetary Mass Dark Matter in the Galactic Halo from Gravitational Microlensing.* Astrophysical Journal, 471:774. astro-ph/9604176. 48

[Ama15] A. Amara (2015). *Gravitational lenses of the universe.* University Lecture. vii, 44

[And14] L. Anderson, E. Aubourg, S. Bailey, et al. (2014). *The clustering of galaxies in the sdss-iii baryon oscillation spectroscopic survey: Baryon acoustic oscillations in the data release 10 and 11 galaxy samples.* Monthly Notices of the Royal Astronomical Society, 441:24–62. http://arxiv.org/abs/1312.4877, arXiv:1312.4877. vi, 18, 19

[Ast06] P. Astier, J. Guy, N. Regnault, et al. (2006). *The supernova legacy survey: Measurement of $\omega_m$, $\omega_\lambda$ and $w$ from the first year data set.* Astron.Astrophys., 447:31–48. http://arxiv.org/abs/astro-ph/0510447, astro-ph/0510447. vi, 25, 37

[Ast12] P. Astier (2012). *The expansion of the universe observed with supernovae.* ArXiv e-prints. http://arxiv.org/abs/1211.2590, arXiv:1211.2590 [astro-ph.CO]. vii, 26, 33

[Ast13] P. Astier, P. E. Hage, J. Guy, et al. (2013). *Photometry of supernovae in an image series : methods and application to the supernova legacy survey (snls).* Astronomy & Astrophysics, 557:A55. http://arxiv.org/abs/1306.5153, arXiv:1306.5153 [astro-ph.IM]. 33, 64, 65

[Bab91] A. Babul and M. H. Lee (1991). *Gravitational lensing by smooth inhomogeneities in the universe.* Monthly Notices of the Royal Astronomical Society, 250:407–413. 81

[Bal09] C. Balland, S. Baumont, S. Basa, et al. (2009). *The eso/vlt 3rd year type ia supernova data set from the supernova legacy survey.* Astronomy & Astrophysics, 507:85–103. http://arxiv.org/abs/0909.3316, arXiv:0909.3316. 32

[Bar01] M. Bartelmann and P. Schneider (2001). *Weak gravitational lensing.* physrep, 340:291–472. http://arxiv.org/abs/astro-ph/9912508, astro-ph/9912508. vii, 43

[Bau13] D. Bauman (2013). *Thermal history.* University Lecture. xi, 9

[Baz11] G. Bazin, V. Ruhlmann-Kleider, N. Palanque-Delabrouille, et al. (2011). *Photometric selection of Type Ia supernovae in the Supernova Legacy Survey.* Astronomy & Astrophysics, 534:A43. arXiv:1109.0948 [astro-ph.CO]. 33

[Beg87] K. G. Begeman (1987). *HI rotation curves of spiral galaxies.* Ph.D. thesis, , Kapteyn Institute, (1987). 51

[Beg91] K. G. Begeman, A. H. Broeils, and R. H. Sanders (1991). *Extended rotation curves of spiral galaxies - Dark haloes and modified dynamics.* Monthly Notices of the Royal Astronomical Society, 249:523–537. viii, 52

[Ben03] C. L. Bennett et al. (WMAP) (2003). *The Microwave Anisotropy Probe (MAP) mission.* Astrophys. J., 583:1–23. astro-ph/0301158. 17

[Ben13] C. L. Bennett, D. Larson, J. L. Weiland, et al. (2013). *Nine-year Wilkinson Microwave Anisotropy Probe (WMAP) Observations: Final Maps and Results.* Astrophysical Journal Supplement Series, 208:20. arXiv:1212.5225. 38

[Ber96] E. Bertin and S. Arnouts (1996). *Sextractor: Software for source extraction.* Astronomy and Astrophysics Supplement Series, 117(2):393–404. 64

[Bet13] M. Betoule, J. Marriner, N. Regnault, et al. (2013). *Improved photometric calibration of the SNLS and the SDSS supernova surveys.* Astronomy & Astrophysics, 552:A124. arXiv:1212.4864. vii, 32, 34, 35, 37

[Bet14] M. Betoule, R. Kessler, J. Guy, et al. (2014). *Improved cosmological constraints from a joint analysis of the sdss-ii and snls supernova samples.* Astronomy & Astrophysics, 568:A22. http://arxiv.org/abs/1401.4064, arXiv:1401.4064. vi, vii, 1, 20, 27, 38, 90

[Bla11] C. Blake, E. A. Kazin, F. Beutler, et al. (2011). *The wigglez dark energy survey: mapping the distance–redshift relation with baryon acoustic oscillations.* Monthly Notices of the Royal Astronomical Society, 418(3):1707–1724. vi, 18, 19

[Boe04] A. Boehm, B. Ziegler, R. Saglia, et al. (2004). *The tully-fisher relation at intermediate redshift.* Astronomy & Astrophysics, 420:97–114. http://arxiv.org/abs/astro-ph/0309263, astro-ph/0309263. ix, 58, 59, 60, 77, 78

[Boh04] R. C. Bohlin and R. L. Gilliland (2004). *Hubble Space Telescope Absolute Spectrophotometry of Vega from the Far-Ultraviolet to the Infrared.* Astronomical Journal, 127:3508–3515. 35

[Boh10] R. C. Bohlin (2010). *Hubble Space Telescope Spectrophotometry and Models for Solar Analogs.* Astronomical Journal, 139:1515–1520. arXiv:1002.4381 [astro-ph.SR]. 35

[Bou03] O. Boulade, X. Charlot, P. Abbon, et al. (2003). *MegaCam: the new Canada-France-Hawaii Telescope wide-field imaging camera.* In M. Iye and A. F. M. Moorwood, editors, *Society of Photo-Optical Instrumentation Engineers (SPIE) Conference Series*, volume 4841 of *Society of Photo-Optical Instrumentation Engineers (SPIE) Conference Series*, pages 72–81. 28

[Bra82] D. Branch, R. Buta, S. W. Falk, et al. (1982). *Interpretation of the maximum light spectrum of a Type I supernova.* Astrophysical Journal Letters, 252:L61–L64. 23

[Bra83] D. Branch, C. H. Lacy, M. L. McCall, et al. (1983). *The Type I supernova 1981b in NGC 4536 - The first 100 days*. Astrophysical Journal, 270:123–125. 23

[Bri13] F. Brimioulle, S. Seitz, M. Lerchster, et al. (2013). *Dark matter halo properties from galaxy-galaxy lensing*. Monthly Notices of the Royal Astronomical Society, 432:1046–1102. arXiv:1303.6287. 52, 59, 60

[Bro08] T. J. Bronder, I. M. Hook, P. Astier, et al. (2008). *SNLS spectroscopy: testing for evolution in type Ia supernovae*. Astronomy & Astrophysics, 477:717–734. arXiv:0709.0859. 32

[Bry98] G. L. Bryan and M. L. Norman (1998). *Statistical Properties of X-Ray Clusters: Analytic and Numerical Comparisons*. Astrophysical Journal, 495:80–99. astro-ph/9710107. 51

[Bul01] J. S. Bullock, T. S. Kolatt, Y. Sigad, et al. (2001). *Profiles of dark haloes: evolution, scatter and environment*. Monthly Notices of the Royal Astronomical Society, 321:559–575. astro-ph/9908159. 55

[Bun09] E. F. Bunn and D. W. Hogg (2009). *The kinematic origin of the cosmological redshift*. American Journal of Physics, 77:688–694. http://arxiv.org/abs/0808.1081, arXiv:0808.1081 [physics.pop-ph]. 6

[Bur97] A. Burkert (1997). *The structure of dark matter halos. observation versus theory*. In H. V. Klapdor-Kleingrothaus and Y. Ramachers, editors, *Dark matter in Astro- and Particle Physics*, page 35. http://arxiv.org/abs/astro-ph/9703057, astro-ph/9703057. 55

[Car89] J. A. Cardelli, G. C. Clayton, and J. S. Mathis (1989). *The relationship between infrared, optical, and ultraviolet extinction*. APJ, 345:245–256. 36

[Car92] S. M. Carroll, W. H. Press, and E. L. Turner (1992). *The cosmological constant*. araa, 30:499–542. 6

[Cla16] J. Clampitt, C. Sánchez, J. Kwan, et al. (2016). *Galaxy-Galaxy Lensing in the DES Science Verification Data*. ArXiv e-prints. arXiv:1603.05790. 59

[Col69] S. A. Colgate and C. McKee (1969). *Early Supernova Luminosity*. Astrophysical Journal, 157:623. 24

[Con08] A. Conley, M. Sullivan, E. Y. Hsiao, et al. (2008). *SiFTO: An Empirical Method for Fitting SN Ia Light Curves*. Astrophysical Journal, 681:482–498. arXiv:0803.3441. 36

[Con11] A. Conley, J. Guy, M. Sullivan, et al. (2011). *Supernova constraints and systematic uncertainties from the first 3 years of the supernova legacy survey*. Astrophysical Journal Supplement Series, 192:1. http://arxiv.org/abs/1104.1443, arXiv:1104.1443 [astro-ph.CO]. 27, 30, 38

[Cou09] Coupon, J. Coupon, O. Ilbert, et al. (2009). *Photometric redshifts for the cfhtls t0004 deep and wide fields "*. Astronomy & Astrophysics, 500(3):18. 74

[Dav03] M. Davis, S. M. Faber, J. A. Newman, et al. (2003). *Science objectives and early results of the deep2 redshift survey*. In P. Guhathakurta, editor, *Discoveries and Research Prospects from 6- to 10-Meter-Class Telescopes II*, volume 4834 of *Society of Photo-Optical Instrumentation Engineers (SPIE) Conference Series*, pages 161–172. http://arxiv.org/abs/astro-ph/0209419, astro-ph/0209419. 70

[Dav07] M. Davis, P. Guhathakurta, N. Konidaris, et al. (2007). *The all-wavelength extended groth strip international survey (aegis) data sets*. Astrophysical Journal Letters, 660:L1–L6. http://arxiv.org/abs/astro-ph/0607355, astro-ph/0607355. 70

[Dem16] C. Demetroullas and M. L. Brown (2016). *Cross-correlation cosmic shear with the SDSS and VLA FIRST surveys*. Monthly Notices of the Royal Astronomical Society, 456:3100–3118. arXiv:1507.05977. 20

[deV82] G. deVaucouleurs and D. W. Olson (1982). *The central velocity dispersion in elliptical and lenticular galaxies as an extragalactic distance indicator*. Astrophysical Journal, 256:346–369. 58

[Djo87] S. Djorgovski and M. Davis (1987). *Fundamental properties of elliptical galaxies*. Astrophysical Journal, 313:59–68. 58

[Dod06] S. Dodelson and A. Vallinotto (2006). *Learning from the scatter in type Ia supernovae*. Physical Review D, 74(6):063515. arXiv:astro-ph/0511086. 61

[Dre87] A. Dressler, D. Lynden-Bell, D. Burstein, et al. (1987). *Spectroscopy and photometry of elliptical galaxies. I - A new distance estimator*. Astrophysical Journal, 313:42–58. 58

[Duf08] A. R. Duffy, J. Schaye, S. T. Kay, et al. (2008). *Dark matter halo concentrations in the wilkinson microwave anisotropy probe year 5 cosmology*. Monthly Notices of the Royal Astronomical Society, 390:L64–L68. http://arxiv.org/abs/0804.2486, 0804.2486. ix, 55, 56, 78, 79

[EH14] P. El Hage (2014). *Measurement of the Dark Energy Equation of State Using the Full SNLS Supernova Sample*. Theses, Université Pierre et Marie Curie - Paris VI. https://tel.archives-ouvertes.fr/tel-01127401. 38

[Ein36] A. Einstein (1936). *Lens-Like Action of a Star by the Deviation of Light in the Gravitational Field*. Science, 84:506–507. 46

[Ein52] A. Einstein (1952). *The foundation of the general theory of relativity*, pages 109–164. 3

[Eis96] D. J. Eisenstein and A. Loeb (1996). *Can the Tully-Fisher Relation Be the Result of Initial Conditions?* Astrophysical Journal, 459:432. astro-ph/9506074. 58

[Eis98] D. J. Eisenstein and W. Hu (1998). *Baryonic Features in the Matter Transfer Function*. Astrophysical Journal, 496:605–614. http://adsabs.harvard.edu/abs/1998ApJ...496..605E, astro-ph/9709112. 13

[Eis05] D. J. Eisenstein, I. Zehavi, D. W. Hogg, et al. (2005). *Detection of the baryon acoustic peak in the large-scale correlation function of sdss luminous red galaxies*. Astrophysical Journal, 633:560–574. http://arxiv.org/abs/astro-ph/0501171, astro-ph/0501171. 37

[Ell08] R. S. Ellis, M. Sullivan, P. E. Nugent, et al. (2008). *Verifying the Cosmological Utility of Type Ia Supernovae: Implications of a Dispersion in the Ultraviolet Spectra*. Astrophysical Journal, 674:51–69. arXiv:0710.3896. 32

[Erb09] T. Erben, H. Hildebrandt, M. Lerchster, et al. (2009). *Cars: the cfhtls-archive-research survey; i. five-band multi-colour data from 37 sq. deg. cfhtls-wide observations*. Astronomy & Astrophysics, 493:1197–1222. http://arxiv.org/abs/0811.2239, arXiv:0811.2239. ix, 68, 69, 76

[Fab76] S. M. Faber and R. E. Jackson (1976). *Velocity dispersions and mass-to-light ratios for elliptical galaxies*. Astrophysical Journal, 204:668–683. 58

[Fer09] M. Fernández Lorenzo, J. Cepa, A. Bongiovanni, et al. (2009). *Evolution of the optical Tully-Fisher relation up to z = 1.3*. Astronomy & Astrophysics, 496:389–397. arXiv:0901.3676 [astro-ph.CO]. 58

[Fie14] B. D. Fields, P. Molaro, and S. Sarkar (2014). *Big-Bang Nucleosynthesis*. ArXiv e-prints. arXiv:1412.1408. v, 8

[Fio99] M. Fioc and B. Rocca-Volmerange (1999). *PEGASE.2, a metallicity-consistent spectral evolution model of galaxies: the documentation and the code*. ArXiv Astrophysics e-prints. arXiv:astro-ph/9912179. 37, 68

[Fix96] D. J. Fixsen, E. S. Cheng, J. M. Gales, et al. (1996). *The Cosmic Microwave Background Spectrum from the Full COBE FIRAS Data Set*. Astrophysical Journal, 473:576. astro-ph/9605054. 3

[Fix09] D. J. Fixsen (2009). *The Temperature of the Cosmic Microwave Background*. Astrophysical Journal, 707:916–920. arXiv:0911.1955. 3

[Fri96] J. A. Frieman (1996). *Weak Lensing and the Measurement of $q_0$; from Type Ia Supernovae*. Comments on Astrophysics, 18:323. arXiv:astro-ph/9608068. 60

[Fri99] A. Friedmann (1999). *On the Curvature of Space*. General Relativity and Gravitation, 31:1991. 3

[Fri08] J. A. Frieman, M. S. Turner, and D. Huterer (2008). *Dark Energy and the Accelerating Universe*. Annual Review of Astronomy & Astrophysics, 46:385–432. arXiv:0803.0982. vi, 19

[Fu14] L. Fu, M. Kilbinger, T. Erben, et al. (2014). *CFHTLenS: cosmological constraints from a combination of cosmic shear two-point and three-point correlations*. Monthly Notices of the Royal Astronomical Society, 441:2725–2743. arXiv:1404.5469. 20

[Gal05] J. S. Gallagher, P. M. Garnavich, P. Berlind, et al. (2005). *Chemistry and Star Formation in the Host Galaxies of Type Ia Supernovae*. Astrophysical Journal, 634:210–226. astro-ph/0508180. 27

[Gal08] J. S. Gallagher, P. M. Garnavich, N. Caldwell, et al. (2008). *Supernovae in Early-Type Galaxies: Directly Connecting Age and Metallicity with Type Ia Luminosity*. Astrophysical Journal, 685:752–766. arXiv:0805.4360. 27

[Gou06] A. Gould, A. Udalski, D. An, et al. (2006). *Microlens OGLE-2005-BLG-169 Implies That Cool Neptune-like Planets Are Common*. Astrophysical Journal Letters, 644:L37–L40. astro-ph/0603276. 46

[Gun04] C. Gunnarsson (2004). *Q-let - quick lensing estimation tool - an application to sn2003es*. Journal of Cosmology and Astroparticle Physics, 3:002. http://arxiv.org/abs/astro-ph/0311380, astro-ph/0311380. 77, 79

[Gun05] C. Gunnarsson (2005). *Supernovae under the gravitational lens*. Ph.D. thesis, University of Stockholm. 52, 56, 77, 79

[Gun06] C. Gunnarsson, T. Dahlen, A. Goobar, et al. (2006). *Corrections for gravitational lensing of supernovae: better than average?* Astrophysical Journal, 640:417–427. http://arxiv.org/abs/astro-ph/0506764, astro-ph/0506764. xi, 58, 60, 61

[Guy07] J. Guy, P. Astier, S. Baumont, et al. (2007). *SALT2: using distant supernovae to improve the use of type Ia supernovae as distance indicators*. Astronomy & Astrophysics, 466:11–21. arXiv:astro-ph/0701828. 25, 36

[Ham00] M. Hamuy, S. C. Trager, P. A. Pinto, et al. (2000). *A Search for Environmental Effects on Type IA Supernovae*. Astronomical Journal, 120:1479–1486. astro-ph/0005213. 27

[Har70] E. R. Harrison (1970). *Fluctuations at the Threshold of Classical Cosmology*. Physical Review D, 1:2726–2730. 13

[Har12] D. Hardin (2012). *Thèse d'habilitation à diriger des recherches*. 64, 68

[Har16a] D. Hardin (2015-2016). *Notes on general cosmology*. 3

[Har16b] D. Hardin (2015-2016). *Notes on gravitationnal lensing*. 41

[Hei03] J. Heidt, I. Appenzeller, A. Gabasch, et al. (2003). *The fors deep field: Field selection, photometric observations and photometric catalog*. Astronomy & Astrophysics, 398:49–61. http://arxiv.org/abs/astro-ph/0211044, astro-ph/0211044. 58

[Hey13] C. Heymans, E. Grocutt, A. Heavens, et al. (2013). *CFHTLenS tomographic weak lensing cosmological parameter constraints: Mitigating the impact of intrinsic galaxy alignments*. Monthly Notices of the Royal Astronomical Society, 432:2433–2453. arXiv:1303.1808. 20

[Hin13] G. Hinshaw, D. Larson, E. Komatsu, et al. (2013). *Nine-year wilkinson microwave anisotropy probe (wmap) observations: Cosmological parameter results*. Astrophysical Journal Supplement Series, 208:19. http://arxiv.org/abs/1212.5226, arXiv:1212.5226. 5

[Hoe96] P. Hoeflich, A. Khokhlov, J. C. Wheeler, et al. (1996). *Maximum Brightness and Postmaximum Decline of Light Curves of Type IA Supernovae: A Comparison of Theory and Observations*. Astrophysical Journal Letters, 472:L81. arXiv:astro-ph/9609070. 24

[Hoe04] H. Hoekstra, H. K. C. Yee, and M. D. Gladders (2004). *Properties of Galaxy Dark Matter Halos from Weak Lensing*. Astrophysical Journal, 606:67–77. astro-ph/0306515. 59

[Hog99] D. W. Hogg (1999). *Distance measures in cosmology*. ArXiv Astrophysics e-prints. http://arxiv.org/abs/astro-ph/9905116, astro-ph/9905116. v, 10, 12

[Hog05] D. W. Hogg, D. J. Eisenstein, M. R. Blanton, et al. (2005). *Cosmic Homogeneity Demonstrated with Luminous Red Galaxies*. Astrophysical Journal, 624:54–58. astro-ph/0411197. 3

[Hol05] D. E. Holz and E. V. Linder (2005). *Safety in numbers: Gravitational lensing degradation of the luminosity distance-redshift relation*. Astrophysical Journal, 631:678–688. http://arxiv.org/abs/astro-ph/0412173, astro-ph/0412173. viii, 60, 61

[How05] D. A. Howell, M. Sullivan, K. Perrett, et al. (2005). *Gemini Spectroscopy of Supernovae from the Supernova Legacy Survey: Improving High-Redshift Supernova Selection and Classification*. Astrophysical Journal, 634:1190–1201. arXiv:astro-ph/0509195. 32

[How09] D. A. Howell, M. Sullivan, E. F. Brown, et al. (2009). *The Effect of Progenitor Age and Metallicity on Luminosity and $^{56}$Ni Yield in Type Ia Supernovae*. Astrophysical Journal, 691:661–671. arXiv:0810.0031. 27

[Hoy60] F. Hoyle and W. A. Fowler (1960). *Nucleosynthesis in Supernovae.* Astrophysical Journal, 132:565. 24

[Hub26] E. P. Hubble (1926). *Extragalactic nebulae.* Astrophysical Journal, 64:321–369. 3

[Hub29] E. Hubble (1929). *A Relation between Distance and Radial Velocity among Extra-Galactic Nebulae*. Proceedings of the National Academy of Science, 15:168–173. 5

[Hub36] E. P. Hubble (1936). *Realm of the Nebulae*. 3

[Ibe83] I. Iben, Jr. and A. Renzini (1983). *Asymptotic giant branch evolution and beyond*. Annual Review of Astronomy & Astrophysics, 21:271–342. 24

[Ilb06] O. Ilbert, S. Arnouts, H. J. McCracken, et al. (2006). *Accurate photometric redshifts for the CFHT legacy survey calibrated using the VIMOS VLT deep survey*. Astronomy & Astrophysics, 457:841–856. astro-ph/0603217. 70

[Ilb09] O. Ilbert, P. Capak, M. Salvato, et al. (2009). *Cosmos Photometric Redshifts with 30-BANDS for 2-deg$^2$*. Astrophysical Journal, 690:1236–1249. 70

[Jha07] S. Jha, A. G. Riess, and R. P. Kirshner (2007). *Improved Distances to Type Ia Supernovae with Multicolor Light-Curve Shapes: MLCS2k2*. Astrophysical Journal, 659:122–148. arXiv:astro-ph/0612666. 25, 36

[Joh07] D. E. Johnston, E. S. Sheldon, R. H. Wechsler, et al. (2007). *Cross-correlation Weak Lensing of SDSS galaxy Clusters II: Cluster Density Profiles and the Mass–Richness Relation*. ArXiv e-prints. arXiv:0709.1159. 56

[Jon06] J. Jonsson, T. Dahlen, A. Goobar, et al. (2006). *Lensing magnification of supernovae in the goods-fields*. Astrophysical Journal, 639:991–998. http://arxiv.org/abs/astro-ph/0506765, astro-ph/0506765. 94

[Jon07] J. Jonsson, T. Dahlén, A. Goobar, et al. (2007). *Tentative detection of the gravitational magnification of Type Ia supernovae*. Journal of Cosmology and Astroparticle Physics, 6:002. arXiv:astro-ph/0612329. 62

[Jon08] J. Jonsson, T. Kronborg, E. Mortsell, et al. (2008). *Prospects and pitfalls of gravitational lensing in large supernova surveys*. Astronomy & Astrophysics, 487:467–473. http://arxiv.org/abs/0806.1387, 0806.1387. 58, 61

[Jon09] D. H. Jones, M. A. Read, W. Saunders, et al. (2009). *The 6df galaxy survey: Final redshift release (dr3) and southern large-scale structures*. Monthly Notices of the Royal Astronomical Society, 399:683–698. http://arxiv.org/abs/0903.5451, arXiv:0903.5451. vi, 18, 19

[Jon10] J. Jonsson, M. Sullivan, I. Hook, et al. (2010). *Constraining dark matter halo properties using lensed snls supernovae*. Monthly Notices of the Royal Astronomical Society, 405:535–544. http://arxiv.org/abs/1002.1374, arXiv:1002.1374. 62, 77, 94, 95

[Kai93] N. Kaiser and G. Squires (1993). *Mapping the dark matter with weak gravitational lensing*. Astrophysical Journal, 404:441–450. vii, 47

[Kas07] D. Kasen and S. E. Woosley (2007). *On the Origin of the Type Ia Supernova Width-Luminosity Relation*. Astrophysical Journal, 656:661–665. arXiv:astro-ph/0609540. 24

[Kas09] D. Kasen, F. K. Röpke, and S. E. Woosley (2009). *The diversity of type Ia supernovae from broken symmetries*. Nature, 460:869–872. arXiv:0907.0708 [astro-ph.HE]. 25

[Kee07] W. Keel (2007). *The Road to Galaxy Formation*. Springer Praxis Books. Springer Berlin Heidelberg. https://books.google.fr/books?id=BUgJGypUYF0C. v, 4

[Kel10] P. L. Kelly, M. Hicken, D. L. Burke, et al. (2010). *Hubble Residuals of Nearby Type Ia Supernovae are Correlated with Host Galaxy Masses*. Astrophysical Journal, 715:743–756. arXiv:0912.0929. 27

[Kel15] P. L. Kelly, S. A. Rodney, T. Treu, et al. (2015). *Multiple images of a highly magnified supernova formed by an early-type cluster galaxy lens*. Science, 347:1123–1126. http://arxiv.org/abs/1411.6009, arXiv:1411.6009. 61

[Ken94] S. M. Kent (1994). *Sloan digital sky survey*. In *Science with astronomical near-infrared sky surveys*, pages 27–30. Springer. 18

[Kle04] M. Kleinheinrich, P. Schneider, H. . Rix, et al. (2004). *Weak lensing measurements of dark matter halos of galaxies from COMBO-17*. ArXiv Astrophysics e-prints. astro-ph/0412615. 59

[Koc94] C. S. Kochanek (1994). *The dynamics of luminous galaxies in isothermal halos*. Astrophysical Journal, 436:56–66. 58

[Koc06] C. Kochanek (2006). *Gravitational Lensing: Strong, Weak and Micro*. In G. Meylan, P. Jetzer, P. North, et al., editors, *The Saas Fee Lectures on Strong Gravitational Lensing*. http://arxiv.org/abs/astro-ph/0407232, astro-ph/0407232. vii, viii, 47, 57

[Kro80] R. G. Kron (1980). *Photometry of a complete sample of faint galaxies*. Astrophysical Journal Supplement Series, 43:305–325. 64

[Kro10] T. Kronborg, D. Hardin, J. Guy, et al. (2010). *Gravitational lensing in the supernova legacy survey (SNLS)*. Astronomy & Astrophysics, 514:A44. arXiv:1002.1249. ix, 2, 58, 62, 63, 68, 86, 95

[Lac93] C. Lacey and S. Cole (1993). *Merger rates in hierarchical models of galaxy formation*. Monthly Notices of the Royal Astronomical Society, 262:627–649. viii, 50

[Lan92] A. U. Landolt (1992). *UBVRI photometric standard stars in the magnitude range 11.5-16.0 around the celestial equator*. Astronomical Journal, 104:340–371. 35

[Lan07] A. U. Landolt and A. K. Uomoto (2007). *Optical Multicolor Photometry of Spectrophotometric Standard Stars*. Astronomical Journal, 133:768–790. arXiv:0704.3030. vii, 34, 35

[Las90] B. M. Lasker, C. R. Sturch, B. J. McLean, et al. (1990). *The Guide Star Catalog. I - Astronomical foundations and image processing*. Astronomical Journal, 99:2019–2058. 68

[Las08] B. M. Lasker, M. G. Lattanzi, B. J. McLean, et al. (2008). *The Second-Generation Guide Star Catalog: Description and Properties*. Astronomical Journal, 136:735–766. arXiv:0807.2522. 68

[Le 04] O. Le Fèvre, Y. Mellier, H. J. McCracken, et al. (2004). *The VIRMOS deep imaging survey. I. Overview, survey strategy, and CFH12K observations*. Astronomy & Astrophysics, 417:839–846. arXiv:astro-ph/0306252. ix, 70, 71

[Lem33] G. Lemaître (1933). *L'Univers en expansion*. Annales de la Société Scientifique de Bruxelles, 53:51. 3

[Lid03] A. Liddle (2003). *An Introduction to Modern Cosmology*. Wiley. https://books.google.fr/books?id=zZg3AQAAIAAJ. 3

[Liu15] J. Liu and J. C. Hill (2015). *Cross-correlation of Planck CMB lensing and CFHTLenS galaxy weak lensing maps*. Physical Review D, 92(6):063517. arXiv:1504.05598. xi, 21

[Lov02] J. Loveday (2002). *The Sloan Digital Sky Survey*. Contemporary Physics, 43:437–449. http://adsabs.harvard.edu/abs/2002ConPh..43..437L, astro-ph/0207189. 18

[Maa15] R. Maartens, F. B. Abdalla, M. Jarvis, et al. (2015). *Cosmology with the ska – overview*. ArXiv e-prints. http://arxiv.org/abs/1501.04076, arXiv:1501.04076. v, 4

[Mac08] A. V. Maccio', A. A. Dutton, and F. C. van den Bosch (2008). *Concentration, spin and shape of dark matter haloes as a function of the cosmological model: Wmap1, wmap3 and wmap5 results*. Monthly Notices of the Royal Astronomical Society, 391:1940–1954. http://arxiv.org/abs/0805.1926, 0805.1926. 56

[Man06] R. Mandelbaum, U. Seljak, G. Kauffmann, et al. (2006). *Galaxy halo masses and satellite fractions from galaxy-galaxy lensing in the Sloan Digital Sky Survey: stellar mass, luminosity, morphology and environment dependencies*. Monthly Notices of the Royal Astronomical Society, 368:715–731. astro-ph/0511164. 59

[Man08] R. Mandelbaum, U. Seljak, and C. M. Hirata (2008). *Halo mass - concentration relation from weak lensing*. Journal of Cosmology and Astroparticle Physics, 8:006. http://arxiv.org/abs/0805.2552, 0805.2552. 55

[Man15] A. B. Mantz et al. (2015). *Weighing the giants – IV. Cosmology and neutrino mass*. Mon. Not. Roy. Astron. Soc., 446:2205–2225. arXiv:1407.4516 [astro-ph.CO]. v, vi, 14, 22

[Mas07] R. Massey, J. Rhodes, R. Ellis, et al. (2007). *Dark matter maps reveal cosmic scaffolding*. Nature, 445:286–290. astro-ph/0701594. vii, 47

[Meg05] MegaCam (2005). *The megaprime/megacam home page*. Website. Http://www.cfht.hawaii.edu/Instruments/Imaging/MegaPrime/. vi, 29

[Mén05] B. Ménard and N. Dalal (2005). *Revisiting the magnification of type ia supernovae with sdss*. Monthly Notices of the Royal Astronomical Society, 358:101–104. http://arxiv.org/abs/astro-ph/0407023, astro-ph/0407023. 62

[Men15] M. Meneghetti (2015). *Introduction to gravitational lensing*. University Lecture. 41

[Met99] R. B. Metcalf (1999). *Gravitational lensing of high-redshift Type IA supernovae: a probe of medium-scale structure*. Monthly Notices of the Royal Astronomical Society, 305:746–754. arXiv:astro-ph/9803319. 61

[Mit05] J. L. Mitchell, C. R. Keeton, J. A. Frieman, et al. (2005). *Improved cosmological constraints from gravitational lens statistics*. Astrophysical Journal, 622:81–98. http://arxiv.org/abs/astro-ph/0401138, astro-ph/0401138. ix, 58, 59, 60, 77, 78

[Mon98] D. Monet, A. Bird, B. Canzian, et al. (1998). *The usno-a2. 0 catalogue*. US Naval Observatory, Washington DC. 68

[Mor01] J. E. Morrison, S. Röser, B. McLean, et al. (2001). *The Guide Star Catalog, Version 1.2: An Astrometric Recalibration and Other Refinements*. Astronomical Journal, 121:1752–1763. 68

[Nar96] R. Narayan and M. Bartelmann (1996). *Lectures on Gravitational Lensing*. ArXiv Astrophysics e-prints. astro-ph/9606001. 41

[Nav95] J. F. Navarro, C. S. Frenk, and S. D. M. White (1995). *The assembly of galaxies in a hierarchically clustering universe*. Monthly Notices of the Royal Astronomical Society, 275:56–66. astro-ph/9408067. 53

[Nav96] J. F. Navarro, C. S. Frenk, and S. D. M. White (1996). *The Structure of Cold Dark Matter Halos*. Astrophysical Journal, 462:563. astro-ph/9508025. 53, 55

[Nav97] J. F. Navarro, C. S. Frenk, and S. D. M. White (1997). *A Universal Density Profile from Hierarchical Clustering*. Astrophysical Journal, 490:493–508. astro-ph/9611107. viii, 53, 54, 55

[Net07] A. F. Neto, L. Gao, P. Bett, et al. (2007). *The statistics of lcdm halo concentrations*. Monthly Notices of the Royal Astronomical Society, 381:1450–1462. http://arxiv.org/abs/0706.2919, 0706.2919. xi, 56

[Nol04] S. Noll, D. Mehlert, I. Appenzeller, et al. (2004). *The fors deep field spectroscopic survey*. Astronomy & Astrophysics, 418:885–906. http://arxiv.org/abs/astro-ph/0401500, astro-ph/0401500. 58

[Nor10] M. L. Norman (2010). *Simulating Galaxy Clusters*. ArXiv e-prints. arXiv:1005.1100 [astro-ph.CO]. viii, 50

[Oka10] N. Okabe, M. Takada, K. Umetsu, et al. (2010). *LoCuSS: Subaru Weak Lensing Study of 30 Galaxy Clusters*. Publications of the Astronomical Society of Japan, 62:811–870. arXiv:0903.1103. 21

[Ost73] J. P. Ostriker and P. J. E. Peebles (1973). *A Numerical Study of the Stability of Flattened Galaxies: or, can Cold Galaxies Survive?* Astrophysical Journal, 186:467–480. 51

[Ost08] L. Ostman, A. Goobar, and E. Mortsell (2008). *Extinction properties of lensing galaxies*. Astron. Astrophys., 485:403–415. arXiv:0711.4267 [astro-ph]. 25

[Pad03] T. Padmanabhan (2003). *Cosmological constant - the weight of the vacuum*. physrep, 380:235–320. http://arxiv.org/abs/hep-th/0212290, hep-th/0212290. 6

[Pag03] L. Page (2003). *The Wilkinson Microwave Anisotropy Probe*. In *Measuring and modeling the universe. Proceedings, Symposium, Pasadena, USA, November 17-22, 2002*, pages 330–348. astro-ph/0306381. 17

[Pea99] J. Peacock (1999). *Cosmological Physics*. Cambridge Uinversity Press. 12

[Pen65] A. A. Penzias and R. W. Wilson (1965). *A Measurement of Excess Antenna Temperature at 4080 Mc/s*. Astrophysical Journal, 142:419–421. 9

[Per97] S. Perlmutter, G. Aldering, S. Deustua, et al. (1997). *Cosmology from type ia supernovae*. In *American Astronomical Society Meeting Abstracts*, volume 29 of *Bulletin of the American Astronomical Society*, page 1351. astro-ph/9812473. 1, 19

[Per99] S. Perlmutter, G. Aldering, G. Goldhaber, et al. (1999). *Measurements of omega and lambda from 42 high-redshift supernovae.* Astrophysical Journal, 517:565–586. http://arxiv.org/abs/astro-ph/9812133, astro-ph/9812133. 6, 11

[Per10] K. Perrett, D. Balam, M. Sullivan, et al. (2010). *Real-time Analysis and Selection Biases in the Supernova Legacy Survey.* Astronomical Journal, 140:518–532. arXiv:1006.2254 [astro-ph.CO]. 32

[Phi93] M. M. Phillips (1993). *The absolute magnitudes of Type IA supernovae.* Astrophysical Journal Letters, 413:L105–L108. 24, 25

[Pie92] M. J. Pierce and R. B. Tully (1992). *Luminosity-line width relations and the extragalactic distance scale. I - Absolute calibration.* Astrophysical Journal, 387:47–55. 58

[Pla14] Planck Collaboration, P. A. R. Ade, N. Aghanim, et al. (2014). *Planck 2013 results. XVI. Cosmological parameters.* Astronomy & Astrophysics, 571:A16. arXiv:1303.5076. 38

[Pre98] P. Premadi, H. Martel, and R. Matzner (1998). *Light Propagation in Inhomogeneous Universes. I. Methodology and Preliminary Results.* Astrophysical Journal, 493:10–27. astro-ph/9708129. viii, 49

[Psk69] Y. P. Pskovskii (1969). *Identification of the Absorption Spectrum of the Type I Supernova.* Soviet Astronomy, 12:750. 23

[Reg09] N. Regnault, A. Conley, J. Guy, et al. (2009). *Photometric calibration of the Supernova Legacy Survey fields.* Astronomy & Astrophysics, 506:999–1042. arXiv:0908.3808 [astro-ph.IM]. 32, 35, 64

[Ric09] J. Rich (2009). *Fundamentals of Cosmology.* École polytechnique. Springer Berlin Heidelberg. https://books.google.fr/books?id=Qf5DAAAAQBAJ. 3, 6, 41, 50

[Rie98] A. G. Riess, A. V. Filippenko, P. Challis, et al. (1998). *Observational evidence from supernovae for an accelerating universe and a cosmological constant.* Astronomical Journal, 116:1009–1038. http://arxiv.org/abs/astro-ph/9805201, astro-ph/9805201. 1, 19

[Rie04] A. G. Riess, L.-G. Strolger, J. Tonry, et al. (2004). *Type ia supernova discoveries at z>1 from the hubble space telescope: Evidence for past deceleration and constraints on dark energy evolution.* Astrophysical Journal, 607:665–687. http://arxiv.org/abs/astro-ph/0402512, astro-ph/0402512. 62

[Rig13] M. Rigault, Y. Copin, G. Aldering, et al. (2013). *Evidence of environmental dependencies of Type Ia supernovae from the Nearby Supernova Factory indicated by local Hα.* Astronomy & Astrophysics, 560:A66. arXiv:1309.1182 [astro-ph.CO]. 27

[Rig15] M. Rigault, G. Aldering, M. Kowalski, et al. (2015). *Confirmation of a Star Formation Bias in Type Ia Supernova Distances and its Effect on the Measurement of the Hubble Constant.* Astrophysical Journal, 802:20. arXiv:1412.6501. 20

[Rix97] H.-W. Rix, P. T. de Zeeuw, N. Cretton, et al. (1997). *Dynamical Modeling of Velocity Profiles: The Dark Halo around the Elliptical Galaxy NGC 2434.* Astrophysical Journal, 488:702–719. astro-ph/9702126. 51

[Rub78] V. C. Rubin, N. Thonnard, and W. K. Ford, Jr. (1978). *Extended rotation curves of high-luminosity spiral galaxies. IV - Systematic dynamical properties, SA through SC.* Astrophysical Journal Letters, 225:L107–L111. 51

[Rug00] S. E. Rugh and H. Zinkernagel (2000). *The quantum vacuum and the cosmological constant problem*. ArXiv High Energy Physics - Theory e-prints. http://arxiv.org/abs/hep-th/0012253, hep-th/0012253. 6

[Ryd03] B. Ryden (2003). *Introduction to Cosmology*. Addison Wesley. 3

[Sad04] S. Sadeh and Y. Rephaeli (2004). *S-Z anisotropy and cluster counts: consistent selection of $\sigma_8$ and the temperature-mass relation*. New Astronomy, 9:373–382. astro-ph/0401033. 15

[Sam04] M. Sami and A. Toporensky (2004). *Phantom field and the fate of universe*. Modern Physics Letters A, 19:1509–1517. http://arxiv.org/abs/gr-qc/0312009, gr-qc/0312009. 7

[San15] M. G. Santos, D. Alonso, P. Bull, et al. (2015). *Hi galaxy simulations for the ska: number counts and bias*. Advancing Astrophysics with the Square Kilometre Array (AASKA14), 21. http://arxiv.org/abs/1501.03990, arXiv:1501.03990. v, 4

[Sar08] D. Sarkar, A. Amblard, D. E. Holz, et al. (2008). *Lensing and Supernovae: Quantifying the Bias on the Dark Energy Equation of State*. Astrophysical Journal, 678:1-5. arXiv:0710.4143. 60, 61

[SC07] M. Sanchez-Conde, J. Betancort-Rijo, and F. Prada (2007). *The spherical collapse model with shell crossing*. Monthly Notices of the Royal Astronomical Society, 378:339–352. http://arxiv.org/abs/astro-ph/0609479, astro-ph/0609479. viii, 51

[Sch92] P. Schneider, J. Ehlers, and E. Falco (1992). *Gravitational lenses*. Springer-Verlag Berlin Heidelberg New York. 41, 48, 49, 81

[Sch98] B. P. Schmidt, N. B. Suntzeff, M. M. Phillips, et al. (1998). *The high-z supernova search: Measuring cosmic deceleration and global cur vature of the universe using type ia supernovae*. Astrophysical Journal, 507:46–63. http://arxiv.org/abs/astro-ph/9805200, astro-ph/9805200. 1, 6, 11

[Sch04] R. T. Schilizzi (2004). *The square kilometer array*. Proc. SPIE, 5489:62–71. http://dx.doi.org/10.1117/12.551206. v, 4

[Sch10] T. Schrabback, J. Hartlap, B. Joachimi, et al. (2010). *Evidence of the accelerated expansion of the Universe from weak lensing tomography with COSMOS*. Astronomy & Astrophysics, 516:A63. arXiv:0911.0053. 20

[Sco07] N. Scoville, R. G. Abraham, H. Aussel, et al. (2007). *COSMOS: Hubble Space Telescope Observations*. Astrophysical Journal Supplement Series, 172:38–45. arXiv:astro-ph/0612306. 70

[Sel00] U. Seljak (2000). *Analytic model for galaxy and dark matter clustering*. Monthly Notices of the Royal Astronomical Society, 318:203–213. astro-ph/0001493. 56

[Ser10] S. Serjeant (2010). *Observational Cosmology*. Cambridge University Press. 45

[She03] R. K. Sheth, M. Bernardi, P. L. Schechter, et al. (2003). *The Velocity Dispersion Function of Early-Type Galaxies*. Astrophysical Journal, 594:225–231. astro-ph/0303092. 58

[Sli13] V. M. Slipher (1913). *The radial velocity of the Andromeda Nebula*. Lowell Observatory Bulletin, 2:56–57. 3

[Sli15] V. M. Slipher (1915). *Spectrographic Observations of Nebulae.* Popular Astronomy, 23:21–24. 6

[Smi14] M. Smith, D. J. Bacon, R. C. Nichol, et al. (2014). *The effect of weak lensing on distance estimates from supernovae.* Astrophysical Journal, 780:24. http://arxiv.org/abs/1307.2566, arXiv:1307.2566 [astro-ph.CO]. viii, 60, 61, 62

[Smo92] G. F. Smoot, C. L. Bennett, A. Kogut, et al. (1992). *Structure in the COBE differential microwave radiometer first-year maps.* Astrophysical Journal Letters, 396:L1–L5. 3

[Sul06] M. Sullivan, D. Le Borgne, C. J. Pritchet, et al. (2006). *Rates and Properties of Type Ia Supernovae as a Function of Mass and Star Formation in Their Host Galaxies.* Astrophysical Journal, 648:868–883. astro-ph/0605455. vi, 26, 27

[Sul10] M. Sullivan, A. Conley, D. A. Howell, et al. (2010). *The dependence of Type Ia Supernovae luminosities on their host galaxies.* Monthly Notices of the Royal Astronomical Society, 406:782–802. arXiv:1003.5119. 27

[Tab10] N. Taburet, M. Douspis, and N. Aghanim (2010). *The Sunyaev-Zel'dovich contribution in CMB analyses.* Monthly Notices of the Royal Astronomical Society, 404:1197–1202. arXiv:0908.1653. 15

[Tak03] M. Takada and B. Jain (2003). *The three-point correlation function in cosmology.* Monthly Notices of the Royal Astronomical Society, 340:580–608. astro-ph/0209167. 79

[Tak14] Y. Takeuchi, K. Ichiki, T. Takahashi, et al. (2014). *Probing quintessence potential with future cosmological surveys.* Journal of Cosmology and Astroparticle Physics, 3:045. http://arxiv.org/abs/1401.7031, arXiv:1401.7031. 7

[Tau04] J. Tauber, ESA, the Planck Scientific Collaboration, et al. (2004). *The planck mission.* Advances in Space Research, 34(3):491–496. 15, 17

[Teg04] M. Tegmark, M. R. Blanton, M. A. Strauss, et al. (2004). *The Three-Dimensional Power Spectrum of Galaxies from the Sloan Digital Sky Survey.* Astrophysical Journal, 606:702–740. astro-ph/0310725. v, 13

[The06] The Planck Collaboration (2006). *The Scientific Programme of Planck.* ArXiv Astrophysics e-prints. astro-ph/0604069. 17

[Tim03] F. X. Timmes, E. F. Brown, and J. W. Truran (2003). *On Variations in the Peak Luminosity of Type Ia Supernovae.* Astrophysical Journal Letters, 590:L83–L86. arXiv:astro-ph/0305114. 27

[Tis07] P. Tisserand, L. Le Guillou, C. Afonso, et al. (2007). *Limits on the Macho content of the Galactic Halo from the EROS-2 Survey of the Magellanic Clouds.* Astronomy & Astrophysics, 469:387–404. astro-ph/0607207. 48

[Tri99] R. Tripp and D. Branch (1999). *Determination of the hubble constant using a two-parameter luminosity correction for type ia supernovae.* Astrophysical Journal, 525:209–214. http://arxiv.org/abs/astro-ph/9904347, astro-ph/9904347. 24

[Tsu13] S. Tsujikawa (2013). *Quintessence: A review.* Classical and Quantum Gravity, 30(21):214003. http://arxiv.org/abs/1304.1961, arXiv:1304.1961 [gr-qc]. 7

[Tul77] R. B. Tully and J. R. Fisher (1977). *A new method of determining distances to galaxies*. Astronomy & Astrophysics, 54:661–673. 56

[Ume14] K. Umetsu, E. Medezinski, M. Nonino, et al. (2014). *CLASH: Weak-lensing Shear-and-magnification Analysis of 20 Galaxy Clusters*. Astrophysical Journal, 795:163. arXiv:1404.1375. 21

[van11] E. van Uitert, H. Hoekstra, M. Velander, et al. (2011). *Galaxy-galaxy lensing constraints on the relation between baryons and dark matter in galaxies in the Red Sequence Cluster Survey 2*. Astronomy & Astrophysics, 534:A14. arXiv:1107.4093. 59

[Van13] L. Van Waerbeke, J. Benjamin, T. Erben, et al. (2013). *CFHTLenS: mapping the large-scale structure with gravitational lensing*. Monthly Notices of the Royal Astronomical Society, 433:3373–3388. arXiv:1303.1806. 20

[Wan99] Y. Wang (1999). *Analytical Modeling of the Weak Lensing of Standard Candles. I. Empirical Fitting of Numerical Simulation Results*. Astrophysical Journal, 525:651–658. astro-ph/9901212. 61

[Wan02] Y. Wang, D. E. Holz, and D. Munshi (2002). *A Universal Probability Distribution Function for Weak-lensing Amplification*. Astrophysical Journal Letters, 572:L15–L18. astro-ph/0204169. 61

[Wan05] Y. Wang (2005). *Observational signatures of the weak lensing magnification of supernovae*. Journal of Cosmology and Astroparticle Physics, 3:005. arXiv:astro-ph/0406635. 62

[Whe73] J. Whelan and I. Iben, Jr. (1973). *Binaries and Supernovae of Type I*. Astrophysical Journal, 186:1007–1014. 24

[Wil04] L. L. Williams and J. Song (2004). *Weak lensing of the high redshift snia sample*. Monthly Notices of the Royal Astronomical Society, 351:1387–1394. http://arxiv.org/abs/astro-ph/0403680, astro-ph/0403680. 62

[Woo02] S. E. Woosley, A. Heger, and T. A. Weaver (2002). *The evolution and explosion of massive stars*. Reviews of modern physics, 74(4):1015. 24

[Yor00] D. G. York, J. Adelman, J. E. Anderson Jr, et al. (2000). *The sloan digital sky survey: Technical summary*. The Astronomical Journal, 120(3):1579. 18

[Zel70] Y. B. Zel'dovich (1970). *Gravitational instability: An approximate theory for large density perturbations*. Astronomy & Astrophysics, 5:84–89. 13

# Abstract

The presence of mass inhomogeneities along the line of sight of propagation of light from distant objects can induce deflection in the flight path of the photon. This phenomenon is called as gravitational lensing. Lensing can have both distortion (shear) and isotropic magnification effects on the source. We studied the effect of lensing magnification on supernova (SN) Ia in this thesis.

Presence of lensing would introduce a source of contamination to the brightness distribution of the source (SN Ia in our case). Thus it also enables one to compute the lensing effect indirectly from the Hubble diagram (i.e. from the residual to the Hubble diagram). In this thesis we computed the correlation between these two effects : the Hubble residual and the computed lensing magnification for the SN by the line of sight foreground dark matter haloes. A detection of positive correlation between these two would signify the positivity of lensing signal detection.

The data sample is the spectroscopic SNe Ia sample from the five years full SNLS data and the Hubble resiudals are those of the preliminary cosmology analysis performed on SNLS5 data. We obtain a signal of $\rho = 0.177$ at $2.51\sigma$. This result is consistent with the previous SNLS three years data lensing analysis results.